\documentclass[11pt]{report}
\usepackage{latexsym,amsfonts,mathrsfs,plain}
\usepackage{graphicx}
\usepackage{cite}
\def\bseq{\begin{subequation}}  
\def\eseq{\end{subequation}}
\def\bsea{\begin{subeqnarray}}  
\def\esea{\end{subeqnarray}}

\newcommand{\beq}{\begin{equation}}
\newcommand{\eeq}{\end{equation}}
\newcommand{\bea}{\begin{eqnarray}}
\newcommand{\eea}{\end{eqnarray}}
\newcommand{\bdm}{\begin{displaymath}}
\newcommand{\edm}{\end{displaymath}}
\newcommand{\ba}{\begin{array}}
\newcommand{\ea}{\end{array}}     
\newcommand{\ben}{\begin{enumerate}}
\newcommand{\een}{\end{enumerate}}
\newcommand{\bde}{\begin{description}}
\newcommand{\ede}{\end{description}}
\newcommand{\nn}{\nonumber}

\renewcommand{\r}{\right}
\renewcommand{\l}{\left}

\newcommand{\complesso}{{\ \hbox{{\rm I}\kern-.6em\hbox{\bf C}}}}
\newcommand{\reale}{{\hbox{{\rm I}\kern-.2em\hbox{\rm R}}}}
\newcommand{\1}{ \,  \raisebox{+0.14em}{{\hbox{{\rm \scriptsize ]}} \raisebox{-0.2em}{\kern-.8em\hbox{1}}}} \, }  

\newcommand{\p}{\partial}

\renewcommand{\a}{\alpha}
\renewcommand{\b}{\beta}
\newcommand{\g}{\gamma}

\renewcommand{\d}{\delta}

\newcommand{\e}{\epsilon}

\renewcommand{\k}{\kappa}

\renewcommand{\l}{\lambda}

\newcommand{\m}{\mu}

\newcommand{\n}{\nu}

\renewcommand{\r}{\rho}

\newcommand{\s}{\sigma}

\renewcommand{\t}{\theta}

\newcommand{\z}{\zeta}

\begin{document}

\begin{titlepage}
\begin{flushright}
IFUM-859-FT\\
January 2006
\end{flushright}
\vspace{2cm}

{\Huge $\mathcal Mechanisms\ for\ Supersymmetry$\\} 

 {\Huge $\mathcal\  Breaking\ in\ Open\ String\ Vacua$ \\ [.5cm]}
\\

 

\vspace{1cm}
{\bf \hrule width 16.cm}
\vspace {1cm}

\noindent{\large \bf Matteo A. Cardella}

\vskip 2mm

{ \small \noindent Dipartimento di Fisica dell'Universit\`a di Milano and

\noindent INFN, Sezione di Milano, Via Celoria 16, 20133 Milano, Italy}
\vfill
\begin{center}
{\bf Abstract}
\end{center}
 We  investigate
  mechanisms that can trigger supersymmetry breaking
 in open string vacua. 
 The focus is on  backgrounds with D-branes and orientifold planes
  that have an  exact string description, and
 allow to study  some of the  quantum effects induced
  by supersymmetry breaking.

\vspace{2mm} 
\vfill \hrule width 6.cm
\begin{flushleft}
e-mail: matteo.cardella@mi.infn.it
\end{flushleft}
\end{titlepage}

\newpage

-------------------------------------------------------------------------------------------------------------------

\newpage

\begin{titlepage}
\begin{Large}
\begin{center}
Universit\`a degli Studi di Milano
\vspace{.3cm}

Facolt\`a di Scienze Matematiche, Fisiche e Naturali



\vspace{.3cm}

\begin{figure}[htpb]
\begin{center}
\includegraphics[scale=0.7,angle=0]{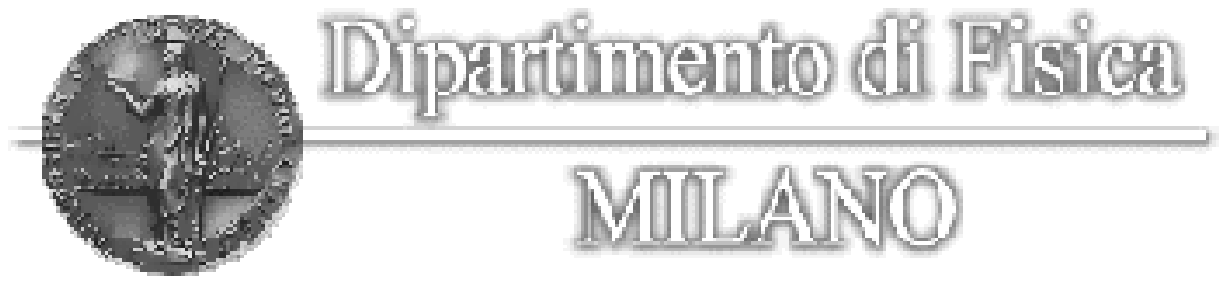}
\end{center}
\end{figure} 

Corso di Dottorato in Fisica, Astrofisica e Fisica Applicata

\vspace{.3cm}

XVIII ciclo

\vspace{.6 cm}

Tesi di Dottorato di Ricerca


\vspace{.6cm}

{\Huge $\mathcal Mechanisms\ for\ Supersymmetry\ Breaking$\\} 
{\Huge $\mathcal\ in\ Open\ String\ Vacua$ \\ [.5cm]}

{\small (settore scientifico disciplinare: FIS/02)}

\end{center}

\vspace{.5cm}


\begin{flushright}

\large Candidato:

\large Matteo Aureliano Cardella

\large matr. R05128

\end{flushright}

\begin{flushleft}

\large Relatore Interno: 

\large Prof. Daniela Zanon

\vspace{.1cm}

\large Relatore Esterno: 

\large Dr. Carlo Angelantonj

\vspace{.1cm}

\large Coordinatore:

\large Prof. Gianpaolo Bellini

\end{flushleft}

\vfill

\begin{center}

Anno Accademico 2004-2005

\end{center}

\end{Large}

\end{titlepage}

\thispagestyle{empty}


\newpage
--------------------------------------------------------------------------------------------------------------\thispagestyle{empty}
\newpage

\tableofcontents

\newpage
--------------------------------------------------------------------------------------------------------------
\newpage

\chapter{Introduction}
\everypar{\hspace{-.6cm}}

String Theory is a quite  interesting framework
 with  the promising  characteristics to provide
  a unified scheme that incorporates
 the fundamental interactions. One of its  ambitious  goals is to
  describe the status of space-time and matter in
the early instants after the Big Bang.

In its perturbative formulation String Theory replaces the concept of point
particles with one-dimensional objects, whose internal
 quantum vibrational modes gives rise to
  a certain number of massless states plus an  infinite tower 
of massive ones. The main result is to identify the different 
 particles of the Standard Model together with
  the graviton  as different quantum  states of a single entity.
Indeed this replacement solves  the UV-divergence problems
  that plague any attempt
 of constructing a consistent quantum field theory for a spin-two particle (the graviton)
 on Minkowski space-time \footnote{It is fair to say that although the perturbative
expansion in Riemann surfaces that replaces Feynman diagrams is conjectured
to be finite term by term, there is no rigorous proof of that.
 Moreover, the full perturbative series has been shown not
to be Borel summable \cite{Gross:1988ib}, and, as a consequence of that,
 it should be considered as an asymptotic series. 
This is one of the arguments that  imply   that
the perturbative formulation cannot be the final form 
of a \emph{fundamental} theory for quantum gravity.}.

The Action for a string propagating on a given space-time has the
peculiar feature of being classically invariant under local rescaling of the
world-sheet metric (2d-conformal transformations). The request for this symmetry
 to  survive  after quantisation puts strong constraints  on the
 number $D$  of space-time dimensions, on the equations 
  for the background fields  and on the spectrum of string vibrations.
In  the critical dimension  $D_{crit}$ the world-sheet scale factor decouples from the
dynamics  after quantisation, while for a different number of dimensions
 the quantum (conformal)  theory acquires an extra-dimension 
 given by the  coupled scale factor mode.  
 For the fermionic string $D_{crit} = 10$, 
 and the only stable solutions on a  ten-dimensional
 Minkowski  space-time enjoy supersymmetry,  which therefore
represents an important ingredient in all the constructions. 
 For supersymmetric string theories  the equation of motions for the background fields
 at leading order in the string length
 agree with Einstein equations in the vacuum,
and consistently  one of the massless states of the closed string 
 can be identified with the graviton.

 
There are  five ten-dimensional perturbative  string   theories: type IIA, type IIB, type I and 
the two Heterotic ones, that, together with eleven-dimensional supergravity,
  are  interconnected  via a rich web of  dualities, relating
 their various  weak  and strong  coupling  regimes. 

 This has led to the conclusion that these  are indeed
 different perturbative corners of a unique conjectured theory called M-theory,
 whose fundamental degrees of freedom and formulation remain so far unknown.
  
A fundamental role in order to establish these dualities and, to gain
some insights  in the non-perturbative regime of the theory,
 has been played
by the Dirichlet p-branes, extended objects that are sources
 for multidimensional generalisations of
 the Maxwell potential called
the Ramond-Ramond forms.
 Dirichlet p-branes are regarded as solitons of the theory,
  carrying a  charge  and a tension proportional to the
inverse of the string coupling constant, and quite remarkably having open strings
as their quantum excitations. 

\begin{figure}  
\begin{center} 
\includegraphics[scale=.5, height=8cm]{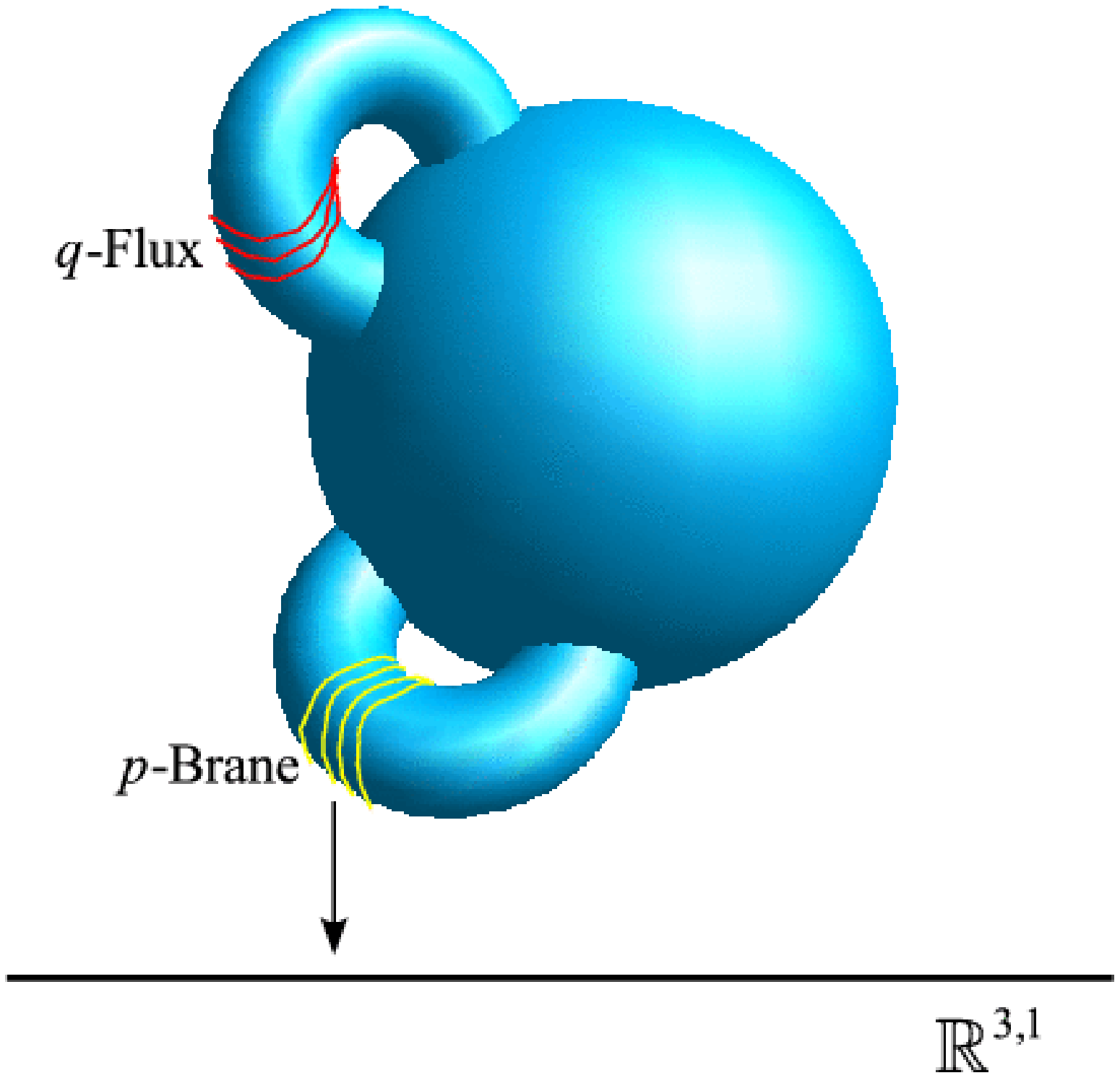}
\caption{An image of the possible structure of space-time after a phase transition from
 an unknown quantum gravity regime: six of the original dimensions are compact
   (in the  figure only two of them  are represented),
 while four are extended, (represented by the straight line).  
Flux lines are trapped in the topology of the
compact space, while p-branes are space-time defects
  that can wrap some of the non-contractible circles and invade the extended dimensions.} 
\label{compactspace}
\end{center}
\end{figure}

 It is fascinating to look at these extended objects
as  space-time defects (figure \ref{compactspace}), remnants of a phase transition
 that might have occurred during  the cooling down of the universe after the 
 Big Bang.  It is remarkable as well how the  mathematical consistency 
of string theory suggests  their existence, by including 
 the multidimensional generalisation of Maxwell potentials (Ramond-Ramond fluxes)
 in one sector of its perturbative closed-string spectra.

The flat ten-dimensional solutions are regarded as a guide 
to produce  a much larger class of new solutions, via 
a process of symmetry breaking called compactification.
 Among them, of particular interest are those   where
 six of the ten dimensions are highly curled up  to form a compact space,
and the full space-time factorises as the product
of the  four-dimensional Minkowski space times this compact space.
All these solutions are considered  as  different vacua
of the same theory and,   
due to the present lack of a  non-perturbative
formulation,   it is not possible to fully
  understand the relations among these different  vacua.
  Therefore it is quite hard 
  to  address the question of  whether  
  the theory prefers a vacuum rather than another,
 an open  issue that goes under the name of  vacuum degeneracy
 problem.
  

 The  perturbative nature of the world-sheet formulation
  results from an initial  splitting of the space-time
 metric in a classical background plus  quantum
 stringy fluctuations. 
 This approach  does not address the problem
 of  finding  the symmetries and principles that need
to replace General Covariance, in order to obtain a fully consistent
   fusion between Quantum Mechanics  and General Relativity. 

In a  \emph{classical} spin-two field theory on Minkowsky space-time,
  if one starts by considering 
  only a   three-points
self-interaction, in order to achieve a ghost-free field theory one
 needs to add higher and higher interaction vertexes to the Lagrangian.
 The final result is 
  a non-polynomial action, from which,  without the knowledge of the underlying  geometrical principles,
   it  would be quite hard  to recognise General Covariance
behind  its  nice properties.

 A similar state of affairs  happens in string theory, with the
advantage, with respect to the field counterpart, of  solving the UV-divergences problem
but with the big drawback of not possessing yet enough clues about the new geometrical ideas.
The description of quantum gravity as string fluctuation on a 
  non-dynamical  space-time  is  intrinsically
 perturbative and not manifestly
 background independent. The non-perturbative  formulation is likely 
 to be found through new
 geometrical  principles,  counterpart
 of  classical  General Covariance, a task that so far has been proved  quite hard.


  Waiting for new insights toward a  non-perturbative formulation,
 one can still investigate possible mechanisms by
  working in particular vacua of the theory.
   There are currently several tasks that
  are object of intense research in following this last
approach, among them of particular relevance is the
 question of finding a controlled description for the process
 of   moduli stabilisation and  (super)symmetry breaking.

If string theory is on the right track it should contain
 the answers to these questions, and vacua with identical features
  of the Standard Model.
In principle, the Standard Model parameters
 could be   reproduced as the
 result of  moduli stabilisation.
 Moduli is the name that indicates    VEVs of background scalar fields, arising from 
 compactification, some of that   describing   properties  of the compact space.

  The second vital question deals with finding   mechanisms  that
 break the original supersymmetries,
  in order to make contact with the Standard Model of particle physics
 and with the observed value of the Cosmological Constant.  
Although a complete  answer might  be found, if ever, from a non-perturbative point
of view, still there are several features that can be observed by studying
 string fluctuations around backgrounds containing non-perturbative
 objects such as Dirichlet branes and orientifold planes.  
 
\vspace{.5  cm}
 Historically the  first attempts  at  looking 
 for semi-realistic vacua have been pursued from
 compactifications of the Heterotic string.
In this case, in order to be in 
a perturbative regime, both  the string scale
 and the compactification scale need to be
not far from 
  the Planck scale.
 The study of the string dynamics for this
 class of compactifications is
 confined to the massless modes and involves 
 the construction of an effective Action. 
  This is  obtained by integrating  out all
the heavy modes, given by string excitations
and  Kaluza-Klein  states, and its
  form  is then  determined by supersymmetry and
topological data of the compact manifold.

\vspace{.1 cm} A different approach for looking
at semi-realistic vacua that offers a more stringy description 
 was born after a development in the understanding 
  of the role of  Dp-branes.
  For example,  in open string vacua 
  in  the presence of some
 \emph{large} extra-dimensions,
 the string scale can be  lowered to
 the TeV range in a perturbative description.
In this case, massive string excitations
 do participate  to  the low-energy
 dynamics, so that 
 their effects can be taken into account
   in all the compactifications
 that allow an exact conformal 
 description.
 In models of brane-worlds D-p branes invade  four extended dimensions, as  
represented in fig. \ref{compactspace},
 and wrap some cycles of the compact space. 
The standard model forces are then  mediated by open strings confined on the 
brane, while only  gravity, mediated by closed strings,
 can experience all the ten-spacetime dimensions.
 This has offered a new  explanation  for the observed hierarchy
 between the electro-weak and the gravitational forces.  
 Gravity is so weak because of the dispersion of its flux lines 
 in a  higher number of dimensions comparing to the other forces.

In summary, the  presence of D-branes opens up the study of 
  new classes of string
 compactifications, where important
 open   issues  such as  moduli 
 stabilisation, breaking of supersymmetry,
  and  the hierarchy problem  can be studied from a 
 more stringy perspective.

\vspace{.4 cm}
This thesis follows this line of thought
 for investigating 
possible mechanisms that can trigger supersymmetry breaking
 in open string vacua. 
 The focus is on backgrounds with D-branes and orientifold planes
  that have an  exact string descriptions,
 which allow to study  some of the  quantum effects
 of supersymmetry breaking.
  
 In the first chapter  I review some of the fundamentals
of perturbative string theory, with a final emphasis on one-loop  
 closed string amplitudes and on the informations that the torus amplitude
 contains about the 
ten-dimensional closed superstring spectra.

  The second chapter  describes  open string excitations from D-p
branes and  the introduction of orientifold planes, often necessary to 
find consistent string backgrounds. The emphasis is on the genus-one
open and closed string diagrams that describe type I superstring vacua.

 The third chapter deals with supersymmetric toroidal compactifications
in the presence of D-branes and orientifold planes    
 and the description of several interesting effects 
associated to different points in the background moduli space. 

 Chapter four  introduces possible mechanisms for supersymmetry breaking
 that originate from various configurations of D-branes and
O-planes, and among them, a novel mechanism for supersymmetry breaking
with a vanishing vacuum energy \cite{Angelantonj:2004cm}. In this case it is possible to
estimate the vanishing of  higher-genus corrections  to the vacuum energy,
even without an explicit computations of the amplitudes that
  is still beyond  the present possibilities.
   The one-loop potential depending 
 on the open string moduli   and  the stability of the vacua 
for this class of solutions is then studied.
 In the second part of the same chapter 
 the Scherk-Schwarz mechanism is considered,
  an alternative way for supersymmetry breaking
originating from the compactification.
In particular,  the computation
 of one-loop potential for a novel Scherk-Schwarz mechanism   
 is presented \cite{inpreparation}
 with a quite interesting behaviour in the background moduli.

Chapter five is introductory on the main features of intersecting
branes vacua with a focus on the genus-one amplitudes,
that allow to determine the open and closed string spectra 
 for these classes of vacua via  tadpole cancellation conditions.
This should  give a  background for the subject of the last chapter.

 The last chapter of this thesis  is devoted to the
 analysis of the effects of the Scherk-Schwarz mechanism
on intersecting branes \cite{Angelantonj:2005hs}. 
 Though interesting in itself,
 a wise use of this mechanism   gives also a solution to one of the long-standing problems
in  intersecting branes vacua. In fact,  in the massless open string
spectra both gauginos and non-adjoint  non-chiral fermions are always present,
and  therefore a mechanism to make them  acquire a mass is needed
in order to agree with observations.   
By using the methods  proposed in  \cite{Angelantonj:2005hs} one  is able to give a tree-level mass to
 these non-chiral fermions with the virtue of not affecting the massless chiral
 fermions living at  branes intersections.

\chapter{$D=10$ superstring theories}

\section{CFT and  the bosonic closed string}
\everypar{\hspace{-.6cm}}

\vspace{1 cm}

\everypar{\hspace{-.6cm}}

 The embedding string coordinates $X^{\m}(\s,\tau)$,  with $\m =0,...,D-1$,
 are maps from a closed Riemann surface  $\Sigma$ of genus $g$
 (the string world-sheet) to a  $D$
 dimensional target space (the spacetime).

 We will consider a  two dimensional massless quantum field theory 
 on  $\Sigma$, with the fields/coordinates $X^{\m}(\s,\tau)$
 playing the double role of   scalar fields on the surface
 and coordinates on the target space.
   The most general Action
 that is both  two-dimensional   and $D$-dimensional general covariant  has the following form
\cite{Brink:1976sc,Deser:1976rb,Polyakov:1981rd}:
\beq
S = \frac{1}{4\pi \a'}\int_{\Sigma}d\s d\tau \sqrt{g} 
\left(G_{\m\n}(X)g^{\a\b} + 
 B_{\m\n}(X)\e^{\a\b} \right)\p_{\a}X^{\m}\p_{\b}X^{\n} + 
 \frac{1}{8\pi}\int_{\Sigma}d\s d\tau \sqrt{g} R^{(2)}\phi(X). \label{bwsaction} 
 \eeq

  The fields $G_{\m\n}(X)$,  $B_{\m\n}(X)$,  $\phi(X)$,
 describe a classical background on which the string
 fluctuates,  $G_{\m\n}(X)$ being the spacetime metric,
 $B_{\m\n}(X)$  an anti-symmetric tensor and   $\phi(X)$
  a scalar called the dilaton.
  From   the
 world-sheet point of view these same fields are  couplings
  for the dynamical scalar bosonic fields $X^{\m}(\s,\tau)$
   in the Lagrangian. If the theory is conformal at 
 quantum level we can expect that the absence of running for 
these couplings  might translate to a condition of stability 
for the spacetime classical background, an idea that we will try to 
describe more preciselely in the following.  

$g_{\a\b}$ is the  metric on  the surface  $\Sigma$,
   that, thanks to a  specific property of two-dimensional
 manifolds, can always be cast  
  in a conformally flat form  $g_{\a\b} = e^{\varphi}\eta_{\a\b}$, 
    for a proper choice of local woldsheet coordinates, with $\varphi = \varphi(\s, \tau)$
 a local scaling factor.
 
In the \emph{quantum} theory of the bosonic string one considers a sum
 over random surfaces by means of  a path integral, where the integrated
 variables are the 
 world-sheet metric  $g_{\a\b}$ and the coordinate fields  $X^{\m}(\s,\tau)$,
 but \emph{not} the classical background fields  $G_{\m\n}(X)$,  $B_{\m\n}(X)$,  $\phi(X)$.

\hspace{.1 cm}

 The action on the world-sheet is  classically  conformal invariant,
  which means  that  for a local rescaling of the metric $ g_{\a\b} \rightarrow  e^{\varphi}g_{\a\b}$
    the conformal factor  $e^{\varphi}$ drops out from the Lagrangian.
  However,  after  quantisation of the two-dimensional theory,
 a conformal anomaly  generally arises,
    and the conditions to
  have \emph{quantum} conformal invariance
  translate into  non trivial  constraints on the  fields  $G_{\m\n}(X)$,  $B_{\m\n}(X)$,  $\phi(X)$
 that describe the classical background.
 These constraints are called the string equations of motion, 
 and are expressed as a perturbative
 expansion in the string scale $\a'$. They 
  select  the 
   class of backgrounds on which the quantum string can  consistently propagate.
 


 \vspace{.5 cm}

  Conformal transformations on $\Sigma$  are defined as  the subgroup of general
 coordinate transformations $x^{\a} \rightarrow x'^{\a}(x)$, ($\a =1,2$)
   that multiply the intrinsic metric
 by a local scale factor
\beq
g'_{\a\b}(x') = \frac{\p x^{\r}}{ \p x'^{\a}} \frac{\p x^{\s}}{\p x'^{\b}}g_{\r\s}(x) 
 = \varphi(x)g_{\a\b}(x). \label{twodiff} 
\eeq
For an infinitesimal  transformation $x^{\a} \rightarrow  x^{\a} + \e^{\a}$   
  the variation of the metric  is
\beq 
 ds'^{2} =  ds^{2} - (\p_{\a}\e_{\b} + \p_{\b}\e_{\a}) dx^{\a} dx^{\b} + O(\e^{2})
 \eeq 
 and  the request for the transformation to be conformal then  gives the following
  conditions on the local parameters  $\e^{\a}$
\beq
\p_{\a}\e_{\b} + \p_{\b}\e_{\a} = (\p \cdot \e ) g_{\a\b}. 
\eeq

  For reasons that we will explain  in the following,
let us consider the special case  $g_{\a\b} = \d_{\a\b}$.
 By acting with $\p^{\a}$
  on the previous condition we recover 
\beq
\p^{2}\e_{\b} = 0 \label{wave}.\label{Laplace}
\eeq 

  The transformation parameters $\e_{\b}$ satisfy the  two dimensional Laplace equation. 
 In  complex coordinates $\e = \e_{1} + i \e_{2}$ \  eq.(\ref{Laplace}) 
 translates  into
\beq
\bar{\p}\e = 0 \qquad \p \bar{\e} = 0,  
\eeq
   showing that infinitesimal coordinate transformations are
 analytic transformation  $ z \rightarrow \e (z)$.

 The generators  $l_n$  of the conformal  transformation
  can be extracted by  expanding  $\e(z)$  in a  power series  $\e(z) = \sum c_{n} z^{n+1}$,
\beq
z \rightarrow z + \sum c_{n} z^{n+1} = (1  + \sum c_{n} z^{n+1}\p )z,   
\eeq
  therefore $l_n = z^{n+1}\p$.

In particular a global scale transformation, $z \rightarrow  \l z$,  $\l \in \mathbb{C}$
is generated by $l_0 = z \p$ and will play a fundamental role in the quantum two-dimensional
 world-sheet theory. 

 The $l_n$ generators   form an infinite dimensional algebra, whose
 commutators are:
\beq 
\left[ l_m , l_m  \right] = ( m - n) l_{m + n}. \label{classconf}
\eeq

\begin{figure}  
\begin{center} 
{\includegraphics[scale=1, height=5cm]{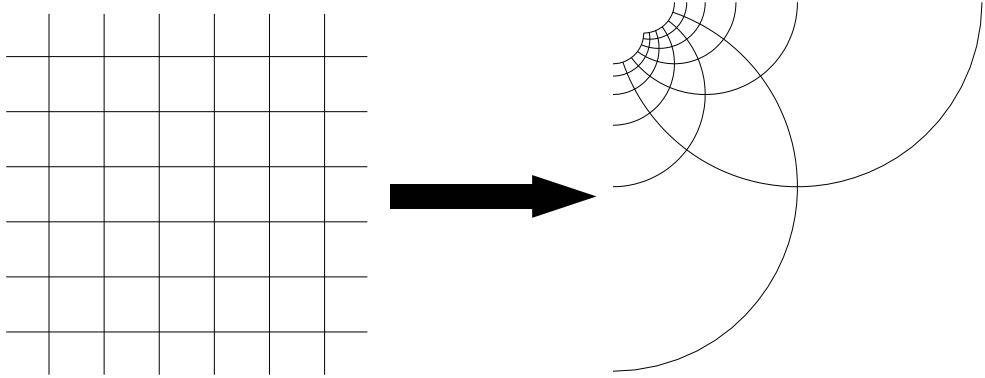}}
\caption{A conformal tranformation changes the distances between  points on the world-sheet
by a local rescaling of the metric without altering the angles between the lines.}                       
\label{ConfTransf}
\end{center}
\end{figure}

 We can now  turn to analyse the conditions for the
 theory to remain conformal at quantum level.
 A preliminary step towards this direction consists in computing
  the response of 
 (\ref{bwsaction}) to a general variation $\d g_{\a\b}$ of the 
 intrinsic metric:
\beq
\d S = \int d\s d\tau \frac{\d S}{\d g_{\a\b}}\d g_{\a\b} = \int d\s d\tau \sqrt{g} T^{\a\b}\d g_{\a\b}.
\eeq
  $\d S$  defines the stress tensor 
(in Euclidean signature for the intrinsic metric):
\beq
 T_{\a\b} = \frac{4\pi}{\sqrt{g}}\frac{\d S}{\d g^{\a\b}}, 
\eeq
 the factor $4\pi$ in the definition simplifies its explicit form, that we are going to obtain. 
  
  Whenever    the action  is invariant under a subgroup of all  possible variations
 of the intrinsic metric, the stress 
   tensor will in correspondence
   enjoy a specific property.
 
  To be specific, for an   infinitesimal
  conformal transformation $\d g_{\a\b} = \e  g_{\a\b}$,
 the variation of the Action yields
\beq
\d S  =  \frac{1}{4\pi} \int d\s d\tau \sqrt{g}  T^{\a\b} \ g_{\a\b} \ \e,
\eeq
 and thus the vanishing of $\d S$ asks for the vanishing of the trace
 of the stress tensor, $T^{\a}_{\a}= 0$.

 The most general local variation of the intrinsic metric
 corresponds   to a  world-sheet diffeomorphism, and if
  the Action is invariant
   under diffeomorphisms,  the  stress tensor is conserved,
 $\nabla^{\a}T_{\a\b}=0$.

 Let us consider the world-sheet Action (\ref{bwsaction}) for the 
  class of backgrounds where 
 $B_{\m\n} = 0$ and  the  dilaton  is constant
 $\nabla_{\m}\phi = 0$, and  compute  the stress tensor.
 The response of the Action to a variation 
  of the intrinsic metric gives:
\beq
\d S = \frac{1}{4\pi \a'}
\int d\s d\tau \left( \d (\sqrt{g}) g^{\a\b} + \sqrt{g} \d  g^{\a\b} \right)G_{\m\n} \p_{\a}X^{\m}  \p_{\b}X^{\n},
\eeq 
 and thus, by using $\d (\sqrt{g}) = \frac{1}{2}\sqrt{g} \d g_{\r\s} g^{\r\s}$ 
one gets
\beq
T_{\a\b} = \frac{4 \pi}{\sqrt{g}} \frac{\d S}{\d g^{\a\b}} = \frac{1}{\a'}\left( \p_{\a}X^{\m}\p_{\b}X^{\n}
 - \frac{1}{2} g_{\a\b} \p_{\g}X^{\m}\p^{\g}X^{\n}\right)G_{\m\n}. \label{stress}
\eeq

    A conformal 
  transformation is a particular kind of
 (world-sheet) diffeomorphism, and  the metric
 on the world-sheet can be chosen to be conformally flat,
  therefore we can always  take a flat metric $\eta_{\a\b}$ as a reference metric
 for the classical action.
  Conformal invariance asks for  $T^{\a}_{\a}= 0$, while, for a flat reference metric, 
 diffeomorphism  invariance imposes $\p^{\a}T_{\a\b}=0$. In complex coordinates
    $T^{\a}_{\a}= 0$  translates into  $T_{z \bar{z}} = 0$
 and in this case    $\p^{\a}T_{\a\b}=0$   yields  $\bar{\p} T_{z z}=0$
  and  $\p T_{\bar{z}\bar{z}}=0$,  which
      mean that  $T_{z z}= T(z)$ is a holomorphic function
 and $T_{\bar{z}\bar{z}} = \bar{T}(\bar{z})$ is  antiholomorphic.

  After quantisations of the two-dimensional
 theory in the flat reference metric, conformal invariance
  will demand the same analiticity properties for
 the stress tensor to hold at the quantum level as well.

\vspace{.5 cm}

In order to detect which are the backgrounds that
preserve conformal invariance at the  quantum level on
 the world-sheet sigma model (\ref{bwsaction}), one can employ a background field
 method in the two dimensional field theory, by writing
 the  string coordinates $X^{\m} = X^{\m}_{cl} + \xi^{\m}$,
 where  $X^{\m}_{cl}$ is the centre of mass string coordinate
 and   $\xi^{\m}$ is a quantum fluctuation of the order of the string length 
 $\sqrt{\a'}$.
  Then the computation of the  beta functions for the two dimensional
 couplings
 gives a perturbative expansion in the string length   $\sqrt{\a'}$ \cite{Friedan:1980jf},
that at the  leading order in  $\a'$ have the form
 
\bea
\b_{\m\n}^{G} &=& \a' \left( R_{\m\n} - \frac{1}{4}H_{\m\r\s}H_{\n}^{\r\s} + \nabla_{\m}\nabla_{\n}\phi\right)            + O(\a'^{2}) \nn \\
\b_{\m\n}^{B} &=& \a' \nabla^{\r} \left(e^{-\phi}H_{\m\n\r}\right) + O(\a'^{2}) 
\nn \\
\b_{\m\n}^{\phi} &=& D - 26 +  3\a' \left[ (\nabla \phi)^{2}  - R + \frac{1}{12}H^{2} \right]   + O(\a'^{2})\label{bsem} 
\eea

where $ H_{\m\n\r} = \p_{\m}B_{\n\r} +   \p_{\n}B_{\r\m} +  \p_{\r}B_{\m\n}$.

The previous equations show the existence of a large class of spacetimes
 that at the  leading order satisfy Einstein vacuum equations
and, quite interestingly, these equations predict the number of spacetime  dimensions 
  where the bosonic string can fluctuate to be $D = 26$, called the critical dimension.
However, the critical string is not the only possibility, since by picking up
 an arbitrary number of dimensions one can still obtain conformal invariance
by adding to the coordinate fields  the world-sheet metric scale factor, that in this case does not
decouple, and it represents an extra coordinate in the quantum theory.
 This last possibility   goes under the name of non-critical string.
 
 Remaining in the  critical dimension  $D = 26$, 
 the two-dimensional theory  (\ref{bwsaction}) for general background fields
 $G_{\m\n}(X)$,  $B_{\m\n}(X)$ and $\phi(X)$ is a non linear sigma model
 for which a perturbative quantisation is hard to obtain.
   Let us therefore  consider
    the simpler case where the 
 action  is free, corresponding to  a flat spacetime metric 
 $G_{\m\n}(X) = \eta_{\m\n}$,   $B_{\m\n} = 0$  and  $\phi$  coordinate independent.

 These  flat spacetimes
  are indeed exact conformal  backgrounds, since higher order corrections in $\a'$
 to the condition for conformal invariance for the sigma model
  depend on complicate expressions of the Riemann
 tensor $R_{\m\n\r\s}$, the field strength
   $ H_{\m\n\r}$ and derivatives of the dilaton $\phi$  \cite{Gross:1986iv,Grisaru:1986px,Grisaru:1986kw}.
 These contributions are therefore vanishing
   for the class of flat spacetimes
 that we are presently considering.  

The  world-sheet action (\ref{bwsaction}) for this choice of background  is given by
\beq
S = \frac{1}{4\pi \a'}\int_{\Sigma}d\s d\tau  
 \eta_{\m\n}\p_{\a}X^{\m}\p^{\a}X^{\n} + 
 \chi_{g} \phi \label{flatg} 
 \eeq

where $\chi_{g}$ is the Euler characteristic
 of the world-sheet
\beq
\chi_{g} = 
\frac{1}{8\pi}\int_{\Sigma}d\s d\tau \sqrt{g} R^{(2)} = 2 - 2g, 
\eeq
 and is an integer number 
 that depends on the topology of the world-sheet,
 $g$ is  the genus of the two-dimensional surface, and
is  given by the number of handles $h$ of the closed surface.

\vspace{.5 cm}

\begin{figure}  
\begin{center} 
{\includegraphics[scale=1, height=5cm]{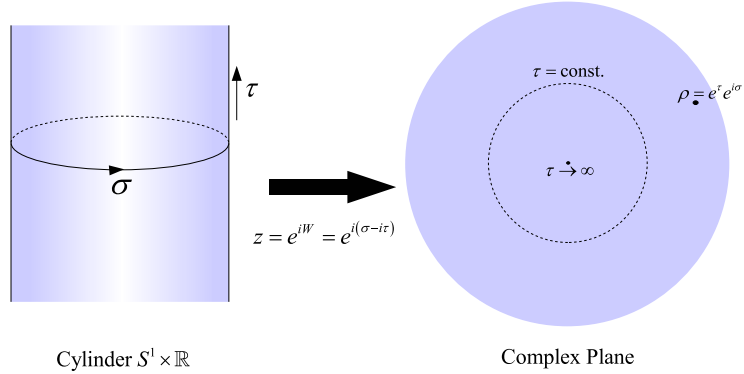}}
\caption{ The conformal transformation
 $z = e^{iw}$  maps the cylinder into the Riemann sphere $\mathcal{S}^{2} = \{ \mathbb{C} \cup \{ \infty \}\}$, the  complex plane together with the points at infinite that have been identified.}                       
\label{CylindtoComplex}
\end{center}
\end{figure}

A closed string  propagating freely
 on a  flat spacetime describes
an infinite  cylinder  $ 0 \le \s < 2\pi$,   $ -\infty < \tau  < \infty$.
Quantisation of the two dimensional theory
 can be worked out  after an Euclidean rotation on the surface,
so that  the coordinate on the cylinder are complexified $w = \s -it$.
  Through the conformal transformation
 $z = e^{iw}$ we can map the cylinder into the Riemann sphere $\mathcal{S}^{2} = \{ \mathbb{C} \cup \{ \infty \}\}$, the  complex plane together with the points at infinite that have been identified, as  shown in fig. \ref{CylindtoComplex}.

 Constant time slices $t = t_{0}$ on the cylinder are conformally mapped
 into  the  circles  $C = \{ z = e^{i\s + t_{0}} \}$  centered in the origin
 of the complex plane and  with radius $e^{t_{0}}$. Hence 
   time grows radially on the plane
  and the origin  corresponds to $t = - \infty$ on the cylinder.

\begin{figure}  
\begin{center} 
{\includegraphics[scale=1, height=5cm]{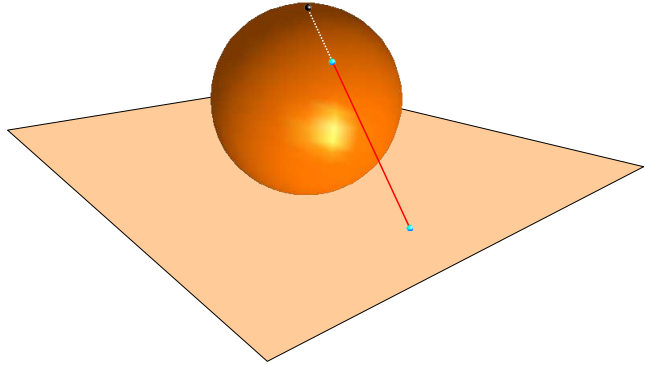}}
\caption{Stereographic projection of a sphere into a plane. The north pole is projected to
 the points at infinite of the plane. Therefore a sphere is equivalent to the complex
plane with the points at infinite identificated (Riemann Sphere).}                       
\label{SpheretoPlane}
\end{center}
\end{figure}

At the tree level we can therefore consider  the 
  world-sheet to be the  Riemann sphere fig. (\ref{SpheretoPlane}) with the action given
 by   the dynamical term
 in (\ref{flatg})
\beq
S = \frac{1}{2\pi \a'}\int dz d\bar{z}  
 \eta_{\m\n}\p X^{\m} \bar{\p} X^{\n},   
\eeq  
whose  classical equations of motion are
\beq
 \frac{\d S}{\d X_{\m}} =  \frac{1}{\pi \a'} \bar{\p}\p X^{\m} = 0.
\eeq
   As a result  $\p X^{\m}$ is a holomorphic function 
 and  $\bar{\p} X^{\m}$ is anti-holomorphic.

  In the quantised theory  the  product of fields at coincident points
  is singular and the    equations of motion are violated
  at  such coincident points.
  This can be easily understood by
  the following formal relation for 
 the Euclidean  path integral on the Riemann sphere
\bea
 0 = \int \mathcal{D}X  \frac{\d}{\d X_{\n}(z,\bar{ z})} \left( X^{\m}(z', \bar{z}') e^{-S} \right)
  \nn  \\ = \int \mathcal{D}X \left(\eta^{\m\n} \d(z - z',\bar{z} - \bar{z}' ) - 
 X^{\m}(z', \bar{z}') \frac{1}{\pi \a'}  \bar{\p}\p X^{\n}(z,\bar{ z})\right)  e^{-S}. \label{path}
\eea
  The equations of motion  are therefore satisfied except at the contact points:
\beq
 X^{\m}(z', \bar{z}')   \bar{\p}\p X^{\n}(z,\bar{ z}) = \pi \a' \eta^{\m\n} \d(z - z',\bar{z} - \bar{z}' ).  
\eeq
 Using the well known  Green Function
 for the Laplace operator in two dimensions  $ \bar{\p}\p \ln (|z|^{2}) =  2\pi \d(z,\bar{z})$, one gets
 
\beq
 X^{\m}(z', \bar{z}') X^{\n}(z,\bar{ z}) = \frac{\a'}{2} \eta^{\m\n} ln (|z  - z'|^{2}). \label{green}
\eeq

 Acting with $\p = \p_{z}$  and  $\p' = \p_{z'}$ on the previous equation one  also obtains
 another  useful relation

\beq
\p' X^{\m}(z') \p \ X^{\n}(z) = - \frac{\a'}{2} \frac{\eta^{\m\n}}{(z  - z')^{2}} \label{corr}.
\eeq

Before showing the use of the above operator product expansion (OPE),
 it is convenient to expand the
 holomorphic function $\p X^{\m}(z)$  in Laurent series:

\beq
\p X^{\m} =   \sqrt{\frac{\a'}{2}} \sum_{n = - \infty}^{\infty} z^{-n-1}\a^{\m}_{n} \label{holcord}
\eeq
and similarly:
\beq
\bar{\p} X^{\m} =   \sqrt{\frac{\a'}{2}} \sum_{n = - \infty}^{\infty} \bar{z}^{-n-1}\tilde{\a}^{\m}_{n}. \label{antiholcord} 
\eeq

The inverse relations are then

\bea
\a^{\m}_{n} &=& \sqrt{\frac{2}{\a'}} \oint_{C} \frac{dz}{2i\pi}    z^{n} \p X^{\m}(z) \nn \\
\tilde{\a}^{\m}_{n} &=&  -  \sqrt{\frac{2}{\a'}}\oint_{C} \frac{d\bar{z}}{2i\pi}\bar{z}^{n} \bar{\p} X^{\m}(\bar{z}),
\eea
  where the  integration contour $C$ winds the origin.

 From the  world-sheet current $1/ \a' \cdot \p_{\a} X^{\m}$,
associated to the  spacetime momentum  operator $\p_{\m} =1/ \a' \cdot \p_{\a} X^{\m} \p_{\a}$,
one can obtain the correspondent  Noether charge by performing contour integral

\beq
 p_{\m} = \frac{1}{\a'}\left( \oint_{C}\frac{dz}{2i\pi}\p X_{\m} 
  - \oint_{C}\frac{d\bar{z}}{2i\pi}\bar{\p} X_{\m} \right) = - \sqrt{\frac{2}{\a'}}\a_{0 \ \m} \label{closedcdmP}
 \eeq

where the  minus sign  takes care of the opposite orientation of  the $d\bar{z}$ the contour.
The independence of the integrals on the choice of the integration
 contour on the complex plane reflects the time independence
 of the Noether charge and thus its conservation.

Due to the singularity 
 of the product  (\ref{corr}), at the coincident points
 the modes $\a^{\m}_{n}$ no longer commute in 
the quantum theory.  Their commutator can be
 easily computed by  performing the following contour integrals:
\bea 
[\a^{\m}_{n}, \a^{\n}_{m}] &=&     \sqrt{\frac{2}{\a'}}      \oint \frac{dz}{2i\pi}    z^{n} \p_{z} X^{\m}(z)
 \sqrt{\frac{2}{\a'}} \oint \frac{dw}{2i\pi}    w^{m} \p_{w} X^{\n}(w) \nn \\  &-&
   \sqrt{\frac{2}{\a'}}\oint \frac{dw}{2i\pi}    w^{m} \p_{w} X^{\n}(w)
  \sqrt{\frac{2}{\a'}} \oint \frac{dz}{2i\pi}    z^{n} \p_{z} X^{\m}(z) \nn \\
 &=& n \eta^{\m\n} \d_{n + m, 0}, \label{ladders}
\eea
 and shows that 
   the operators $a^{\m}_{n} = \a^{\m}_{n}/ \sqrt{n}$ and  $(a^{\m}_{n})^{\dag} = \a^{\m}_{-n}/ \sqrt{n}$
  represent  an infinite set 
of ladder  harmonic oscillator operators. 

Integration of (\ref{holcord}) and (\ref{antiholcord}) then gives the normal mode expansion
 for the coordinates
\beq
X^{\m}(z,\bar{z}) =  x^{\m} +  \frac{\a'}{2}p^{\m}\ln(z\bar{z}) -  \sqrt{\frac{\a'}{2}} \sum_{n \ne 0} \left( \frac{z^{-n}}{n} \a^{\m}_{n}
+   \frac{\bar{z}^{-n}}{n} \tilde{\a}^{\m}_{n} \right),\label{coordexpansion}
\eeq
where $p^{\m}$ is the spacetime momentum of the centre of mass of the string.

Let us  return to the stress tensor that reads
\beq 
 T_{z z}= T(z) = \frac{1}{\a'} \p X^{\m}(z,\bar{ z})  \p X_{\m}(z,\bar{ z}).
\eeq
 Eq. (\ref{corr}) says that this expression is ill-defined 
 and needs to  be regularized    by subtracting
 its divergence at the coincidence points
\beq
 T(z) = \frac{1}{\a'}: \p X^{\m}(z,\bar{ z})  \p X_{\m}(z,\bar{z}):
 =    \frac{1}{\a'} \p X^{\m}(z,\bar{ z})  \p X_{\m}(u,\bar{u}) 
 -  \frac{D}{2(z  - u)^{2}} \label{regul}
\eeq
We are interested at the OPE of the stress tensor with itself since,
 as we will show,   the 
  knowledge of the singularities in the product $T(z)T(w)$
  is crucial  in order to
 reconstruct the quantum version of the
 classical conformal algebra (\ref{classconf}).
 
In order to compute
  $T(z)T(u)$ we use its regularized  definition (\ref{regul}) 
 and the OPE in (\ref{corr}). The computation gives 
\bea
T(z)T(u) =  :\frac{1}{ \a'} \p X^{\m}(z,\bar{ z})  \p X_{\m}(z',\bar{z}'):
 : \frac{1}{\a'} \p X^{\r}(u,\bar{ u})  \p X_{\r}(u',\bar{u}'): \nn \\
 = \frac{\eta^{\m}_{\m}/2}{(z-u)^{4}} +  \frac{2 T(u)}{(z-u)^{2}}
+ \frac{\p T(u)}{(z-u)}
=  \frac{D/2}{(z-u)^{4}} +  \frac{2 T(u)}{(z-u)^{2}}
+ \frac{\p T(u)}{(z-u)}. \label{ope stress}
\eea

  The  Laurent coefficients $L_{n}$ in the stress tensor expansion
   define the Virasoro operators  
  \beq 
 T_{zz}(z) = T(z) = \sum_{n} z^{-n-2}L_{n} \qquad 
   T_{\bar {z}\bar{z}}(\bar{z}) = \bar{T}(\bar{z}) = \sum_{n} \bar{z}^{-n-2}\bar{L}_{n}.
\eeq

 Since the stress tensor is the Noether current for the conformal symmetry
 we can compute the generators of the conformal transformations by
 integrating the current on a constant time slice
 
\beq
L_n = \frac{1}{2\pi i} \oint_{C} dz z^{n+1} T(z), \label{mode}
\eeq 

and similarly for the generators $\bar{L}_n$ of the antiholomorphic 
copy of the conformal algebra
\beq
\bar{L}_n = - \frac{1}{2\pi i} \oint_{C} d\bar{z} \bar{z}^{n+1} \bar{T}(\bar{z}). 
\eeq 

 Finally, through the OPE (\ref{ope stress}) and the relation (\ref{mode})
  we can compute  the quantum version of the
 the  conformal algebra
\beq
[ L_{n}, L _{m} ] = (n-m) L_{n+m}
  +\frac{D}{12} n(n-1)(n+1) \d_{n+m,0}. \label{Virasoro}
\eeq
 An identical relation holds for the anti-holomorphic Virasoro operators $\bar{L}_{n}$. 
 This is called the  Virasoro algebra $\mathcal{V}$.
 Comparison  with  eq. (\ref{classconf}) for the classical commutators  shows
 that the  quantisation of the theory
 introduces  a central extension in the conformal algebra.
 Indeed the number of spacetime dimensions $D$ that appears
  on the right hand-side
   corresponds to what is called  the central 
charge of the  the Virasoro algebra,
 and
  signals the presence of a  conformal anomaly. 
 In  the covariant path integral quantisation for the bosonic string,
 the symmetries ask for the introductions of woldsheet ghost
fields that contribute themselves to the stress tensor
  and therefore  to the central charge in the Virasoro algebra\footnote{See \cite{Green:1987sp,Green:1987mn,Lust:1989tj,Kiritsis:1997hj,Polchinski:1998rq,Polchinski:1998rr,Kaku:1999yd,Kaku:2000dq,Johnson:2003gi,Zwiebach:2004tj} for more details and a more comprehensive introduction to string theory.}.
   In this approach, imposing conformal invariance
 at quantum level asks for the cancellation of the
  conformal anomaly, a condition that constraints
   the number of  coordinate fields to be  $D_{crit} = 26$,
   this number is called the spacetime critical dimension.

\vspace{.5 cm}

In the two dimensional quantum CFT there is an important correspondence
 between states and operators that allows to classify  the representations of the Virasoro
algebra. At the end of the day,  the string  excitations 
   carry a representation of the Virasoro algebra,
 so this analysis is useful to classify the states in the string spectrum.
 
Once one has identified an asymptotic inner
 ($z=0$ on the sphere or $t= - \infty$ on the cylinder)  ground state $|0\rangle$, one can   
  obtain  a generic  asymptotic inner state
   $|f\rangle$  by acting with a field $f(0)$ 
 on the ground state  $|f\rangle = f(0)|0\rangle$.

 For example, the string ground state with center of mass  momentum $k_{\m}$ is given by
 $e^{ik_{\m}X^{\m}(0)}|0\rangle = |0, k_{\m}\rangle$.
 
 Similarly, for an excited state  
 $ \a^{\m}_{-n} \tilde{\a}^{\n}_{-n}|0\rangle $ one has 

\beq
 \a^{\m}_{-n} \tilde{\a}^{\n}_{-n}|0\rangle =  \frac{2}{\a'} \oint \frac{dz}{2i\pi}    z^{n} \p X^{\m}(z) 
 \oint \frac{d\bar{z}}{2i\pi}\bar{z}^{n} \bar{\p} X^{\n}(z)|0\rangle. \label{sate-ope}
\eeq
 Expanding  $\p X^{\m}(z)$($\bar{\p} X^{\n}(\bar{z})$) around $z=0 (\bar{z} = 0)$,
 \beq
\p X^{\m}(z) = \p X^{\m}(0) + ... + \frac{z^{n-1}}{(n-1)!} \p^{n} X^{\m}(0) + ...,
\eeq
 and performing the  contour integrals, one gets
\beq
\a^{\m}_{-n} \tilde{\a}^{\n}_{-n}|0\rangle = \frac{2}{\a' (n-1)!(n-1)!} \p^{n} X^{\m}(0)
 \bar{\p}^{n} X^{\n}(0)|0\rangle,
\eeq
that shows a general correspondence between  inner states and  fields at $z=0$.

  States  created at a  generic point $z_{0}$ on the sphere
 and carring  momentum $k$ are then associated to the operator
\beq
\frac{2}{\a'} e^{ik_{\m}X^{\m}(z_{0},\bar{z}_{0})}  \p^{n} X^{\m}(z_{0})\bar{\p}^{n} X^{\n}(\bar{z_{0}}).
\eeq

Let us see which kind of constraints  the conformal algebra imposes
  on an operator of this kind and therefore on the correspondent quantum state.

 From the Virasoro algebra  in
 (\ref{Virasoro}), it is clear  that the only subalgebra that survives
quantisation is generated by $(L_{1}, L_{0}, L_{-1})$, because for 
 $n = -1,0,1$ the central charge contribution vanishes.
This algebra is isomorphic to $SL(2,\mathbb{R})$ and, together with the
 anti-holomorphic counterpart,  forms the algebra
 $SL(2,\mathbb{C}) = SL(2,\mathbb{R}) \oplus  SL(2,\mathbb{R})$.

 $SL(2,\mathbb{C})$  is the only subalgebra of the conformal algebra whose
 transformations map the Riemann sphere into itself,
in particular
${L}_{0} = z\p$ generates  a constant scale
 transformations $z \rightarrow \l z$ for $\l \in \mathbb{C}$.
  Therefore  to respect the unbroken  $SL(2,\mathbb{C})$
  symmetry  a generic operator
\beq
\frac{2}{\a'} \int dz d\bar{z} e^{ik_{\m}X^{\m}(z,\bar{z})}  \p^{n} X^{\n}(z)\bar{\p}^{n} X^{\m}(\bar{z})
\label{intsphere}
\eeq
  must be ${L}_{0}$ invariant.

  Since the integration measure on the sphere scales as $dzd\bar{z} \rightarrow |\l|^{2} dzd\bar{z}$,
   the operator itself needs to scale  oppositely 
  for the integral  (\ref{intsphere}) to be $SL(2,\mathbb{C})$ 
 invariant.

The operators $f(z,\bar{z})$  that create physical states need to have
  definite global scaling  properties. This kind of operators  are called  primaries
 with conformal weights $(h,\bar{h})$, ($h$ and $\bar{h}$ real numbers), if 
   under the dilatation $z \rightarrow \l z$
 they transform as
  $f(\l z, \bar{\l}\bar{z}) = \l^{-h} \bar{\l}^{-\bar{h}}f(z)$.

With this definition  in mind it is clear that
\beq
\int  dz d\bar{z} V(z,\bar{z}), 
\eeq
 is $L_{0}$ invariant only if
$ V(z,\bar{z})$ has conformal weights $(1,1)$.
This implies that  physical asymptotic states are created by  $(1,1)$ operators, 
i.e.
\beq
L_{0}|phys\rangle = |phys\rangle  \qquad  \bar{L}_{0}|phys\rangle = |phys\rangle.
\eeq

 Let us consider the rank-two tensor operator 
\beq
\int dz d\bar{z} e^{ik_{\m}X^{\m}(z,\bar{z})}
\p X^{\m}(z) \bar{\p} X^{\n}(\bar{z}) \label{ranktwo}
\eeq
$\p X^{\m}(z) \bar{\p} X^{\n}(\bar{z})$ has \emph{classical} $(1,1)$  scaling properties
 due to the presence of  the derivatives $\p$ and $\bar{\p}$ 
 and  $(1,1)$ conformal weights, as we can check by computing
 $\left[ L_{0},\p X^{\m}(0)\right] = \a'\p X^{\m}(0)$ 
with the help of (\ref{mode}) and (\ref{corr}).  A similar computation
 shows instead  that   $e^{ik_{\m}X^{\m}(z,\bar{z})}$
 has conformal weights $(\a'k^{2}/2,\a'k^{2}/2)$
( note that classically this operators has 
zero scaling dimension a property that we recover
 in the limit $\a' \rightarrow 0$).
 $SL(2,\mathbb{C})$ invariance for the rank-two tensor operator
 therefore implies
 that   the sum of the conformal weights
 of  $\p X^{\m}(z) \bar{\p} X^{\n}(\bar{z})$ and  $e^{ik_{\m}X^{\m}(z,\bar{z})}$  
 be equal to $(1,1)$
\beq
(\a'k^{2}/2 + 1,  \  \a'k^{2}/2 +1) = (1,1),
\eeq
which happens only if $k^{2} = - m^{2} = 0$ \  i.e. only if the operator creates 
 massless states.

 These states need to carry an  irreducible 
 representation of the spacetime Lorentz group and the decomposition
 of  the rank two
tensor gives  the spin two symmetric traceless part
 that we identify with the graviton $G_{\m\n}$,
  the antisymmetric
part $B_{\m\n}$  and the trace part $\phi$, the dilaton.
  
     Note that  operators  like  $e^{ik_{\m}X^{\m}(z,\bar{z})}\p^{n} X^{\m} \bar{\p}^{m} X^{\n}$
 for $m$ different from $n$ \emph{cannot}
 create physical states since even at non zero mass they cannot respect 
  the $SL(2,\mathbb{C})$ symmetry,
whence for $n$ equal to $m$ and with a non vanishing mass they can respect
 this symmetry. This is an example of a
   level matching constraint, which in general means that
the  state
\bea
e^{ik_{\m}X^{\m}(0,0)}  \p^{n_1} X^{\m_{1}}(0) \bar{\p}^{\tilde{n}_1}   X^{\n_{1}}(0)
... \p^{n_p} X^{\m_{p}}(0) \bar{\p}^{\tilde{n}_p}  X^{\n_{p}}(0)|0\rangle \nn \\
  \sim  \a^{\m_{1}}_{-n_{1}}\tilde{\a}^{\n_{1}}_{- \tilde{n}_{1}}...
 \a^{\m_{p}}_{-n_{p}}\tilde{\a}^{\n_{p}}_{-\tilde{n}_{p}}|k,0\rangle \label{generic}
\eea
needs to satisfy the level matching condition
\beq
n_{1} + ... n_{p} = \tilde{n}_{1} + ... + \tilde{n}_{p},
\eeq
in order to  respect the  $SL(2,\mathbb{C})$ and therefore
to by physical.

In this case,
 the physical states created by (\ref{generic})
have mass
\beq
m^{2} = - k^{2} = \frac{2}{\a'}( n_{1} + ... + n_{p} - 1).
\eeq


%

To summarise,   $SL(2,\mathbb{C})$ invariance implies 
that only operators with well defined scaling properties (conformal weights) can create  asymptotic physical
 states. 
 There is a precise relation between the rank
 of  tensor operators and the mass of the corresponding physical states.
 In particular,
 the scalar groundstate  $e^{ik_{\m}X_{\m}(z,\bar{z})}|0 \rangle$,
  has conformal weights $(\a'k^{2}/2, \a' k^{2}/2)$ and thus it represents a tachyon 
  $m^{2}  = - 2/ \a' $.
 The only massless states are those created by the rank-two tensor: 
  the metric $G_{\m\n}$ the antisymmetric tensor $B_{\m\n}$
and the dilaton $\phi$.
 All  higher rank tensors that respect  level matching
create massive states.



\vspace{.5 cm}

 The quatization of the bosonic string
  presents some subtleties
 related to the Lorentzian signature
of the space-time metric. 
In fact, one can see  from  (\ref{ladders}),
 that  the oscillator  modes associated with the time coordinate
 $X^{0}$ satisfy a \emph{wrong} sign commutation relation,
 and thus  create negative-norm quantum states.

 There  are various routes to deal with this problem 
that correspond to different, but at the end equivalent,
 ways to quantise the theory. At the classical level,
  the equations of motion   
   for the world-sheet  metric  yield the constraint $T_{\a\b} = 0$.
 There are  two alternative ways  to proceed with the quantisation:
 one can  impose the $T_{\a\b} = 0$ constraint before quantisation,
 reducing in this way
the number of variables that describe the string degrees of freedom,
  or  one can quantise  the whole  $X^{\m}$  and impose
 the constraint in a operator form on the quantum states.
 In the first case  the quantisation is worked out
  in a 
 gauge   in which the  oscillators
  satisfying  \emph{wrong} sign commutation relations
  are not dynamical, because they are expressed in terms
of the transverse oscillators via the solution of the classical constraint.
  However D-dimensional  Lorentz covariance is not manifest and it is
 recovered only for  $D=26$.
  This needs to be checked 
by verifying  the proper commutation relations for the Lorentz generators,
   constructed with a reduced number of oscillators.

In the second case all  coordinates are quantised in
a manifestly covariant way, but the absence of negative norm states in the spectrum
  then follows by imposing
 the constraints  in a operatorial
 form on the physical states.
Also in this case the Hilbert space is indeed free of nagative-norm states
again only if $D=26$.

Here we follow the most economic way to derive
 the  physical spectrum
  by introducing light cone  coordinates
\beq
X^{\pm}= \frac{1}{\sqrt{2}}(X^{0} \pm X^{1}),
\eeq
 a  choice  manifestly  non covariant.
  One can  choose  a gauge that allows to solve the constraints in terms
of the dangerous longitudinal oscillators.
 In such a gauge
$X^{+}$ is identified with the world-sheet time $\tau$,
 a choice that is indeed possible because  every conformal transformation
is induced by a change of coordinates on the world-sheet that satisfies the 
 Laplace equation as we have shown in eq. (\ref{wave})
 and the equation of motion for $X^{+}$ is precisely the same equation:
\beq
\p^{\a}\p_{\a}X^{+} = 0.
\eeq

With this choice  
\beq 
X^{+} = x^{+} + \frac{\a'}{2} p^{+}\ln(z\bar{z})  =  x^{+} - \sqrt{\frac{\a'}{2}} \a^{+}_{0}\ln(z\bar{z})\label{blcgauge} 
\eeq
 
 we have set to zero the oscillations
 along the $X^{+}$ directions,  $\a_{n}^{+} =0$.
 This is quite natural since
  oscillations along the world-sheet are expected not to be physical due
to (world-sheet) reparametrization invariance. 

 The classical  constraints $L_{n} = 0$ then
  eliminate the non-dynamical variables. Indeed
\bea
L_n &=&  \oint_{C}  \frac{dz}{2\pi i}  z^{n+1} T(z)= 
 \frac{1}{\a'} \oint_{C}  \frac{dz}{2\pi i}z^{n+1}:\p X^{\m}\p X_{\m}(z):\nn \\     
 &=& \frac{1}{2} \oint_{C}  \frac{dz}{2\pi i} z^{n+1} : \sum_{m = - \infty}^{\infty} z^{-m-1}\a_{m}^{\m}
 \sum_{l = - \infty}^{\infty} z^{-l-1}\a_{l \ \m}: \nn \\
 &=&  \frac{1}{2} \sum_{m = - \infty}^{\infty}\a_{-m + n }^{\m}\a_{m \ \m}, \label{Ln}
\eea 
and in  the lightcone gauge (\ref{blcgauge})

\beq
0 =  L_n =   \sum_{m = - \infty}^{\infty}\a_{-m + n }^{+}\a_{m}^{-} -  \frac{1}{2} \sum_{m = - \infty}^{\infty}\a_{-m + n }^{i}\a_{m}^{i} =   \frac{1}{2} \a_{0}^{+}  \a_{n}^{-} -  \frac{1}{2} \sum_{m = - \infty}^{\infty}\a_{-m + n }^{i}\a_{m}^{i}
\eeq
allow to express the $\a_{n}^{-}$ in terms of the physical transverse oscillators:

\beq
\a_{n}^{-}= \frac{1}{2\a_{0}^{+}}\sum_{m = - \infty}^{\infty}\a_{-m + n }^{i}\a_{m}^{i}
 = - \frac{1}{\sqrt{2\a'}p^{+}}\sum_{m = - \infty}^{\infty}\a_{-m + n }^{i}\a_{m}^{i}.
\eeq
 Thus showing  that  only the oscillations transverse to
the world-sheet are physical with a positive-definite
scalar product.  A closed string state
 is therefore described in  the lightcone gauge by a $D$-dimensional center of mass momentum $p^{\m}$,
and its oscillations are created by 
 $D-2$ transverse bosonic oscillators $\a^{i}_{n<0} = \sqrt{-n}(a^{i}_{-n})^{\dag}$

\bea
e^{ik_{\m}X^{\m}(0,0)}  \p^{n_1} X^{i_{1}}(0) \bar{\p}^{\tilde{n}_1}   X^{j_{1}}(0)
... \p^{n_p} X^{i_{p}}(0) \bar{\p}^{\tilde{n}_p}  X^{j_{p}}(0)|0\rangle \label{tensorop} \\
  \sim  \a^{i_{1}}_{-n_{1}}\tilde{\a}^{j_{1}}_{-\tilde{n}_{1}}...
 \a^{i_{p}}_{-n_{p}}\tilde{\a}^{j_{p}}_{-\tilde{n}_{p}}|k^{\m},0\rangle. \nn
\eea
 These states satisfy the level matching condition
\beq
n_{1}+ ... + n_{p} = \tilde{n}_{1} + .... + \tilde{n}_{p}
\eeq

 and have a  mass 
\beq
 m^{2} =  -k^{2} = \frac{1}{\a'} \left[ 2( n_1 +...+ n_{p}) - 2 \right].   
\eeq

 Consistency with  D-dimensional Lorentz symmetry requires
  for the   tensor field  operator  (\ref{tensorop}) to  create
    states  carrying  a  representation of the  little group of  $SO(1,D-1)$.
  Massless states needs to be in irreducible
  representation of   $SO(D-2)$, while   massive states in $SO(D-1)$ ones.
 
  As a  consequence of $SL(2, \mathbb{C})$ invariance 
  the only massless  states in the closed string spectrum
  are those created  by the rank-two $(1,1)$
 tensor field 

\beq
 \frac{2}{\a'} e^{ik_{\m}X^{\m}(0,0)}  \p X^{i}(0) \bar{\p} X^{j}(0) |0\rangle \label{ranktwo2}
= \a^{i}_{-1}\tilde{\a}^{j}_{-1}|k_{\m},0\rangle     \qquad i = 1,...,D-2.
\eeq
These states  correspond to the decomposition of the rank-two tensor field into $SO(D-2)$ irreducible representations  
\beq
(D - 2)^{2}  =  \left(\frac{(D - 2)(D - 1)}{2} - 1 \right) + \frac{(D - 2)(D - 3)}{2} + 1.
\eeq
 Clearly,  the $(D - 2)^{2}$ degrees of freedom
    do not fit into a sum of  two-tensor irreducible representations of  $SO(D-1)$
\beq
(D - 1)^{2}  =  \left(\frac{(D - 1)D}{2} - 1 \right) + \frac{(D - 1)(D - 2)}{2} + 1.
\eeq
In other worlds, consistency with Lorentz covariance requires  for the states created
by the rank-two tensor to be massless.

 For  massive states, created by Higher-rank tensors operators,
 the situation is different and the number of degrees of freedom
 fits  in producing a decomposition in terms of $SO(D - 1)$ irreducible representations.
 This is possible because
 there is more then one operator that coresponds to a given conformal weight.
 For example, there are two operators with   $(2,2)$ weights
\beq
   \p X^{i_1}(0) \p X^{i_2}(0) \bar{\p} X^{j_1}(0)  \bar{\p} X^{j_2}(0) \qquad
  \p^{2} X^{i_1}(0) \bar{\p}^{2} X^{j_1}(0),                 
\eeq
 which give rise to $(D-2)^{4} + (D-2)^{2}$ degrees of freedom for the first
 closed string massive level.
  It is possible to show  that this number  
  equals a sum 
    of   $SO(D-1)$-irreducible representations, and that actually 
this is the case for the full list of primary fields,  in agreement with the Lorentz
symmetry.





 
The condition dictated by  both  $SL(2, \mathbb{C})$ and  $SO(1,D-1)$ invariance
  on the states   created by the 
rank-two tensor operator actually is satisfied only 
  if  the number of spacetime dimension is $D=26$.

  Let us consider the expression for $L_0$
 in terms of lightcone oscillators eq. (\ref{Ln}) 
\beq
 L_0 = \frac{1}{2} \sum_{m = - \infty}^{\infty}\a_{-m}^{\m}\a_{m \ \m} = 
    \frac{1}{2} \sum_{m = - \infty}^{-1}\a_{-m}^{i}\a_{m}^{i}  +  \frac{\a_{0}^{2}}{2}
 + \frac{1}{2} \sum_{m = 1}^{\infty}\a_{-m}^{i}\a_{m}^{i} 
\eeq
with $\a_{0}^{2} = 2\a_{0}^{+}\a_{0}^{-} - \a_{0}^{i}\a_{0}^{i} =  \a'/2 \cdot p^{2}$.    
The first term of the last r.h.s.  needs to be normal ordered,
so that
\beq 
\sum_{m = - \infty}^{-1}\a_{-m}^{i}\a_{m}^{i} =  \sum_{m = 1}^{\infty}\a_{-m}^{i}\a_{m}^{i}
+ \frac{D-2}{2} \sum_{m=1}^{\infty}m
\eeq
and 
\beq
 L_0 =   \frac{\a_{0}^{2}}{2}
 +  \sum_{m = 1}^{\infty}\a_{-m}^{i}\a_{m}^{i} +  \frac{D-2}{2} \sum_{m=1}^{\infty}m . 
\eeq
  The divergent sum can be regularised by inserting a smooth cutoff
 
\beq 
 \sum_{m=1}^{\infty}m \rightarrow \sum_{m=1}^{\infty}m e^{-m\e} = \frac{1}{\e^{2}} - \frac{1}{12} + O(\e)
\eeq
In this way the normal ordered expression for $L_0$ reads:
\beq
 L_0 =   \frac{\a_{0}^{2}}{2}
 +  \sum_{m = 1}^{\infty}\a_{-m}^{i}\a_{m}^{i} -  \frac{D-2}{24}  =
    \frac{\a'}{2}  p^{2}
 +  \sum_{m = 1}^{\infty}m (a_{m}^{i})^{\dag} a_{m}^{i} -  \frac{D-2}{24}  \label{L0bos}
\eeq
 and thus the first excited states are massless only if  $D = 26$.

\vspace{.5 cm}

We have seen that conformal invariance at
tree level
 requires that  operators associated to physical string excitations 
  have
  conformal weights equal to $(1,1)$. This in turn determines the mass of the
 associated states.
   In the CFT language operators with conformal weights  $(1,1)$ are called \emph{marginal}, since 
 at linear order a perturbation induced by them does not break conformal invariance
   (see below). It follows that all the tree level string states are created by marginal operators.

 However, whenever a string state flows in a loop diagram,
 its momentum can be off-shell.
In the UV regime the fluctuations are
  regulated by  the finiteness of the string length, 
 and thus are not UV  dangerous. 
 On the other hand,  the IR regime is much more
  subtle, since the    fluctuations can 
 induce effects on the background.
  As we will mention below,  zero momentum   fluctuations 
 can  break conformal invariance
 and change the infrared (background) properties of the theory.

In the infrared limit $k^{2} \rightarrow 0$, 
   the effect of a   fluctuation can be taken into account
   by adding to the original  CFT 
 a perturbation term  which corresponds  to a variation
 of the couplings  $ \l_{k}$ (the string background) of  primary operators $\mathcal{O}_{h}^{k}$
 (the fields whose mode is the fluctuation)
 of conformal weights $h = \bar{h}$ 
 
\beq
\d S = \int dz d \bar{z}\  \mathcal{O}_{h}^{k}\d \l_{k}. 
\eeq
The beta functions $\beta_{k}$ associated to  the couplings  $\l_{k}$
 can be obtained  by a perturbative expansion
 in an increasing number of tree level correlators of the two dimensional theory.

  Up to second order in the fluctuations $\d \l_{k}$,  the 
 $\beta_{k}$ are calculated to be \cite{Polchinski:1998rr}:
\bea
\beta_{k}^{(1)} &=& 2(h-1)\d \l_{k}, \nn \\
\beta_{k}^{(2)} &=& c_{k}^{ij} \d \l_{i}  \d \l_{i}. \label{betas}  
\eea

 The first order $\beta_{k}^{(1)}$ contribution emerges from the OPEs 
between the primary operators $ \mathcal{O}_{h}^{k}$
 and the stress tensor $T$ and therefore depends on the 
 conformal  weight $h$ of the field.  The second order $\beta_{k}^{(2)}$  contribution
 emerges  instead from the OPEs
 between two primary fields, i.e. from  the two point function
\beq
 \mathcal{O}_{h}^{i}(z,\bar{z})     \mathcal{O}_{h}^{j}(w,\bar{w})   \sim
 \frac{c^{ij}_{k}}{|z - w|^{2}} \mathcal{O}_{h}^{k}(w,\bar{w})
\eeq
and is therefore quadratic in the perturbation.

From  equation  (\ref{betas}) it is then clear  the role played by
  irrelevant $h > 1$, marginal $h = 1$ and $h < 1$ irrelevant perturbations
 on the background.
 Massive string modes with   $h > 1$ are irrelevant perturbations, since they
 induce a positive $\beta_{k}^{(1)}$, and therefore their perturbation 
 disappears in the IR with no effect on the background. 
The ground state  $h = -1$ is instead relevant, the beta function is
 negative so that a perturbation to the original CFT
grows in the infrared and changes drastically the
 background configuration.

For marginal perturbations $h =1$,
 such as those induced from the infrared
 fluctuations of the string massless modes,  at first order in the fluctuation
 the beta function vanishes.
 As an example  consider
 a fluctuation  of the metric at zero momentum  $\d G_{\mu\nu}$.
At linearised level the beta function is zero, (first equation of (\ref{betas})),
 so that the theory remains at its critical point.
 
 From the second equation in  (\ref{betas}),
we see that the perturbation can in principle
   break
 conformal invariance.
 The running of the background metric is
\beq
\frac{\p}{\p (ln \Lambda)} G_{\mu\nu} = \beta_{\mu\nu}^{(2)} = C_{\m\n}^{\a\b\g\d} 
  \d G_{\a\b} \d G_{\g\d} + O(G^{3}) \label{beta2}  
\eeq
where the coefficients  $C_{\m\n}^{\a\b\g\d}$ are 
 those in the OPE
\beq
:\p X^{\a}\bar{\p}X^{\b}(z,\bar{z})  : :\p X^{\g}\bar{\p}X^{\d}(w,\bar{w}): \sim
 \frac{C^{\a\b\g\d}_{\m\n}}{|z - w|^{2}} :\p X^{\m}\bar{\p}X^{\n}(w,\bar{w}):
\eeq
that gives the propagator of the graviton on a given background.

 In order to have stability for the background
 one needs therefore to check that the
 second order beta
function is \emph{positive} so that  the
perturbation does not affect
 the IR properties of the theory.
Of course for consistency one should check 
 the value of the beta at every order
 in this non linear perturbative expansion. 

  String backgrounds  compatible with
  conformal invariance must
  satisfy the  field equations  (\ref{bsem})
 that are expressed as a  perturbative expansion
 in the  string scale $\a'$, whose terms
  involve  an increasing number of derivatives
  of the background fields.
 %
 In a slow varying field regime the   leading order in  $\a'$
 can be already a good approximation. This corresponds
  for example to the case where 
  the string length
 is much  smaller than the scale of variation of the background metric,
 described  by the curvature radius of the manifold on which
 the string propagates.

 Conformal invariance allows to  split  the string between
 on-shell classical backgrounds  and quantum fluctuations.
 In general quantum effects spoil the aproximation
  and infrared  tachyonic fluctuations or  tadpoles
  ask for background redefinition.
This might be cured if we knew how to shift the vacuum, but
   the first quantised formulation
 is restricted on-shell and the off-shell continuation remains unknown.
 Spacetime supersymmetry represents a partial way out for this problem  
 since it exclude  the presence of these destabilization effects. The first step to achieve specetime supersymmetry
 can be via the introduction of world-sheet spinors and a two dimensional
world-sheet supersymmetry.
 The conformal symmetry gets  enlarged into  a superconformal symmetry a subject 
   that will be  discussed in   the next two  sections.

\vspace{3 cm}

\section{$N =2$ superconformal symmetry and spacetime supersymmetry}

 \everypar{\hspace{-.6cm}}


   
  
 It is interesting to see under which
  assumptions  the closed string spectrum  
  exhibits  \emph{spacetime} supersymmetry.
   
   On $D=10$  Minkowski spacetime,
   as we will discuss in the next section, 
  there are several closed string spectra
  that exhibit modular invariance, a fundamental constraint
 that singles out the consistent string spectra
 discussed in section four.
  In particular, type IIA and type IIB are
  supersymmetric, while there are  two more
   theories that have a  modular invariant spectrum 
    but with no supersymmetries: type 0A and type 0B.
   The non-supersymmetric type 0
  theories are purely bosonic
    and  are unstable on $D=10$ flat spacetime
  as the presence of a  tachyonic excitation in their spectra suggests.

  After compactification the preservation
  of some spacetime supersymmetries depends on
    the nature of the compact space.
     Actually spacetime supersymmetry does not seem
    to be required by any fundamental
    principle, rather
    is the absence of backreaction on
    the background
     that singles out the supersymmetric
   solutions as those that are really tractable in
   a first quantised formulation.
    A  background
   giving rise to \emph{spacetime} supersymmetry
    and satisfying the $\a '$ leading order
    string equations of motion
     will  generally   not be
   destabilised by classical stringy corrections
   (Higher order in $\a'$),
  in the sense that 
   either it remains a solution to \emph{all}
 orders of the  $\a '$ corrected background equations of motion    
  or,  more generally,
   some of its
 geometrical data needs to be  corrected
  in order to keep up with the perturbative correction 
 from the sigma model \cite{Gross:1986iv,Grisaru:1986px,Grisaru:1986kw}, but these
 changes do not modify the form
 of the low energy effective action \cite{Witten:1981nf,Witten:1985bz,Nemeschansky:1986yx}.
  Moreover, in the  presence of supersymmetry the  background
   cannot be destabilised by loop quantum corrections,
  since  both vacuum diagrams
   and tadpoles vanish  for supersymmetric solutions  \cite{Martinec:1986wa}  \cite{Friedan:1985ge}.
 
   Conversely,  for non-supersymmetric
   compactifications
   both classical  string effects
  (in  $\a '$) and quantum  corrections
     to the vacuum energy (in $g_{s}$) 
 give  perturbative corrections to the equations
 of motion that in most of the cases 
    invalidate
  the original background as a solution.
    This cases are rather more complicate then
the supersymmetric ones, since order by order
 in perturbation theory the background receives substantial
 corrections from the fluctuations.
   The  hope is  that taking 
   the supersymmetric solutions
  as a guide one  could  find solutions
  where supersymmetry is hidden (spontaneously broken)
  with the analytic control
   of the unbroken phase  still present.

\vspace{1 cm}

   Let us study under which conditions the spectrum of
   states in a world-sheet SCFT (superconformal theory)
   enjoys \emph{spacetime} supersymmetry.
   We will focus to the case of $\mathcal{N} = 2$ superconformal symmetry on the world-sheet,
   actually $\mathcal{N} = (2,2) $ taking into account the holomorphic
   and anti-holomorphic closed string sectors.
 As we shall see  this is the symmetry possessed 
    by the  Neveu-Schwarz Ramond  string  in the light cone gauge.
  
  The aim of this section is to present the construction
   of a superconformal algebra and
   to  show which kind
   of  projection on the Hilbert space of states (GSO projection \cite{Gliozzi:1976qd}
  and its generalisations) yields \emph{spacetime} supersymmetry.

  To construct  an $\mathcal{N} = 2$ superconformal theory, (fixing the attention to
  the holomorphic sector), besides a conformal symmetry 
  with a stress tensor $T(z)$ that satisfies all the properties
 illustrated in the previous section,   we need  to introduce  two
supercurrents $G^1 (z)$ and $G^2 (z)$, that satisfy the  OPE \cite{Gepner:1989gr}
\bea
T(z)G^{i}(w) &\sim& \frac{3/2}{(z-w)^2}G^{i}(w) +  \frac{\p_w G^{i}(w)}{(z-w)}, \nn \\ 
G^{1}(z)G^{2}(w) &\sim&  \frac{2 c/3}{(z-w)^3} \frac{2 T(w)}{(z-w)} + i\left( \frac{2 j(w)}{(z-w)^2}+ \frac{\p_{w} j(w)}{(z-w)}\right) \label{OPEn=2,1}
\eea
 In turn it, this  introduces a  new $U(1)$ current $j(z)$ 
 appearing  in the last  of the previous OPEs.\\
The new current  $j(z)$ satisfies
\bea
T(z)j(w) &\sim& \frac{j(w)}{(z-w)^2} +  \frac{\p_w j(w)}{(z-w)} \nn \\ 
j(z)G^{1}(w) &\sim&   \frac{iG^{2}(w)}{(z-w)} \nn \\
j(z)G^{2}(w) &\sim&   \frac{-iG^{2}(w)}{(z-w)}  \nn \\
j(z)j(z)&\sim& \frac{c/3}{(z-w)}  \label{OPEn=2,2}
\eea


The important point is that the $\mathcal{N} = 2$ theories 
 are  one-parameter  families
 of  isomorphic algebras.
The continuous parameter corresponds to the 
choice of boundary conditions for the supercurrents
 \beq
 G^{\pm}(e^{2 i\pi}z) = - e^{ \mp 2 i\pi a}  G^{\pm}(z), \label{twistbc} 
\eeq
  with  $G^{\pm}(z) = \frac{1}{2}( G^{1}(z) \pm i G^{2}(z))$.

For every value of $a = \eta + \frac{1}{2}$, with  $0 \le \eta \le \frac{1}{2}$,  
 the algebras are all isomorphic. 
 This isomorphism  can be checked directly
 by  showing that  the new operators
\bea
L_{n}' &=& L_n + \eta j_n + \frac{c}{6} \eta^{2}\d_{n,0} \nn \\
j_{n}' &=& j_n + \eta j_n + \frac{c}{3} \eta \d_{n,0} \nn \\
G_{r}^{\pm '} &=& G_{r \pm \eta}^{\pm},
\eea
 generate the same $\mathcal{N}  = 2$ superconformal
 algebra.

 One can extend the isomorphism to the representations as well, 
the operation that
 connects two representation in two different 
twisted sectors is called spectral flow.

A unitary map that realizes the spectral flow 
is given explicitly by: $U_{\eta} = e^{-i\sqrt{\frac{c}{3}} \eta \Phi}$ 
when the U(1) current is bosonised $j(z) = i \sqrt{\frac{c}{3}}\p_{z}\Phi$.
 A  generic  field with a  U(1) charge $q$
 under $j(z)$  can then be written as
\beq
f = \hat{f}e^{-i\sqrt{\frac{3}{c}}\Phi}, 
\eeq
where $\hat{f}$ is U(1) neutral. $f$ creates 
the state $|f \rangle$, while the twisted state $|f_{\eta}\rangle$
 is created by the same field through the map $U_{\eta}f U^{-1}_{\eta} $.

 Now,  the R and NS sectors can be defined to  correspond to the choices $a=0$
 and $a=\frac{1}{2}$,
 for the (\ref{twistbc}) 
 boundary conditions for the supercurrent.
 Therefore the states in the two sectors are connected
 by $U_{1/2}$ spectral flow.
Of course  the OPE between the operator 
$U_{1/2}$ and $f$ needs to be singular
 in order to connect  $f$ to the corresponding 
 twisted field $f_{\eta}$,
otherwise the action of the operator $U_{1/2}$
  on the field $f$ would be trivial. 
  From the  explicit form of the OPE

\beq
f(z,\bar{z})   U_{\eta}(z,\bar{z}) = 
 \hat{f}e^{-i\sqrt{\frac{3}{c}}\Phi(z,\bar{z})}
 e^{-i\sqrt{\frac{c}{3}} \frac{1}{2}\Phi(w,\bar{w})} \sim
 |w|^{\frac{q}{2}} e^{-i\left( \sqrt{\frac{3}{c}} - 
\sqrt{\frac{c}{3}}\right)\Phi(w,\bar{w})} \left[1+ O(w,\bar{w}) \right],
\eeq

 one can  see that demanding   semilocality in the above
 expansion corresponds to the
 restriction  $q \in 2 \mathbb{Z} +1$.
 In this case $U_{1/2}$  creates a square-root
branch cut on the world-sheet and interchanges 
spacetime bosons and fermions.

The projection of the $\mathcal{N}=2$ spectrum  of the
 superconformal theory that include both
 R and NS sectors  into states with integer 
and odd $U(1)$ charge gives rise
 to spacetime supersymmetry 
 as more  careful investigations had proved \cite{Sen:1986mg,Banks:1987cy}.  
  This truncation of the spectrum  is  called the GSO projection.
We will reconsider this projection in the specific example
 of $D = 10$ superstring on flat spacetime in the next section
and further in the section dedicated to the torus vacuum amplitudes.

In the more general case of orbifold theories,
 where various twisted sectors may
 coexist, one needs to workout a similar condition
 of semilocality for the action of the
spectral flow operator. This   restricts the allowed
 $U(1)$ charges for the states, and 
  the  spectrum may enjoy spacetime supersymmetry in
 all    cases where 
 a solution of the semilocality condition  exists.
  The corresponding restriction on the $q$ charges 
 represents  a generalised GSO projection.
   
In the above discussion we focused only on the holomorphic
 sector.  Indeed, the \emph{bulk}
SCFT posses  $ N = (2,2)$ superconformal symmetry.
The four world-sheet supercharges, $Q_{L1}(z)$,  $Q_{L2}(z)$
from the holomorphic sector and   $Q_{R1}(\bar{z})$,   $Q_{R2}(\bar{z})$
from the anti-holomorphic one
 are  the zero modes of the $G^{\pm}_{L}(z)$ and $G^{\pm}_{R}(\bar{z})$ 
supercurrents.
If  an  appropriate  GSO projection on the spectrum is performed,
   than the states of $N =(2,2)$ superconformal theory     
 realize $N = 2$ spacetime supersymmetry.
 The spacetime supercharges are the spectral flow
 operators (holomorphic and antiholomorphic),
 constructed by  world-sheet quantities, as discussed above. 

\vspace{.5 cm}

\section{R-NS $D=10$ superstring in light cone gauge}

 \everypar{\hspace{-.6cm}}

 Now we turn to a more concrete description, by considering  a world-sheet Action
 with two dimensional
 fermionic degrees of freedom  on a flat spacetime background. 
 The main motivation is that  two-dimensional  anticommuting  fermions $\psi^{\m}$
  with target space vector index $\m = 0, ... ,D - 1$, 
 after quantisations, have the nice virtue of generating states that carry spinorial representations
 of the $SO(1,D - 1 )$ target Lorentz group. Under some 
circumstances,  together with the bosonic states, they realise spacetime supersymmetry. 

 The introduction of $\psi^{0}$ brings
 a new  infinite discrete set of
 wrong sign oscillators besides those introduced by  $X^{0}$.
   In order to eliminate them we need therefore a
 larger set of constraints, which 
  can be obtained from the Noether currents
   corresponding  to the  extension of  the
 original conformal symmetry into a superconformal symmetry.

 The general covariant two dimensional Action that includes  $\psi^{\m}$ is
\beq
S = \frac{1}{4\pi \a'}\int d \s  d \tau \sqrt{g}  \left[ g^{\a\b}\p_{\a}X^{\m}\p_{\b}X^{\n}
  + \frac{i}{2} \bar{\psi}^{\m}\gamma^{\a}\nabla_{\a}\psi^{\n} 
+ \frac{i}{2}\left( \bar{\chi}_{\a}\gamma^{\b}\gamma^{\a}\psi^{\m} \right)
 \left(\p_{\b}X^{\n} - \frac{i}{4}\bar{\chi}_{\b}\psi^{\nu} \right) \right] \eta_{\m\n} \label{wssaction}
\eeq
where $\chi_{\a}$ is a two dimensional  gravitino.

This action enjoys $N=(1,1)$ superconformal  symmetry,
\bea
\d g_{\a\b} = i\e (\gamma_{\a}\chi_{\b} + \gamma_{\b}\chi_{\a}), \qquad \d \chi_{\a} = 2\nabla_{\a}\e, \nn \\
\p X^{\m} = i\e \psi^{\m}, \qquad  \d \psi^{\m} = \gamma^{\a}(\p_{\a}X^{\m} - \frac{i}{2}\chi_{\a}\psi^{\m})\e, \qquad \p \bar{\psi}^{\m} = 0.
\eea    
where $\e$ is a right moving two dimensional   spinor and similarly for a left moving supersymmetry.

 In the superconformal gauge:
\beq
g_{\a\b}= e^{\varphi}\eta_{\a\b}, \qquad \chi_{\a} = \gamma_{\a}\chi,  \label{scgauge} 
\eeq
where $\chi$ is a constant Majorana spinor, 
 both the conformal factor $\varphi$ and
 the spinor $\chi$ decouple from the action,
  that reduces to a   free two dimensional theory \cite{Polyakov:1981re} 
\beq
S =  \frac{1}{4\pi \a'} \int d\s d\tau \left( \p_{\a}X^{\m}\p^{\a}X^{\n} + i\bar{\psi}^{\m}\gamma^{\a}\p_{\a}\psi^{\n} \right)\eta_{\m\n}. \label{scgAction}
\eeq
On the infinite cylinder  we need to assign periodicity
 conditions along $0 \le \s < 2\pi$. For fermion fields  $\psi^{\m}(\s,\tau)$,
 we have two choices compatible with maximal target space Poincar\`e
 invariance:
 they can be either periodic (Ramond spinors)\cite{Ramond:1971gb} or anti-periodic (Neveu-Schwarz spinors)\cite{Neveu:1971rx}.
  As we will
see in following sections, it is possible to consider
 more general twisted periodicity condition for $\psi^{\m}(\s,\tau)$
 and for the coordinates $X^{\m}(\s,\tau)$  at 
 the price of breaking  the maximal spacetime symmetry.
  In this section, however,  we restrict the analysis to 
 a flat Minkowski spacetime so that 
   $X^{\m}$ are periodic  and  $\psi^{\m}$  periodic (Ramond)
and anti-periodic (Neveu-Schwarz).

\bea  
\psi_{R}^{\m}(w + 2\pi) &=& \psi_{R}^{\m}(w + 2\pi)  \qquad Ramond, \nn \\
 \psi_{R}^{\m}(w + 2\pi) &=& - \psi_{R}^{\m}(w + 2\pi)  \qquad Neveu-Schwarz, 
 \eea
  with $w = \s - i \tau$. Of course,
 similar conditions hold for  the left mover fields

\bea  
\psi_{L}^{\m}(\bar{w} + 2\pi) &=& \psi_{L}^{\m}(\bar{w} + 2\pi)  \qquad Ramond, \nn \\
\psi_{L}^{\m}(\bar{w} + 2\pi)  &=& - \psi_{L}^{\m}(\bar{w} + 2\pi)  \qquad Neveu-Schwarz. 
\eea
The Fourier expansions
\bea  
\psi_{R}^{\m}(w) &=& \sqrt{\frac{\a'}{2}}  \sum_{r \in \mathbb{Z}} d_{r}e^{irw}  \qquad Ramond, \nn \\
\psi_{R}^{\m}(w) &=&  \sqrt{\frac{\a'}{2}}\sum_{r \in \mathbb{Z} + 1/2} b_{r}e^{irw}            \qquad Neveu-Schwarz 
\eea
have then integer modes in the R sector and half-integer in the NS one.

As usual, it is convenient  to go from the cylinder to the Riemann sphere 
 through the conformal transformation $z = e^{iw}$
\beq
S =  \frac{1}{2\pi \a'} \int dz d\bar{z} \left( \p X^{\m} \bar{\p} X^{\n} + i \l^{\m} \bar{\p} \l^{\n}  +   i\bar{\l}^{\m} \p \bar{\l}^{\n}   \right)\eta_{\m\n}. \label{sswsaction}
\eeq
where  $\psi^{\m} = (\l^{\m}, \bar{\l}^{\m})^{T}$.

 The equations of motion for the fields are  then
\beq
 \bar{\p}\p  X^{\m} = 0,  \qquad   \bar{\p} \l^{\m} = 0,  \qquad  \p  \bar{\l}^{\m} = 0,
\eeq
 implying that
 $\p X^{\m}(z)$ and  $\l^{\m}(z)$ are   holomorphic,
 while  $\bar{\p} X^{\m}(\bar{z})$ and $\bar{\l}^{\m}(\bar{z})$ are anti-holomorphic.
To avoid repetitions we only discuss
 holomorphic quantities, similar expressions
 holding for the anti-holomorphic counterparts.

The Laurent expansion  for the holomorphic
 spinors are
\bea  
\l^{\m}(z) &=& \sqrt{\frac{\a'}{2}} \sum_{r \in \mathbb{Z}} d_{r}z^{-r - 1/2 } \qquad Ramond, \nn \\
\l^{\m}(z) &=& \sqrt{\frac{\a'}{2}} \sum_{r \in \mathbb{Z} + 1/2} b_{r} z ^{-r - 1/2 }\qquad Neveu-Schwarz.
\label{fermoscillators} 
\eea
Notice that now Ramond fermions have a square-root brunch cut on the complex plane
 since they are properly defined on the 
 double covering of the sphere,
 while the Neveu-Schwarz have now integer modes. 
  $\l^{\m}$ and $\bar{\l}^{\m}$   have conformal weight $1/2$
 as a consequence of scale invariance
 $z \rightarrow \l z$  of  the action  (\ref{sswsaction}).

As usual the inverse relations are given by contour integrals
\bea
 d_{r}^{\m} &=& \sqrt{\frac{2}{\a'}}\oint \frac{dz}{2\pi i} z^{ r - 1/2} \l_{R}^{\m}(z) \qquad r \in \mathbb{Z}, \nn \\
  b_{r}^{\m} &=& \sqrt{\frac{2}{\a'}}\oint \frac{dz}{2\pi i} z^{r - 1/2} \l_{NS}^{\m}(z) \qquad r \in \mathbb{Z}+ 1/2.
\eea

The supercurrent associated with the  holomorphic  supersymmetry is
\beq
G(z) = \frac{i}{\a'} \l^{\m}\p X_{\m}(z), \label{T}
\eeq
while the stress tensor now includes
 also the contribution from the world-sheet fermions
\beq
T(z)  = -\frac{1}{\a'}\left( \p X^{\m}\p X_{\m} -  \l^{\m}\p \l_{\m}\right).\label{G}
\eeq

  We can now  use the same argument as in (\ref{path}) for the bosonic string
  and consider the  path integral on the Riemann sphere
   for the Action  in superconformal gauge (\ref{scgAction}).
 The equations of motion are not satisfied by  the quantised fermionic fields
  at the coincident points, the divergence being
\beq
\l^{\m}(z')\bar{\p}\l^{\n}(z) = \a' \pi \eta^{\m \n}\d(z - z'). \label{corrferm}
\eeq
  The OPE between  two fermionic fields is then
\beq
\l^{\m}(z')\l^{\n}(z) =  \frac{\a'}{2}\frac{ \eta^{\m\n}}{(z - z')}. \label{singferm}
\eeq

 Due to the above singularity  the oscillation modes of the fermonic fields do
 not anticommute any more:

\bea
\{ d_{r}^{\m}, d_{s}^{\n} \} =  \frac{2}{\a'} \oint \frac{dz}{2\pi i} z^{r - 1/2} \l^{\m}(z) 
 \oint \frac{dw}{2\pi i} w^{ s  - 1/2} \l^{\n}(w) \nn \\
 +  \frac{2}{\a'} \oint \frac{dw}{2\pi i} w^{ s  - 1/2} \l^{\n}(w)
 \oint \frac{dz}{2\pi i} z^{r - 1/2} \l^{\m}(z)  = \eta^{\m\n}\d_{r+s,0} \qquad r,s \in \mathbb{Z} \label{Roscill}
\eea
 for Ramond oscillators. Identical relations  
\beq
\{ b_{r}^{\m}, b_{s}^{\n} \} =  \eta^{\m\n}\d_{r+s,0} \qquad r,s \in \mathbb{Z} + 1/2
\eeq
  hold for Neveu-Scwharz oscillators.

\vspace{.5 cm}

 The next step is to compute the OPE between  the  currents in the quantised theory:
 
\bea
T(z) &=& \frac{1}{\a'}\left(   :\p X^{\m}\p X_{\m}: +  :\l^{\m} \p \l_{\m}: \right), \nn \\
G(z)  &=& \frac{1}{\a'} \l^{\m}\p X_{\m},
\eea 
 then  by using  the OPEs that we have already obtained in  (\ref{corr}) and in  (\ref{singferm})
 \bea
\p X^{\m}(z') \p \ X^{\n}(z) &=&  \frac{\a'}{2} \frac{\eta^{\m\n}}{(z  - z')^{2}}, \nn \\
\l^{\m}(z')\l^{\n}(z) &=&  \frac{\a'}{2}\frac{ \eta^{\m\n}}{(z - z')}, 
\eea
 the OPEs between the currents are:
\bea
T(z)T(w) &=& \frac{3/2\cdot D}{2(z-w)^{4}} + \frac{2T(z)}{(z-w)^{2}} + \frac{\p T(z)}{z-w}, \nn \\
T(z)G(w)  &=& \frac{3/2 G(w)}{(z-w)^{2}} + \frac{\p G(w)}{z-w}, \nn \\ 
G(z)G(w) &=& \frac{D}{(z-w)^{3}} + \frac{2 T(w)}{z-w}.\label{superOPE}  
\eea
Once we expand the currents in Laurent series
\beq
T(z) = \sum_{n} L_{n}z^{-n-2},  \qquad G(z) = \sum_{r} G_{r}z^{-n-3/2}, 
\eeq 
\beq
L_{n} = \oint \frac{dz}{2\pi i}z^{n + 1}T(z), \qquad G_{r} = \oint \frac{dz}{2\pi i}z^{r + 1/2}G(z), \label{LnGr},
\eeq
the knowledge of the singularities  (\ref{superOPE})
 in the  OPEs  between currents given in (\ref{superOPE})
  allows to  compute via contour integrals the (anti)commutators relation for
 the (light-cone) superconformal algebra

\bea
[L_{n}, L_{m}] &=& (n-m)L_{n + m} + \frac{3/2\cdot D}{12} n(n-1)(n+1)\d_{n + m,0}, \nn \\
\{ G_{r}, G_{s}\}   &=&  2L_{r + s} +  \frac{3/2\cdot D}{12} (4r^{2}-1)\d_{r + s,0},  \nn \\
\left[L_{n}, G_{s}\right]  &=& \frac{m-2r}{2} G_{n+s}.
\eea

\vspace{.1 cm}
The most economical way to obtain the 
spectrum is to solve the  constraints
 $T(z) = 0$ and $G(z) = 0$ at the  classical level.
 With lightcone spacetime coordinates,
  we can choose a gauge  where
  the solutions of the constraints allow to express
 the oscillators along the lightcone directions,
 as a function of those along the transverse directions.
  Although after  this gauge choice 
 Lorentz covariance is not manifest, we can remove in this way
  \emph{wrong-sign} oscillators that arise from
 $X^{0}$ and $\psi^{0}$ and
 obtain a system of dynamical   variables  with a positive definite scalar product.
 
\vspace{.1 cm}
 
The lightcone coordinates are $X^{\pm} = \frac{1}{\sqrt{2}}( X^{0} \pm  X^{1})$
 and  $\psi^{\pm} = \frac{1}{\sqrt{2}}( \psi^{0} \pm  \psi^{1})$
and the lightcone gauge is given by:
\beq 
X^{+} = x^{+} + \frac{\a'}{2} p^{+}\ln(z\bar{z})  \qquad \l^{+} = \bar{\l}^{+} =  0. \label{sstlightcone}
\eeq



As  in the bosonic string case,  one needs after quantisation to
 construct spacetime charges with the reduced
 set of variables and verify that indeed
they satisfy the Lorentz algebra, a condition 
that is satisfied only for $D_{crit} = 10$.

In the covariant approach, that we will not discuss here, one needs instead  to 
cancel the conformal anomaly between the matter fields
$ X^{\m}$ and $\psi^{\m}$ and the system of
 anticommuting  ghost necessary for the commuting
coordinates that contribute to $-26$ to the central charge and new commuting ghosts
 whose presence is due to the world-sheet fermions,
 that give a central charge of  $+11$.  Since a boson $X$ contributes  $+1$ and
a fermion  $+1/2$, the cancellation
 for the total  central charge gives $D + D/2 - 26 + 11 = 0$
that fixes the critical dimension to be $D_{crit} = 10$.
\vspace{.1 cm}

We start by obtaining the spectrum in the NS sector. In this case we want to write
 the charges $L_{n}$ and $G_{s}$, defined in (\ref{LnGr}),  in terms of  bosonic oscillators
 $\a^{+}_{0}$, $\a^{-}_{n}$,  $\a^{i}_{n}$
 of the Laurent expansion for $\p X^{\m}$ eq.(\ref{holcord}) 
and transverse fermionic oscilators $b^{i}_{r}$
 in the expansion for the NS spinor $\l_{NS}$,  second relation
 in   (\ref{fermoscillators}):
\beq
L_{n} =     \frac{1}{2} \left( - 2\a^{+}_{0}\a^{-}_{n} +   \sum_{m  \in \mathbb{Z}  } \a^{i}_{n-m} \a^{i}_{m} +
 \sum_{r \in \mathbb{Z} + 1/2} \left( r - \frac{n}{2} \right) b^{i}_{n-r} b^{i}_{r}\right),  
\eeq
 in particular $L_{0}$ has the form
\beq
L_{0} =     \frac{1}{2} \left( - 2\a^{+}_{0}\a^{-}_{0} +   \sum_{i = 2}^{D}(\a^{i}_{0})^{2}
    +    \sum_{m \in \mathbb{Z} - \{0 \}} \a^{i}_{-m} \a^{i}_{m} +
 \sum_{r \in \mathbb{Z} + 1/2}  r   b^{i}_{-r} b^{i}_{r}\right),  
\eeq
 and needs to be normal ordered,

\beq
L_{0} =     \frac{1}{2} \left( \frac{\a'}{2}p^{2}
    +   2 \sum_{m = 1}^{\infty} \a^{i}_{-m} \a^{i}_{m} + (D -2) \sum_{m = 1}^{\infty}m
 + 2\sum_{r = 1/2}^{\infty}  r   b^{i}_{-r} b^{i}_{r}
 - (D -2) \sum_{r = 1/2}^{\infty}r \right),  
\eeq
with $r$ always half-integer since we are in the NS sector. 

 We need to regularize the divergent sums in $L_{0}$
\bea
\sum_{m = 1}^{\infty}m - \sum_{m = 1}^{\infty}\left(m - 1 /2\right) &\rightarrow&
 \sum_{m = 1}^{\infty}me^{-\e m} - \sum_{m = 1}^{\infty}\left(m - 1/2 \right)e^{-\e (m  - 1/2)} \nn \\
 &=& -1/8 + O(\e)
\eea
where we have used the $\e$-expansion:
\beq
\sum_{m = 1}^{\infty}\left(m - \d  \right)e^{-\e (m  - \d )} = 1/\e^{2} - 1/12 + \d /2 - \d^{2} /2 + O(\e).
\eeq
 
The regular expression for $L_{0}$ is therefore
\beq
L_{0} =   \frac{\a'}{4}p^{2} +  \sum_{m = 1}^{\infty} \a^{i}_{-m} \a^{i}_{m}
 +  \sum_{r = 1/2}^{\infty}  r   b^{i}_{-r} b^{i}_{r} 
  -  \frac{D-2}{16}.
\eeq
For closed strings we have actually  two copies of the superconformal algebra, and
every closed-string excitation has the form $|phys \rangle \otimes |phys \rangle$,
with the first factor from the holomorphic sector
and the second from the anti-holomorphic one.
The left-mover counterpart for $L_{0}$ has the form
\beq
\bar{L}_{0} =   \frac{\a'}{4}p^{2} +  \sum_{m = 1}^{\infty} \tilde{\a}^{i}_{-m} \tilde{\a}^{i}_{m}
 +  \sum_{r = 1/2}^{\infty}  r   \tilde{b}^{i}_{-r} \tilde{b}^{i}_{r} 
  -  \frac{D-2}{16},
\eeq 
the same expression but involving anti-holomorphic quantities.

 The constraints  $L_{0}|phys \rangle = 0 $ and $\bar{L}_{0}|phys \rangle = 0$ 
  contain both the square $p^{2}$ of the D-dimensional momentum,
  yielding  the mass-shell condition
\beq
m^{2}|phys \rangle  = \frac{4}{\a '} \left(  \sum_{m = 1}^{\infty} m(a^{i}_{m})^{\dag} a^{i}_{m}
+ \sum_{r = 1/2}^{\infty}  r  ( b^{i}_{r})^{\dag} b^{i}_{r} - \frac{D-2}{16} \right)|phys \rangle, \label{NSmassshell}  
\eeq
together with the Level Matching condition
  $(L_0 - \bar{L}_0) |phys \rangle = 0$. 

In particular, the state  $( b^{i}_{1/2})^{\dag} ( \tilde{b}^{j}_{1/2})^{\dag}|0 \rangle$
   can be decomposed into   irreducible representations of  the rotation group $SO(D-2)$,
 but with its $(D - 2)^{2}$ degrees of freedom
it is not possible to fit a sum of irreducible representation  of $SO(D-1)$.
 Therefore, since $SO(D-2)$ is the little group for \emph{massless}
 D-dimensional Lorentz representation, this state needs to be massless. 
\beq
m^{2}(b^{i}_{1/2})^{\dag} ( \tilde{b}^{j}_{1/2})^{\dag}|0 \rangle = \frac{4}{\a '} \left(\frac{1}{2}  -  \frac{D-2}{16}   \right)( b^{i}_{1/2})^{\dag} ( \tilde{b}^{j}_{1/2})^{\dag}|0 \rangle,   
\eeq
consistency with Lorentz symmetry therefore demands $D = 10$.
These massless states correspond to o the graviton $G_{\m\n}$, 
  the antisymmetric tensor $B_{\m\n}$ and the dilaton $\phi$.

 
 The $NS \otimes NS$  ground state  is therefore tachyonic with a mass
\beq
m^{2}|0, NS \rangle \otimes |0, NS \rangle= - \frac{2}{\a '},
\eeq
but we will see how  the GSO projection  can eliminate it from the spectrum.  
\vspace{.2 cm}

Turning to the Ramond sector,

\beq
L_{n} =     \frac{1}{2} \left( - 2\a^{+}_{0}\a^{-}_{n} +   \sum_{m  \in \mathbb{Z}  } \a^{i}_{n-m} \a^{i}_{m} +
 \sum_{r \in \mathbb{Z}} \left( r - \frac{n}{2} \right) d^{i}_{n-r} d^{i}_{r}\right),  
\eeq
 and for $n = 0$,
\beq
L_{0} =     \frac{1}{2} \left( - 2\a^{+}_{0}\a^{-}_{0} +   \sum_{i = 2}^{D}(\a^{i}_{0})^{2}
    +    \sum_{m \in \mathbb{Z} - \{0 \}} \a^{i}_{-m} \a^{i}_{m} +
 \sum_{r \in \mathbb{Z}}  r   d^{i}_{-r} d^{i}_{r}\right).  
\eeq
  Normal ordering gives two divergent sums that cancel one-another

\beq
L_{0} =      \frac{\a'}{4}p^{2}
    +   \sum_{m = 1}^{\infty} \a^{i}_{-m} \a^{i}_{m} + \frac{(D -2)}{2} \sum_{m = 1}^{\infty}m
 + \sum_{r = 1}^{\infty}  r   d^{i}_{-r} d^{i}_{r}
 - \frac{(D -2)}{2} \sum_{r = 1}^{\infty}r,  
\eeq
 so that the mass-shell condition in the R sector then reads
\beq
m^{2}|phys \rangle  = \frac{4}{\a '} \left(  \sum_{m = 1}^{\infty} m(a^{i}_{m})^{\dag} a^{i}_{m}
+ \sum_{r = 1/2}^{\infty}  r  ( d^{i}_{r})^{\dag} d^{i}_{r} \right)|phys \rangle.   
\eeq
The Ramond groundstate $|0, R \rangle$  is therefore massless and  degenerate,  since
 the states $d_{0}^{i}|0, R \rangle$ $i= 2,...,D-2$ are all massless.  
Actually, the zero modes  $d_{0}^{i}$, due to the anti-commutators relations in (\ref{Roscill}),
satisfy the Clifford algebra for the rotation group $SO(D-2)$ 
\beq
\{ d_{0}^{i}, d_{0}^{j} \} = \d^{ij}.
\eeq
 By defining 
\beq
 d_{0}^{i \pm} = \frac{1}{2}(d_{0}^{2i} \pm d_{0}^{2i +1}), \qquad  i = 2,...,D/2 - 1,
\eeq
we obtain in a standard way a set of ladder operators, with whome one can 
   construct   a basis for the $2^{D/2 - 1} = 16$ dimensional linear space of $SO(D-2)$ spinors
\footnote{In the appendix of \cite{Polchinski:1998rr}  there is a quite useful report about  Spinors and SUSY in various dimensions.} 
\beq 
\z^{\mathbf{s}} = (d_{0}^{2 +})^{S_2 - 1/2}...(d_{0}^{D/2 +})^{S_{D/2} - 1/2}\z,  \qquad  (d_{0}^{2i -})\z = 0.
\eeq 

The Ramond (holomorphic) groundstates is therefore massless and carries a spinorial 
representation of the rotation group $SO(D-2)$.

\vspace{.2 cm}

To eliminate the tachyon in the NS sector one can use the operator $(-)^{G}$, where $G$ counts the
 number of holomorphic NS raising operators $(d_{r}^{i})^{\dag} = d_{-r < 0}^{i}$, $r \in \mathbb{Z} + 1/2$.
Since the tachyon $|0, NS \rangle$
  is the groundstate, it has  $(-)^{G} = +$,
 while the holomorphic massless vector  $d_{-1}^{i}|0, NS \rangle$,
   has  a  negative holomorphic world-sheet fermion number $(-)^{G} = -$.   
Therefore one might  retain in the  NS sector only  $(-)^{G} = -$ states,
thus constructing the tachyon-free subsector $NS_{-} = NS/(-)^{G + 1}$ and similarly for the
anti-holomorphic sector by using the index  $(-)^{\bar{G}} = -$.
 In this way the tachyon is projected out
from the closed string spectrum.

In the R sectors (holomorphic and anti-holomorphic) one considers a similar projection,
  but this time the operators is $(-)^{G}\Gamma_{9}$.  
 At massless level
it singles out one $SO(D-2)$ chirality,
thanks to the chirality matrix $\Gamma_{9}$. 
\vspace{.1 cm}

 The GSO projection 
   is  the key to  construct consistent  closed
 and  open string theories.
 Modular invariance at one-loop, a subject that we will  consider in the following section,
 plays a fundamental role in  this respect
  and clarifies   the necessity of using the GSO projections.

\vspace{.1 cm}

 Spacetime supersymmetry
  imposes   an equal number of bosonic and fermionic
degrees of freedom at every mass level.
 In the closed string case the spectrum originates from 
 four sectors $NS \otimes NS$, $NS \otimes R$, $R \otimes NS$ and $R \otimes R$.
   After the tachyon is projected out from the holomorphic NS sector 
  the massless states $b_{-1}^{i}|0 ,NS \rangle$ describe the  eight physical degrees
of freedom of an $SO(1,9)$ vector.  Without  the analogous
 projection by $(-)^{G}\Gamma_{9}$ that reduces the sixteen
 components of the (holomorphic) R groundstate to eight  one would not achieve a pairing 
  of states in the NS and R sectors,
 and consequently  spacetime bosons from  $NS\otimes NS$ and   $R\otimes R$
 and spacetime spinors from  $NS\otimes R$ and $R\otimes NS$
 would not have the same number of degrees of freedom, even at massless level.

 The more general question of the pairing of fermionic and bosonic
 degrees of freedom   at every mass level
  can be checked  by computing the  string partition function and by using identity involving
 Jacobi theta functions, a matter considered in the following section.  

\vspace{.5 cm}

In the $NS_{-} \otimes NS_{-}$  the massless states originate from
 the two tensor $d_{-1}^{i}|0,NS \rangle \otimes \tilde{d}_{-1}^{j}|0,NS \rangle$,
 that decompose into 
 $SO(8)$ irreducible components as  $\mathbf{8}_{V}\times \mathbf{8}_{V} =  35 + 28 + 1$.

 $35 = (8\cdot 9)/2 - 1$ is the number of components of the symmetric traceless part
 that correspond to the $G_{\m\n}$ graviton,  $28 = (8\cdot 7)/2$ is the
 number of components of the antisymmetric part $B_{\m\n}$ and $1$ is the trace
 that corresponds to the dilaton $\phi$. 

For the holomorphic and anti-holomorphic Ramond sectors one  has two possible choices to
enforce the GSO projection. One can select states with opposite chirality 
  or with the same chiralty,
 we denote these two possibilities as   $(R_{+},R_{-})$ and  $(R_{+},R_{+})$,
where the first entry denotes the choice for the holomorphic sector and the 
second for the anti-holomorphic one.
 
 The first choice corresponds to type IIA superstring,
 while the second to   type IIB superstring:

\bea
type IIA:  &\hspace{.1cm}& (NS_{-},R_{+}) \otimes  (NS_{-},R_{-}) = (NS_{-}\otimes NS_{-}),   (NS_{-}\otimes R_{-}),
 (R_{+}\otimes  NS_{-}),  (R_{+}\otimes  R_{-}) \nn \\  
type IIB:  &\hspace{.1 cm}&        (NS_{-},R_{+}) \otimes  (NS_{-},R_{+}) = (NS_{-}\otimes NS_{-}),   (NS_{-}\otimes R_{+}),  (R_{+}\otimes  NS_{-}),   (R_{+}\otimes  R_{+}) \nn  
\eea

\vspace{.5 cm}
The massless content in the    $NS_{-} \otimes R_{-}$ sector can be obtained 
 by the decomposition  $\mathbf{8}_{V}\times \mathbf{8}_{S} = \mathbf{56}_{S} + \mathbf{8}_{C}$
 where the subscripts $S$ and $C$ distinguish the two opposite chiralities of an  $SO(8)$ spinor.  
$\mathbf{56}_{S}$ are the number of physical degrees of a massless
 ten-dimensional  spin $3/2$ field
 $\psi^{\m}_{\a}$
that corresponds to the gravitino,
while the  $\mathbf{8}_{C}$ corresponds to  the dilatino  $\bar{\z}_{\dot{\a}}$.

 For  $NS_{-} \otimes R_{+}$ we would have instead 
the opposite chiralities for the gravitino $\bar{\psi}^{\m}_{\dot{\a}}$
 and the dilatino  $ \z_{\a}$.

\vspace{.5 cm}  
 In  the Ramond-Ramond $R \otimes R$ sectors, the product of two spinors can be decomposed in terms
 of p-forms with  odd-degree if the spinors have opposite chirality as in type IIA
and with even degree if the chirality of the spinors is the same as in type IIB.
All the form satisfy an Hodge duality condition through their fieldstrength,
so that a p-form is dual to an (8 - p)-form.

Therefore  $(R_{+}\oplus  R_{-})$ in type IIA
 contains a one form $C_{1}$  (with its dual  $C_{7}$),
 and  a three form $C_{3}$  (with its dual to $C_{5}$).
   $(R_{+}\oplus  R_{+})$ in type IIB
 contains instead a zero form  $C_{0}$  (and its dual  $C_{8}$),
 a two form $C_{2}$  (and its  dual  $C_{6}$)
 and a selfdual four form $C_{4 +}$

\vspace{.5 cm}

All these  RR forms represent a multindex generalisation of the Maxwell potential
 $A_{\m}$, and  in the effective supergravity Lagrangians they satisfy a generalisation
 of the local $U(1)$ invariance $A \rightarrow A + d\Lambda$,
 in the form  $C_{p} \rightarrow C_{p} + d\Lambda_{p - 1}$.

  A (zero dimensional) point-like electric charge $q_{e}^{(0)}$ is a source 
  for the electromagnetic field $dA$ and it couples to the vector
 potential through the term 
 \beq
 q_{e}^{(0)}  \int_{\mathcal{\gamma}} dx^{\m}A_{\m}(x),
\eeq
 where the integral is performed along the worldline $\mathcal{\gamma}$  of the point-like electric charge.
A natural electric source for a form
 $C_{p + 2}$ is  a (p + 1)-dimensional
 object called p-brane, with  electric charge  $q_{e}^{(p+1)}$ and coupling
\beq
 q_{e}^{(p+1)}  \int_{\mathcal{M}_{p+1}} C_{p + 2}.
\eeq
where $\mathcal{M}_{p+1}$ is the p-brane world-volume.
 
In four dimension, if one introuduce magnetic point-like charges,
  Maxwell equations $\p^{\m}F_{\m\n} = J^{e}_{\n}$ and  $\p^{\m \ *}F_{\m\n} = J^{m}_{\n}$ and  ( with $J^{e}$   ($J^{m}$)  the electric(magnetic) current, and $F^{*}$ the Hodge dual of $F$),
 are symmetric under the substitution $ F \rightarrow F^{*}$. 

The electric charge contained in a region  surrounded by a  two-sphere $S^{2}$
 is given by
\beq 
 q_{e}^{(0)}  = \int_{S^{2}} d\vec{S} \cdot  \vec{J}^{e} = \int_{S^{2}} F^{*},
\eeq

 while the magnetic charge is given  by
\beq 
 q_{m}^{(0)}  = \int_{S^{2}} d\vec{S} \cdot  \vec{J}^{m} = \int_{S^{2}} F. \label{magncharge}
\eeq

Now in all the non-trivial topological situations on which
 $F \ne dA$ globally \footnote{which means that the relation $F = dA$ holds 
 separatley  on different coordinate patches and the different values of the
 vector potential are connected through gauge transformations, so that
$F$ is globally defined.},  the integral  (\ref{magncharge})
 can be non-vanishing and we can have magnetic monopoles  in four dimensions.
\vspace{.1 cm}

\begin{figure}  
\begin{center} 
{\includegraphics[scale=1, height=5cm]{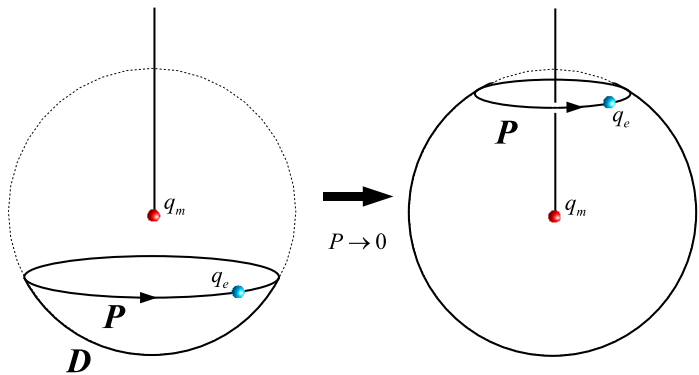}}
\caption{We can shrink the path  $\mathcal{P}$ to zero around the Dirac String so that the surface $\mathcal{D}$
 on the left closed to a two sphere $S^{2}$.}                       
\label{DiracString}
\end{center}
\end{figure}

The electric and magnetic charges need to satisfy a Dirac quantisation condition.
 Indeed, consider a   the two-sphere $S^{2}$ surrounding  a 
  non vanishing  magnetic charge   $q_{m}^{(0)}$, 
 and suppose that $F = dA$ everywhere except on
a one dimensional line that starts on the charge and extends to infinity,
 (called the Dirac string). If one considers a path  $\mathcal{P}$ for an electric
 charge  $q_{e}^{(0)}$   surrounding  the line (see fig.  \ref{DiracString}), after a tour one
 would get a phase due to the coupling of $q_{e}^{(0)}$ to the potential
 $A_{\m}$
\beq
 e^{i q_{e}^{(0)} \oint_{\mathcal{P}} A_{\m}dx^{\m}} =  e^{i q_{e}^{(0)} \int_{\mathcal{D}} dA_{\m}}. 
\eeq
 For the previous relation we have used Stockes theorem and $\mathcal{D}$ is a portion of two-sphere whose  boundary
 is  $\mathcal{P}$. Now if $\mathcal{P}$ shrinks to zero, $\mathcal{D}$ becomes a two-sphere
 and since the Dirac string must invisible, the phase needs to be equal to one
\beq
 1 =  e^{i q_{e}^{(0)} \int_{\mathcal{S}^{2}} dA } =  e^{i q_{e}^{(0)}  q_{m}^{(0)}}.
\eeq
Therefore we have the Dirac quantisation condition
\beq
 q_{e}^{(0)}  q_{m}^{(0)} = 2 \pi n \qquad n \in \mathbb{Z}, \label{Diracquantpoint} 
\eeq
which states 
 that whenever
 electric charged particles are weakly coupled,
 then the magnetic monopoles have an high charge  $q_{m}^{(0)} = 2 \pi n / q_{e}^{(0)}$
and therefore  represent non-perturbative objects that can be described as
topologically non-trivial classical solutions of the
 equations of motion, called solitons.

\vspace{.1 cm}
In $D = 10$ a similar phenomenon occurs
 for  the extended objects that are sources for  the RR p-forms,
with the  difference that  electric and magnetic duals objects generally
have a different number of dimensions.

For example we have seen that type IIA contains $C_{1}$, $C_{3}$ and 
 and their respective (fieldstrength) Hodge duals $C_{5}$, $C_{7}$.
If   $C_{1}$, $C_{3}$ couple electrically
  with respectively a  0-brane,   2-brane,
 $C_{7}$, $C_{5}$  couple magnetically to their dual 
   6-branes, 4-branes. 
In a regime where the electric coupling is weak, the  6-branes, and  4-branes 
 are thus  non-perturbative objects that can wrap some of the compact directions
 of ten-dimensional spacetime, while invading the extended directions,
 so that they do not break Poincar\`e invariance of the uncompactified spacetime.

The presence of p-branes  therefore enrich the possibilities for string background,
in fact for some configurations, they can preserve some of the  supersymmetries
  of the original theory on the vacuum,  moreover these solitons are indeed Dirichlet p-branes
 (D-p branes) hypersurfaces where closed strings can open up \cite{Polchinski:1995mt},
and therefore their quantum excitations can be described by open strings
with end-points  confined on their world-volumes.
 
 Backgrounds with the presence of D-p branes are of main interest for the present discussion.
Together with other objects (orientifold planes),
 they allow to find consistent solutions for the string
equations of motion  with an exact world-sheet description, 
involving closed and open strings.
 One of the  focus in the following will be the study of possible mechanisms for
supersymmetry breaking that can be suggested by the presence of D-branes and
 orientifold planes. 

\vspace{.5 cm}
We want to finish this section by analysing   the relation between  $\mathcal{N} =(1,1)$
superconformal symmetry of the fermionic string Action (\ref{scgAction})
and the extended algebras  $\mathcal{N}=(2,2)$    described in a more abstract
 way in the previous section.
We have seen that by  choosing a lightcone gauge (\ref{sstlightcone})
 one eliminates the longitudinal oscillators from the dynamical
 degrees of freedom.
 If one picks up  the superconformal gauge  (\ref{scgauge}), the
 leftover bosonic and fermionic fields represent a system of $2\cdot (D-2)$-free fields.
Actually, an Action for their oscillating degrees of freedom
enjoies  $\mathcal{N}=(2,2)$ superconformal symmetry, as one can check
 by coupling the coordinates in complex pairs
  $X^{I} = X^{i} + i  X^{i+1}$,  $\l^{I} = \l^{i} + i  \l^{i+1}$ and
  $\bar{\l}^{I} = \bar{\l}^{i} + i  \bar{\l}^{i+1}$ with $i = 1,3,5,7,9$.

The action for the free system of oscillators degrees of freedom  then will look like 
\beq
S =  \frac{1}{2\pi \a'} \int dz d\bar{z} \left( \p X^{I}\bar{\p} X^{I} + i \l^{I}\p \l^{I}
 + i \bar{\l}^{I}\bar{\p} \bar{\l}^{I}   \right),
\eeq 
 where $I =1,2,3,4.$. It is easy then to check that
  the holomorphic $\mathcal{N} =2$ supercurrents

\bea T(z)  &=& - \p X^{I}\p X^{I *} + \frac{1}{2}\psi^{I *}\p \psi^{I}
 + \frac{1}{2}\psi^{I}\p \psi^{I *}, \nn \\
G^{+}(z) &=& \frac{1}{2}\psi^{I *} \p X^{I}, \nn \\
G^{-}(z) &=& \frac{1}{2}\psi^{I} \p X^{I *}, \nn \\
j(z) &=& \frac{1}{4}\psi^{I *}\psi^{I},
\eea
  satisfy the
 OPE  given in (\ref{OPEn=2,1}) and  (\ref{OPEn=2,2}),
 that define the $\mathcal{N}=2$ superconformal symmetry.

We have seen in the previous section that for $\mathcal{N} = (2,2)$ superconformal algebra
  different  choices  of periodicity conditions for the supercurrents
  give rise  to  sets of charges that  generate 
 isomorphic algebras.
 The realisation of this isomorphism, called spectral flow, on the representation
 of the algebras is obtained  by an operator playing on the world-sheet
  the role of a spacetime supersymmetry charge.

 Indeed spacetime supersymmetry is gained after imposing the existence
of a square root brunch cut in the OPE between the spectral flow operator
 and the allowed primary fields of the theory, a condition that translates
 into a projection on their  $U(1)$ charges (generalised GSO projections \cite{Sen:1986mg,Banks:1987cy}).

 We therefore expect 
from the critical $D = 10$ theory    solutions with some ammount of spacetime
supersymmetry, 
by taking sets of fields with different twisted periodicity conditions
more general than the Ramond and the Neveu-Schwarz ones.
 This actually is the case  if a correct $U(1)$ projection on the states is performed.
 The price for introducing these twisted sectors is 
 to break some of the original $SO(1,9)$ spacetime invariance.
 This is quite natural, if one considers  some of the extra-dimensions 
  to be compact, so that the original symmetry breaks to $SO(1,9) \rightarrow SO(1,9 - k) \otimes SO(k)$.
 Since the original $D = 10$ flat spacetime type II solutions possess
 32 supercharges (the number of components of the two ten-dimensional Majorana-Weyl spinors),
  a simple toroidal compactification preserving  all the supercharge
 gives rise to a by far too  large number of supersymmetries.

 Theories  with  $\mathcal{N}=(2,2)$ allow  to introduce different twisted boundary conditions
    and through a generalised GSO can give rise to a  reduced number of supersymmetries
in the compactifications.
 These solutions are called orbifold compactifications \cite{Dixon:1985jw,Dixon:1986jc},
a subject on which   we will return  in chapter five
   by describing  some special examples.

\vspace{2 cm}
\section{The torus partition function}\label{sectiontoro}

 \everypar{\hspace{-.6cm}}

As a further step one can consider   topologies different then  the sphere for the closed  (super)string
 world-sheet.
These two dimensional surfaces are called closed Riemann surfaces and are
classified by their genus $g$ that corresponds  to the
  number $h$  of handles. The Riemann sphere has $g = h = 0$,
    the torus $ g = h = 1$ and so on.
   The Euler characteristic of the  surface $\chi$ is a topological invariant
 \beq
\chi = \frac{1}{8\pi} \int \sqrt{g}R^{(2)} = 2 - 2g,  
\eeq
 with $R^{(2)}$ the scalar curvature of the surface.

\begin{figure}  
\begin{center} 
\includegraphics[scale=1, height=3cm]{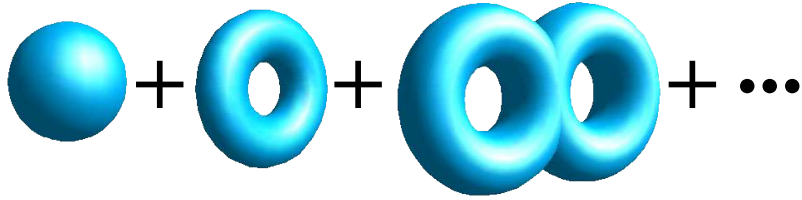}
\caption{The perturbative series of vacuum fluctuations for the closed string is ordered by  
   incrising genus.}                       
\label{vacuumexpansion}
\end{center}
\end{figure}

The interest of genus $g$ string world-sheets follows from the the Polyakov integral \cite{Polyakov:1981rd,Polyakov:1981re}
\beq
\int \mathcal{D}g_{ab} \mathcal{D}X^{\m} e^{-S},
\eeq
that allows to compute  in a perturbative regime  string amplitudes by
    summing over all the  possible  topologies of the world-sheet diagrams (fig.\ref{vacuumexpansion}).

For a constant  dilaton background,  the   dilaton term
 of the world-sheet action
 becomes proportional to the Euler characteristic of the surface
\beq
 \frac{\phi}{8\pi} \int \sqrt{g}R^{(2)} = (2 - 2g)\phi,
\eeq
 and therefore the sum over all the topologies can be expressed as a power expansion
in the string coupling constant $g_{s} = e^{-\phi}$
\beq
\int \mathcal{D}g_{ab} \mathcal{D}X^{\m} e^{-S} = \sum_{g = 0}^{\infty}g_{s}^{2g -2}\int_{\Sigma_{g}} \mathcal{D}g_{ab} \mathcal{D}X^{\m} e^{-S'},\label{gexpansion}
\eeq
 $S'$ being the world-sheet action without the dilaton term.

 For every term  in the above sum, the path integral is now performed only on  surfaces of fixed genus $g$.
 If the string coupling is small,
  eq. (\ref{gexpansion}) represents   a perturbative expansion that gives a diagrammatic definition
for the theory, where Riemann surfaces replace ordinary  Feynman diagrams, with a single 
 string diagram for every perturbative order in $g_{s}$.
 
For every term in the series the integration over the metric involves an integration over
the moduli, complex parameters that describe the shape of the surface.
 The sphere has no moduli, the torus has a complex modulus and a $g \ge 2$   
surface has $3(g - 1)$ complex moduli.

 The moduli are defined on special regions,
 called fundamental regions $\mathcal{F}_{g}$, of  $\mathbb{C}^{n}$  where every
point describes an inequivalent shape of the  surface. The  points outside of
 the region  $\mathbb{C}^{n} - \mathcal{F}_{g}$   are redundant since are connected
to those in $\mathcal{F}_{g}$  via  modular transformations.
   Modular transformations
 form a group $\mathcal{G}_{g}$  called the modular group so that a fundamental region
is  the quotient $\mathcal{F}_{g} = \mathbb{C}^{n}/ \mathcal{G}_{g}$   of the complex plane by the modular group.  

An important point is that modular invariance for $g \ge 1$
 world-sheets  path integrals is an essential requirement for  the
consistency of the
 closed string theory, in the   $g = 1$ torus case modular invariance
 singles out the
 consistent  closed string spectra.
 
 Higher genus  $g > 1$ modular invariance is satisfied by the theory
 once   modular invariance on the torus  and spin-statistics are satisfied,  \cite{Witten:1985mj,Vafa:1986wx,Antoniadis:1985az}.
  In the following of this section  we will therefore focus on the torus vacuum
amplitude.

\begin{figure}  
\begin{center} 
\includegraphics[scale=1, height=3cm]{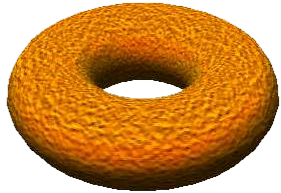}
\caption{ The Torus represents the lowest genus topology
 that gives the lowest order quantum correction to the classical string amplitudes.}
\label{PertTorus}
\end{center}
\end{figure}

\vspace{.3 cm}

\begin{figure}  
\begin{center} 
{\includegraphics[scale=1, height=5cm]{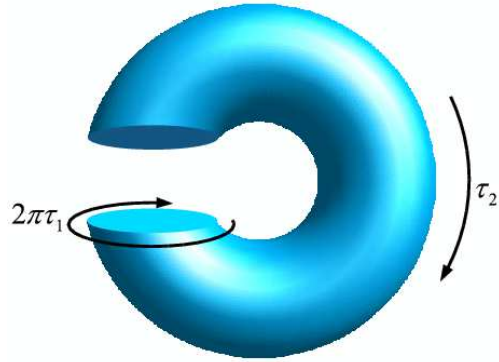}}
\caption{ A torus can be obtained from a cylinder of height 
$Im(\tau)$ by  gluing together
its two   boundaries  after a 
relative twist of an amount  $2\pi Re(\tau)$}
\label{TwistedTorus}
\end{center}
\end{figure}

\begin{figure}  
\begin{center} 
{\includegraphics[scale=1, height=5cm]{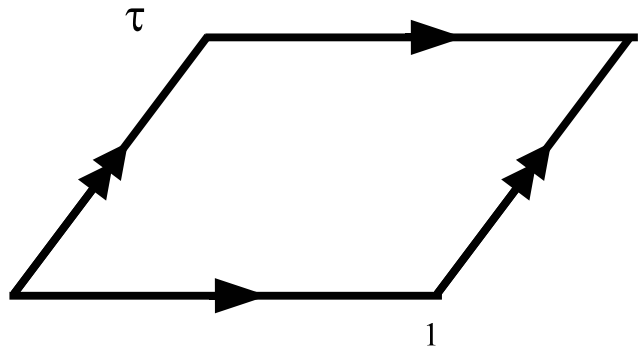}}
\caption{ By cutting a torus along its two homology cycles, on can be represent
 this surface  on a complex plane
 by a parallelogram with opposite sides identified.}
\label{torusident}
\end{center}
\end{figure}

The shape of a torus can be described by a complex number $\tau$.
 This  surface can be obtained from a cylinder of height 
$Im(\tau)$ by  gluing together
its two   boundaries  after a 
relative twist of an amount  $Re(\tau)$ (see  figure  \ref{TwistedTorus}).
 We can cut the surface along the two homology cycles of length
 $Im(\tau)$ and $2\pi$ so that it is represented on a complex plane
 by a parallelogram with opposite sides identified, fig. \ref{torusident}.
 The torus partition function,
that gives the amplitude for  a closed string  to be created by the vacuum
 and  to describe a fluctuation
 with a torus  world-sheet of  shape   $\tau$,
  is given by the following  trace on the closed string  Hilbert space 

\beq
Z = tr_{\mathscr{H}} e ^{-Im(\tau)H +i Re(\tau)P} \label{parttoro}
\eeq

The Hamiltonian is given by the generator of translations along $\tau$
     
\beq 
H  =  2\pi ( L_{0} +  \bar{L}_{0}), 
\eeq
while   
\beq
P =  2\pi( L_{0} -  \bar{L}_{0})
\eeq
  generates    translations along $\s$.

 $e ^{-Im(\tau)H}$ realizes   
 a translation 
 equal to the height $Im(\tau)$ of the cylinder, while
  $e ^{i Re(\tau)P}$ gives the   $Re(\tau)$  relative twist
between the two boundaries 
 and the trace takes care of the
 gluing between the two boundary
 of the cylinder  so that  the initial and final state of the fluctuation
are indeed the same.

With the definition $q = e^{2\pi i\tau}$, the partition function can be rewritten as
\beq
Z = tr_{\mathscr{H}} q ^{L_{0}} \bar{q}^{\bar{L}_{0}}. \label{parttoro2}
\eeq

 In order to obtain the full  vacuum amplitude
 we need to integrate  (\ref{parttoro2}) over
 all the possible shapes \footnote{The origin of the measure $d^{2}\tau/\tau_{2}$ in the torus amplitude  (\ref{parttoro2}) is due to 
the computation of the path integral as a determinant of a kinetic operator.
For example, in the easier case of a scalar free field theory 
 the path integral $\int \mathcal{D}\varphi e^{\int\ \varphi(\Delta - m^{2})\varphi}$
is Gaussian and equals $\sqrt{ det(\Delta - m^{2})}$.
 The vacuum energy is the logaritm of the above quantity
 $\log \sqrt{ det(\Delta - m^{2})} = - \int_{\e}^{\infty} \frac{dt}{t}Tr(e^{-t(\Delta - m^{2})})$,
  $\e$ being  a UV cutoff that
reguralise this quantity.
 Written in this last form, it is clear the analogy with the  expression used for the torus vacuum  amplitude (\ref{torusvacuum}). In going from point particles to  strings    one replaces the circle proper
time $t$ of a vacuum Feynaman diagram  with the torus modular parameter $\tau$.}$\tau$
\beq
 \mathcal{T} = - \int_{\mathcal{F}}
 \frac{d^{2}\tau}{\tau_{2}}
    tr_{\mathscr{H}}  \left( q ^{L_{0}} \bar{q}^{\bar{L}_{0}}\right).\label{torusvacuum}
 \eeq

\begin{figure}  
\begin{center} 
\includegraphics[scale=1, height=5cm]{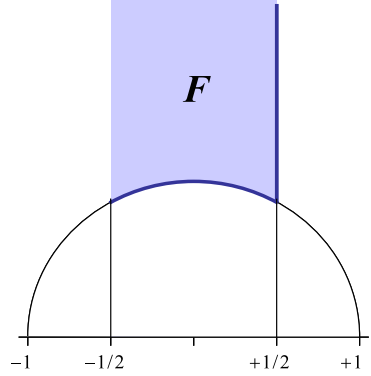}
\caption{Fundamental domain $\mathcal{F} = \mathbb{C} /PSL(2,\mathbb{Z})$ for $\tau$.}                          
\label{FundamentalD}
\end{center}
\end{figure}

The integration domain is given by the fundamental
 region  $\mathcal{F} = \{ \tau \in \mathbb{C} \ \  |\tau| > 1, \  -1/2 < \tau_{1} \le 1/2 \}$
 of the torus modular group, shown in fig. \ref{FundamentalD}. 
 The modular  transformations  are the big diffeomorphisms
on the torus metric that are not connected
to the identity,
 they form a discrete group, whose elements
 are finite sequences of two basic
operations: $T$ and $S$ that leave the shape of the torus unchanged.

$T :   \tau  \rightarrow  \tau + 1$  corresponds
to add a $2  \pi $-twist  between the two boundaries of the cylinder  before
gluing them together, while $S :   \tau  \rightarrow  - 1/ \tau$
exchanges the two homology cycles of the torus. 

 The modular group is isomorphic to
 $PSL(2,\mathbb{Z})$, since a generic sequence of these two basic
 transformations affects the modular parameter as follows
\beq
\tau \rightarrow \frac{a\tau + b}{c\tau + d},
\eeq
with $a,b,c,d$ such that $ad - bc = 1$.
 
 The integral over the modular parameter
 $\tau$ that gives  the full torus amplitude, 
 needs to be evaluated on the fundamental region 
 $\mathcal{F}=  \mathbb{C} /PSL(2,\mathbb{Z})$
 of the complex plane, 
  in order not to  overcount
 the  contributions from different shapes. 
The measure $d\tau_{1}d\tau_{2}/\tau_{2}$ that appears in the torus amplitude
 is not in itself  modular invariant \footnote{The modular invariant measure
 is actually $d\tau_{1}d\tau_{2}/\tau_{2}^{2}$, as one can check directly by
acting with a  a $T$ transformation and a  $S$ transformation.
 As a pure curiosity, by using this proper invariant measure one can compute
 the area of the fundamental region $\mathcal{F}$,  $\int_{\mathcal{F}}d\tau_{1}d\tau_{2}/\tau_{2}^{2} = \pi/3$.  It is less easy then one at first sight might expect. \cite{Lerche:1987qk,Dixon:1990pc}
}, but we will check that
the  complete final result does.

\vspace{.2 cm}
 We want to compute
the torus amplitude for the $D = 26$  bosonic string and then for 
the $D = 10$  closed superstring theories.

Let us start with the bosonic string in the light cone gauge. Expressions for  $L_{0}$ and $\bar{L}_{0}$
  in terms of the transverse oscillators have been given in (\ref{L0bos}),
  and in the critical dimension read
\bea
 L_{0} &=&    \frac{\a'}{2}  p^{2}
 +  \sum_{m = 1}^{\infty}m (a_{m}^{i})^{\dag} a_{m}^{i} -  1,  \nn \\ 
 \bar{L}_{0} &=&   \frac{\a'}{2}  p^{2}
 +  \sum_{m = 1}^{\infty}m (\tilde{a}_{m}^{i})^{\dag} \tilde{a}_{m}^{i} - 1.  \label{L0bosII}  
\eea 
  A generic closed string state has the form 
\beq
(a_{1}^{i_{1} \dag})^{n_1}... (a_{k}^{i_{k} \dag })^{n_k}
(\tilde{a}_{1}^{j_{1} \dag })^{\tilde{n}_{1}}...(\tilde{a}_{k}^{j_{1} \dag})^{\tilde{n}_{k}}|p, 0 \rangle
 \label{bosstate}
\eeq
with 
\beq 
n_1 + .... + n_k = \tilde{n}_1 + .... + \tilde{n}_k
\eeq
to satisfy the level matching condition, where  $n_i$ and  $\tilde{n}_i$ are non negative integers
 as well as $k$, and $p$ is the spacetime momentum of the centre of mass of the bosonic string.

We first take care of the trace over the continuum of spacetime momenta $p$
 in (\ref{torusvacuum}) that needs to be evaluated  over
 a complete  set of momentum eigenstates $\{ |p \rangle \}$

\bea
 \mathcal{T} &=& - \int_{\mathcal{F}}
 \frac{d^{2}\tau}{\tau_{2}^{2}}
    Tr_{\mathscr{H}}  \left( q ^{L_{0}} \bar{q}^{\bar{L}_{0}}  \right) \nn \\ 
&=& -  \int \frac{d^{26}p}{(2\pi)^{26}} e^{-\pi \tau_{2}\a' 2p^{2}} \int_{\mathcal{F}}
\frac{d^{2}\tau}{\tau_{2}^{2}} Tr \left( q^{  \left( \sum m (a_{m}^{i})^{\dag} a_{m}^{i} -  1  \right)}
 \bar{q}^{  \left( \sum (\tilde{a}_{m}^{i})^{\dag} \tilde{a}_{m}^{i} -  1  \right)}\right). \label{bostorus}    \nn
 \eea
The Gaussian integral over the momenta then gives
\beq
\mathcal{T}= -\frac{2^{13}}{(\pi \a')^{13}}       \int_{\mathcal{F}}
\frac{d^{2}\tau}{\tau_{2}^{14}} Tr \left( q^{  \left( \sum m (a_{m}^{i})^{\dag} a_{m}^{i} -  1  \right)}
 \bar{q}^{  \left( \sum (\tilde{a}_{m}^{i})^{\dag} \tilde{a}_{m}^{i} -  1  \right)} \right). \label{bostor}
\eeq
 The trace over the oscillators yields
\beq
 Tr \left[ q^{  \left( \sum_{m = 1}^{\infty} m (a_{m}^{i})^{\dag} a_{m}^{i} -  1  \right)} \right]
 = \prod_{1 = 2}^{26} Tr \left[  q^{  \left( \sum_{m = 1}^{\infty} m (a_{m})^{\dag} a_{m} -  \frac{1}{24}  \right)} \right] = q^{-1} \sum_{k = 0}^{\infty} d_{k} q^{k}, \label{characterbos}
\eeq
 from which one can read the number of  closed string states  $d_{k}$  that have
 a   mass proportional to $k - 1$. In fact the mass shell condition 
 given by $L_{0}|phys \rangle = 0$  is
\beq
m^{2} |phys \rangle = \frac{4}{\a'}\left( \sum_{m = 1}^{\infty}m (a_{m}^{i})^{\dag} a_{m}^{i} -  1  \right),
\eeq
so that the exponent of $q$ in the l.h.s. of (\ref{characterbos})
 is proportional  to the masses of the states. 
  The coefficients
 in front of the  various powers of $q$  precisely count the degeneration
of every mass level, i.e. the number of closed string states that have a given mass.

Now, in order to obtain the degeneracy coefficients it  is enough to think at the structure
of the holomorphic sector of the bosonic   Fock  space. The generic closed 
 string excitation (\ref{bosstate}) has the form $|n_1 \rangle \otimes....\otimes|n_k \rangle \otimes...$
 where $n_{i}$ counts the number of bosonic excitation for the mode $i$,
 whose mass 
 is proportional to $m^{2} \sim n_1 + 2n_2 + ....+ k n_k + ... - 1$.

 Let us  consider first  the following set of states of the holomorphic
 part of the Fock space 
\beq 
  \mathcal{S}_{1} =  \{ |i \rangle \otimes |0 \rangle  ....\otimes|0 \rangle \otimes..., \qquad i = 0,1,2,... \} \label{seti}
\eeq

  where the first mode has an arbitrary number $i$ of  excitations, while \emph{all} the other modes are not excited. 
If we compute the trace of $q^{L_{0}}$ over  $\mathcal{S}_{1}$
 we obtain:
\beq
 Tr_{\mathcal{S}_{1}}  \ q^{\left(\sum_{m = 1}^{\infty}m (a_{m})^{\dag} a_{m} -  1 \right)} =  q^{-1}(1 +  q + q^{2} + ....+  q^{i} + ...) =  \frac{q^{-1}}{1 - q},
\eeq
while for the subset where only the second mode  can be  excitated 
\beq 
  \mathcal{S}_{2} =     \{ |0 \rangle \otimes |j \rangle  ....\otimes|0 \rangle \otimes..., \qquad j = 0,1,2,... \}
\eeq
 the trace of $q^{L_{0}}$ over $\mathcal{S}_{2}$    gives
\beq
  Tr_{\mathcal{S}_{2}}  \ q^{\left(\sum_{m = 1}^{\infty}m (a_{m})^{\dag} a_{m} -  1 \right)}  =   q^{-1}( 1 + q^{2\cdot 1} + q^{2\cdot 2} + ....+  q^{2\cdot i} + ...) =  \frac{q^{-1}}{1 - q^{2}}.
\eeq
If we consider the subset where both the  the first two modes are excited but not the others
\beq 
  \mathcal{S}_{12} =     \{ |i \rangle \otimes |j \rangle  ....\otimes|0 \rangle \otimes..., \qquad i,j = 0, 1,2,... \} \label{setij}
\eeq
 the trace of $q^{L_{0}}$ over    $\mathcal{S}_{12}$ will   be given by the following product

\bea 
  Tr_{\mathcal{S}_{12}}   \ q^{\left(\sum_{m = 1}^{\infty}m (a_{m})^{\dag} a_{m} -  1 \right)}                      & =&    q^{-1}( 1 + q + q^{2} + ....+  q^{i} + ...)\cdot ( 1 + q^{2\cdot 1} + q^{2\cdot 2} + ....+  q^{2\cdot j} + ...) \nn  \\ & =&  \frac{q^{-1}}{(1 - q)(1 - q^{2})} 
\eea

because once we compute the product we saturate all the
 possibilities of exciting independently the first two modes.
Therefore the trace over the full holomorphic sector
 where all the infinite tower of modes can be independently excited is given
  by the following infinite product
\beq
 Tr    \ q^{\left(\sum_{m = 1}^{\infty}m (a_{m})^{\dag} a_{m} -  1 \right)}  =  \frac{q^{-1}}{(1 - q)(1 - q^{2})...(1 - q^{k})...} =  
\frac{q^{-1}}{\prod_{n = 1}^{\infty}(1 - q^{n})}.
\eeq
If one expands the above product up to some order in $q$,
  $d_{0} + d_{1}q + ... + d_{k}q^{k}$,  the coefficients $d_{i}$
 will correspond to  number of  states
 that have a mass equal to the corresponding power of $q$.

 Finally, when we compute the trace over the full Fock space of the bosonic string 
 we have to remember that we have 24 identical copies of the Fock space,
one for each transverse coordinate, so that

\bea
 Tr \left[ q^{  \left( \sum_{m = 1}^{\infty} m (a_{m}^{i})^{\dag} a_{m}^{i} -  1  \right)} \right]
 = \prod_{1 = 2}^{26} Tr \left[  q^{  \left( \sum_{m = 1}^{\infty} m (a_{m})^{\dag} a_{m} -  \frac{1}{24}  \right)} \right] \nn \\
= \left(\frac{1}{q^{1/24} \prod_{n = 1}^{\infty}(1 - q^{n})} \right)^{24} = \left(\frac{1}{\eta(q)}\right)^{24},\eea
where $\eta(q)$ is the Dedekind eta function.

Altogether  the torus amplitude in (\ref{bostor}) reads
 
\bea
\mathcal{T}= -\frac{2^{13}}{(\pi \a')^{13}}       \int_{\mathcal{F}}
\frac{d^{2}\tau}{\tau_{2}^{14}} Tr \left( q^{  \left( \sum m (a_{m}^{i})^{\dag} a_{m}^{i} -  1  \right)}
 \bar{q}^{  \left( \sum (\tilde{a}_{m}^{i})^{\dag} \tilde{a}_{m}^{i} -  1  \right)} \right)\\
=   -\frac{2^{13}}{(\pi \a')^{13}}       \int_{\mathcal{F}}\frac{d^{2}\tau}{\tau_{2}^{14}}
 \frac{1}{(\eta(q)\eta(\bar{q}))^{24}} =   -\frac{1}{(\pi \a')^{13}}       \int_{\mathcal{F}}\frac{d^{2}\tau}{\tau_{2}^{2}}
 \frac{1}{ \tau_{2}^{12}(\eta(q)\eta(\bar{q}))^{24}}.
\eea

  One can check the   modular invariance of the torus vacuum amplitude
by using the modular properties of the Dedekind function
\beq
T: \ \eta(\tau + 1) = e^{i\pi/12}\eta(\tau), \qquad S: \  \eta(- 1/\tau) = \sqrt{-i\tau}\ \eta(\tau).
\eeq  

The measure  $d^{2}\tau / \tau_{2}^{2}$ is by itself  modular invariant,
 while for the rest of the integrand function
\bea
T:  \ \ \frac{1}{ \tau_{2}^{12}(\eta(\tau + 1)\eta(\bar{\tau} + 1 ))^{24}} = 
 \frac{1}{ \tau_{2}^{12}(\eta(\tau)\eta(\bar{\tau}))^{24}} \nn \\
S: \ \  \frac{|\tau|^{24}}{ \tau_{2}^{12}(\eta(- 1/\tau   )\eta( - 1/\bar{\tau} ))^{24}} =
   \frac{1}{ \tau_{2}^{12}(\eta(\tau)\eta(\bar{\tau}))^{24}}.
\eea
thus showing the invariance of the integrand  under a generic modular
 transformation.

\vspace{.5 cm}

The torus partition function  corresponds to the torus vacuum amplitude with
 the omission of the integration in $\tau$ and the normalization coefficient
\beq
Z =  \frac{1}{(\eta(q)\eta(\bar{q}))^{24}} = \frac{1}{q\bar{q} \prod_{n = 1}^{\infty}(1 - q^{n})^{24} \prod_{n = 1}^{\infty}(1 - \bar{q}^{n})^{24}},
\eeq
by expanding the infinite products we can read the number of states corresponding to a given mass
 as the coefficient in front of the powers $(q\bar{q})^{k}$.
 Since the background for the closed string where we have computed the spectrum is 
$D = 26$ Minkowski spacetime we expect that every string state carry a representation
 of $SO(1,25)$. In particular since we are working in the light cone gauge we expect
to read from the partition function only the physical degrees of freedom, which
implies that massless states should carry a representation of $SO(24)$, while massive states 
should come in representation of $SO(25)$.

Up to massive states the expansion gives
\beq
Z = \frac{1}{q\bar{q} \prod_{n = 1}^{\infty}(1 - q^{n})^{24} \prod_{n = 1}^{\infty}(1 - \bar{q}^{n})^{24}},
  = (q\bar{q})^{-1}(1 + 24^{2}q\bar{q} + ...) = (q\bar{q})^{-1} +  24^{2}(q\bar{q})^{0} + O(q\bar{q})
\eeq
the first term $(q\bar{q})^{-1}$ corresponds to one degree of freedom (its coefficient) i.e. a Lorentz scalar
 with a negative mass (its exponent). This state is the infamous closed string tachyon.
 The second term  $24^{2}(q\bar{q})^{0}$ indicates  $24^{2} = 576$ massless degrees of freedom.
 These are the massless states in the bosonic string spectrum previously described
that carry a representation of the massless little group $SO(24)$ of the $D =26$ Lorentz group.
 The decomposition in irreducible representation gives
\beq
 24^{2} =  \left(\frac{24\cdot 25}{2} - 1 \right) +  \left(\frac{24\cdot 23}{2}\right) + (1)
 = (299) + (276) + (1)
\eeq

the traceless symmetric part $(299)$ of the two tensor corresponds to the physical degrees of freedom
of the metric $G_{\m\n}$, the antisymmetric part $(276)$  to those of the $B_{\m\n}$
 and the trace part $(1)$  to the dilaton $\phi$.

\vspace{1 cm}

\begin{figure}  
\begin{center} 
\includegraphics[scale=1, height=5cm]{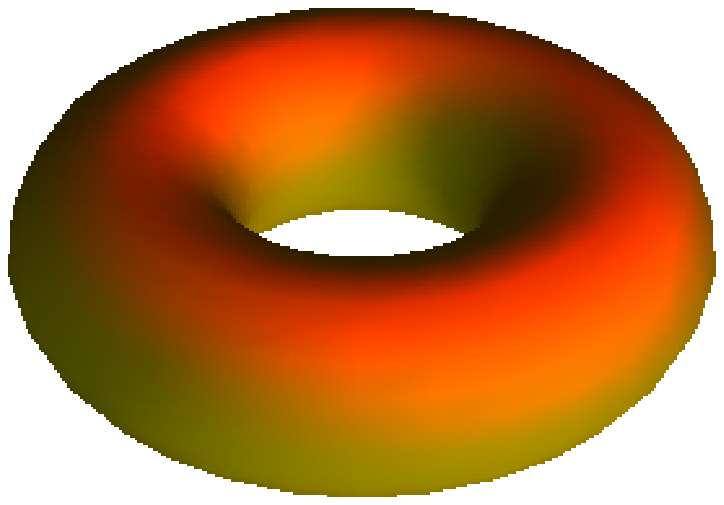}
\caption{}                          
\label{RedTorus}
\end{center}
\end{figure}

 Now  we turn to consider the $D = 10$ closed string theories and compute their torus partition
functions.
We will compute here the $L_{0}$ traces in the diverse sectors,
 introduce the notion of characters and show how consistent superstring theories
can be obtained by imposing modular invariance.
 As before, the general form for the torus  amplitude is 
\beq
 \mathcal{T} = - \int_{\mathcal{F}}
 \frac{d^{2}\tau}{\tau_{2}}
    tr_{\mathscr{H}}  \left( q ^{L_{0}} \bar{q}^{\bar{L}_{0}}\right). \label{tor}
 \eeq
Where now we have  to consider separately  the NS sector, where $L_{0}$ reads
\beq
L_{0} =   \frac{\a'}{4}p^{2} +  \sum_{m = 1}^{\infty} m (a^{i}_{m})^{\dag} a^{i}_{m}
 +  \sum_{r = 1/2}^{\infty}  r  ( b^{i}_{r})^{\dag} b^{i}_{r} 
  -  \frac{1}{2}, \label{L0NS}
\eeq
and the R sector, where 
\beq
L_{0} =   \frac{\a'}{4}p^{2} +  \sum_{m = 1}^{\infty} (a^{i}_{m})^{\dag} a^{i}_{m}
 +  \sum_{r = 1}^{\infty}  r   (d^{i}_{r})^{\dag} d^{i}_{r}.  \label{L0R}
\eeq

 As for the bosonic string,    the trace over the ten-dimensional 
momentum    gives a Gaussian integral,

\bea
 \mathcal{T} &=& - \int_{\mathcal{F}}
 \frac{d^{2}\tau}{\tau_{2}}
    Tr_{\mathscr{H}}  \left( q ^{L_{0}} \bar{q}^{\bar{L}_{0}}  \right) \nn \\ 
&=& -  \int \frac{d^{10}p}{(2\pi)^{10}} e^{-\pi \tau_{2}\a'p^{2}/2} \int_{\mathcal{F}}
\frac{d^{2}\tau}{\tau_{2}} Tr \left( q^{L_{0}}
 \bar{q}^{ \bar{L}_{0}}\right)     \nn \\
 &=&  -\frac{2^{5}}{(\pi \a')^{5}}       \int_{\mathcal{F}}
\frac{d^{2}\tau}{\tau_{2}^{6}}  Tr \left( q^{L_{0}}
 \bar{q}^{ \bar{L}_{0}} \right),\label{supertoro}
\eea
 where   $L_{0}$   does not contain anymore the momentum $p$.


In calculating the contribution from the world-sheet fermions in the NS sector
we need to  take into account the Pauli exclusion principle, so that every
  world-sheet fermionic  mode can be   excited at most once.
 For example, the subset  $\mathcal{S}_{1}$  of the  holomorphic Fock space
 that we were considering in (\ref{seti})
  contains in this case only two states

\beq 
 \mathcal{S}_{1} =  \{ |i \rangle \otimes |0 \rangle  ....\otimes|0 \rangle \otimes..., \qquad i = 0,1 \}.
\eeq

 The first mode can be excited ($i =1$) or not ($i =0$), while \emph{all} the other modes are not. 
 The trace of $q^{L_{0}}$ over  $\mathcal{S}_{1}$, with $L_{0}$ given by  (\ref{L0NS}), then yields
\beq
  Tr_{\mathcal{S}_{1}}  \ q^{ \left( \sum_{r = 1/2}^{\infty}  r  ( b_{r})^{\dag} b_{r} 
  -  \frac{1}{2}\right)} =   q^{-1/2}(1 +  q^{1/2}),
\eeq

 For  the subset  
\beq 
 \mathcal{S}_{12} =      \{ |i \rangle \otimes |j \rangle \otimes|0 \rangle \otimes..., \qquad i,j = 0, 1 \} 
\eeq
  that contains four states,  since the first two modes can be independently excited  but not the others,
the trace of $q^{L_{0}}$ gives
\beq 
  Tr_{\mathcal{S}_{12}}  \ q^{ \left( \sum_{r = 1/2}^{\infty}  r  ( b_{r})^{\dag} b_{r} 
  -  \frac{1}{2}\right)}   =     q^{-1/2}( 1 + q^{1/2})( 1 + q). 
\eeq

 Iterating, the full trace of $q^{L_{0}}$ in the NS sector  reads
 
\bea
 Tr_{NS} \  q^{L_{0}} &=&  Tr_{NS} \  q^{  ( \sum_{m = 1}^{\infty} m (a^{i}_{m})^{\dag} a^{i}_{m}
 +  \sum_{r = 1/2}^{\infty}  r  ( b^{i}_{r})^{\dag} b^{i}_{r} 
  -  \frac{1}{2})}  \nn  \\
 &=& q^{-1/2} \prod_{i = 2}^{10} Tr \  q^{  ( \sum_{m = 1}^{\infty} m (a_{m})^{\dag} a_{m}
 +  \sum_{r = 1/2}^{\infty}  r  ( b_{r})^{\dag} b_{r})}\nn \\
 &=& \frac{q^{-1/2}\prod_{n}^{\infty}(1 + q^{n - 1/2})^{8}}{\prod_{n}^{\infty}(1 - q^{n})^{8}}
 = \frac{1}{\eta^{8}}\left(\frac{\theta_{3}}{\eta} \right)^{4}, \label{trNS}
\eea
 where  we have included also the contributions from
 the eight transverse bosonic excitations,
and in the last equality 
 $\theta_{3}$ is one of the Jacobi theta constants,  defined by
\beq
\theta_{3} = \prod_{n}^{\infty}(1 + q^{n - 1/2})^{2}(1 - q^{n}).\label{theta3}
\eeq

In computing the trace  in (\ref{trNS})  we need to give  boundary conditions
 to the spinor fields along the 
 two cycles $\s \sim \s + 2\pi$ and  $t \sim t + 2\pi$
 of the worlsheet torus. 
In the NS sector  these fields are anti-periodic along the $\s$-cycle,
while a single choice between periodic and anti-periodic boundary 
   conditions along the $t$-cycle  is inconsistent with  modular invariance.
To see this  one can consider    $t$-antiperiodic NS fermions for definiteness, 
 then (\ref{trNS}) is invariant under an $S$ modular transformation that
exchanges the two cycles, since the fields have the same boundary conditions.
A $T$ modular transformation adds a $2\pi$ twist to $Re(\tau)$, the real part 
of the torus modulus,
 so that the fields now get a further minus sign in moving along the 
 $t$ cycle and therefore   (\ref{trNS}) is not $T$ invariant.  
The  $T$ transformation of  (\ref{trNS}) is equivalent to the choice
of \emph{perodic} boundary conditons along the $t$-cycle, 
 the alternative choice respect to the one we were considering
\beq
 Tr_{NS} \  q^{L_{0}} \stackrel{T}{\leftrightarrow}  Tr_{NS} \ (-)^{G} \  q^{L_{0}}
\eeq
 where the insertion of the operator  $(-)^{G}$ in the trace exchange
 the fermionic  boundary conditions along the $t$-cycle.

  An $S$ transformation on the r.h.s in
the previous relation
 interchanges the $\s$ with the $t$ cycles,
 the resulting expression results in a   trace  computed with  $\s$-\emph{periodic}
   and $t$-\emph{antiperiodic} boundary conditions, that corresponds
to the Ramond sector.
\beq 
 Tr_{NS} \ (-)^{G} \  q^{L_{0}}
  \stackrel{S}{\leftrightarrow}  Tr_{R} \  q^{L_{0}}. 
\eeq

\begin{figure}  
\begin{center} 
\includegraphics[scale=.5, height=8cm]{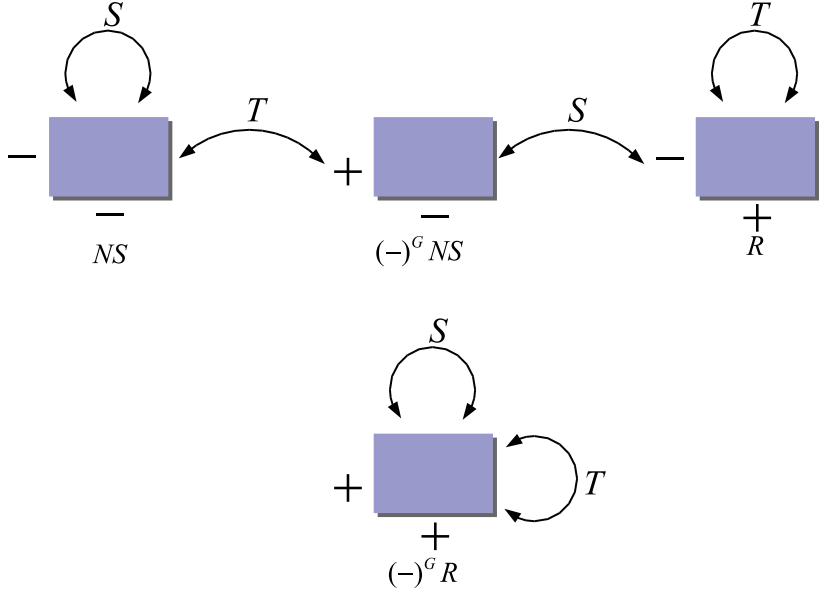}
\caption{The fermionic spin structures on the torus are organised in two disjoint modular orbits.
 An insertion of the world-sheet holomorphic fermion index $(-)^{G}$ in the trace flips the periodicity
conditions along the time direction (vertical side of the torus).} 
\label{spinstructures}
\end{center}
\end{figure}

To summarise, a closed modular orbit is given by, (see fig. \ref{spinstructures})  
\beq
 Tr_{NS} \  q^{L_{0}} \stackrel{T}{\leftrightarrow}  Tr_{NS} \ (-)^{G} \  q^{L_{0}}
  \stackrel{S}{\leftrightarrow}  Tr_{R} \  q^{L_{0}}, \label{orbitI} 
\eeq
and therefore all the above sectors need to be considered in order to construct
modular invariant amplitudes.
There is a  fourth possibility in prescribing  the boundary
 conditions that we did not considered yet. It corresponds  to periodicity along both  cycles
\beq
 Tr_{R} \ (-)^{G} \  q^{L_{0}}.
\eeq
This last quantity is modular invariant in itself, therefore it represents a closed modular
orbit, disjoint from (\ref{orbitI}).

\hspace{.5 cm}

We now compute the trace  in the  NS  sector  with the  operator $(-)^{G}$ inserted.
Notice  that $(-)^{G}$ simply gives a minus sign to states with an odd number
 of $b$ excitations, so that  
\bea
 Tr_{NS} \  \left( (-)^{G} \  q^{( \sum_{m = 1}^{\infty} m (a^{i}_{m})^{\dag} a^{i}_{m}
 +  \sum_{r = 1/2}^{\infty}  r  ( b^{i}_{r})^{\dag} b^{i}_{r} 
  -  \frac{1}{2})}    \right)
&=& \frac{q^{-1/2}\prod_{n}^{\infty}(1 - q^{n - 1/2})^{8}}{\prod_{n}^{\infty}(1 - q^{n})^{8}} \nn \\
&=& \frac{1}{\eta^{8}}\left(\frac{\theta_{4}}{\eta} \right)^{4},
\eea
where $\theta_{4}$, another Jacobi theta constant, is defined by
 
\beq
\theta_{4} = \prod_{n}^{\infty}(1 - q^{n - 1/2})^{2}(1 - q^{n}).\label{theta4}
\eeq
 
With the final goal of obtaining modular invariant torus  amplitudes, we are
therefore led to consider the following two choices for the NS sector

 \bea
 Tr_{NS} \left( \frac{1 \pm (-)^{G}}{2} \ q^{L_{0}} \right)  
 &=& \frac{q^{-1/2}}{\prod_{n}^{\infty}(1 - q^{n})^{8}}\left( \frac{\prod_{n}^{\infty}(1 + q^{n - 1/2})^{8} 
 \pm \prod_{n}^{\infty}(1 - q^{n - 1/2})^{8}}{2} \right) \nn \\
 &=& \frac{1}{\eta^{8}} \frac{\theta_{4}^{4} \pm \theta_{3}^{4}}{2\eta^{4}}, 
\eea
  which single out the two subsets $NS_{-}$ and $NS_{+}$ of the $NS$ sector.


The lowest state in   $NS_{-}$ is a massless vector and all the other states are massive tensors,
while the lowest state in $NS_{+}$ is the tachyonic scalar and all the other states are
massive tensors. With this in mind, we define the following characters
that correspond to the contribution form holomorphic spinor oscillators
 for the two different GSO projections  \cite{Bianchi:1990yu,Angelantonj:2002ct}
\bea
V_8 &=&  \frac{\theta_{4}^{4} - \theta_{3}^{4}}{2\eta^{4}}, \nn \\
O_8 &=&  \frac{\theta_{4}^{4} + \theta_{3}^{4}}{2\eta^{4}}.
\eea
Their names recall the lowest mass state that they contain, $V$ for the vector and $O$ for the scalar.  

 Turning to the R sector, let us start with
 
\bea
 Tr_{R} \  q^{L_{0}} &=&  Tr_{R} \  q^{  ( \sum_{m = 1}^{\infty} m (a^{i}_{m})^{\dag} a^{i}_{m}
 +  \sum_{r = 1}^{\infty}  r  ( d^{i}_{r})^{\dag} d^{i}_{r})}  \nn  \\
 &=& 2^{4} \prod_{i = 2}^{10} Tr \  q^{  ( \sum_{m = 1}^{\infty} m (a_{m})^{\dag} a_{m}
 +  \sum_{r = 1}^{\infty}  r  ( b_{r})^{\dag} b_{r})}\nn \\
 &=&  2^{4} \frac{\prod_{n}^{\infty}(1 + q^{n})^{8}}{\prod_{n}^{\infty}(1 - q^{n})^{8}}
 = \frac{1}{\eta^{8}}\left(\frac{\theta_{2}}{\eta} \right)^{4}
\eea
 where one has  to take into account that the holomorphic R ground state is a 
ten dimensional massless   spinor, whose physical degrees of freedom
 correspond to the 16 components of an $SO(8)$ spinor,  while 
  the  Jacobi  $\theta_{2}$ constant 
   is defined by
\beq
\theta_{2} = 2q^{1/8} \prod_{n}^{\infty}(1 + q^{n})^{2}(1 - q^{n}).\label{theta2}
\eeq
We can consider two different GSO projections for the  Ramond sector
with the insertion of the  operator $(-)^{G}\Gamma_{9}$ that selects one of the two chiralities
for the spinors.

If one computes  the trace over the Ramond sector with the  $(-)^{G}\Gamma_{9}$
inserted, one finds
\beq
 Tr_{R} \ \left( (-)^{G}\Gamma_{9} q^{L_{0}} \right) =
  \frac{1}{\eta^{8}}\left(\frac{\theta_{1}}{\eta} \right)^{4} = 0. \label{theta1}
\eeq
 This  is due to the equal number of  opposite chirality states at every mass level.
 We define the following characters
that correspond to the contribution form holomorphic Ramond fermionic  oscillators
 for the two different spacetime chiralities  \cite{Bianchi:1990yu,Angelantonj:2002ct}
\bea
S_8 &=&  \frac{\theta_{2}^{4} - \theta_{1}^{4}}{2\eta^{4}}, \nn \\
C_8 &=&  \frac{\theta_{2}^{4} + \theta_{1}^{4}}{2\eta^{4}}.
\eea

 So far we have introduced 
 a basis of holomorphic characters that correspond
 to   combinations of the two possible twists
 along  the $\s$-cycle (R and NS) and
along the $\tau$-cycle ($(-)^{G}$R and $(-)^{G}$NS)

\bea
O_8 (q) &=& tr_{NS} \left(\frac{1+(-)^{G}}{2}\right)q^{ L_{0}}, \qquad
V_8 (q) = tr_{NS} \left(\frac{1-(-)^{G}}{2}\right)q^{ L_{0}}, \nn \\  
C_8 (q) &=& tr_{R} \left(\frac{1+ \Gamma_{9}(-)^{G}}{2}\right)q^{ L_{0}}, \qquad
S_8 (q) = tr_{R} \left(\frac{1-  \Gamma_{9}(-)^{G}}{2}\right)q^{ L_{0}}. \label{GSOcharact}   
\eea

 These projections, from the spacetime point of view,
 single out the conjugacy classes
 for the  $SO(8)$  little group.
 Indeed these classes are the scalar  $O$, the vector $V$,
 and the two  spinors $S$, and $C$  with opposite chirality.

  The explicit computation of the traces in the above formulae has shown
 that the SO(8) characters can be written in terms
 of the four Jacobi theta constants and the Dedekind eta function 
\bea
O_8  &=&  \frac{\theta^{4}_{3} + \theta^{4}_{4}}{2 \eta^{4}},
\qquad V_8  =  \frac{\theta^{4}_{3} - \theta^{4}_{4}}{2 \eta^{4}}     \nn  \\  
C_8  &=&   \frac{\theta^{4}_{2} + \theta^{4}_{1}}{2 \eta^{4}}, 
\qquad S_8  =   \frac{\theta^{4}_{2} - \theta^{4}_{1}}{2 \eta^{4}}. \label{so8}  
\eea

From the modular properties of the theta constants
 it    can be derived that  a $T$ modular transformation 
acts on the  column vector $(O_8, V_8,  S_8, C_8)$
of  (holomorphic) characters with the  matrix
\beq
T = diag(-1,1,1,1) \label{interniT},
\eeq
while an $S$ modular  transformation acts as

\bea
 \mathcal{S} = 
\frac{1}{2} \left( \begin{array}{cccc} 
1 & 1 & 1 & 1  \\ 
1 & 1 & -1 & -1  \\ 
1 & -1 & 1 & -1  \\
1 & -1 & -1 & 1 \end{array} \right). \label{interniS}    
\eea
Actually there is a subtlety that one needs to 
take into account in order to obtain  the proper modular transformation
properties of the \emph{spacetime} characters.
 Due to the opposite statistic,
 spacetime bosonic and fermionic excitations
 contribute with an opposite sign to the
 vacuum energy.
This can be taken into account by giving an opposite sign
 to the holomorphic spacetime fermionic characters
 $(O_{8},V_{8}, - S_{8}, -C_{8})$ (and in the anti-holomorphic sector as well).
 In this case spacetime
 closed string bosonic and fermionic states that are
 encoded in products of an holomorphic and one anti-holomorphic
characters have opposite sign.

 
Moreover, we need  to ensure that  all  the interactions  respect the
GSO projections stated in (\ref{GSOcharact})
  via  conditions on the
 fusion rules coefficients $\mathcal{N}_{h k}^{i}$
\beq
 [f_{h}] \times  [f_{k}] = \sum_{h k} \mathcal{N}_{h k}^{i} [f_{i}]. \label{fusion} 
 \eeq
 
 Those  encode  superselection rules
 coming from the OPEs between fields belonging to 
 different conformal families  $[f_{h}]$ and  $[f_{k}]$.
 The fusion rules coefficients $\mathcal{N}_{h k}^{i}$ 
 are connected to the elements of the $S$
 modular matrix that acts on the characters
 via the Verlinde formula \cite{Verlinde:1988sn}
\beq
{N}_{h k}^{i} = \sum_{l} \frac{S_{i}^{l} S_{j}^{l}  S_{l}^{k \dag}}{S_{1}^{l}}.
\eeq

 It can be proved \cite{Englert:1986na} that a
 proper account for the spin statistic
 and  the request that all the interactions  respect the
GSO projections stated in (\ref{GSOcharact}),
 interchange the roles of $O_8$ and $V_8$
 in the $S$ and $T$ matrixes.
 Therefore for the $SO(8)$ \emph{spacetime} characters $(O_8 , V_8 , - S_8 , - C_8 )$
   the correct form for the $S$  matrix is 

\bea
 \mathcal{S} = 
\frac{1}{2} \left( \begin{array}{cccc} 
1 & 1 & -1 & -1  \\ 
1 & 1 & 1 & 1  \\ 
1 & -1 & 1 & -1  \\
1 & -1 & -1 & 1 \end{array} \right). \label{spacetimeS}    
\eea
 
This is indeed a crucial point to keep in mind in order to recover 
 all the possible ten dimensional
 closed string modular invariant  partition functions.

\vspace{.5 cm}
 Knowing  the modular 
 properties of the characters
 we can search for  the possible modular invariant sesquilinear combinations
 of characters. 
 Actually there are four  possible choices, 
 the  Type IIA and Type IIB combinations \cite{Dixon:1986iz,Seiberg:1986by}
\bea
Z_{IIA} &=& (V_8 - S_8)(\bar{V}_8 - \bar{C}_8) = V_8 \bar{V}_8 +   S_8 \bar{C}_8 - V_8 \bar{C}_8 - \bar{V}_8 S_8, \nn \\
Z_{IIB} &=& (V_8 - S_8)(\bar{V}_8 - \bar{S}_8) = V_8 \bar{V}_8 +   S_8 \bar{S}_8 - V_8 \bar{S}_8 - \bar{V}_8 S_8, \label{IIaIIb}
\eea
 and  the  Type 0A and Type 0B combinations \cite{Bianchi:1990yu}
 \bea
Z_{0A} &=& O_8 \bar{O}_8 +    V_8 \bar{V}_8 +   S_8 \bar{C}_8 + C_8 \bar{S}_8,  \nn \\
Z_{0B} &=&  O_8 \bar{O}_8 +    V_8 \bar{V}_8 +   S_8 \bar{S}_8 + C_8 \bar{C}_8  \label{0a0b}.
\eea

In the above formulae we have displayed only the contributions from world-sheet fermions excitations.
The complete torus amplitudes including  
 also  the contributions from the transverse
bosonic oscillators and the integration over the torus modular parameter are given by
(see \ref{supertoro})
\bea
\mathcal{T}_{IIA} =  -\frac{2^{5}}{(\pi \a')^{5}}   \int_{\mathcal{F}}
\frac{d^{2}\tau}{\tau_{2}^{6}}\frac{1}{(\eta(q)\eta(\bar{q}))^{8}} 
(V_8 - S_8)(q)(V_8 - C_8)(\bar{q}) \nn \\
 = -\frac{2^{5}}{(\pi \a')^{5}}   \int_{\mathcal{F}}
\frac{d^{2}\tau}{\tau_{2}^{6}}\frac{1}{4(\eta(q)\eta(\bar{q}))^{12}} 
(\theta_{3} - \theta_{4} - \theta_{2} + \theta_{1})(q)
(\theta_{3} - \theta_{4} - \theta_{2} - \theta_{1})(\bar{q}), \nn 
\eea
and
\bea
\mathcal{T}_{IIB} =  -\frac{2^{5}}{(\pi \a')^{5}}   \int_{\mathcal{F}}
\frac{d^{2}\tau}{\tau_{2}^{6}}\frac{1}{(\eta(q)\eta(\bar{q}))^{8}} 
(V_8 - S_8)(q)(V_8 - S_8)(\bar{q}) \nn \\
 = -\frac{2^{5}}{(\pi \a')^{5}}   \int_{\mathcal{F}}
\frac{d^{2}\tau}{\tau_{2}^{6}}\frac{1}{4(\eta(q)\eta(\bar{q}))^{12}} 
(\theta_{3} - \theta_{4} - \theta_{2} + \theta_{1})(q)
(\theta_{3} - \theta_{4} - \theta_{2} + \theta_{1})(\bar{q}) \nn 
\eea
for the supersymmetric closed string theories and similar expression for the type 0 theories.
A necessary conditions for supersymmetry on Minkowski space is that  at
every mass level  there is an  equal number of bosonic and fermionic 
degrees of freedom. Since spacetime bosons and fermions contribute with  opposite signs
 in a loop diagram, the supersymmetric vacuum amplitudes $\mathcal{T}_{IIA}$ and $\mathcal{T}_{IIB}$ 
 are expected to vanish.
Indeed they vanish   as a consequence of the Abstrusa Jacobi Aequatio \cite{jacobi}
\beq
 \theta_{3}^{4} - \theta_{4}^{4} - \theta_{2}^{4} = 0, \label{Abstrusa}
\eeq
 together with $\theta_{1} = 0$. 
Let us  extract now 
 the massless spectrum of  Type IIA and Type IIB superstrings,
by expanding the characters in their corresponding partition functions.
  Since the excitations of the  transverse
bosons oscillators  are always massive, we use the lighter formulae
that involve only contributions from world-sheet spinor oscillators and expand their
characters.
 We start with type IIA,  where  the holomorphic and anti-holomorphic Ramond
 sectors  have \emph{opposite} GSO projections
\beq
Z_{IIA} = (V_8 - S_8)(\bar{V}_8 - \bar{C}_8) = V_8 \bar{V}_8 +   S_8 \bar{C}_8 - V_8 \bar{C}_8 - S_{8}\bar{V}_8. \label{IIA}
\eeq
 We have  the bosonic sectors  $NS_{-}NS_{-}$ and   $R_{-}R_{+}$     encoded in $V_8(q) V_8(\bar{q})$,
 and     $S_8(q) C_8(\bar{q})$, and the fermionic sectors
  $NS_{-}R_{+}$  and   $NS_{-}R_{-}$, encoded in    $- V_8(q) C_8(\bar{q})$
   and  $ -   S_8(q) V_8(\bar{q})$.
 
From the expansion of the $NS_{-}$ character

 \bea
 Tr_{NS} \left( \frac{1 - (-)^{G}}{2} \ q^{L_{0}} \right)  
 &=& \frac{q^{-1/2}}{\prod_{n}^{\infty}(1 - q^{n})^{8}}\left( \frac{\prod_{n}^{\infty}(1 + q^{n - 1/2})^{8} 
 \pm \prod_{n}^{\infty}(1 - q^{n - 1/2})^{8}}{2} \right) \nn \\
&=&  \frac{V_{8}}{\eta^{8}}\sim  8q^{0} + O(q)
\eea
  and similarly for the expansions of  two Ramond sectors,  $R_{-}$
\bea
 Tr_{R} \left( \frac{1 - (-)^{G}}{2} \ q^{L_{0}} \right) 
&=& \frac{2^{4}}{\prod_{n}^{\infty}(1 - q^{n})^{8}}\left( \frac{\prod_{n}^{\infty}(1 + q^{n})^{8}}{2} \right) \nn \\
  &=&  \frac{S_{8}}{\eta^{8}} \hspace{.1 cm} \sim  \hspace{.1 cm} 8q^{0} + O(q),
\eea
and $R_{+}$
\bea
 Tr_{R} \left( \frac{1 + (-)^{G}}{2} \ q^{L_{0}} \right) 
&=& \frac{2^{4}}{\prod_{n}^{\infty}(1 - q^{n})^{8}}\left( \frac{\prod_{n}^{\infty}(1 + q^{n})^{8}}{2} \right) \nn \\
  &=&  \frac{C_{8}}{\eta^{8}} \hspace{.1 cm} \sim  \hspace{.1 cm} 8q^{0} + O(q),
\eea
we  can read the massless spectrum of type IIA.

 Indeed in  the $NS \otimes NS$ sector
\beq 
V_{8}(q)V_{8}(\bar{q}) \sim 64 (q\bar{q})^{0} + O(q\bar{q}) = (35) + (28) + (1)
\eeq
the  $(64)$  massless degrees of freedom correspond to a two-tensor  of $SO(8)$,
whose irreducible part are the symmetric traceless part $(35) = (8\cdot 9)/2 - 1 $ (spin two
 ten dimensional graviton $G_{\m\n}$),
 the antisymmetric part $(28) = (8\cdot 7)/2$ that correspond to the excitations of the two form
 $B_{\m\n}$ and the trace $(1)$ excitation of the dilaton $\phi$.  
\vspace{.1 cm}

From the $R_{-}R_{+}$ sector we have 
\beq
S_{8}(q)C_{8}(\bar{q}) \sim  64 (q\bar{q})^{0} + O(q\bar{q}) = (8) + (56)
\eeq
the physical  degrees of freedom of a one form $C_{1}$ and a tree form $C_{3}$.
 $(8)$ are the light component of the vector $C_{1}$,
 while  $(56) = (8\cdot 7 \cdot 6)/(2 \cdot 3)$ are those of $C_{3}$.

 $NS_{-}R_{+}$ gives
\beq
 V_{8}(q)C_{8}(\bar{q}) \sim  64 (q\bar{q})^{0} + O(q\bar{q}) =  (8_{S}) + (56_{C}) 
\eeq
where  we have used the decomposition for the product of the vectorial
and (one chirality) spinorial representations
 $\mathbf{8}_{V} \times \mathbf{8}_{C} = \mathbf{56}_{C} + \mathbf{8}_{S}$ 
( the subscripts $C$ and $S$ distinguish the two chiralities).
  $(56_{C})$  is a left-handed gravitino $\bar{\psi}^{\m}_{\dot{\a}}$ 
 and $(8_{S})$ is a right-handed dilatino $\l_{\a}$.  

While from  $R_{-}NS_{+}$
\beq
 S_{8}(q)V_{8}(\bar{q}) \sim  64 (q\bar{q})^{0} + O(q\bar{q}) =  (8_{C}) + (56_{S}) 
\eeq
we have   $\mathbf{8}_{V} \times \mathbf{8}_{S} = \mathbf{56}_{S} + \mathbf{8}_{C}$, where
$(56_{S})$ is a right handed gravitino $\psi^{\m}_{\a}$ and $(8_{C})$ is  a left-handed  
 dilatino  $\l_{\dot{\a}}$.
\hspace{.1 cm}

To summarise, for type IIA superstring we have found the following massless spectrum:
\vspace{.5 cm}

bosons:
\bea
NS_{-}NS_{-}  &:&   (G_{\m\n}, \  B_{\m\n}, \ \phi) \qquad from \ \  V_{8}(q)V_{8}(\bar{q}) \nn \\
R_{-}R_{+}  &:&   (C_{\m}, \  C_{\m\n \r} ) \qquad   from \ \  S_{8}(q)C_{8}(\bar{q}) \nn
\eea 
fermions:
\bea
NS_{-}R_{+}  &:&   ( \bar{\psi}^{\m}_{\dot{\a}}, \  \l_{\a}) \qquad  from \ \  V_{8}(q)C_{8}(\bar{q}) \nn \\
R_{-} NS_{-}  &:&   ( \psi^{\m}_{\a}, \   \l_{\dot{\a}} ) \qquad  from \ \   S_{8}(q)V_{8}(\bar{q}).  \nn
\eea
This spectrum  corresponds to type IIa  $\mathcal{N} = (1,1)$ supergravity that is not chiral.

\vspace{.1 cm}

We now  consider the type IIB string, where we employ the same GSO projection
 in the holomorphic and anti-holomorphic sectors.  The partition function thus reads 
\beq
Z_{IIB} = (V_8 - S_8)(\bar{V}_8 - \bar{S}_8) = V_8 \bar{V}_8 +   S_8 \bar{S}_8 - V_8 \bar{S}_8 -  S_{8} \bar{V}_8.
\eeq
The $NS \otimes NS$ sector is the same as in type IIA case
\beq 
V_{8}(q)V_{8}(\bar{q}) \sim 64 (q\bar{q})^{0} + O(q\bar{q}) = (35) + (28) + (1)
\eeq
and at the massless level comprises
$ (G_{\m\n}, \  B_{\m\n}, \ \phi) $,
while the     $R_{-}R_{-}$ sector at massless level now  containes  even-degrees forms
\beq 
S_{8}(q)S_{8}(\bar{q}) \sim 64 (q\bar{q})^{0} + O(q\bar{q}) = (1) + (28) + (35)
\eeq
$(1)$ is a zero form $C_{0}$,  $(28)$ a two form $C_{2}$ and  
 $(35) = \frac{1}{2}\frac{8\cdot 7 \cdot6 \cdot 5}{ 2 \cdot 3 \cdot 4}$ corresponds 
 to the physical degrees of freedom of   a  four form $C_{4+}$,
 whose field strength  satisfies a self duality conditions
\beq 
\p_{\m}C_{\n \r \s \t}^{+} = \e_{\m\n\r\s\t}^{ \ \ \ \ \ \ \a\b \gamma \d \eta} \p_{\a}C_{\b \gamma \d \eta}^{+}.
\eeq

The RNS spinors from  $V_8 \bar{S}_8$ and  $S_{8} \bar{V}_8$
give two right-handed gravitini and two left-handed dilatini
$( \psi^{\m}_{\a}, \   \l_{\dot{\a}} )$.

To summarise, the massless spectrum of type IIB superstring is:

\vspace{.5 cm}
bosons:
\bea
NS_{-}NS_{-}  &:&   (G_{\m\n}, \  B_{\m\n}, \ \phi) \qquad from \ \  V_{8}(q)V_{8}(\bar{q}) \nn \\
R_{-}R_{+}  &:&   (C, \  C_{\m\n}, \   C_{\m\n\r\s}^{+}) \qquad   from \ \  S_{8}(q)S_{8}(\bar{q}) \nn
\eea 
fermions:
\bea
NS_{-}R_{-}  &:&   ( \psi^{\m}_{\a}, \   \l_{\dot{\a}} ) \qquad  from \ \   V_{8}(q)S_{8}(\bar{q}) \nn \\
R_{-} NS_{-}  &:&   ( \psi^{\m}_{\a}, \   \l_{\dot{\a}} ) \qquad  from \ \   S_{8}(q)V_{8}(\bar{q}).  \nn
\eea
This spectrum  corresponds to type IIb  $\mathcal{N} = (2,0)$ supergravity
that although chiral is free of   anomalies.


\vspace{.1 cm}



\newpage
--------------------------------------------------------------------------------------------------------------
\newpage

\chapter{$D = 10$  Open string Vacua}

\section{Open strings and the Annulus vacuum amplitude}
 \everypar{\hspace{-.6cm}}

\begin{figure}  
\begin{center} 
\includegraphics[scale=1, height=5cm]{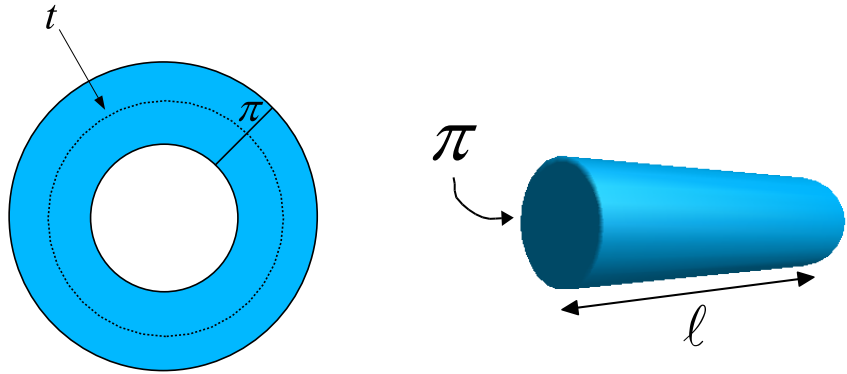}
\caption{The double role of the cylinder as a closed string tree diagram and a one loop open string
 one. The two possibilities are related by the choice of \emph{vertical}, along the boundary or \emph{horizontal}
proper time, along the axis.}
\label{cylinderfig}
\end{center}
\end{figure}

 A second option    for constructing a quantum theory
of a one-dimensional object is to consider the topology of a segment
 instead of that of a circle.
 This second possibility gives rise to a description of a quantum open string propagating
on spacetime, in many respects  related to the
closed string theory. At an intuitive level, the relation can be understood 
by considering that  loop open string diagrams, as  the one loop annulus for example,
  represent also  closed string tree level diagram (cylinder), see fig. (\ref{cylinderfig}).
 Indeed, a theory of open strings needs for its consistency
 the presence  of a 
closed string sector, while the viceversa is not true.
The segment  $\{ 0 \le \s \le \pi \}$  is related to the circle  $S = \{ \s \sim \s + 2\pi \}$
 by one of the simplest  examples of orbifold, the 
 quotient  $S / \mathbb{Z}_{2}$ with  $\mathbb{Z}_{2}: \s \rightarrow - \s$ as shown in fig (\ref{Orbifold}).
 This already  suggests that closed and open string might be at times related
in a sort of world-sheet orbifold \cite{cargese}, an idea that we will make precise
in the following.
 
\begin{figure}  
\begin{center} 
\includegraphics[scale=1, height=5cm]{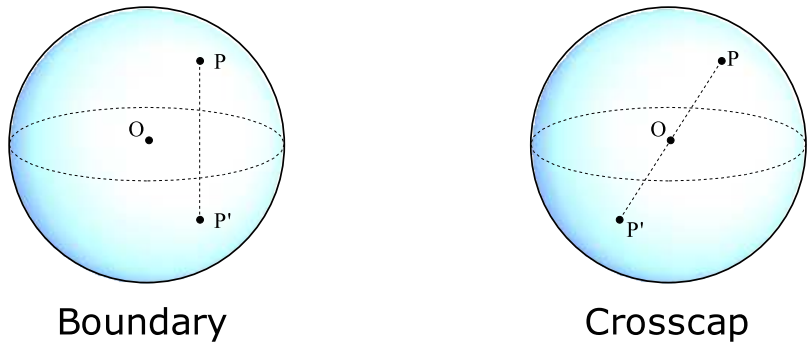}
\caption{The tree-level oriented and unoriented open string diagrams are obtained
from two different involutions on a two sphere, (the latter represents  a closed string tree-level
vacuum diagram).   On the left side,  the identification
 of points at opposite latitude  gives rise to a disk, whose boundary is the equator.
 On the right side, the identification of antipodal points
on a two sphere produces a semisphere whose equator has opposite points
identified (crosscap). The crosscap is isomorphic to the real projective plain.}
\label{boundcross}
\end{center}
\end{figure}

The involution $\mathbb{Z}_{2}$ creates two boundaries in the infinite
 cylinder, giving  an infinite strip,  the world-sheet that the open string describes in its
motion.
If we introduce complex  coordinates $w = \s - i\tau$ on the  closed string cylinder,
so that the world-sheet metric acquires an Euclidean signature, 
and  conformally map this surface into the Riemann sphere $z = e^{iw} = e^{i\s}e^{\tau}$,
we see that the involution $\mathbb{Z}_{2} : \s \rightarrow - \s$
 translates into $z \sim \bar{z}$, the open string world-sheet becomes  the (upper) half complex plane
  with all the points at  infinity  identified. This set can also be obtained
 from the stereographical projection
 of a disk, whose boundary is projected on the real axis, shown in  figure \ref{diskhalfplane}. 
 
\begin{figure}  
\begin{center} 
\includegraphics[scale=1, height=5cm]{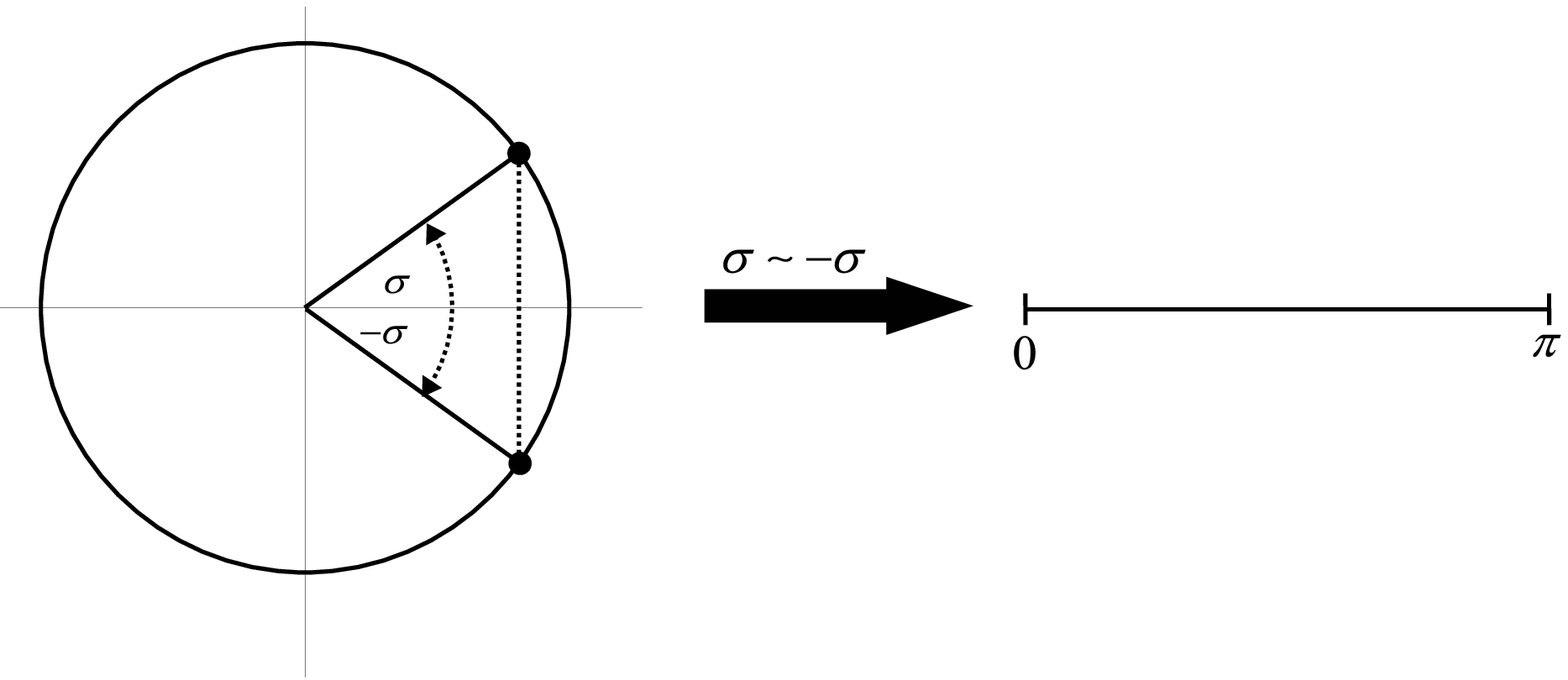}
\caption{The segment  $\{ 0 \le \s \le \pi \}$  is related to the circle  $S = \{ \s \sim \s + 2\pi \}$
 by one of the simplest  example of orbifold, the 
 quotient  $S / \mathbb{Z}_{2}$ with  $\mathbb{Z}_{2}: \s \rightarrow - \s$.                                 .}
\label{Orbifold}
\end{center}
\end{figure}

All the considerations about the quantisation
 of the two dimensional world-sheet closed string theory  
 go through for open strings as well, keeping in mind of course that there is only one holomorphic sector
 and so only one copy of the Virasoro Algebra.  
This is the result of the boundary conditions on the real axis that 
impose for the world-sheet fields 
\beq
\p X^{\m}(z) = \bar{\p}X^{\m}(\bar{z}) \qquad  on  \qquad  z = \bar{z}, 
\eeq
which implies the existence of only one set of oscillators in the open string coordinate mode expansion
\beq
 X^{\m}(z,\bar{z}) = \a' p^{\m} \ln |z|^{2} + \sqrt{\frac{\a'}{2}}\sum_{m \ne 0}\frac{\a_{m}^{\m}}{m}
\left( z^{-m} + \bar{z}^{-m} \right). \label{opencoord} 
\eeq
Notice that the open string center of mass momentum is one-half with respect to the 
  closed string case,
 since the contour integral to obtain it (\ref{closedcdmP}) needs in this case
to be performed over a semicircle.

In particular tree level asymptotic states are created on the boundary of the disk
by  operators that  need to be $SL(2,\mathbb{R})$ invariant.
  Global scale invariance, in perfect analogy to the
 holomorphic closed string sector,  fixes their   conformal weight to be equal to one,
 so that for the tree level open string spectrum
  we find the same connections between  rank of tensor operators
   and the masses of the corresponding open string excitations.
  The open string ground state   is
necessarily tachyonic,  the Lorentz   vector is necessarily  massless and higher rank tensor states are all
necessarily massive.

An important feature, peculiar to  open strings,  is the possibility of supporting
\emph{quarks}, since their endpoints can be dressed with the so called Chan-Paton
factors \cite{Paton:1969je}, that give rise to the possibility of describing non-Abelian gauge symmetries.
  The presence of colours carried by the open string endpoints 
  reflects in the open string amplitudes
  in  the presence of  traces of group valued matrices, a feature compatible with the  symmetry
  under cyclic permutation of external legs for these diagrams.
 A careful investigation of the consistency for the open string amplitudes
with unitarity and factorisation at intermediate poles has shown \cite{Marcus:1982fr,Marcus:1986cm}
that the allowed open string gauge groups are the classical groups $U(n)$, $SO(n)$ and $USp(2n)$. 
 
We want to focus on  open strings in $D = 10$ critical dimension,
 so we consider the $\mathcal{N} = (1,1)$  two dimensional
action  (\ref{wssaction}) in the  superconformal gauge (\ref{scgauge})
  on the infinite strip $\mathcal{S} = \{ 0 \le \s \le \pi, \ -\infty < \tau
 < \infty \}$ 
  
\beq
S = \frac{1}{4\pi \a'} \int d\s d\tau \p^{\a}X^{\m} \p_{\a}X_{\m} 
+ \frac{i}{4\pi \a '}\int d\s d\tau \bar{\psi}^{\m}\gamma^{\a}\p_{\a}\psi_{\m}.
\eeq
 The boundary conditions 
 for the  world-sheet fields at the open  string endpoints $\s = 0$ and  $\s = \pi$ are
\bea
\d X^{\m} \p_{\s}X^{\m}&=& 0  \\
\d \psi^{\m}_{L}  \psi^{\m}_{L} + \d \psi^{\m}_{R}  \psi^{\m}_{R} &=& 0 \qquad \m = 0,...,D - 1,\qquad (not \  summed) \label{fbc}
\eea
where   $\psi^{\m}_{L}$, $\psi^{\m}_{R}$ are (left)right handed two dimensional
 spinors.
For each of the two endpoint coordinates we can either have   $\p_{\s}X^{i} = 0$
(Neumann boundary condition) or  a $\d X^{i} = 0$ (Dirichlet boundary condition).
In the latter  case, the endpoint is stuck at one point along the direction given by the coordinate,
for example  $ X^{i}(0,\tau) = x^{i}_{0}$ for the left endpoint or  $X^{i}(\pi,\tau) = x^{i}_{\pi}$  
for the right one.
  Open strings  can have one or both  endpoints constrained to move 
 on a submanifold of the $D = 10$  space that is called a Dirichlet p-brane,
where $p + 1$ is the number of its spacetime dimension. The dimension of a p-brane corresponds to the number 
 of  open string  coordinates that satisfy Neumann boundary conditions.
 
 For a coordinate that satisfies N-N  (Neumann- Neumann) boundary condition the solution of the two dimensional
Laplace equation is
\beq
X^{\m} = x^{\m} +  \a ' p^{\m} \tau + \sqrt{\frac{\a'}{2}}\sum_{n \ne 0} \a^{\m}_{n}e^{i n\tau} \cos(n\s).
\eeq
In case of mixed N-D  (Neumann- Dirichlet) conditions we have instead an  expansion
in half integer modes
\beq
X^{\m} =   \sqrt{\frac{\a'}{2}}\sum_{n} \a^{\m}_{n - 1/2}e^{i(n - 1/2)\tau}\cos((n - 1/2) \s),
\eeq
and for  D-D  (Dirichlet- Dirichlet) boundary conditions
\beq
X^{\m} =  x^{\m}\frac{\s}{\pi} +    \sqrt{\frac{\a'}{2}}\sum_{n \ne 0 } \a^{\m}_{n}e^{i n\tau}\sin(n \s).
\eeq

\begin{figure}  
\begin{center} 
\includegraphics[scale=1, height=5cm]{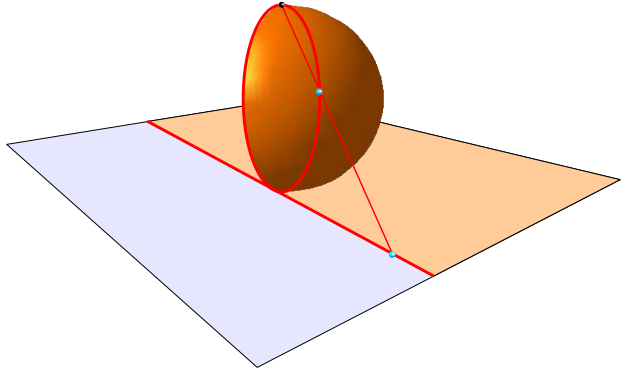}
\caption{Stereographic projection of a disk into the half complex plane, the boundary 
of the disk is mapped by the projection into the real axis.}                   
\label{diskhalfplane}
\end{center}
\end{figure}

 Turning to the world-sheet fermions, we have two inequivalent  options
       for satisfying  their  boundary conditions (\ref{fbc})
\bea
\psi^{\m}_{L}(0,\tau) &=& \psi^{\m}_{R}(0,\tau) \nn \\
\psi^{\m}_{L}(\pi,\tau) &=& (-)^{a} \psi^{\m}_{R}(\pi ,\tau), 
\eea
 where $a=0$ defines  the Ramond sector, while $a=1$  the Neveu-Schwarz  sector.

By using the basis 
\begin{displaymath}
 \psi_{r} = \left( \begin{array}{c} 
 \psi_{r, L} \\ 
 \psi_{r, R}    \end{array} \right) = \frac{1}{\sqrt{2}} \left( \begin{array}{c} 
 e^{- ir(\tau + \s)  } \\ 
 e^{- i r(\tau - \s)}\end{array} \right),   
\end{displaymath}
of solutions to the two-dimensional Dirac equation, one can write
 write a generic solution as the expansion
\beq
\psi = \sum_{r} \left(\b_{r} \psi_{r, R} +  \tilde{\b}_{r} \psi_{r, L} \right).
\eeq
 After imposing the boundary conditions, one finds  that  $\b_{r} = \tilde{\b}_{r}$
 and $r \in \mathbb{Z}$ for Ramond, while  $r \in \mathbb{Z} + 1/2$ for Neveu-Schwarz. 
Therefore the mode expansions for the two sectors are 
\bea
\psi &=& \sqrt{\frac{\a'}{2}}\sum_{r \in \mathbb{Z}}d_{r}\psi_{r}, \qquad Ramond, \nn \\
\psi &=& \sqrt{\frac{\a'}{2}}\sum_{r \in \mathbb{Z} + 1/2}b_{r}\psi_{r}, \qquad Neveu-Schwarz.
\eea

 Following the same route discussed for one sector of the closed string,
 one can pick up  a light cone gauge
 and solve the classical constraints for the superconformal algebra
 for the lightcone oscillators in the various sectors.
 In this case, longitudinal oscillators
 are expressed in terms of the physical transverse oscillators and
   the Hilbert space is manifestly  free of negative norm states, because constructed by
quantising only transverse modes.

 In particular, in order to construct the one-loop open string
annulus amplitude we need  the  generator  $L_{0}$  for  time translations  that
 in the R sector reads  
\beq
L_{0} = \left(  \a' p^{2} +  \sum_{m = 1}^{\infty} m (a^{i}_{m})^{\dag} a^{i}_{m}
 +  \sum_{r = 1/2}^{\infty}  r  ( b^{i}_{r})^{\dag} b^{i}_{r} 
  -  \frac{1}{2} \right), 
\eeq
 while in the NS sector reads 
\beq
L_{0} = \left(  \a' p^{2} +  \sum_{m = 1}^{\infty} (a^{i}_{m})^{\dag} a^{i}_{m}
 +  \sum_{r = 1}^{\infty}  r   (d^{i}_{r})^{\dag} d^{i}_{r} 
  \right). 
\eeq
 Notice the different normalisation of the ten-dimensional momentum $p$ from 
 the closed string case eq. (\ref{L0NS}) and  (\ref{L0R}) as a consequence of
 (\ref{opencoord}).

\begin{figure}  
\begin{center} 
\includegraphics[scale=1, height=5cm]{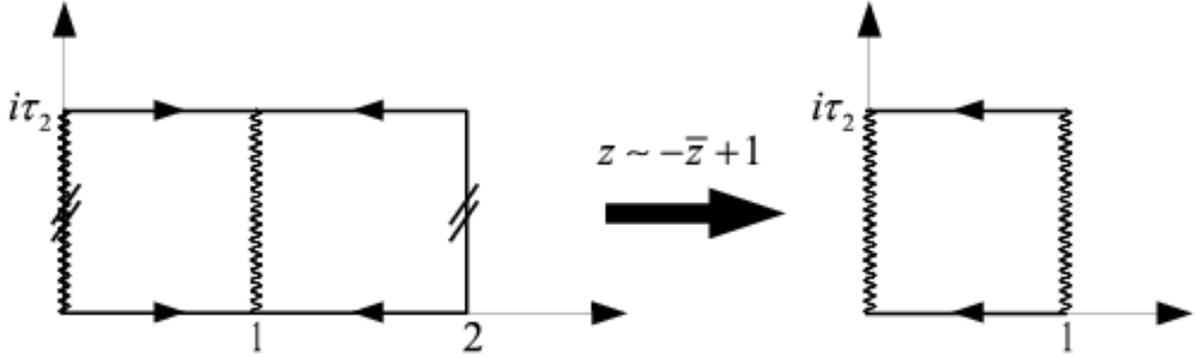}
\caption{The annulus can be obtained from a double covering torus (left side) by the involution
  $z \sim  -\bar{z} + 1$ that creates the two boundaries, (represented by the wavy lines).
The modulus of the double covering torus  $i\tau_{2}/2$, that appears naturally in the 
Annulus vacuum amplitude (\ref{anello10}), 
  is obtained by rescaling to one its horizontal side.}
\label{anellofigura}
\end{center}
\end{figure}

 The Annulus is a surface that can be obtained from a double covering square torus with
 a  pure imaginary  modulus $\tau = i\tau_{2}$ after an anti-holomorphic involution (see fig. \ref{anellofigura}) that creates two boundaries on the covering torus.
This amplitude is then given by
\bea
\mathcal{A} &=& -N^{2}\int \frac{d\tau_{2}}{\tau_{2}}Tr_{\mathcal{H}} e^{-2\pi \frac{\tau_{2}}{2}L_0} \nn \\
 &=& -N^{2}\int \frac{d\tau_{2}}{\tau_{2}}Tr_{\mathcal{H}}q^{\frac{1}{2}L_0}.
\eea
   The factor $N^{2}$ takes
into account the  Chan-Paton multiplicities.
After computation of the trace over the continuum of the ten dimensional momenta,
one recovers
\beq  
\mathcal{A} = -N^{2}\int \frac{d\tau_{2}}{\tau_{2}^{6}}Tr_{\mathcal{H}}q^{\frac{1}{2}L_0},
\eeq
where in the last expression to lighten the notation
  we omitted the   normalisation coefficient  that originates from the Gaussian 
integral over the ten dimensional momentum $p$.

Similarly to one sector of the $D = 10$ closed string, the ground state in the NS sector
is tachyonic, while the first excited state is a massless vector.
 In the R sector the ground state is instead a ten-dimensional Dirac spinor,
while all the excited states are massive.

In order to project out the tachyon
 and  to obtain a supersymmetric  open string spectrum, one can consider 
  the various  GSO projection on  the $SO(8)$ little group characters

\bea
O_8 (q) &=& tr_{NS} \left(\frac{1+(-)^{G}}{2}\right)q^{ \frac{1}{2} L_{0}} \qquad
V_8 (q) = tr_{NS} \left(\frac{1-(-)^{G}}{2}\right)q^{ \frac{1}{2}  L_0} \nn \\  
C_8 (q) &=& tr_{R} \left(\frac{1+ \Gamma_{9}(-)^{G}}{2}\right)q^{\frac{1}{2}     L_{0}} \qquad
S_8 (q) = tr_{R} \left(\frac{1-  \Gamma_{9}(-)^{G}}{2}\right)q^{ \frac{1}{2} L_{0}} \label{GSOchar}   
\eea
where $(-)^{G}$ is the world-sheet open string fermion number and $q = e^{- 2\pi \tau_{2}}$.

For a  tachyon-free  and spacetime supersymmetric spectrum the right
  choice in the NS sector is to select the GSO that  gives the vector character $V_8$,
 while in the R sector is to select one spinorial character of definite  chirality
 
\beq
\mathcal{A} = N^{2}\int \frac{d\tau_{2}}{\tau_{2}^{6} \eta^{8}(i\tau_{2}/2)} (V_{8} - S_{8})(i\tau_{2}/2).\label{anello10}
\eeq

 
The opposite sign between the NS  and the R characters
 takes into account
that, due to their opposite statistic, bosons and fermions give opposite 
contributions to the vacuum energy.


The integral in eq. (\ref{anello10}) needs to be performed on the region $0 < \tau_{2} < \infty $
 since there is not a modular group for the annulus, in particular there
is not an $S$ transformation that cuts off the UV part
of the integration domain, and  the amplitude is
divergent in the UV limit  $\tau_{2} \rightarrow 0$. 

One can study this divergence from the closed string point of view.
 In fact, the annulus can be considered as an amplitude for a 
tree level propagation for closed strings.
In order to obtain the correct closed string amplitude,
 that corresponds to a different choice of proper time,
 one first makes the change of integration variable $\tau_{2} = 2t$
 
\bea
\mathcal{A} = N^{2}\int \frac{d\tau_{2}}{\tau_{2}^{6}}Tr_{\mathcal{H}}q^{\frac{1}{2}L_0} 
=    2^{ - 5} N^{2}\int \frac{d t}{ t^{6}}  Tr_{\mathcal{H}}q^{L_0}\nn \\
 =  2^{- 5} N^{2}\int \frac{d t}{t^{6}\eta^{8}(i t)} (V_{8} - S_{8})(i t),
\eea 

 in order to obtain the closed string normalisation for the characters,
as it appears in the second of the previous  equations.
 ($L_{0}$ does not contain the momentum $p$
 as we can see    from the power of $t$ in the expression). 

Then  with the  change of variable $t = 1/l$
one obtains the correct closed string length  in the diagram
\beq
\tilde{\mathcal{A}} = 2^{ - 5} N^{2}\int_{0}^{\infty} \frac{dl \  l^{4}}{\eta^{8}\left(i /l \right)}(V_{8} - S_{8})\left(i/ l\right)  =   2^{ - 5} N^{2}\int_{0}^{\infty} \frac{dl}{\eta^{8}(i l)}(V_{8} - S_{8})(i l).                         \label{treeanul}
\eeq 
 In the last equality we used  the transformations properties
  of the Dedekind eta function under an $S$ transformation
   $\eta(- 1/\tau) = \sqrt{- i \tau} \eta(\tau)$ and the modular invariance
 of the supercharacter $V_{8} - S_{8}$.
 
The change of variable $t = 1/l$ has mapped
  the  $t \rightarrow 0$ UV  divergence  into a
 $l \rightarrow \infty$ IR one.

By expanding the characters in  (\ref{treeanul}) in powers of $q = e^{-2\pi l}$
and recalling that the exponents are proportional to the masses of the 
closed string states flowing in the tube, one can see that the IR divergence of the 
cylinder is due only to the propagation of massless closed string states
\bea
\tilde{\mathcal{A}}  &=& 2^{ - 5} N^{2}\int_{0}^{\infty} \frac{dl}{\eta^{8}(i l)}(V_{8} - S_{8})(i l) \nn \\
 &=&  2^{ - 5} N^{2}\int_{0}^{\infty} dl\left[ (8 - 8) + (d_{1} - d_{1}) e^{-2\pi l}
 +...+ (d_{k} - d_{k}) e^{-2\pi lk}+ ... \right].
\eea
The origin of the IR divergence is clear by considering that the closed string states
flowing on the tube are on-shell
\beq
\int_{0}^{\infty} dl e^{-2\pi lk} = \frac{1}{2\pi k} = \frac{1}{2\pi \a' p^{2}} \qquad for \ \  p^{2} = m^{2} = \frac{k}{\a'}.
\eeq

 The tree level cylinder amplitude is therefore a  sum of on-shell  propagators of  the
 $NS \otimes NS$ and RR states flowing in the diagram, multiplied by a coefficient
 which is the square of the amplitude for emission of the state by a D-p brane (as shown in the 
upper part in figure \ref{transverse}).



\begin{figure}  
\begin{center} 
{\includegraphics[scale=1, height=6cm]{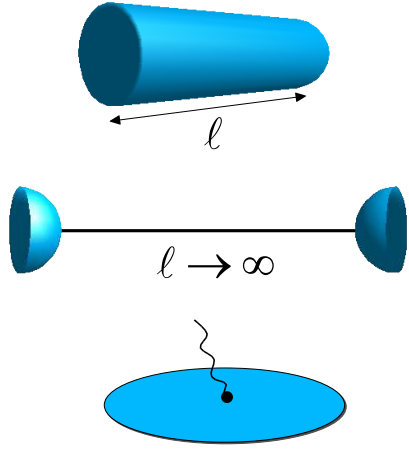}}
\caption{Above: The origin of the divergence in the cylinder diagram can be recognised by the 
factorisation of the amplitude into  reflection coefficients at the boundary
times on shell propagators. In the limit of infinite closed string 
proper time $l$ is quite visible the presence of disk tadpole diagrams, that represent
the amplitude for emission of closed string from the internal of the disk into the vacuum.
 These tadpole diagrams   gives rise  to a  divergence  for massless states, since the external  propagators are forced to be on mass shell.}
\label{tadpoles}
\end{center}
\end{figure}

These IR divergences are called tadpoles, the $g = 1$ cylinder diagram contains tadpole contributions
 arising from both the  $NS \otimes NS$ sector  and the RR one, that are indeed  $g = 1/2$ disk diagrams,
as shown in figure \ref{tadpoles},
  and that
correspond  to the emission of massless $NS \otimes NS$ and  RR states into the vacuum.
They arise in the cylinder diagram in the limit of infinite proper time $ l \rightarrow \infty$.

\begin{figure}  
\begin{center} 
\includegraphics[scale=1, height=5cm]{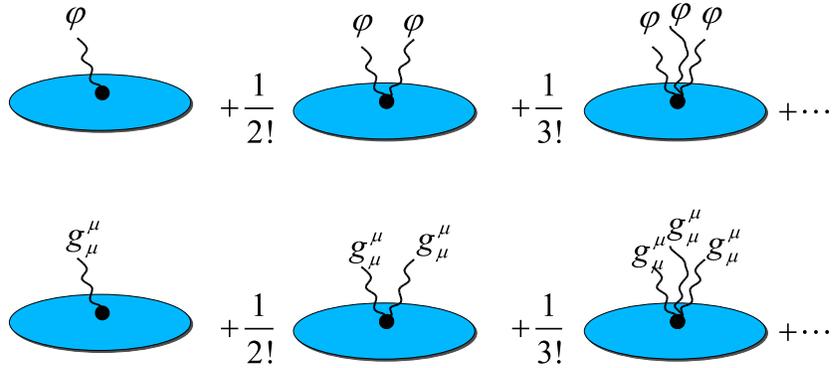}
\caption{The $NS \otimes NS$ tadpole series. Due to Lorentz invariance the only closed string states that can be 
emitted by a disk are $g^{\m}_{\m}$ and $\phi$. The two series can be resummed to obtain
  the correction  term (\ref{NSaction}) for the background fields Action. This corrective term is
proportional to the tension of the D-branes present in the background and represents a non vanishing
 vacuum energy.} 
\label{tadpoleexpansion}
\end{center}
\end{figure}

 In the $NS \otimes NS$ sector, the only non-vanishing 
 disk-tadpole diagrams compatible with Lorentz symmetry are  those  involving the emission of one or more dilatons $\phi$
into the vacuum and the trace $g_{\m}^{\m}$ of the graviton.
 All these diagrams are generated by the following Action term \cite{Green:1984ed}

\beq
 S_{\mathcal{B}} = \mathcal{B}\int d^{10}X \sqrt{G}e^{-\phi}, \label{NSaction}
\eeq
 that, after  splitting the fields  $G_{\m\n} = \eta_{\m\n} + g_{\m\n}$ and $\phi =  \phi + \varphi$
 into the background plus a fluctuation,
   computes   the amplitude for the emission of $n$ dilatons from the disk 
\beq  
\frac{\d^{n} S_{\mathcal{B}}}{\d \varphi^{n}} = \frac{1}{n !}\mathcal{B},
 \eeq
 reproducing the correct  combinatory factor $1/n!$ for this diagram, see fig. \ref{tadpoleexpansion}.

 The emission for the trace  $g_{\m}^{\m}$ from the disk can be obtained 
 by expanding the square root  $\sqrt{G} = \sqrt{det(\eta_{\m\n} + g_{\m\n})} = 1 + g_{\m}^{\m} + O(g^{2})$
and by the functional derivarive
\beq
\frac{\d }{\d g_{\m}^{\m}} \left( \mathcal{B}\int d^{10}X \sqrt{G}e^{-\phi}\right) =
  \frac{\d }{\d g_{\m}^{\m}}\left( \mathcal{B}\int d^{10}X e^{-\phi} ( 1 + g_{\m}^{\m}) \right) = \mathcal{B},\eeq
and similarly for the amplitude for emission of a generic number of 
states $g_{\m}^{\m}$ from the disk.

\begin{figure}  
\begin{center} 
\includegraphics[scale=1, height=5cm]{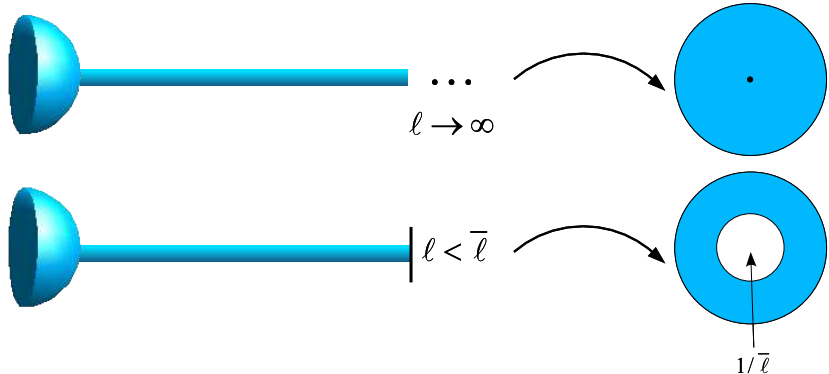}
\caption{One way to regularising the $l \rightarrow \infty$ divergence due to massless modes
 propagating in the tube is to put a IR cut off $\bar{l}$, that breaks conformal invariance on the
disk. A careful inspections shows that  conformal symmetry is restored by the cancellation
 of this effect with the shift on the vacuum energy given  by the correction term in eq. (\ref{NSaction}).}
\label{tadpolelimit}
\end{center}
\end{figure}

The action term induced by (\ref{NSaction}) describes a background with a vacuum energy density
proportional to $\mathcal{B}$ that is not Ricci flat and therefore  seems incompatible with
 the string equation of motions (\ref{bsem}),  that guarantee conformal invariance on the world-sheet theory.
Actually this is not the case, since the tadpole diagram breaks conformal invariance
on the world-sheet. An intuitive way to see this is to consider the tadpole  diagram,
 a disk
 (semisphere) with a tube of infinite length.  One way to regularise the divergence
 is to cut off the length to the tube to a finite size $\bar{l}$. If the insertion of the infinite tube is
  conformally  mapped to a puncture in the interior of the semi-sphere,
 the presence of a cutoff for its length will correspond to substituting the puncture with
a finite size region on the sphere, whose presence breaks conformal invariance
   since two points on the disk cannot get closer than the size of the region (fig. \ref{tadpolelimit}).

The correction  of the string equation of motions from the conformal anomaly described above
can be connected to the action term describing the tadpole, such that via a redefinition of
the background metric from the  flat Minkowski to the curved metric
  satysfing the corrected equations of motions, one obtains conformal invariance
again, as first pointed out by Fischler and Susskind 
\cite{Fischler:1986ci,Fischler:1986tb}.
The conformal background,
 solution of the tadpole corrected string equation of motions, describes a spacetime with a
 non vanishing Cosmological Constant.

The origin of a nonvanishing vacuum energy is  actually connected
  to the presence
 of D-branes, whose tension induces a back-reaction on the background 
   generating a curvature. In fact through the cylinder diagram
one can measure the tension of a brane, since the $NS \otimes NS$ closed string excanged
by  two  D-branes produces their gravitational interaction and therefore
the cylinder diagram is proportional to the square of the tension of a D-brane.
     
The tension of a D-9 brane, (we are considering open string with Neumann-Neumann boundary
conditions on D = 10 spacetime), is proportional to $e^{-\phi} = 1/g_{s}$
as one can read from (\ref{NSaction}).

It is worth to mention also that
 a non vanishing  $NS \otimes NS$ tadpole breaks space-time supersymmetry.
  This happens because  supersymmetry
 guaranties the vanishing of all the
 tadpoles  diagrams.
  For example, a dilaton tadpole diagram is given by 
the insertion of a dilaton vertex operator
  into a point of a Riemann surface. 
If supersymmetry is present the dilaton  vertex operator
 can be written as the commutator of 
the supercharge and a \emph{fermionic}
vertex operator. On the world-sheet its 
local form is given by a contour integral
 of the above commutator that vanishes
 for analyticity \cite{Martinec:1986wa,Friedan:1985ge}. 

\vspace{.5 cm}

  The $RR$  tadpole  has a rather different meaning since induces a breaking
 of gauge invariance.

  Let us focus on the type IIB. In this case, the only RR form that can propagate
  in a   disk tadpole diagram
 without breaking ten dimensional  Lorentz invariance is the ten form $C_{10}$.
 This form is not dynamical since its field strength 
 vanishes identically in ten dimensions.

 The effective Action term that reproduces the tadpole diagrams
 for the ten form by functional derivation is 
\beq
 S_{C_{10}} = \mathcal{B}_{C_{10}} \int C_{10},
\eeq
  which is however  incompatible with its  equation of motion that states
\beq 
\mathcal{B}_{C_{10}} = 0.
\eeq

Therefore an uncancelled RR tadpole from the type IIB brings  actually to an inconsistency,
 since it violates the equation of motion for the RR  ten form.
  A background redefinition that cancels the tadpole in this case
does not exist.
The inconsistency of an uncancelled RR tadpole manifests itself into an anomaly
in the massless spectrum associated to the closed string type IIB excitations and  
 the open string  gauge excitations of super Yang Mills.
The RR tadpole is associated to the RR charge of the D-9 brane 
that is measured by the RR contribution to the closed string cylinder diagram.
A way to cure this inconsistency is to consider negatively charged extended objects
 called orientifold planes, whose presence is able
 to cancel the RR tadpoles and alllow  to obtain  consistent ten-dimensional
 vacua, containing  a precise number of D9 branes that is called type I superstring.

\newpage

\section{Unoriented world-sheets and type I superstring}
 \everypar{\hspace{-.6cm}}


 The involution  $ \Omega: \s \sim -\s$ 
  creates two boundaries in the infinite cylinder $   \mathbb{R} \times S^{1}$
 turning it into the infinite strip  $\mathbb{R} \times \left[0, \pi \right]$,
 the surface  that topologically an open string describes in its motion.
This involution through the   boundaries  identifies the 
 holomorphic and antiholomorphic  sectors
 on the world-sheet.
 However, the involution has been performed only
 on the classical world-sheet fields 
 and not on the spectrum of their quantum states.

A  projection  of the  open and closed spectra
  by an operator that realises  $ \Omega$ on the quantum states
 will ensure the symmetry at the quantum level, giving rise to
an unoriented string theory \cite{Bianchi:1990yu,Bianchi:1990tb,gianf,Pradisi:1996yd,Pradisi:1995pp,Pradisi:1995qy,Angelantonj:2002ct,Dudas:2000bn,Dabholkar:1997zd}.
 It turns out that this projection of the spectrum
 into unoriented states is the cure for the problems
 that we have seen in the last section, namely 
  on  $D = 10$ spacetime the presence of D-9 branes
 in the background gives rise to an uncancelled
 RR tadpole which violates the  RR  ten-form
 equation of motion.
 The RR disk tadpole present in the $g = 1$ cylinder diagram
   needs to be cancelled by some other contributions, in order to
obtain a consistent string vacuum.

In the unoriented theory there are two more $g = 1$ 
 surfaces, besides the torus and the annulus,
 which are their unorientable cousins  the Klein Bottle and the Moebuis strip.

The operator  $\Omega: \  \s \sim - \s$ in the closed string sector 
 interchanges holomorphic and anti-holomorphic oscillators
  $\Omega: \  \a_{-n}^{\m} \rightarrow \tilde{\a}_{-n}^{\m}$,   while in the open sector
 $\Omega: \  \s \sim \pi  - \s$  acts on the oscillators as   $\Omega: \  \a_{-n}^{\m} \rightarrow (-)^{n} \a_{-n}^{\m}$.\\

As an example, the  $NS \otimes NS$
antisymmetric two tensor
\beq
B_{\m \n}: \qquad  \left( \a^{-1 \ \m} \tilde{\a}^{-1 \ \n} - \a^{-1 \ \n} \tilde{\a}^{-1 \ \m} \right)
 |0\rangle,
\eeq
  is odd under  $\Omega$
 \beq
 \Omega: B_{\m \n} \rightarrow  -B_{\m \n}.
 \eeq

$\Omega$ indeed splits the string spectrum into left-right symmetric states that are even
under its action and left-right asymmetric ones that are odd.
 

If  we want  $\Omega$ to be a symmetry in the $D = 10$ closed superstring spectrum
 we need to consider type IIB whose right and left moving sectors are identical,
(in type IIA the different GSO projections in the R sectors produce
only left-right asymmetric states),
and project out all the odd states under the $\Omega$ involution\footnote{See \cite{Angelantonj:2002ct,Dudas:2000bn,Dabholkar:1997zd} for more  general reviews
 on open-string constructions}. 

This can be done by inserting a projector
 into the trace that  computes the closed string  one loop vacuum amplitude
\beq
  \int
 \frac{d^{2}\tau}{\tau_{2}^{6}}
  Tr  \left(\frac{\1 + \Omega}{2}\right) q^{ L_0} \bar{q}^{ \bar{L}_{0}} 
= \frac{1}{2} \mathcal{T} + \mathcal{K}.
\eeq

 The number of  states in $\mathcal{T}$ have been halved by the projection,
  while $\mathcal{K}$ is the Klein bottle vacuum
diagram whose states complete the projection.

The computation of the trace with the operator $\Omega$
 inserted gives

\bea
\mathcal{K} &=& \frac{1}{2} \int_{0}^{\infty}  \frac{d\tau_{2}}{\tau_{2}^{6}} Tr\left(\Omega q^{L_{0}} \bar{q}^{ \bar{L}_{0}}\right) = 
 \frac{1}{2} \int_{0}^{\infty}  \frac{d\tau_{2}}{\tau_{2}^{6}}
  Tr   (q \bar{q})^{L_0 } \nn \\ &=& 
\frac{1}{2} \int_{0}^{\infty}  \frac{d\tau_{2}}{\tau_{2}^{6} \eta^{8}(q \bar{q})}
\left( V_{8}- S_{8} \right)(q \bar{q}) \label{vackb}
\eea

 The  trace is not vanishing only on the $NS \otimes NS$ and RR sectors,
 therefore only states of these sectors flow in the Klein Bottle.
 The argument of the characters is $q \bar{q} = e^{2\pi i(2i\tau_{2})}$
 so that the modulus 
 of the Klein bottle is  purely imaginary, and equals  $2i\tau_{2}$ .

\begin{figure}  
\begin{center} 
\includegraphics[scale=1, height=12cm]{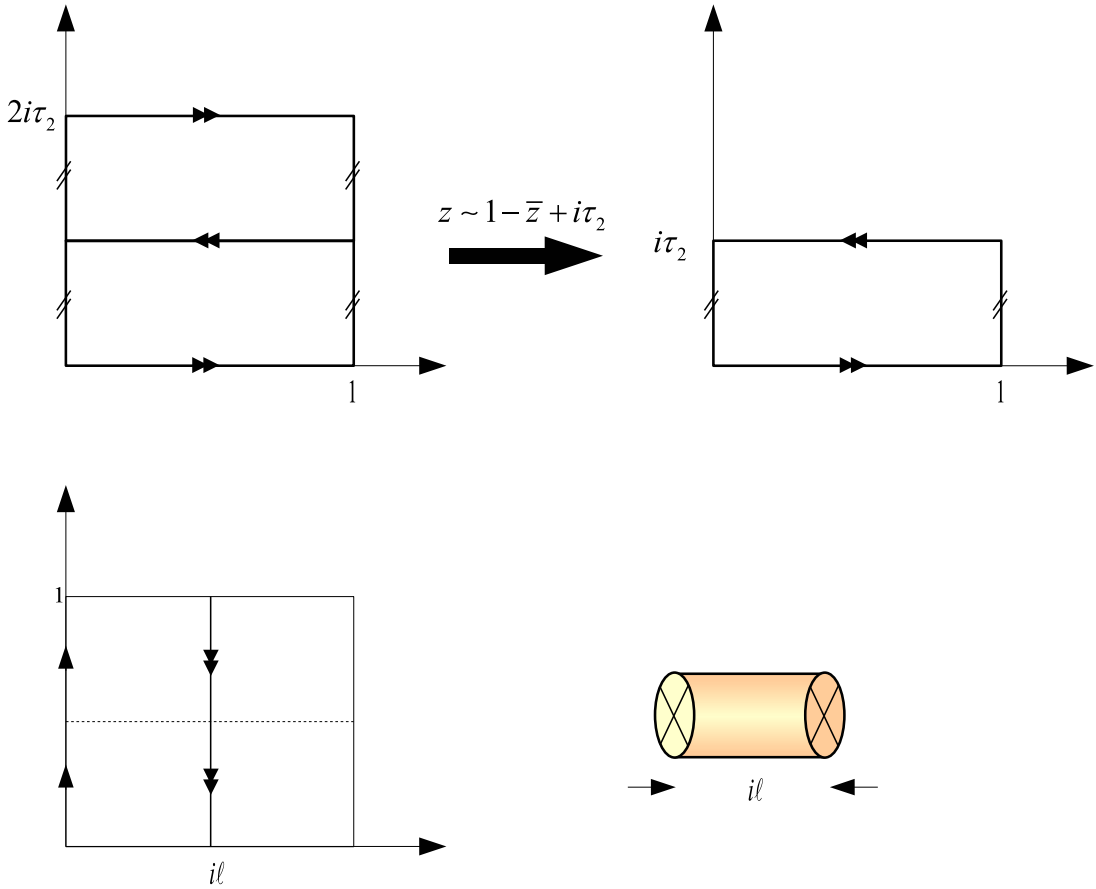}
\caption{Above:  the Klein Bottle (on the right side) 
  can be obtained as an involution from a 
 double covering  \emph{square} torus (on the left side) with a
 pure imaginary modulus $2i\tau_{2}$.
 This representation corresponds to a choice of \emph{vertical} proper time.
 Below:
 A  second inequivalent  choice for the   proper time for the  
  Klein bottle closed string diagram. The involution from the double covering
 torus (on the left) shows  a representation for this surface as a tube terminating
  with two crosscaps (on the right). In this second case, we choose \emph{horizontal} time $l$,
 and in order to obtain the correct closed string length 
 one needs to impose  $l = 1/2 \tau_{2}$.}                          
\label{KBfig}
\end{center}
\end{figure}

  This has a simple explanation since  this unorientable
 surface can be obtained as an involution from a 
 double covering  \emph{squared} torus with a
 purely imaginary modulus $2i\tau_{2}$ (see fig. \ref{KBfig}).
 After the involution, one recovers the unorientable
 world-sheet that represents 
a closed string vacuum fluctuation of proper time  $\tau_{2}$.
 The integral over the Klein Bottle proper time is UV divergent because, as in the 
case of the open string annulus diagram,
 there is not any modular symmetry that is able to cutoff the divergence.

One has a second inequivalent  choice for the   proper time in  the
 Klein bottle, that represents a closed string tree-level
 tube diagram with two crosscaps (see fig. \ref{KBfig}).
 This second choice  is connected to the
 one loop diagram through the change of variable
 $l = 1/ 2 \tau_{2}$ in the integral (\ref{vackb}),
  necessary to properly  rescale the closed string length.

 The  amplitude for tree level closed string propagation     
   between two crosscaps $\tilde{\mathcal{K}}$, corresponding to the choice of
 horizontal proper time is
\beq 
  \tilde{\mathcal{K}}=  2^{4} \int_{0}^{\infty}  \frac{dl}{ \eta^{8}(il)}
\left( V_{8}- S_{8} \right)( il).  \label{treekb}
\eeq 

After the change of variable  $l = 1/ 2 \tau_{2}$,
 in eq. (\ref{treekb}) 
the original UV divergence in the
one loop amplitude has been  mapped into an IR divergence,
due to the massless closed string modes flowing in the tube.

 As for the cylinder amplitude, the divergence in the transverse Klein Bottle
 is due to $NS \otimes NS$ and RR tadpole diagrams, but this time  correspond
 to  the emission of massless closed string states
 from the projective plane  into the vacuum.

 

 
 
  

\vspace{.3 cm}

 We now turn   to the open string spectrum, where  for consistency with the closed sector
 we need  to project the states by $\Omega$.
The aim is to check if after having performed a complete projection,
 both in the closed and in the open sectors, 
 it is possible to find in the $g = 1$ amplitudes  a cancellation between the 
 disk and the croscap tadpoles.
 
 As for the closed string sector, we project the Fock space of open string excitations
  by inserting the $\Omega$ projectior into   the trace
\beq
  \int_{0}^{\infty}
 \frac{d\tau_{2}}{\tau_{2}^{6}}
Tr  \left(\frac{\1 + \Omega}{2}\right) q^{L_0 -1} 
=  \mathcal{A} + \mathcal{M}.
\eeq

 Computation of the above expression gives
\beq
\mathcal{A} = \frac{N^{2}}{2}\int_{0}^{\infty} \frac{d\tau_{2}}{t_{2}^{6}
 \eta^{8}\left(i\frac{\tau_{2}}{2}\right)}
(V_{8} - S_{8})\left(i\frac{\tau_{2}}{2}\right) \label{vacanulII}
\eeq 
for the annulus amplitude, while   the trace  on the open spectrum with $\Omega$ inserted
 gives the M\"obius strip amplitude
\bea
\mathcal{M} &=&  \int_{0}^{\infty}
 \frac{d\tau_{2}}{\tau_{2}^{6}}
Tr  \left( \Omega q^{\frac{1}{2}L_{0}} \right) 
= \int_{0}^{\infty}
 \frac{d\tau_{2}}{\tau_{2}^{6}}
Tr  \left(  (-q)^{\frac{1}{2}L_0}\right) \nn \\
&=& \e\frac{N}{2}\int_{0}^{\infty} \frac{d\tau_{2}}{t_{2}^{6}
 \hat{\eta}^{8}\left(\frac{1}{2}+ i\frac{\tau_{2}}{2}\right)}
(\hat{V}_{8} - \hat{S}_{8})\left(\frac{1}{2} + i\frac{\tau_{2}}{2}\right). \label{vacMoeb}
\eea
 where $\e$ takes into account for a  global sign ambiguity in this amplitude,
  whose origin  will appear clear  when we will discuss the corresponding
 transverse closed string amplitude.
 The factor $N$ indicates the presence of a single boundary in the surface,
(the annulus is proportional to $N^{2}$ due to the two boundaries),
 while  
the minus sign in the second equality,  that produces an alternating sign
 in the $q$-expansion of this amplitude, follows
 from the effect of the reflection $\Omega: \s \rightarrow \pi - \s$ on the open strings
modes
 \beq
\Omega \a_{-n}|0 \rangle = (-)^{n}\a_{-n}|0 \rangle.
\eeq
 In the last 
equality of (\ref{vacMoeb}), we have  taken into account
  the alternating sign $(-)^{n}$
  by  giving a real part equal to $1/2$
 to the pure imaginary argument $i \tau_{2}/ 2$ in the M\"obius characters.
  
 This follows from the definition of the  character that is  computed on a 
Verma modulus 
 from a primary state of weight $h$, i.e. the tower of states created
 by a primary field on the ground state 
\beq
\chi(q) = q^{h - \frac{c}{24}}\sum_{n=0}^{\infty} c_{n}q^{n},
\eeq
 with $q= e^{-2\pi \tau_{2}}$ and $c$ the central charge.  By adding the real part  in the argument of the character it is then necessary  to
 multiply  it for a phase 
 in order to recover the original character with the alternating signs
\beq
  q^{h - \frac{c}{24}}\sum_{n=0}^{\infty}(-)^{n}  c_{n}q^{n}  =
  e^{-i\pi( h - c/24)}\chi\left( \frac{1}{2} + i\tau_{2} \right) = \hat{\chi}\left( \frac{1}{2} + i\tau_{2} \right).  \label{hat}
\eeq
  This last relation defines the hatted characters that appear in the loop  M\"obius amplitude
eq. (\ref{vacMoeb}).

\vspace{.5 cm}

An inequivalent  choice for the proper time is possible also for  the M\"obius strip.
 This different representation for the world-sheet
 is given by a  tube terminating with a  boundary
and a crosscap
and corresponds to a  tree level  closed string  amplitude.

\begin{figure}  
\begin{center} 
\includegraphics[scale=.5, height=8cm]{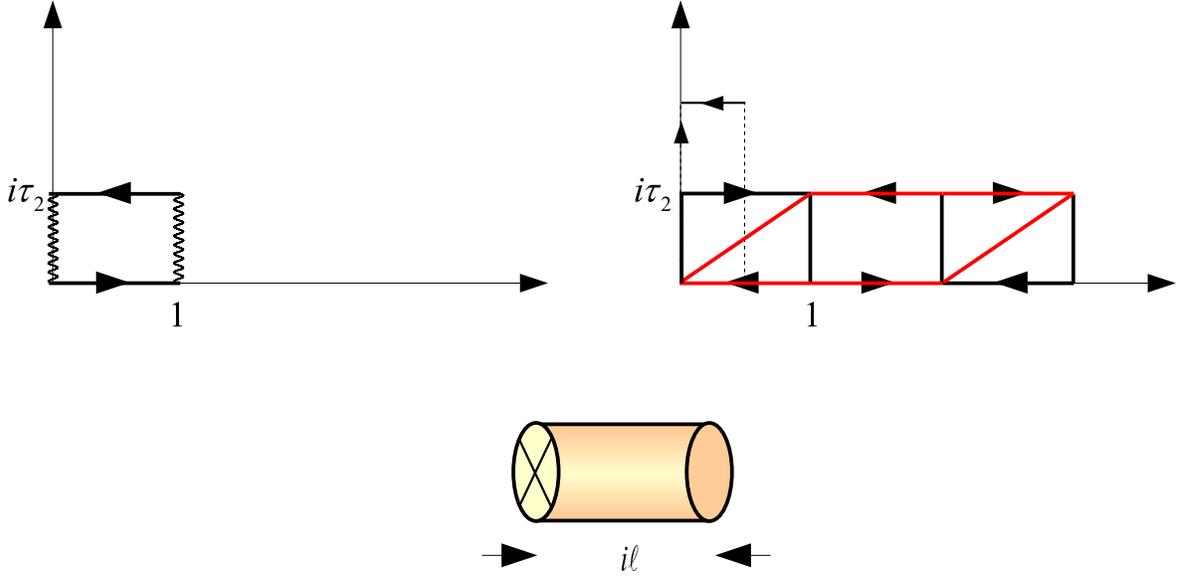}
\caption{On the left above the fundamental polygon for the M\"obius strip: following the
 identifications given by the arrows we can recognise that the wavy lines represent two
 portions of a single boundary. On the right above, the double covering tilted torus (in red)
  for the M\"obius strip, and the two inequivalent  choices for the string proper  time. For the horizontal
time (below),  this surface corresponds to a tube terminating with a boundary and a crosscap.} 
\label{Moebiusfig}
\end{center}
\end{figure}

In this case the double covering surface is not a torus
but an annulus with modulus $i \tau_{2}/2$, fig. \ref{Moebiusfig}.
 There is a double covering for the M\"obius strip
as shown in fig.  \ref{Moebiusfig}, but it is a tilted torus with a real part equal to $1/2$ and it is worth
noticing that the argument of the characters in the M\"obius amplitude
 correspond to its double covering tilted torus.

  The change of integration  variable $\tau_{2} = 1/t$  
 in (\ref{vacMoeb}) is necessary to properly  rescale the length of the closed string
 states  flowing in the tube.

The transverse tree level amplitude $\tilde{\mathcal{M}}$  therefore reads
\beq
\tilde{\mathcal{M}} = \e \frac{N}{2}\int_{0}^{\infty} d t
 \frac{t^{4}}{\hat{\eta}^{8}\left(\frac{1}{2}+ \frac{i}{2t}\right)}
(\hat{V}_{8} - \hat{S}_{8})\left(\frac{1}{2} + \frac{i}{2t}\right). \label{treeMoeb}
\eeq 
 In order to be able to confront it
 with $\mathcal{A}$ and $\mathcal{K}$ displaied in (\ref{treeanul}) and (\ref{treekb}),
 we need to rewrite this closed string amplitude in terms of characters depending
 on $1/2 + it/2$. 
 
 The proper modular transformation that does the job is
 given by the sequence    $P = TST^{2}S$ \cite{Bianchi:1990yu,Bianchi:1990tb,gianf}

\beq
\frac{1}{2} + \frac{i}{2t} = P \left(\frac{1}{2} + \frac{it}{2}\right). \label{PMoebius}
\eeq
 
 Since in the M\"obius amplitude we are using  a  basis of hatted characters
 defined in (\ref{hat}), the actual representation of the above
 modular transformation in terms of the matrices $S$ and $T$ is actually
 slightly modified and is given by  
 $P' = T^{1/2}ST^{2}ST^{1/2}$, so that
\beq
\hat{\chi}_{i}\left( \frac{1}{2} + i \frac{1}{2t} \right) = 
      P'_{ij} \hat{\chi}_{j}\left( \frac{1}{2} + i\frac{t}{2}\right), \label{P'Moebius}
\eeq

 where $ T_{ij}^{1/2} =  \d_{ij} e^{ i\pi( h - c/24)}$.


  By using the $P'$ matrix on the hatted characters and the following
  property for the  hatted $\hat{\eta}$ Dedekind function
\beq
\hat{\eta}\left(\frac{1}{2} + \frac{i}{2t}\right) = \sqrt{t} \  \hat{\eta}\left(\frac{1}{2} + \frac{it}{2}\right),
\eeq
one can re-express the transverse M\"obius amplitude (\ref{treeMoebII})
 in terms of modular functions with
   argument  $1/2 + it/2$ 
 \beq
\tilde{\mathcal{M}} = \e \frac{N}{2}\int_{0}^{\infty}  \frac{dt}{\hat{\eta}^{8}\left(\frac{1}{2}+ \frac{it}{2}\right)}
(\hat{V}_{8} - \hat{S}_{8})\left(\frac{1}{2} + \frac{it}{2}\right). \label{treeMoebII}
\eeq 
Finally, in  this amplitude the proper time $t$ needs to be
 rescaled in order to agree with the proper time 
 of the other two tree amplitudes $\tilde{\mathcal{A}}$ and 
 $\tilde{\mathcal{K}}$,  $l = t/2$
 \beq
\tilde{\mathcal{M}} = \e 2 \times \frac{N}{2}\int_{0}^{\infty}  \frac{dl}{\hat{\eta}^{8}\left(\frac{1}{2}+ il\right)}
(\hat{V}_{8} - \hat{S}_{8})\left(\frac{1}{2} + il \right). \label{treeMoebIII}
\eeq 
  After this rescaling the right factor of 2 appears in front of $\tilde{\mathcal{M}}$ that takes into account the correct combinatory factor for this diagram with respect to $\tilde{\mathcal{A}}$ and 
 $\tilde{\mathcal{K}}$.
 
\vspace{2 cm}

\begin{figure}  
\begin{center} 
\includegraphics[scale=1, height=12cm]{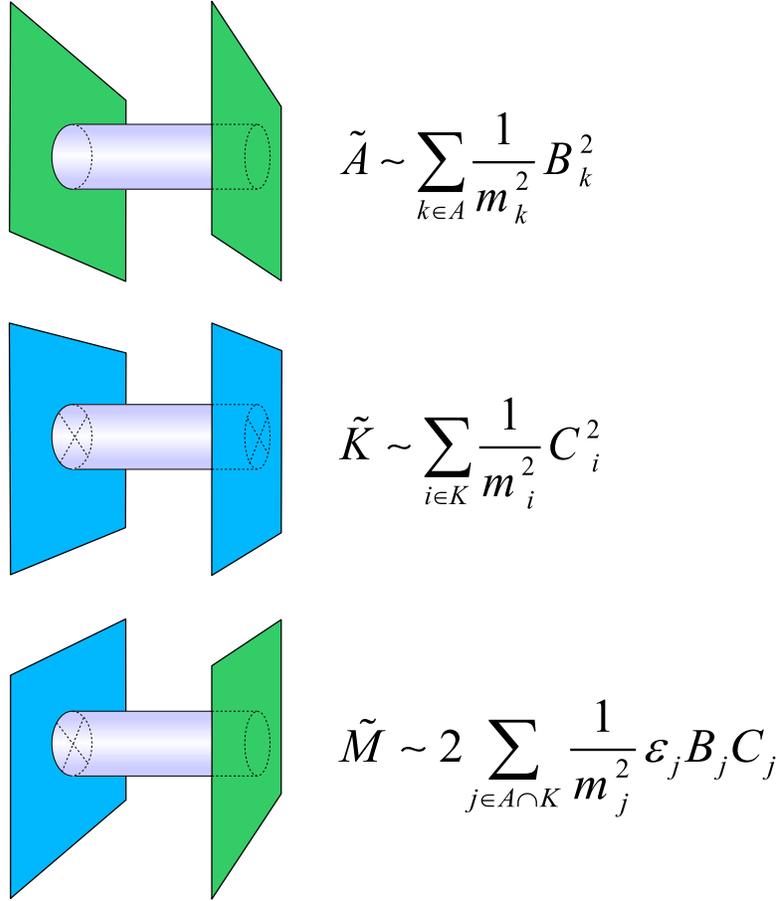}
\caption{The structure of the transverse closed string amplitudes.
 D-branes  are represented   in green  while O-planes  in blue.
$\tilde{\mathcal{A}}$ is the amplitude for a closed string two propagate between two D-branes, while $\tilde{\mathcal{K}}$ is the amplitude for propagation between two O-planes.
$\tilde{\mathcal{M}}$ describes the amplitude for a closed string to propagate between a D-brane
and  a O-plane and therefore it  contains  only the closed string states flowing in \emph{both}
  $\tilde{\mathcal{A}}$ and  $\tilde{\mathcal{K}}$.} 
 \label{transverse}
\end{center}
\end{figure}

    The transverse  M\"obius
 surface, corresponding to a choice of horizontal
 proper time, is a tube terminating with a crosscap and a boundary,
 as it is clear from fig. \ref{Moebiusfig}.
Since the transverse annulus is a tube with two
 boundaries and the transverse klein bottle
  a tube with two crosscap, it is then clear that 
in  $\tilde{\mathcal{M}}$ can flow
 only  those closed string states that are common to the other
 two transerves diagrams, as shown in fig. \ref{transverse}.
 Moreover, the coefficient of the $q = e^{- 2\pi l}$ expansion
 of the amplitude are given by the geometrical
 mean $\sqrt{\mathcal{B}^{2}\mathcal{C}^{2}} = \pm \mathcal{B}\mathcal{C}$,
which is the origin of the overall sign ambiguity taken into account by the sign $\e$.

Also in the case of $\tilde{\mathcal{M}}$  the massless states that flow in
 the diagram generate $NS \otimes NS$ and a RR tadpoles.

\begin{figure}  
\begin{center} 
\includegraphics[scale=1, height=4cm]{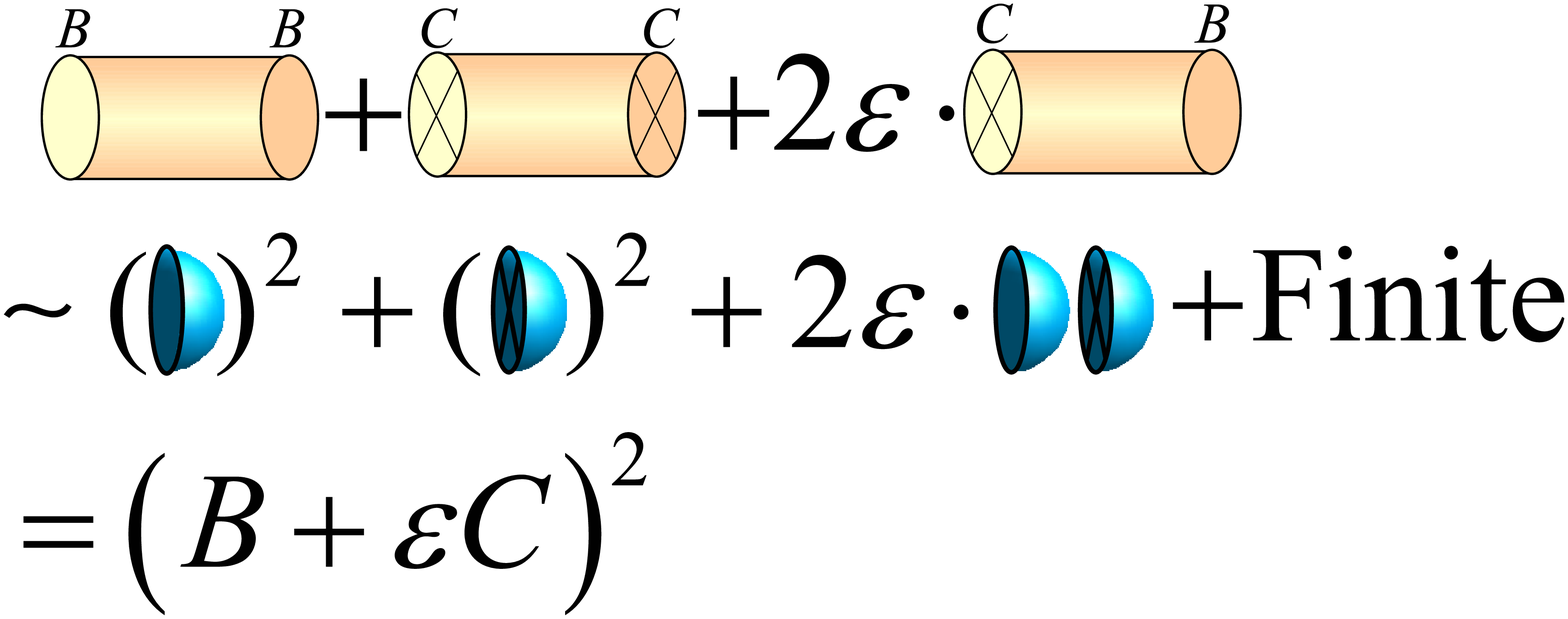}
\caption{Tadpole cancellation is possible among the $g = 1$ transverse diagrams $\tilde{\mathcal{K}}$,
  $\tilde{\mathcal{A}}$ and  $\tilde{\mathcal{M}}$, whose divergent part is due to the propagation
 of closed string massless states.  $\tilde{\mathcal{K}}$ is  proportional to the square of the
   crosscap    reflection coefficient $\mathcal{C}$, while  $\tilde{\mathcal{A}}$ to the square of the boundary reflection coefficient
 $\mathcal{B}$.  $\tilde{\mathcal{M}}$  is proportional to  $2\e \mathcal{C}\mathcal{B}$,
 being the amplitude  for  closed strings to  propagate
in a tube  diagram terminating  with a  boundary  and a  crosscap.  
 The sign ambiguity  $\e$ 
   follows  by extracting the square root of  $(\mathcal{C}\mathcal{B})^{2}$.} 
 
\label{tadpoleconditions}
\end{center}
\end{figure}

We are ready to collect all  tadpole 
contributions from the three diagrams   $\tilde{\mathcal{K}}$,
  $\tilde{\mathcal{A}}$ and  $\tilde{\mathcal{M}}$,
 by considering the constant term in their $q$ expansion, i. e.   the coefficients  in front of the massless states
  flowing in the three diagrams.  Their total contribution both to the   $NS NS$
 and  to the $R R$ sectors, give (see fig. \ref{tadpoleconditions})
\beq
 \left(2^{4} + 2^{-6}N^{2} + \e N \right) = \frac{1}{64}\left(N + \e 32 \right)^{2}.
\eeq

 Therefore the sum of the $NS \otimes NS$  tadpoles contained in the transverse diagrams
 induces a correction to the  flat background given by
\beq
\d S_{\Phi} \sim
             \left(N + \e 32 \right) \int d^{10}x e^{-\Phi} \sqrt{g},  \label{dilatontadpole} 
\eeq
  while RR tadpoles  correct the equation of motion  for $C_{10}$ by adding 
the following term to the Action for the massless modes 
\beq
 S_{C_{10}}  \sim \left(N + \e 32 \right)      \int  C_{10}.
\eeq

This last contribution cannot be cancelled by a modification
 of the background fields and leads to an inconsistency unless $\e = -$ and $N=32$,
 a  condition that in this case actually cancels the dilaton
 tadpole (\ref{dilatontadpole}) as well.

  The cancellations of all the tadpoles is necessary for spacetime
supersymmetry and indeed one can prove that for $\e = -$ and $N=32$ the full spectrum 
enjoies  $N =1$ spacetime supersymmetry.
 This background is given by $D=10$ Minkowski with $16$
 spacefilling  D-9 branes and one O-9 plane.
 $16$ is actually the number of \emph{physical}   D-9 branes
  and not $32$ as one might expect.
  This comes in order to  follow a convention
 consistent with backgrounds with lower dimensional orientifold planes,
 that we shall discuss in the following. In these cases the lower dimensional
planes induce also a reflection symmetry along directions of the target space
 orthogonal to them which forces to introduce image branes in order to respect 
this symmetry. In this case the number of physical D-branes is actually one half
of the factor $N$ in front of the M\"obius.
 
\vspace{2 cm}
 
The configuration of D-9 branes and O-9 planes  preserves one-half  of the original type IIB supersymmetries,
 since under world-sheet parity only the sum of holomorphic and anti-holomorphic 
  supercharges survive but not their difference.
This solution is called type I superstring whose closed massless  spectrum
 comprises in the closed string $NS \otimes NS$   sector the metric $g_{\m\n}$ and the dilaton $\phi$,
 in the $RR$ sector a two form $C_{2}$ and the  (non-dynamical) ten form $C_{10}$.

 Finally,  the  $RNS$ sector yields a gravitino $\Psi_{\a}^{\m}$ and the dilatino $\bar{\zeta}_{\dot{\a}}$.
The number of physical  degrees of freedom in the closed string sectors
can be read by expanding the closed string amplitude
\bea
 \frac{1}{2}\mathcal{T}  + \mathcal{K} &=& \frac{V_{8}(q)V_{8}(\bar{q})+ V_{8}(q\bar{q})}{2} +  \frac{S_{8}(q)S_{8}(\bar{q})- S_{8}(q\bar{q})}{2} - \frac{V_{8}(q)S_{8}(\bar{q})+ S_{8}(q)V_{8}(\bar{q})}{2} \nn \\
 &=& \frac{8^{2} + 8}{2}(q\bar{q})^{0} + \frac{8^{2} + 8}{2}(q\bar{q})^{0}  - \frac{8^{2} + 8}{2}(q\bar{q})^{0}  - \frac{8^{2} + 8}{2}(q\bar{q})^{0} + O(q\bar{q})
 \eea
In the NSNS sector we have
\beq
\frac{V_{8}(q)V_{8}(\bar{q})+ V_{8}(q\bar{q})}{2} = \left( \frac{8^{2} + 8}{2} - 1 \right) + 1 +  O(q\bar{q})
= (35) + (1) + massive
\eeq
 the first contribution $(35)$ are the physical degrees of freedom of the graviton $g_{\m\n}$,
 while the second $(1)$ is the dilaton $\phi$.

In the RR sector
\beq
\frac{S_{8}(q)S_{8}(\bar{q}) -  S_{8}(q\bar{q})}{2} =  \frac{8^{2} - 8}{2} = (28)  
\eeq
 corresponds to the two form $C_{\m\n}$.

 Finally,  the RNS and NSR sectors gives one-half  of the degrees of freedom of
type IIB
\beq
\frac{V_{8}(q)S_{8}(\bar{q})+ S_{8}(q)V_{8}(\bar{q})}{2} = (64) = \bf{56}_{S} + \bf{8}_{S}, 
\eeq
 which indicates the presence of only one right-handed gravitino $\psi^{\m}$ and
one left-handed dilatino $\bar{\z}$, as one can expect from the existence of one spacetime
supersymmetry.

Turning to the open string massless spectrum, one can read the physical degrees of freedom from
\beq
\mathcal{A} + \mathcal{M} = \left(\frac{N^{2}}{2} - \frac{N}{2}\right)V_{8} - \left(\frac{N^{2}}{2} - \frac{N}{2}\right)S_{8}. 
\eeq
In the NS  open sector we have a massless vector in the adjoint of $SO(32)$, while in the R sector
we have the gaugino.
 The massless open string excitations correspond then to $N =(1,0)$ $SO(32)$ pure super Yang Mills
 that comprises  the vector gauge boson $A_{\m}^{a}$ and the gaugino $\l_{\a}^{a}$.
 The full massless spectrum $(g_{\m\n},\phi, C_{\m\n},\psi^{\m}, \bar{\z}, A_{\m}^{a}, \l_{\a}^{a})$
corresponds to type I supergravity plus super Yang-Mills  in $D = 10$, which      is free of gravitational, gauge and mixed anomalies for the gauge group $SO(32)$.

Due to the sign ambiguity  in front of  states flowing in  the
 transverse  M\"obius amplitude one can consider a different solution by relaxing
the NSNS tadpole.
 This corresponds to   a different choice of sign for the NSNS and RR states flowing in the M\"obius
 $\e_{NS} = +$,  $\e_{R} = -$.
 
In this case the background is not supersymmetric due to the uncancelled NSNS tadpole
 and the gauge group becomes $USp(32)$ \cite{Sugimoto:1999tx}.
The NSNS tadpole  generates the dilaton potential
\beq
 V =  \left(N +  32 \right) \int d^{10}x \ \sqrt{g} e^{- \langle \phi \rangle}, 
\eeq
with  a runaway behaviour in the dilaton VEV  $\langle \phi \rangle$.

\newpage
--------------------------------------------------------------------------------------------------------------
\newpage

\chapter{Compactification and Supersymmetric Open String Vacua}

 \everypar{\hspace{-.6cm}}

One of the main tests for string theory is to check whether
 on one of its vacua it provides   excitations 
  that agree with the standard model of particle physics. 
If   future generation high-energy accelerators
  will supply  evidences for the existence
 of  the superpartners, this would represent
  an important indirect indication in favour of superstring theory.

 One of the main problems is vacuum degeneracy. In the compactification scheme
 that we are going to discuss the classical spacetime on which
 the string  propagates is the product of four dimensional
 Minkowski and a compact space.
 The classical spacetime needs to satisfy the string equations of motion
 but besides that, a  dynamical mechanism that produce the background  is not known.
  In this state of affairs,
  the vacuum  is chosen a priori by hand, and 
  this choice 
  is among an incredible huge spectrum of possibilities.
 Some of the vacua resemble the properties and the spectrum of
  the standard model, while other have completely different
 features.
 The Standard Model parameters can be in principle reproduced  by
  stabilising string moduli,   VEVs of background scalar fields,
 that describe the compactification.
 The stabilisation can be achieved, as recent and present investigations
 have been  showing, by intricate combinations of various backgrounds (fluxes) 
 predicted by the superstring. In this way the problem of explaining
 the SM parameters is translated into the problem of explaining
 why the fluxes have the values that reproduce 
correctly  the SM parameters.
 This is another face of the vacuum degeneracy
 problem.

 To find vacua that have features identical to the Standard
 Model is still an open problem and the precise structure
   of the low energy particle world  poses 
   several constraints and non trivial tasks 
that at least one of the  string solutions need to satisfy.

One of the first criteria that one can follow 
is to study vacua that preserve some  supersymmetries,
  since  the absence of quantum effects
 on the  background allow a good analytic control.
Even if the background preserves supersymmetry the choice is still
 among an amazing huge number of possibilities.

In general a  spontaneous breaking of all the  supersymmetries 
  lifts  the moduli by creating a quantum potential,   
  and the effects  of the  quantum fluctuations induce on
the background  are not under control.   
 The background acquires a dynamics such that
 the initial assumptions are no longer true and one should be able
 to have an approximation scheme to describe quantum gravity 
  without spacetime supersymmetry and then, probably
  without
the assumption of  splitting the spacetime into an inert
  flat background and  
  quantum fluctuations.

 

  The  understanding of some nonperturbative properties of string
  theory has enlarged the spectrum of the  possibilities
  for the structure of string vacua.
  Particularly  interesting are the D-p branes that are solitons of the 
  theory, a sort of multidimensional generalisation of  
  magnetic monopoles. They are spacetime defects on which closed strings can
  open up and, quite interestingly  they have  perturbative excitations
  that are open  strings, so they can be thought of  as
  backgrounds, whose configurations can preserve some amount
  of the original supersymmetries of the  \emph{empty} spacetime solutions.  
 
\begin{figure}  
\begin{center} 
\includegraphics[scale=.5, height=8cm]{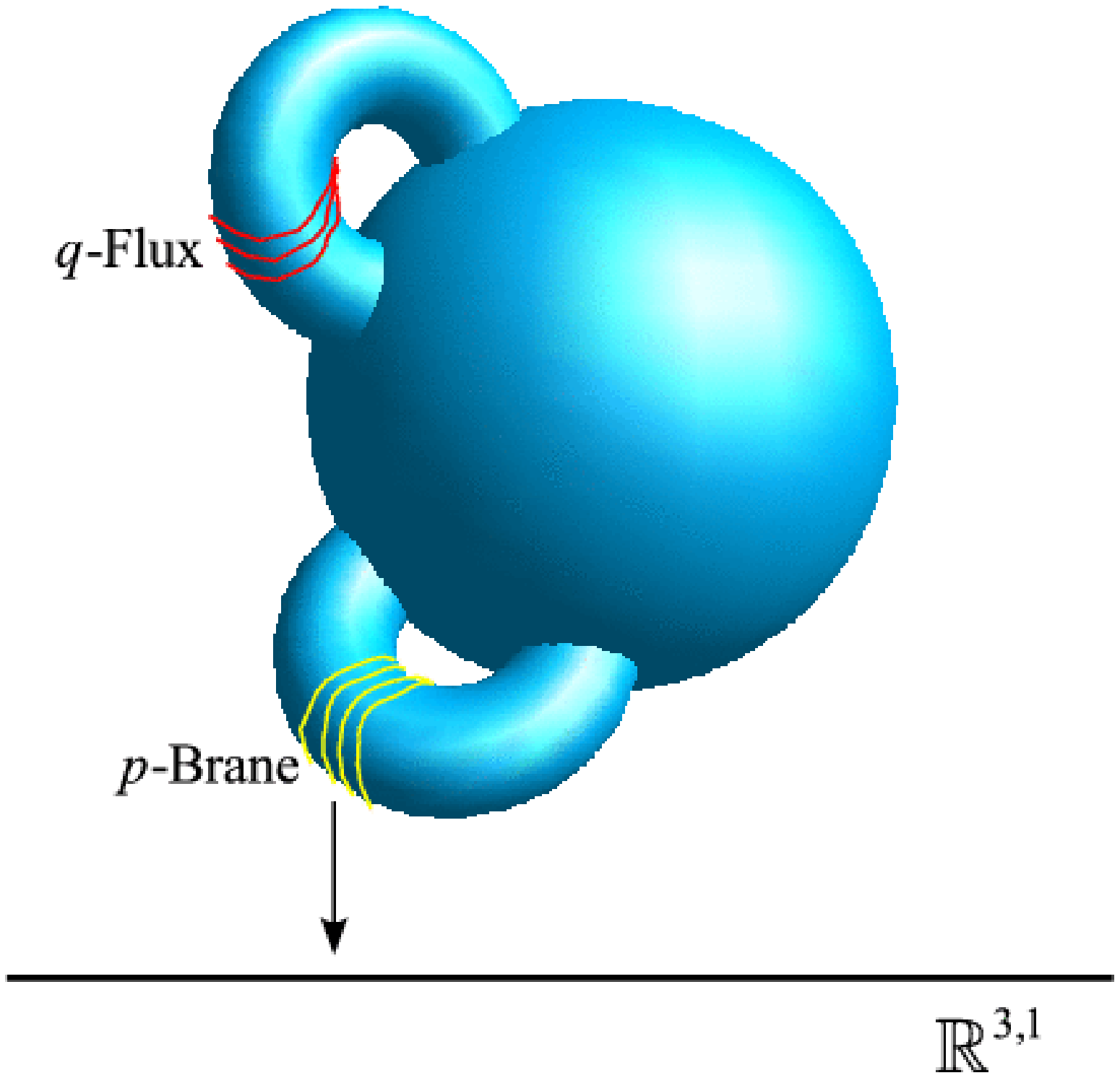}
\caption{} 
\label{compactbranes}
\end{center}
\end{figure}

In models of brane-worlds the branes invade the four extended dimensions, as  
represented in fig. \ref{compactbranes},
 and wrap some cycles of the compact space. 

 The standard model forces are mediated by open strings confined on the 
brane while only  gravity, mediated by closed strings, can experience all the ten spacetime dimensions.
 This has given new possibilies for explaining   the observed hierarchy
 between the electro-weak and the gravitational forces.  
 Gravity is so weak because of the dispersion of its flux lines 
 in a  higher number of dimensions comparing to the other forces.

 As recent observations have confirmed, the Cosmological Constant
 is slightly positive so that, at least in the part of universe
we live in,  the extended space is  a de Sitter spacetime with a  positive curvature.


\vspace{3 cm}

\section{Compactifications}
 \everypar{\hspace{-.6cm}}


Since in general consistent string models require the presence of extra-dimensions
 and, in particular, in its
 perturbative formulation, critical superstrings predict ten space-time dimensions,
  string vacua that claim to describe low energy physics  need to contain compact dimensions.
 Due to the perturbative stability of the ten dimensional supersymmetric  
  vacua, it seems necessary to consider mechanisms that induce a
compactification and, in connection with that,  the breakdown of the original $D = 10$ supersymmetry
 .

Let us for the moment put aside  the question of finding
  mechanisms  that induce compactification of some of the extra dimensions,
 such as those induced by the breaking of supersymmetry, and describe the propagation of
 strings on \emph{a priori}  compactified backgrounds.
 This becomes easier if some of the original ten-dimensional spacetime supersymmetries 
  are preserved in the compactified solution and 
 it is therefore worth to begin with such situation. 

The simplest ansatz for compactification is to consider  backgrounds 
of the form $\mathcal{M}_{d} \otimes \mathcal{X}_{10 - d}$, tensor
product of a $d$-dimensional Minkowski space with a $(10 - d)$-dimensional  compact space.
 Such solutions are generically named compactifications
 to $d$ dimensions.
 For this class  of vacua one can consider the breaking $SO(1,9) \rightarrow SO(1,d - 1)\times SO(10 -d)$
  of the original ten-dimensional Lorentz symmetry
  and a consequent  Kaluza Klein reduction of the fields that describe string excitations.

 With the above breaking of the Lorentz group,
  the range for the indexes of a tensor field 
   pointing  to  compact directions
   counts the number of  copies  of  $SO(1,d - 1)$ fields
 present in the spectrum.
  This dimensional reduction is actually encoded
  by the breaking
of the original ten dimensional spacetime characters into
 sum of  products
 of lower dimensional spacetime and  internal ones,
 that at massless level reproduce the expected group theoretical decomposition
  for the Lorentz irreducible representations.
For a compactification  to $2d + 2$  extended dimensions
 $SO(1,9)\rightarrow  SO(1,2d +1)\times SO(2k)$
 with $2d + 2 + 2k = 10$,
         we have the following decomposition

\bea
O_8  &=&  O_{2d} O_{2k}  +  V_{2d} V_{2k},
\qquad V_8  =  V_{2d} O_{2k} +   O_{2d} V_{2k},                      \nn  \\  
S_8  &=&       S_{2d} C_{2k} +   C_{2d} S_{2k}, 
\qquad C_8  =   S_{2d} S_{2k} +   C_{2d} C_{2k}.  \label{decomposizioni}  
\eea

 
  The  left factors in the above relation
 indicate \emph{spacetime} characters,  while the right
factors are \emph{internal} characters.

 As usual, the lower-dimensional characters are expressed in terms of combinations
 of Jacobi Theta Constants

\bea
O_{2n}  &=&  \frac{\theta^{n}_{3}(0|\tau) + \theta^{n}_{4}(0|\tau)}{2 \eta^{n}},
\qquad V_{2n}  =  \frac{\theta^{n}_{3}(0|\tau) - \theta^{n}_{4}(0|\tau)}{2 \eta^{4}},     \nn  \\  
C_{2n}  &=&   \frac{\theta^{n}_{2}(0|\tau) + i^{-n}\theta^{n}_{1}(0|\tau)}{2 \eta^{4}}, 
\qquad S_{2n}  =   \frac{\theta^{n}_{2}(0|\tau) - i^{-n}  \theta^{n}_{1}(0|\tau)}{2 \eta^{4}}.  
\eea

 One can   verify that at every mass level the decompositions (\ref{decomposizioni})
hold, by  using the representations of the Jacobi functions as infinite products
given in eqs. (\ref{theta3}), (\ref{theta4}), (\ref{theta2}) and (\ref{theta1}).

It is worth to notice that the modular properties of the internal characters
 $(O_{2n} , V_{2n} ,  S_{2n} ,   C_{2n} )$
follow from those of the Theta functions, and are given by the following matrices
\beq
 T = e^{-i\pi/12}diag \left( 1, -1, e^{in\pi/4},  e^{in\pi/4} \right), 
\eeq 
and
\bea
 \mathcal{S}_{2n}  = 
\frac{1}{2} \left( \begin{array}{cccc} 
1 & 1 & 1 & 1  \\ 
1 & 1 & -1 & -1  \\ 
1 & -1 & i^{-n} & -  i^{-n} \\
1 & -1 & - i^{-n} & i^{-n} \end{array} \right). \label{S2n}    
\eea
While for  spacetime characters $(O_{2n} , V_{2n} ,  -S_{2n} ,   -C_{2n} )$, 
 for a proper account of spin-statistic, the correct matrices that realise
 the modular transformations are obtained by interchanging the role of the vector
 $V_{2n}$ and the scalar  $O_{2n}$, as already discussed in section \ref{sectiontoro}
 for the special case of the $SO(8)$
 ten-dimensional characters. One has
\beq
 T = e^{-i\pi/12}diag \left(-1, 1, e^{in\pi/4},  e^{in\pi/4} \right), 
\eeq 
\bea
 \mathcal{S}_{2n}  = 
\frac{1}{2} \left( \begin{array}{cccc} 
1 & 1 & -1 & -1  \\ 
1 & 1 &  1 &  1  \\ 
1 & -1 & i^{-n} & -  i^{-n} \\
1 & -1 & - i^{-n} & i^{-n} \end{array} \right). \label{S2nII}    
\eea

\vspace{.3 cm}
  Compactification implies also a  Kaluza-Klein  reduction 
 for the various fields in the spectrum.
  By  Fourier expanding   a given field, periodic   along the compact directions,
 one obtains  from the lower dimensional point of view    a massless field
  plus an infinite tower of massive ones.

 As an example, after a
 circle compactification $\mathcal{M}_{9} \times S^{1}$ a scalar field $\phi$
\beq
\phi(\mathbf{x},y) = \sum_{m \in \mathbb{Z}} \phi_{m}(\mathbf{x})e^{2\pi i y m/L},
\eeq

 gives rise to a tower of modes  $\phi_{m}(\mathbf{x})$
\beq
 \Box \phi_{m}(\mathbf{x})e^{2\pi i m/L}  = (\Box_{\mathbf{x}} +  \p_{y}^{2})\phi_{m}(\mathbf{x})e^{2\pi i m/L} 
= \left(\Box_{\mathbf{x}} - \left(2\pi  m/L \right)^{2} \right)\phi_{m}(\mathbf{x})e^{2\pi i m/L},
\eeq
 with  masses given by
\beq
M^{2}_{m} = \left(\frac{m}{R}\right)^{2} \qquad m \in \mathbb{Z},
\eeq
$R = L/2 \pi $ being the radius of $S^{1}$.

 A further tower of massive modes are  given by topological non-trivial string configurations
 in which  a closed string winds around a compact direction \cite{narain,cs}.
 Or, for open strings, if two D-branes have a different position on the circle
 such that the string can stretch between them, partially or completely  winding the circle.
 
 In both cases the mass of the topologically non trivial
 modes  is proportional to  an integer number,  the winding   string number and the
effective length of the string.

The zero mode part of the  closed string  coordinate expansion along the circle direction is then
given by
\beq
X_{zm} = x + 2\a' \frac{m}{R}\tau + 2n R \s \qquad m, n \in \mathbb{Z}, \label{zmcircle}
\eeq
where we have changed in the closed string expansion the normalisation in front of the momentum with
respect to eq. (\ref{coordexpansion}) and  we have  halved the closed string length for convenience in later formulae.
To take into account for the change of normalisation it is enough to replace $\a' \rightarrow 4\a'$
in all the previous formulae.
The corresponding   right and left moving zero-mode expansions are
\bea
X_{zm}(\tau - \s) &=& \frac{x}{2} +  \a'\left(  \frac{m}{R} - \frac{n R}{\a'}  \right)(\tau - \s)
 = \frac{x}{2} +   \a'p_{R}(\tau - \s), \nn \\ 
X_{zm}(\tau  + \s) &=& \frac{x}{2} +  \a'\left(  \frac{m}{R} + \frac{n R}{\a'}  \right)(\tau + \s) = 
 \frac{x}{2} + \a'p_{L}(\tau + \s),
\eea
 with  right and left moving momenta
\beq
p_{R} = \frac{m}{R} - \frac{nR}{\a'} \qquad   p_{L} =  \frac{m}{R} + \frac{nR}{\a'}.
\eeq
 The  zero modes therefore give  the following contribution  to the closed string mass formula
\beq
M^{2} = \frac{1}{2}(p_{R}^{2} + p_{L}^{2}) = \frac{m^{2}}{R^{2}} +  \frac{(nR)^{2}}{(\a')^{2}}.
\eeq
It is then  clear the invariance of the closed string spectrum under the
T-duality transformation $ R \rightarrow \a' / R$,
  a symmetry of the full closed string theory \cite{cs}
 that corresponds to the interchange  $\s \leftrightarrow \tau$ between the  two world-sheet coordinates.
 Its effect on  the bosonic coordinates  corresponds to the asymmetric transformation
 $X_{L}^{\m} \rightarrow X_{L}^{\m}$,  $X_{R}^{\m} \rightarrow - X_{R}^{\m}$, 
which leaves the components of the stress tensor invariant.
 This same operation induces a reflection on the right-mover   world-sheet
  fermions, thus changing the chirality of the  spacetime fermionic
 states in the holomorphic  Ramond sector.
 Therefore  a  T-duality
 exchanges type IIA  and type IIB in  nine dimensions.
 However, in the case of circle compactification and for the more general
case of higher-dimensional tori, the spacetime spinors
  get decomposed into lower dimensional  spinors of both  chiralities
and therefore type IIA and type IIB after compactification give rise to the same spectrum.
 
In the presence of D-branes, a T-duality along a direction exchanges
    Neumann and Dirichlet boundary conditions \cite{dlp,horava2,green}, thus 
increasing or decreasing by one the dimensionality of a Dp-brane \cite{pcj},
 whether the T-dualised coordinate
is orthogonal or parallel  to it.
The same phenomenon of dimensional transmutation occurs also for  orientifold planes,
  whose action on the spectrum of
 string  states is turned  by a  T-duality,  into  a combination of
  world-sheet parity $\Omega$ and spacetime reflection $I: X \rightarrow - X $ along the T-dualised direction $X$ 
 \beq
 \Omega \stackrel{T}{\rightarrow} I \Omega.
\eeq

\vspace{3 cm}

\section{Circle Compactifications}
\everypar{\hspace{-.6cm}}
We now turn  to study some properties of closed string theory
  after compactification, by   computing the torus amplitude for
  type IIB  on the circle $S^{1}$. 
   
 The general form for the torus amplitude    eq.(\ref{bostorus}) is given by  
\beq
 \mathcal{T} = \int_{\mathcal{F}}
 \frac{d^{2}\tau}{\tau_{2}}
    Tr_{\mathscr{H}}  \left( q ^{\frac{1}{4}L_{0}} \bar{q}^{\frac{1}{4}\bar{L}_{0}}  \right), \nn \\ 
\eeq
but  in this case
 along the compact coordinate the momenta $p_{R}$ and  $p_{L}$ are quantised.

Therefore  performing  the trace over the ten dimensional  momentum $p$

\beq
 q ^{\frac{1}{4}L_{0}} \bar{q}^{\frac{1}{4}\bar{L}_{0}} = e^{-\pi \tau_{2}\frac{1}{2}(L_{0} + \bar{L}_{0})}  e^{i\pi \tau_{1}\frac{1}{2}(L_{0} - \bar{L}_{0})},
\eeq

\beq
 e^{- \frac{\pi \tau_{2}\a'}{2}(p^{2}_{L} + p^{2}_{R})} 
 e^{\frac{ i\pi \tau_{1} \a'}{2}(p^{2}_{L} - p^{2}_{R})}= 
 e^{-\pi \tau_{2} \a'\left(\frac{m^{2}}{R^{2}} +  \frac{(nR)^{2}}{(\a')^{2}} \right)} 
   e^{ 2 i\pi \tau_{1}mn},
\eeq

one needs to replace the Gaussian integral 
 for the compact direction  with a discrete sum over the Kaluza-Klein $m$
 and winding $n$ quantum numbers 
\beq
\int \frac{dp}{2\pi} e^{-\pi \tau_{2} \a'p^{2}} \rightarrow \sum_{m\in \mathbb{Z}}\sum_{n \in \mathbb{Z}} e^{-\pi \tau_{2} \a'\left(\frac{m^{2}}{R^{2}} +  \frac{(nR)^{2}}{(\a')^{2}} \right)} 
   e^{ 2 i\pi \tau_{1}mn}. \label{latticesum} 
\eeq
The torus amplitude therefore reads
\bea
\mathcal{T} &=& \int_{\mathcal{F}}\frac{d^{2}\tau}{\tau_{2}^{6} (\eta \bar{\eta})^{8}}|V_{8} - S_{8}|^{2}\ \cdot  \sqrt{\tau_{2}} \sum_{m\in \mathbb{Z}}\sum_{n \in \mathbb{Z}} e^{-\pi \tau_{2} \a'\left(\frac{m^{2}}{R^{2}} +  \frac{(nR)^{2}}{(\a')^{2}} \right)} 
   e^{ 2 i\pi \tau_{1}mn} \nn \\  &=& 
 \int_{\mathcal{F}}\frac{d^{2}\tau}{\tau_{2}^{6} (\eta \bar{\eta})^{8}} |V_{8} - S_{8}|^{2} \cdot  \sqrt{\tau_{2}}\sum_{m,n} \Lambda_{m,n}, \label{toruscircle} 
\eea
where 
\beq
\sum_{m,n}\Lambda_{m,n} = \sum_{m, n \in \mathbb{Z}} e^{-\pi \tau_{2} \a'\left(\frac{m^{2}}{R^{2}} +  \frac{(nR)^{2}}{(\a')^{2}} \right)}e^{2i\pi \tau_{1} mn},\label{Lambdalatticesum} 
\eeq
 indicates a double lattice sum.

To check modular invariance, it is easy to see that   $T: \tau_{1} \rightarrow  \tau_{1} + 1$ 
 leaves the above lattice sum invariant, while to check invariance under  $S: \tau \rightarrow - 1/ \tau $ one needs to use the Poisson resummation formula
\beq
\sum_{m \in \mathbb{Z}} e^{-\pi m^{2}a + 2\pi i b} = \frac{1}{\sqrt{|a|}} \sum_{\m \in \mathbb{Z}} e^{-\pi a^{-1}(\mu + b)^{2}}.\label{Poisson}
\eeq
 For example a resummation  on $m$ gives
\beq
\sum_{m,n} \Lambda_{m,n} = \frac{R}{\sqrt{\a ' \tau_{2}}} 
\sum_{\m,n}  e^{-\frac{\pi R^{2}}{\a ' \tau_{2}}|n\tau+\m |^{2}}. \label{mulatticesum}
\eeq
 A  generic modular transformation in $PSL(2, \mathbb{Z})$, represented  by the integral 
matrix 
 \bea
M_{n,\mu}  =  \left(\ba{cc} a &  b \\ n & \mu  \ea \right),    
\eea
transforms the torus parameter as
\beq
\tau \rightarrow \frac{a\tau + b}{n\tau + \mu},
\eeq
  and thus  the imaginary part $\tau_{2}$ transforms as 
\beq
\tau_{2} \stackrel{M_{n ,\mu}}{\longrightarrow} \frac{\tau_{2}}{|n\tau+\m |^{2}}.
\eeq

 We can then  rewrite the lattice sum (\ref{mulatticesum}) as

\beq
\sum_{m,n} \Lambda_{m,n} = \frac{R}{\sqrt{\a ' \tau_{2}}} 
\sum_{\m,n}  e^{-\frac{\pi R^{2}}{\a ' \tau_{2}}|n\tau+\m |^{2}}
 =  \frac{R}{\sqrt{\a ' \tau_{2}}}\sum_{M_{n,\mu}}  e^{-\frac{\pi R^{2}}{\a 'M_{n,\mu}\cdot \tau_{2}}}, 
\eeq
  so that  the torus amplitude (\ref{toruscircle}) becomes
\beq
 \mathcal{T} =  \frac{R}{\sqrt{\a '}} \int_{\mathcal{F}}\frac{d\tau^{2}}{\tau_{2}^{6} (\eta \bar{\eta})^{8}}
|V_{8} - S_{8}|^{2}   \ \cdot  \sum_{M_{n,\mu}}  e^{-\frac{\pi R^{2}}{\a 'M_{n,\mu} \tau_{2}}}.
\eeq
 From the previous equation it becomes quite visible the modular invariance of the integrand,
since  it  contains the image   $M_{n,\mu} \tau_{2}$ of the imaginary
part of the torus modulus under the full modular group.

Moreover, one can use the fact that the above amplitude contains a sum over the terms
 $M_{n,\mu} \tau_{2}$ to unfold the integration domain from the fundamental region $\mathcal{F}$
 to the half-strip $\mathcal{S} = \left[-1/2, 1/2 \right] \times [0, \infty )$.
 This last region is the image of the fundamental domain through
the modular group, modulo  unitary translation on the real axis, i.e. \cite{apostol}
\beq
 \mathcal{S} = PSL(2,\mathbb{Z})(\mathcal{F})/T.
\eeq

\begin{figure}  
\begin{center} 
\includegraphics[scale=1, height=5cm]{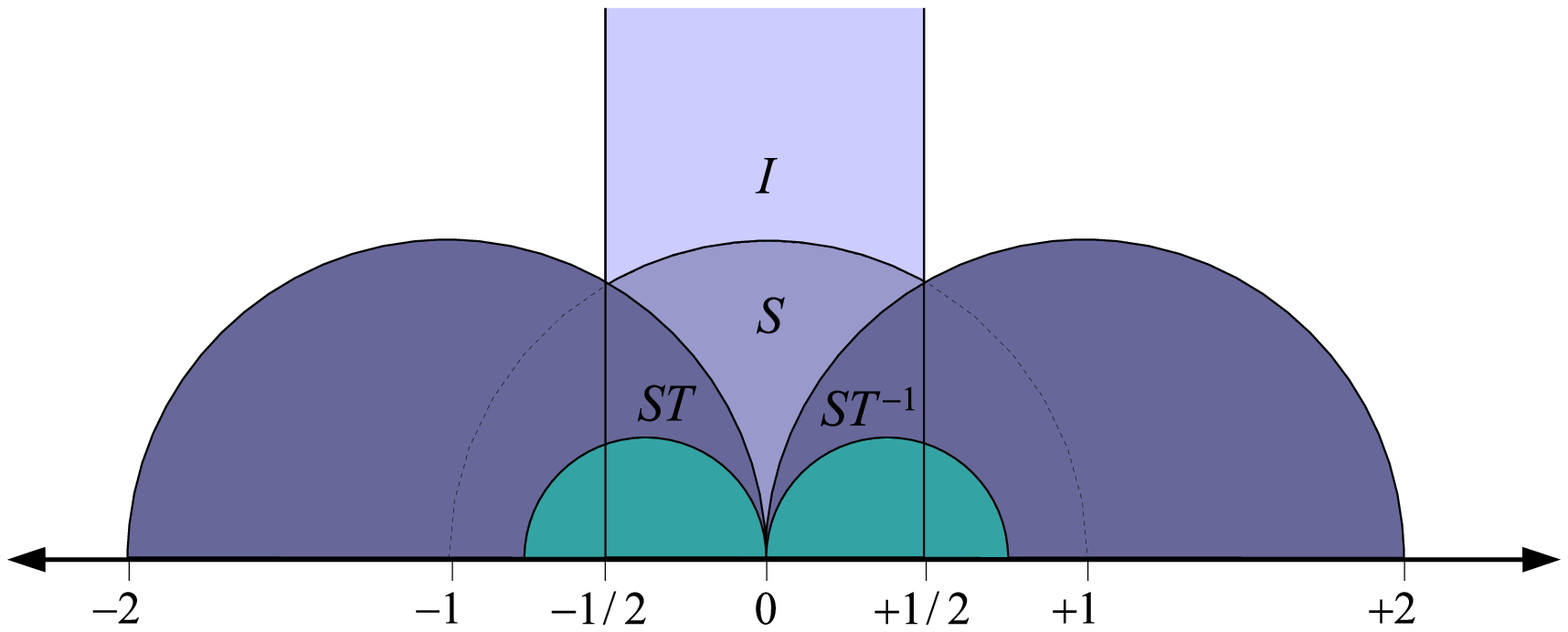}
\caption{  }
\label{}
\end{center}
\end{figure}

\bea
 \mathcal{T} &=&  \frac{R}{\sqrt{\a '}} \int_{\mathcal{F}}\frac{d\tau^{2}}{\tau_{2}^{6} (\eta \bar{\eta})^{8}}|V_{8} - S_{8}|^{2}\ \cdot  \sum_{M_{n,\mu}}  e^{-\frac{\pi R^{2}}{\a 'M_{n,\mu} \tau_{2}}} \nn \\
&=&  \frac{R}{\sqrt{\a '}}   \int_{\cup M_{n, \mu}(\mathcal{F})}\frac{d\tau^{2}}{\tau_{2}^{6} (\eta \bar{\eta})^{8}} |V_{8} - S_{8}|^{2} \ \cdot e^{-\frac{\pi R^{2}}{\a ' \tau_{2}}} \nn \\
&=&  \frac{R}{\sqrt{\a '}} \int_{-1/2}^{1/2} d\tau_{1} \int_{0}^{\infty} \frac{d\tau_{2}}{\tau_{2}^{6} (\eta \bar{\eta})^{8}} |V_{8} - S_{8}|^{2}\ \cdot e^{-\frac{\pi R^{2}}{\a ' \tau_{2}}}.
\eea
In this way we have reduced the integration domain for the torus amplitude
into a simpler one, which allows a  direct computation of the integral.
 Of course, for this supersymmetric case this reduction is not of big interest
since the integrand is  itself vanishing as a result of Jacobi Abstrusa Identity.
However we will come back in the following  to this \emph{unfolding technique}\cite{O'Brien:1987pn,Dixon:1990pc,Mayr:1993mq,Borunda:2002ra,Trapletti:2002uk}
 for non-supersymmetric vacua where it gives a way to compute  one-loop torus potentials.
   
This last expression shows in particular the presence of an exponential factor
 that acts as a regulator  by cutting off the  UV closed string modes
  in the region $\tau_{2} < \pi R^{2}/ \a'$.

\vspace{3 cm}
\section{Rational Conformal Field Theory Point}
\everypar{\hspace{-.6cm}}
Let us return to the general torus amplitude for the circle compactification.
We want to show an interesting phenomena that occurs for a particular value
 of the compactification radius and, for more general compactifications,
for particular values of the background fields
\beq
\mathcal{T} = \int_{\mathcal{F}}\frac{d^{2}\tau}{\tau_{2}^{6} (\eta \bar{\eta})^{8}}|V_{8} - S_{8}|^{2}\ \cdot  \sqrt{\tau_{2}} \sum_{m\in \mathbb{Z}}\sum_{n \in \mathbb{Z}}
 q^{ \a'\left(\frac{m}{R} -  \frac{nR}{\a'}\right)^{2}} 
\bar{q}^{ \a'\left(\frac{m}{R} +  \frac{ nR}{\a'}\right)^{2}}. \label{toruscirc} 
\eeq
If one considers a special value for the background metric, the self-dual radius $R = \sqrt{\a'}$,
 the lattice sum  becomes
\beq
 \sum_{m\in \mathbb{Z}}\sum_{n \in \mathbb{Z}} q^{\frac{1}{4}(m + n)^{2}}\bar{q}^{\frac{1}{4}(m - n)^{2}},
\eeq
 that can be rewritten as
\beq
|\sum_{r}q^{r^{2}}|^{2} + |\sum_{r}q^{(r + 1/2)^{2}}|^{2} = |\chi_{0}|^{2} + |\chi_{1/2}|^{2},
\eeq
 with $\chi_{0}$ and $\chi_{1/2}$ the  characters relative to the scalar and
the spinor $SU(2)$ conjugacy classis.
 At the self-dual radius the conformal field theory becomes \emph{rational} \cite{Verlinde:1988sn},
 in the sense that the torus partition function contains a finite number of characters.
 Actually, this same phenomenon occurs  for a generic  \emph{rational} value of the radius in
 string unites  $R = \sqrt{\a'}p/q$, for coprime integers $p$ and $q$,
while for an  irrationl value of the radius  (\ref{toruscirc})
 contains an infinite number of characters. 
 
 Similar facts  occour for compactification on 
 higher dimensional tori, where  infinite sets   of rational points  can be acheived.
  For example, if   the  background metric $G_{ij}$ is the Cartan matrix of the enanchement
symmetry group and the $B_{ij}$ is the adjacency matrix, as discussed in sec. \ref{A Prototype Six-Dimensional Example}.
 We will come back in the next chapter  to the
 argument where, in order to illustrate 
  a novel mechanism for supersymmetry breaking,
   we will consider  a $T^{4}$ compactification
   in the  $SO(8)$ conformal rational point.
 
\vspace{3 cm}

\section{Circle Compactification with D-8 Branes and O-8 Planes in the Background} 
\begin{figure}  
\begin{center} 
\includegraphics[scale=1, height=5cm]{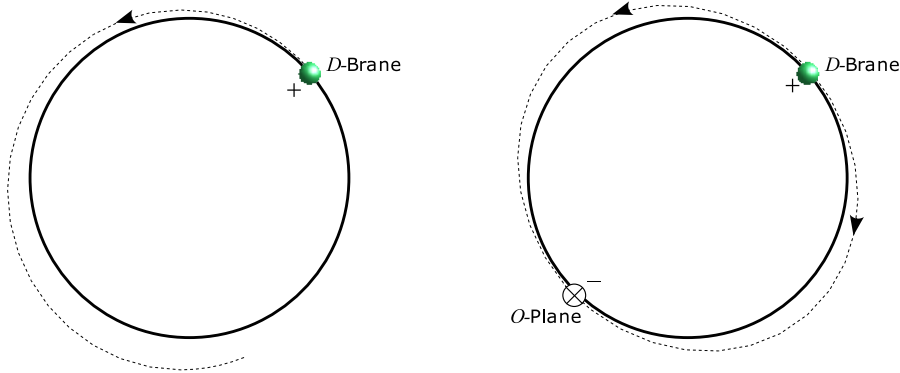}
\caption{ On the left side: A  single D-brane (in green) transverse to a compact direction
  violates the p-form  equation of motions since its Faraday flux lines need to close
to a negative charged object of the same dimension. If an orientifold plane is also present,
 as shown in the right side, the configuration can respect gauge invariance by guaranteeing
 a total vanishing RR charge.  Whenever the total charge is zero, the  massless
 spectra of D-brane 
excitations coupled to  closed strings is free of gravitational, gauge and mixed anomalies.}
\label{Faraday}
\end{center}
\end{figure}

\everypar{\hspace{-.6cm}}
We want now to study the closed and open string spectra \cite{Bianchi:1991eu}
 for  backgrounds corresponding to a  circle compactification $\mathcal{M} \times S^{1}$ 
 with D8 branes orthogonal to the compact direction.
 This corresponds to type IIA with D8 branes connected by T-duality to type IIB with D9
 branes compactificated on the dual circle. 
 However, the only presence of  D8 branes violates the string equations of motions,
 since the RR Faraday flux lines of the charged  D8 brane that extend along the compact direction
  need  to close  on a negatively charged object as shown in fig. \ref{Faraday}.

 One is therefore led to consider O-8 planes, which in turn impose the involution
$ I \cdot \Omega$ on the spectrum of  string excitations, $I$ being a reflection along the compact coordinate.
To take into account of the effects of the O-8 planes,
 one can insert a projector in the trace that computes the one loop closed string amplitudes
\beq
 Tr_{\mathscr{H}}\left[\left(\frac{1 + I\Omega}{2}\right)  q ^{\frac{1}{4}L_{0}} \bar{q}^{\frac{1}{4}\bar{L}_{0}} \right] = \frac{1}{2}\mathcal{T} + \mathcal{K},
\eeq
with the Klein Bottle amplitude given by
\beq
\mathcal{K} = \frac{1}{2} \int_{0}^{\infty} \frac{d\tau_{2}}{\tau_{2}^{11/2}\eta^{8}(q\bar{q})}(V_{8} - S_{8})(q\bar{q})W_{n}.
\eeq
 Now the lattice sum      is restricted only to  winding states
\beq
 W_{n} = \sum_{n \in \mathbb{Z}} e^{-\pi \tau_{2}   \frac{(nR)^{2}}{\a'}}=
 \sum_{n \in \mathbb{Z}}(q\bar{q})^{\frac{1}{4\a'}n^{2}R^{2}},
\eeq
  indeed the only states allowed to flow in  $\mathcal{K}$  by the involution  $I\Omega$,
that selects $p_{L} = - p_{R}$.
\vspace{.5 cm}

\begin{figure}  
\begin{center} 
\includegraphics[scale=1, height=7cm]{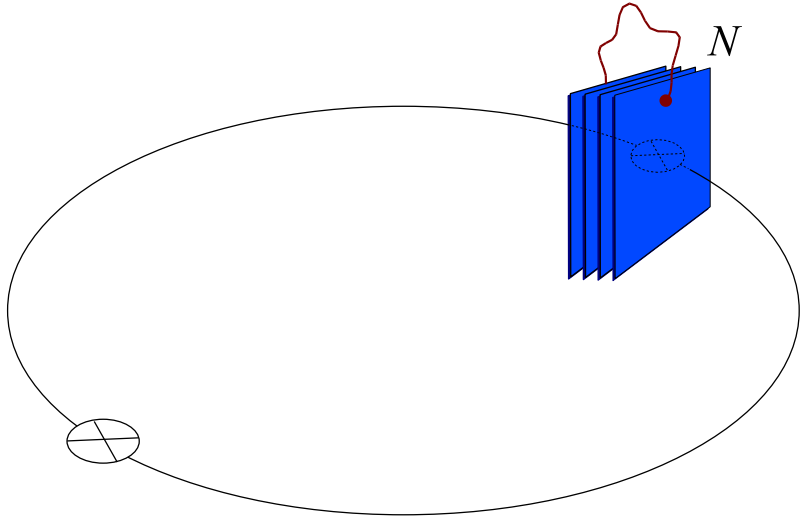}
\caption{A stack of $N$ D-8 branes orthogonal to the circle. Open strings can completely wind
 around the circle giving rise to a tower of massive states, encoded in the lattice sum
 in the annulus amplitude (\ref{anellowinding}).}
\label{onestack}
\end{center}
\end{figure}

 The  change of integration variable $\tau_{2} = 1/2l$ gives the transverse Klein Bottle
 amplitude 
\beq
\tilde\mathcal{K} = \frac{2^{9/2}}{2} \int \frac{dl \ l^{7/2}}{\eta(i/l)^{8}}(V_{8} - S_{8})(il)
 \sum_{n \in \mathbb{Z}}e^{-\pi    \frac{(nR)^{2}}{2l\a'}},
\eeq
  which after a  Poisson resummation on the lattice sum
\beq
 \sum_{n \in \mathbb{Z}}e^{-\pi    \frac{(nR)^{2}}{2l\a'}} = 
\frac{\sqrt{2\a' l}}{R}\sum_{m \in \mathbb{Z}}e^{- \frac{2\pi l\a'}{R^{2}} m^{2}},
\eeq
and  by using the modular property of the eta function $\eta(- 1/ \tau) = \sqrt{ - i \tau_{2}}\eta(\tau)$,     can be rewritten as follows
 
\beq
\tilde\mathcal{K} = \frac{2^{5}}{2} \int \frac{dl}{\eta(il)^{8}}(V_{8} - S_{8})(il)
 \sum_{m \in \mathbb{Z}}e^{- 2\pi l \frac{\a'}{4} \frac{(2m)^{2}}{R^{2}}} = 
  \frac{2^{5}}{2} \int \frac{dl}{\eta(il)^{8}}(V_{8} - S_{8})(il) P_{2m}. \label{circlektilde}
\eeq    
The open string excitations  for a single stuck of D-8 branes  on top of 
  an O-8 plane
 gives rise to the following  annulus amplitude
\beq
\mathcal{A} =  \frac{N^{2}}{2}\int \frac{d \tau_{2}}{\tau_{2}^{11/2}\eta^{8}(i\tau_{2}/2)}
(V_{8} - S_{8})(i\tau_{2}/2)\sum_{n}q^{\frac{1}{2\a'}n^{2}R^{2}}, \label{anellowinding}
\eeq
where the  lattice states correspond to open strings that wind
 around the circle, as shown in figure \ref{onestack}.

  The different normalisation in the lattice sum takes into account the 
different overall coefficient in the open string mass formula. 
With the  change of integration variable $\tau_{2} = 2/l$ we can obtain the cylinder
amplitude for  closed strings to propagate between two D-8 branes
\bea
\tilde{\mathcal{A}} &=& \frac{N^{2}}{2}2^{-9/2} \int  \frac{dl \ l^{7/2}}{\eta(i/l)^{8}}(V_{8} - S_{8})(il)
 \sum_{n} e^{-\frac{2\pi}{l}\frac{n^{2}R^{2}}{\a'}} \nn \\ 
 &=&  2^{-5}\frac{\sqrt{\a'}}{R}  \frac{N^{2}}{2} \int \frac{dl}{\eta(il)^{8}} 
(V_{8} - S_{8})(il)\sum_{m} e^{-2\pi l \frac{\a'}{4}\frac{m^{2}}{R^{2}}} \nn \\ 
 &=& 2^{-5}\frac{\sqrt{\a'}}{R}  \frac{N^{2}}{2} \int \frac{dl}{\eta(il)^{8}} 
(V_{8} - S_{8})(il)\sum_{m} e^{-2\pi l \frac{\a'}{4}\frac{m^{2}}{R^{2}}} \nn \\
 &=& 2^{-5}\frac{\sqrt{\a'}}{R}  \frac{N^{2}}{2} \int \frac{dl}{\eta(il)^{8}} \label{Anellocircle}
(V_{8} - S_{8})(il)P_{m}.  
\eea
The loop M\"obius strip amplitude is then  given by the  symmetrisation under $I\Omega$
 of the open string spectrum
\beq
\mathcal{M} = - \frac{N}{2}\int \frac{d \tau_{2}}{\tau_{2}^{11/2}\hat{\eta}^{8}(i\tau_{2}/2 + 1/2)}
(\hat{V}_{8} - \hat{S}_{8})(i\tau_{2}/2 + 1/2 )\sum_{n}e^{-\pi \tau_{2}\frac{1}{\a'}n^{2}R^{2}},
\eeq 
while the transverse $\tilde{\mathcal{M}}$ corresponds to the amplitude for closed strings
to propagate between a D-8 brane and a O-8 plane and is obtained from (\ref{Anellocircle})
 by a change of integration variable $\tau_{2} = 1/t$
\bea
\mathcal{M} &\rightarrow& - \frac{N}{2}\int \frac{d t \ t^{7/2}}{\hat{\eta}^{8}(i/2t  + 1/2)}
(\hat{V}_{8} - \hat{S}_{8})(i /2t  + 1/2 )
 \sum_{n} e^{-\frac{\pi}{t}\frac{n^{2}R^{2}}{\a'}} \nn \\
 &=& - \frac{N}{2}\int \frac{d t \ t^{7/2}}{t^{4} \hat{\eta}^{8}(it/2  + 1/2)}
(\hat{V}_{8} - \hat{S}_{8})(i t/2  + 1/2 )
   \frac{\sqrt{t}R}{\sqrt{\a'}}  \sum_{m} e^{-\frac{\pi t}{t}\frac{\a'}{R^{2}}m^{2}}  \nn \\
& =&  -  \frac{R}{\sqrt{\a'}} \frac{N}{2}\int \frac{d t}{\hat{\eta}^{8}(it/2  + 1/2)}
(\hat{V}_{8} - \hat{S}_{8})(i t/2  + 1/2 )
   \sum_{m} e^{- \pi t\frac{\a'}{R^{2}}m^{2}},
\eea
followed by the rescaling $t = 2l $  of the horizontal proper time 
\bea
\tilde{\mathcal{M}} &=&  -  2\frac{R}{\sqrt{\a'}} \frac{N}{2}\int \frac{d l}{\hat{\eta}^{8}(il  + 1/2)}
(\hat{V}_{8} - \hat{S}_{8})(il  + 1/2 )
   \sum_{m} e^{- 2\pi l \frac{\a'}{4}\frac{(2m)^{2}}{R^{2}}} \nn \\
 &=&  -  2\frac{R}{\sqrt{\a'}} \frac{N}{2}\int \frac{d l}{\hat{\eta}^{8}(il  + 1/2)}
(\hat{V}_{8} - \hat{S}_{8})(il  + 1/2 )
   \sum_{m} q^{\frac{\a'}{4}\frac{(2m)^{2}}{R^{2}}}.
\eea
 As usual, the factor of two in the above relation  gives the correct multiplicity for the
 transverse diagram, while the normalisation coefficient comes out to be correctly  
the geometric mean between those in $\tilde{\mathcal{A}}$ and $\tilde{\mathcal{K}}$.

It is worth to notice that for consistency only those closed string states that can flow \emph{both}
in $\tilde\mathcal{K}$ and  $\tilde\mathcal{A}$ contribute to this diagram.

Tadpole conditions can be read, as usual, by summing the coefficient 
 in front of the massless states,  flowing in the transverse diagram.
  Also in this case the cancellation of both  RR and NSNS tadples
asks for $N = 32$, so that the gauge group is $SO(32)$, as for the uncompactified
type I solution.

\section{Open String Wilson Lines}
\label{section Wilson Lines}
\everypar{\hspace{-.6cm}}
  Circle compactification corresponds to a non-vanishing VEV for the internal component
 of the metric $\langle G_{99} \rangle = R^{2}$, which is called  a closed string \emph{modulus},
 that
connects continuously  a set of supersymmetric vacua preserving the original gauge group.  
  When D-branes are present one has also the option of switching on
 a VEV for  the internal components of  the gauge boson  $\langle A_{9}^{a} \rangle$ \cite{Bianchi:1991eu}.
 These scalars control
the displacement of D-8 branes along  the circle \cite{horava2,pcj}  and, if they point into the Cartan subalgebra
of the gauge group,  \footnote{Strictly speaking open string Wilson lines  are the corresponding
 effect on the T-dual picture, first analysed in \cite{Bianchi:1991eu}, of what we are describing in this section.
 After a T-duality along the circle coordinate, one recovers a circle compactification with D-9 branes.
 Due to the non-trivial topology of the circle, even pure gauge configuration for a  U(1)
 potential in the background  are able to affect the quantised momenta along the compact direction.
 The Wilson lines are the gauge invariant quantity  $W_{q} = e^{iq\oint dx^{i}A_{i}}$.
 For a pure gauge background $A = -i\Lambda^{-1}\p_{y}\Lambda = \theta/ 2\pi R$, corresponding
 to the choice $\Lambda = e^{-i \theta y/ 2\pi R}$, the Wilson line is $W_{q}= e^{iq\theta}$.
  The momenta are given by $p = \frac{m}{R} + \frac{q\theta}{2\pi R}$, this shift  in the T-dual  picture 
 that we are considering in this section,  translates geometrically into 
a partial wrapping of the open string on the circle due to brane displacement.}, 
  they do not break supersymmetry since the  effective-action  superpotential
is proportional to the commutator of two gauge bosons.

\begin{figure}  
\begin{center} 
\includegraphics[scale=1, height=8cm]{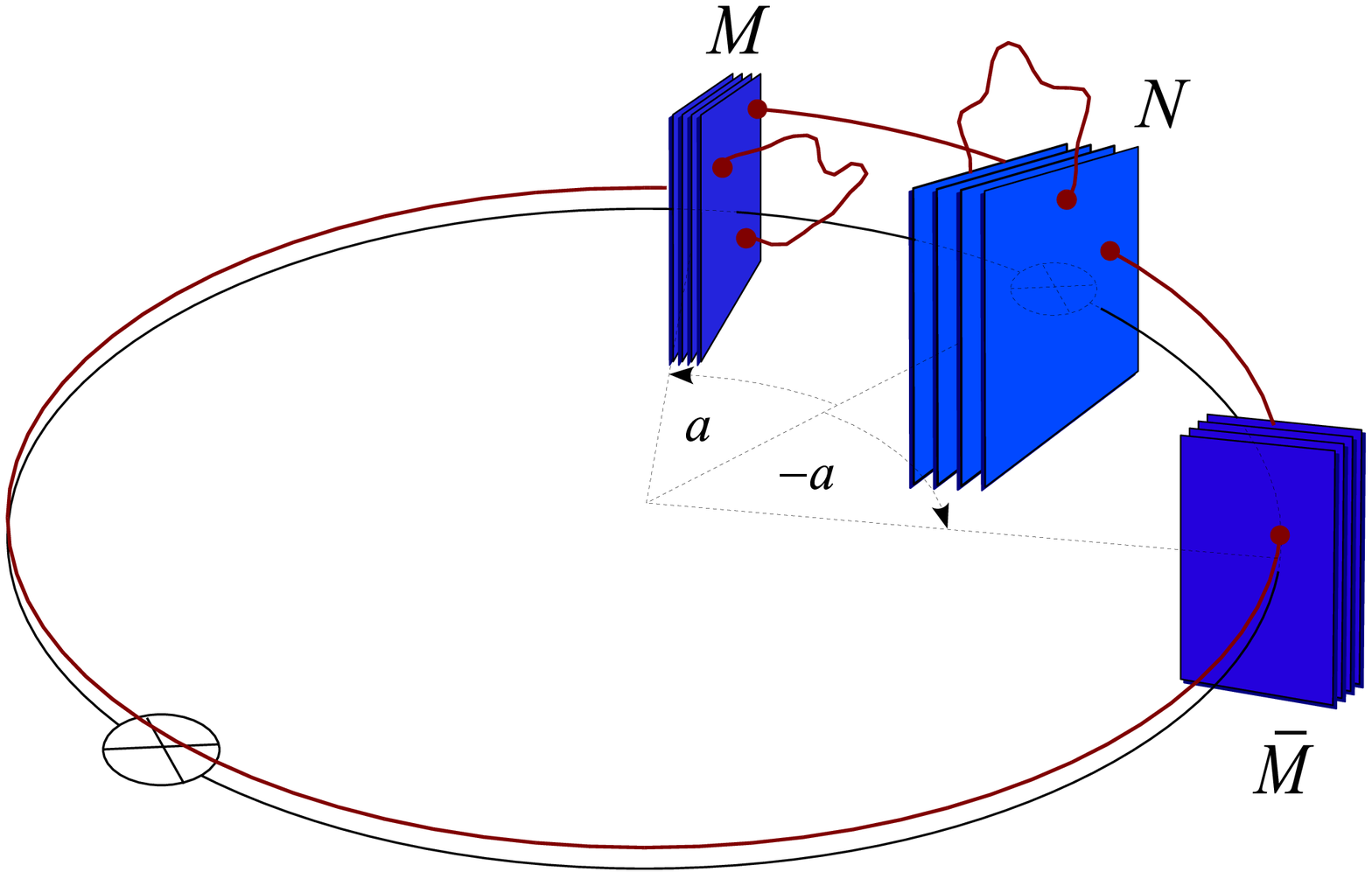}
\caption{}
\label{continuosWL}
\end{center}
\end{figure}

For the configuration in fig. \ref{continuosWL},  the Annulus loop amplitude is
\beq
\mathcal{A} = (V_{8} - S_{8})\left[ \left(\frac{N^{2}}{2} + M\bar{M}\right)W_{n}+
 NMW_{n + a}    + N\bar{M}W_{n - a} +  \frac{M^{2}}{2}W_{n - 2a}  + \frac{\bar{M}^{2}}{2}W_{n - 2a}   \right], \label{Awilsonline}  \eeq
where,  to lighten the notation we omit the integral over the annulus modulus $i\tau_{2}$
in the above equation.

 The  M\"obius loop  diagram is obtained by taking the unoriented open string states
in the previous amplitude
\beq
\mathcal{M} = (\hat{V}_{8} - \hat{S}_{8}) \left[\frac{N}{2} W_{n} + \left( \frac{M}{2}W_{n + 2a}
   + \frac{\bar{M}}{2} W_{n - 2a} \right) \right]. \label{Mwilsonline} 
\eeq 

The corresponding  closed string amplitudes $\tilde{\mathcal{A}}$
 and  $\tilde{\mathcal{M}}$     are given by a proper  change of integration
variables. 
The cylinder amplitude reads
\beq
\tilde{\mathcal{A}} = \frac{2^{-5}\sqrt{\a'}}{2R} (V_{8} - S_{8})\left(N + Me^{2\pi i a m} 
+ \bar{M}e^{-2\pi i a m}\right)^{2}P_{m},  
\eeq
  showing clearly the displacement of the D-8 branes on the circle \footnote{In this case
 the background value $\langle A_{9} \rangle = a/R$ controlls the branes displacement
 on the circle through the scalar parameter $a$, which plays the role of an Higgs field,
though for this supersymmetric vacua there is a flat potential associated to it.
 Notice how the phases that appears in the Annulus amplitude are powers of the
 U(1)-gauge invariant   Wilson line $e^{2\pi i a}$, defined in the previous footnote.}
 (see fig. \ref{continuosWL}).

The transverse M\"obius amplitude
\beq
\tilde{\mathcal{M}} = \frac{\sqrt{\a'}}{R} (\hat{V}_{8} - \hat{S}_{8})\left(N + Me^{4\pi i a m} 
+ \bar{M}e^{-4\pi i a m}\right)P_{2m}  
\eeq
 contains  closed string states with  even $m$, in agreement with
 $\tilde{\mathcal{K}}$ in (\ref{circlektilde}).

In particular,  massless states flowing in the open string diagrams
can be read from (\ref{Awilsonline}) and (\ref{Mwilsonline}) by selecting the
 the constant  terms in the $q$ expansion of the characters 
\bea
\mathcal{A}_{0} &=& (8 - 8)\left( \frac{N^{2}}{2} + M\bar{M} \right) \nn \\
\mathcal{M}_{0} &=& -(8 - 8)\frac{N}{2}.
\eea

This shows  that  the original gauge group $SO(32)$ has been broken to
 $SO(N)\times U(M)$ with $N + 2M = 32$, in a way that preserves the total rank,
since the total number of D-8 branes needed  to cancel
the total tension and charge of the O-8 planes remains the same.
 This phenomenon  represents a nice  geometrical realisation for a
 supersymmetry preserving Higgs mechanism: some of the originally massless gauge bosons
 have acquired a mass due to the minimal stretching induced by the brane displacement, as shown in the 
following figures.


\begin{figure}  
\begin{center} 
\includegraphics[scale=1, height=7cm]{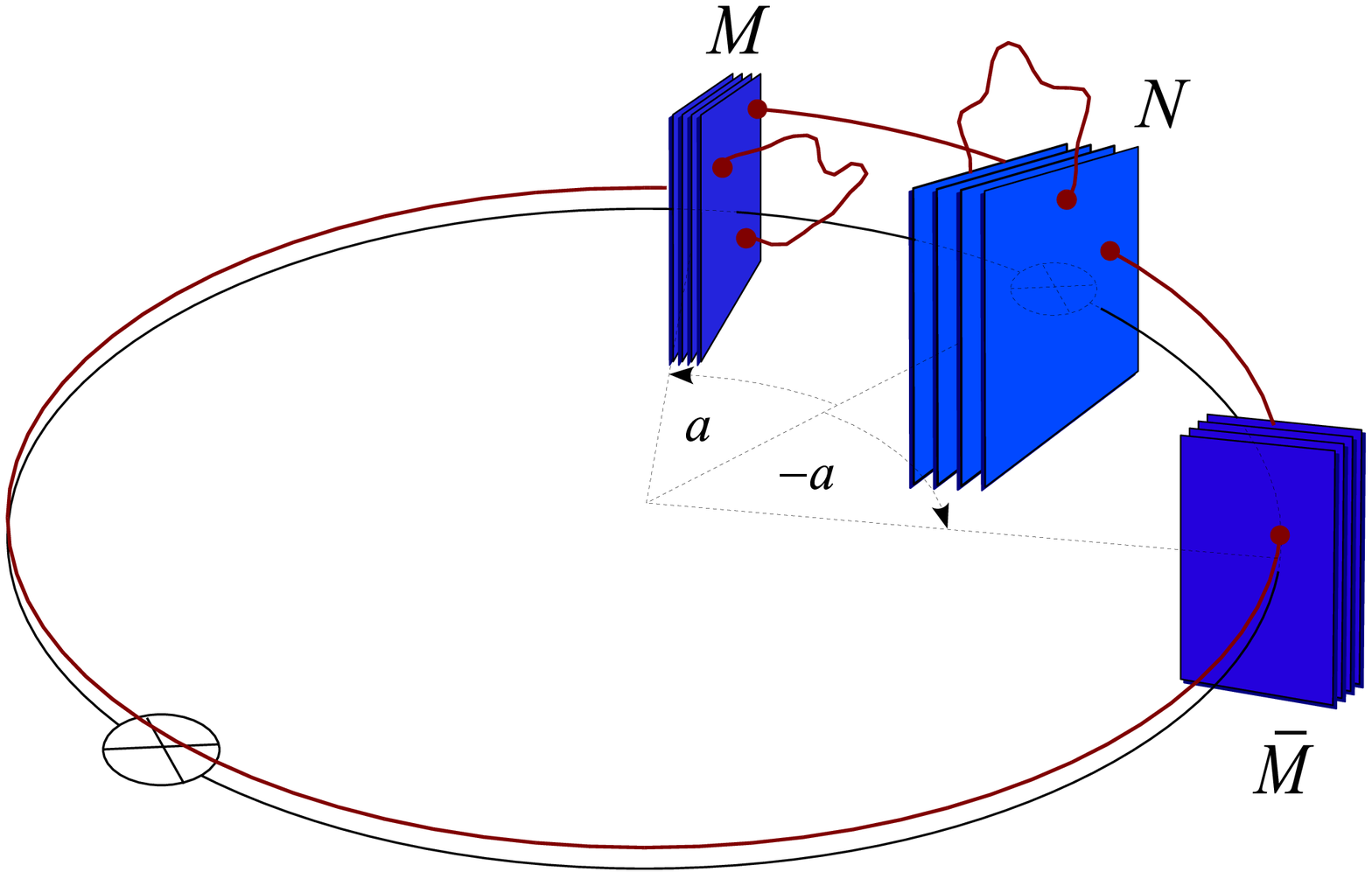}
\caption{A D-brane realisation of the Higgs mechanism:  brane displacement is controlled
 by the internal component of the gauge vector along  Cartan directions (Higgs fields).
 If some brane is moved away from the original stuck, the open strings whose endpoints
 are on  this brane and on  a brane at the original position acquire a mass due to its minimal stretching.
As a consequence the original tadpole-free gauge group is spontaneously broken
 $SO(32) \rightarrow SO(N)\times U(M)$, with $N + 2M = 32$ and $M = \bar{M}$.
 However, since this is a supersymmetric deformation there is not a potential 
 for the Higgs fields, which thus  represent  flat directions in the moduli space.}
\end{center}
\end{figure}

\vspace{3 cm}                                                                                                               
\section{Higher Dimensional Tori and $B_{ij}$ NSNS  Discrete Moduli}
\label{sectionDiscreteModuli}
\everypar{\hspace{-.6cm}}
 We now  turn  to compactifications on higher  dimensional tori $T^{d}$, $d > 1$,
    corresponding  to the case of  non vanishing VEVs 
  for  the internal background metric  components $G_{ij}$, $i,j = 1,...,d$,   
   describing the shape and the size of the target torus.
 Non vanishing   $B_{ij}$ background values have also non-trivial effects 
on the zero modes of the closed string compact coordinate expansion.
In the presence of orientifold planes some of the internal components of the background
fields become discrete moduli giving rise to interesting 
effects on the open string spectra \cite{Bianchi:1991eu,ssethi,bianchitor,wittor,Angelantonj:1999jh,discrete,Kakushadze:1998bw}.

\vspace{.2 cm}
We choose  internal coordinates along  the two  homology cycles of the torus
$X^{i} \sim X^{i} + 2\pi$, of course only in the case the  cycles
are mutual orthogonal, the torus metric $G_{ij}$ is diagonal in
 this coordinate system. With this choice  of torus 
coordinates   we know that the 
conjugate momenta are quantised, being integer numbers.  
The part of the world-sheet action that involves the compact bosonic coordinates $X^{i}$
reads
\beq
I = \frac{1}{2\pi \a'} \int d \tau \int_{0}^{\pi}d\s \left( G_{ij}\p^{\a}X^{i}\p_{\a}X^{j} + B_{ij}\p_{\a}X^{i}\p_{\b}X^{j}\e^{\a\b} \right).    
\eeq
One can make an ansatz for the zero mode in the internal coordinate expansion \cite{narain}
\beq
X^{i}_{ZM} = 2n^{i}\s + q^{i}(\tau)
\eeq
such that the  same action restricted  to the zero modes becomes
\beq
 \frac{1}{2\a'} \int  d\tau \left( G_{ij}\dot{q}^{i}\dot{q}^{j} + 2B_{ij}\dot{q}^{i}n^{j} -  4G_{ij}n^{i}n^{j} \right),
\eeq
with momenta given by
\beq
p_{i} =    G_{ij}\dot{q}^{j} +  2B_{ij}\dot{q}^{j} = m_{i} \in \mathbb{Z}.
\eeq

Therefore 

\beq
\dot{q}^{i}(\tau) =  G^{ij}(m_{j} - 2 B_{ik}n^{k}),
\eeq

and we obtain 
\beq
X^{i}_{ZM} =  x^{i} + \frac{1}{\a'} G^{ij}(m_{j} - 2 B_{ik}n^{k})\tau + 2n^{i}\s.
\eeq
 The left and right moving momenta are then

\bea
p^{i}_{L} &=&   G^{ij}(m_{j} - B_{jk}n^{k}) + n^{i} \nn \\
p^{i}_{R} &=&   G^{ij}(m_{j} + B_{jk}n^{k}) - n^{i}
\eea
i.e.
\bea
p_{i,L} &=&  m_{i} +   (G_{ik} -  B_{ik})n^{k}  \nn \\
p_{i,R} &=&  m_{i} -   (G_{ik} +  B_{ik})n^{k}. \label{PLPR}
\eea
In this case the computation of the torus amplitude gives
\beq
\mathcal{T} = \int_{\mathcal{F}}\frac{d^{2}\tau}{\tau_{2}^{6} (\eta \bar{\eta})^{8}}|V_{8} - S_{8}|^{2}\ \cdot \tau_{2}^{d/2} \sum_{\vec{m},\vec{n}} \Lambda_{\vec{m},\vec{n}}^{(d)},
\eeq
where the  sum is now extended to a d-dimensional lattice
\beq
 \Lambda_{\vec{m},\vec{n}}^{(d)}(\tau) = \sum_{\vec{m},\vec{n}}q^{\frac{1}{4\a'}P^{2}_{R}}
\bar{q}^{\frac{1}{4\a'}P^{2}_{L}}.\label{dlattice}
\eeq

 The simplest  case where we can observe the effects of a non-vanishing 
$B$ NSNS modulus corresponds to a compactification over a two-torus $T^{2}$ in  the presence
of D-7 branes transverse to the compact directions.
O-7 planes, necessary to neutralise the D-7  RR charge, induce a $I_{2}\Omega (-)^{F_{L}}$
involution on the closed and open  string spectra, where $I_{2}$ is a reflection along the
 two compact   coordinates, transverse
to the O-7 planes, and $F_{L}$ is the left-moving spacetime fermion numbers, which
 is necessary in order to have a $\mathbb{Z}_{2}$ involution \cite{dpSO,Sen:1996vd}.

 The closed string amplitudes are computed by
\beq
 Tr_{\mathscr{H}}\left[\left(\frac{1 + I_{2}\Omega(-)^{F_{L}}}{2}\right)  q ^{\frac{1}{4}L_{0}} \bar{q}^{\frac{1}{4}\bar{L}_{0}} \right] = \frac{1}{2}\mathcal{T} + \mathcal{K}.
\eeq

In particular the combined effect of $\Omega$ and $I_{2}$  imposes the following
 condition involving the left and right momenta 
 \beq
P_{L,i} = - P_{R,i}
\eeq
for the states that can flow in the Klein bottle, which via eq.(\ref{PLPR})  gives  
\beq
B_{ij}n^{j} = b\e_{ij}n^{j} =  m_{i}.
\eeq
From the above condition we see that $B$ can  only be  integer or half-integer,
in this last case  the $n^{i}$ need to be even.
\vspace{.5 cm}

We see therefore that due to the presence of the O-7 planes the $B_{ij} = b\e_{ij}$ NSNS field
becomes a discrete modulus \cite{Bianchi:1991eu,ssethi,bianchitor,wittor}.
There are  two possible inequivalent values for $b$, $b = 0$ and $b = 1/2$,
 since the mass formula is invariant under the transformation $b \rightarrow b + 1$.

\begin{figure}  
\begin{center} 
\includegraphics[scale=1, height=8cm]{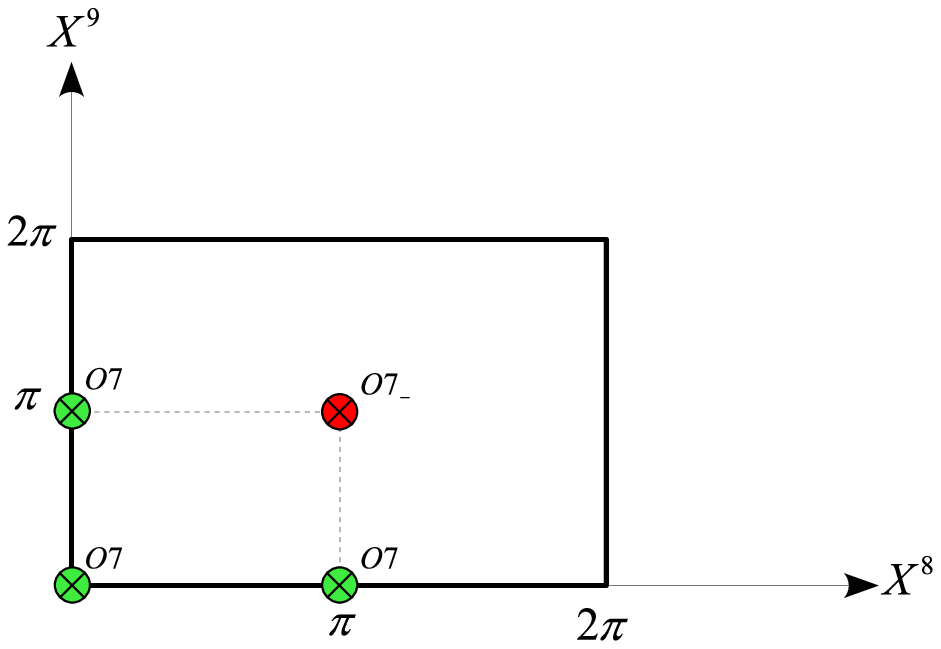}    
\caption{Three $O7_{+}$ planes (in green) with negative tension and charge occupy three of the four  fixed point of the involution $I_{2}$,
while an exotic   $O7_{-}$ plane (in red) with \emph{positive} tension and charge is located in the 
fourth fixed point.}                       
\label{O7distribution}
\end{center}
\end{figure}

Let us consider the case of non vanishing   $b = 1/2$.
The Klein bottle amplitude is given by 
\beq
\mathcal{K} = \frac{1}{2}(V_{8} - S_{8})W_{2n_{1},2n_{2}}, 
\eeq
where to lighten the notation we did not include the integral over the modulus
 of the double covering torus.
A change in the integration variable gives $\tilde{\mathcal{K}}$,  the
amplitude for closed string states to propagate between two O-7 planes

\bea
\tilde{\mathcal{K}} &=&  \frac{2^{5}\a'}{2v_{2}}(V_{8} - S_{8})\frac{1}{4}P_{m_{1},m_{2}} \nn \\
 &=&  \frac{2^{3}\a'}{2v_{2}}(V_{8} - S_{8})\left( \frac{1 + (-)^{m_{1}} + (-)^{m_{2}} -  (-)^{m_{1} +  m_{2}}}{2}\right)^{2}  P_{m_{1},m_{2}}, \label{tildeKB}
 \eea
 that is obtained  via a Poisson resummation
 on the lattice sum, after which the volume   $v_{2}$ of the $T^{2}$ appears.
 In the second line the quantity that looks like a projector actually 
is a fancy way to write $1$, but it is quite instructive since it 
 displays  neatly the geometry of the O-7 planes, 
 which  in the compact space correspond to the four fixed point of the involution $I_{2}: X^{i} \rightarrow
- X^{i}$\\
$(0,0)$, $(0,\pi)$, $(\pi,0)$ and $(\pi,\pi)$. The minus sign infront of the $(-)^{m_{1} + m_{2}}$
 also indicates that the O-plane in  $(\pi,\pi)$ has a reversed tension and charge  with respect
to the other  three O-planes \cite{wittor}, as shown in figure (\ref{O7distribution}).

\vspace{.2 cm}
We consider first  a single stack of D-7 branes  on the $(0,0)$
 fixed point and lately the interesting effects arising after brane displacement on different
fixed points.

\begin{figure}  
\begin{center} 
\includegraphics[scale=1.5, height=7cm]{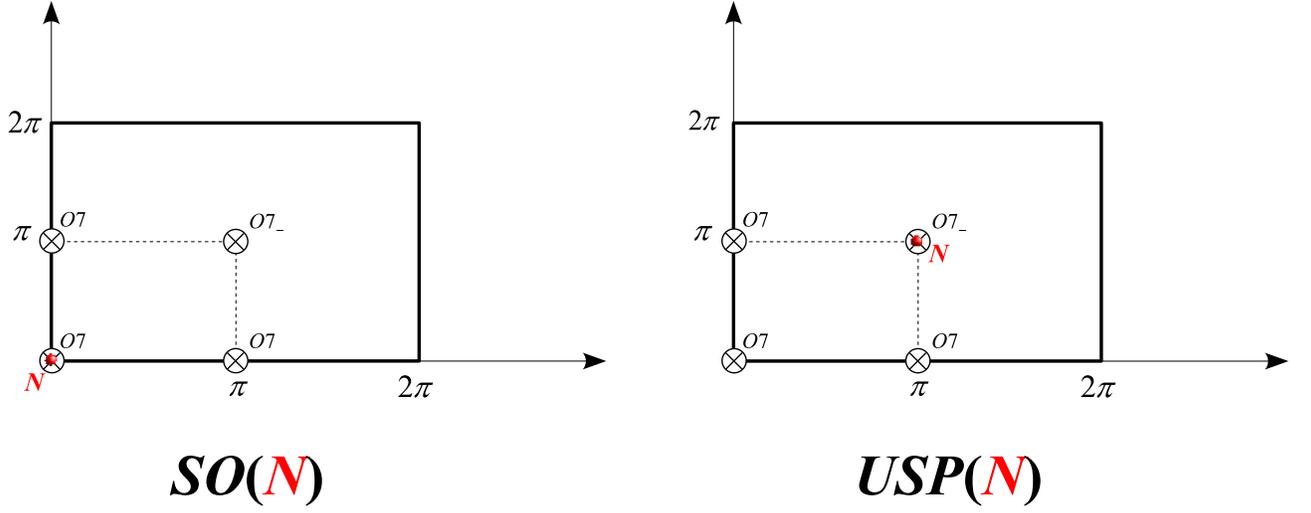}
\caption{Configuration of O-planes at the fixed points of the $I_{2}$ involution. On the left side
a single stuck of D-7 (red point) is on top of a O7 plane, the gauge group for the open string
living on this stuck is $SO(N)$.  
On the right side
 the same stuck is on top of an exotic $O7_{-}$ plane, with  postive tension and charge and
 thus the gauge group is $USp(N)$.} 
\label{discretewilsonlines}
\end{center}
\end{figure}

  In this first case, (represented in the left side of fig.\ref{discretewilsonlines}) ,  the Annulus amplitude is given by
\beq
\mathcal{A} = \frac{N^{2}}{2} (V_{8} - S_{8})W_{n_{1},n_{2}}  \label{A00B}
\eeq
In the transverse channel the cylinder represent the amplitude for closed strings to
propagate between two D-7 branes
\beq
\tilde{\mathcal{A}} = 2^{-5}\frac{\a' }{v_{2}} \frac{N^{2}}{2} (V_{8} - S_{8})P_{m_{1},m_{2}}, 
\eeq

while the transverse M\"obius $\tilde{\mathcal{M}}$ gives  the amplitude 
 for closed string states to propagate between a D-7 brane and a O-7 plane
 and therefore it contains only the closed string states that flow  \emph{both} in
 $\tilde{\mathcal{K}}$ and  $\tilde{\mathcal{A}}$, with reflection coefficients
 in front of the on-shell propagators given  by the geometric mean between
those of the other two transverse amplitudes
\beq 
\tilde{\mathcal{M}} = - 2 \frac{\a' }{v_{2}}\frac{N}{4}\left(\hat{V}_{8} - \hat{S}_{8}\right)\left( \frac{1 + (-)^{m_{1}} + (-)^{m_{2}} -  (-)^{m_{1} +  m_{2}}}{2}\right) P_{m_{1},m_{2}}.
\eeq
From this last amplitude we can read a quite interesting phenomenum that is originated by 
the presence of a non vanishing $B$ discrete NSNS $B$ modulus.
 Namely the O-7 plane located at $(\pi,\pi)$ has a reversed tension and charge ( fig. \ref{O7distribution})
 respect to the other three O-7 planes,
  as displayed by the negative sign in front the phase $(-)^{m_{1} +  m_{2}}$.
 In fact  $\tilde{\mathcal{M}}$  is the amplitude for propagation of closed string 
between a D-brane and a O-plane thus sensitive in the NSNS sector  to the relative
 sign between the tension of these objects, while in the RR sector to the sign between their charge.

 The total tension and charge of the O-7 planes is therefore lower than for a vanishing $B = 0$,
a fact that can be read from the coefficient in front of the closed string
states flowing in the transverse Klein bottle  eq.(\ref{tildeKB}),
that in this case reads
\bea
\tilde{\mathcal{K}}_{0}  &=&    \frac{2^{3}\a'}{2v_{2}}(V_{8} - S_{8})\left( \frac{1 + (-)^{m_{1}} + (-)^{m_{2}} -  (-)^{m_{1} +  m_{2}}}{2}\right)^{2}  P_{m_{1},m_{2}}, \qquad  (m_{1},m_{2}) = (0,0) \nn \\
 &=&  \left(\frac{2 \sqrt{\a'}}{\sqrt{v_{2}}} \right)^{2}(8 - 8). 
\eea
 Therefore tadpole cancellation now requires
\beq
\tilde{\mathcal{K}}_{0} + \tilde{\mathcal{A}}_{0} + \tilde{\mathcal{M}}_{0} = \left( 2^{2} + 2^{-6}N^{2}
- \frac{N}{2} \right)\frac{\a'}{v_{2}}(8 - 8)
 \eeq
 i. e. $N = 16$.

\vspace{.2 cm}

 For a vanishing   $B = 0$    in $\tilde{\mathcal{K}}$ flow only even K.K. closed string states
 
\bea
\tilde{\mathcal{K}}_{0}  &=&    \frac{2^{3}\a'}{2v_{2}}(V_{8} - S_{8})\left( \frac{1 + (-)^{m_{1}} + (-)^{m_{2}} +  (-)^{m_{1} +  m_{2}}}{2}\right)^{2}  P_{m_{1},m_{2}}, \qquad  (m_{1},m_{2}) = (0,0) \nn \\
 &=&  \left(\frac{2^{2}\sqrt{\a'}}{\sqrt{v_{2}}} \right)^{2}(8 - 8), 
\eea 
so that  tadpole cancellation
\beq
\tilde{\mathcal{K}}_{0} + \tilde{\mathcal{A}}_{0} + \tilde{\mathcal{M}}_{0} = \left( 2^{4} + 2^{-6}N^{2}
- N \right)\frac{\a'}{v_{2}}(8 - 8)
 \eeq
 requires  $N = 32$.

Of course the reduction of the  rank of the tadpole-free    gauge group is  the effect
of a lower NSNS and RR charge due to the presence of the inverted orientifold plane.

 By changing the proper time in the M\"obius we recover the loop open string amplitude
\beq
\mathcal{M} = -  \frac{N}{2}\left(\hat{V}_{8} - \hat{S}_{8}\right)\left( W_{2n_{1},2n_{2}}+ W_{2n_{1} + 1,2n_{2}}+ W_{2n_{1}, 2n_{2}+ 1} -  W_{2n_{1}+1, 2n_{2}+ 1 } \right).
\eeq
From this expression and the Annuls vacuum amplitude  (\ref{A00B}) 
we can read $N^{2}/2 - N/2$ states for the gauge bosons so that in this case the gauge group
is $SO(16)$.

Let us now  consider the case where a single stack of D-7 branes sits instead on the $(\pi,\pi)$
 O-7 plane that has reversed tension and charge, as shown in the right side of figure
\ref{discretewilsonlines}.

In this case the transverse Annulus reads

\beq
\tilde{\mathcal{A}} = 2^{-5}\frac{\a' }{v_{2}} \frac{(N (-)^{m_{1} + m_{2}})^{2}}{2} (V_{8} - S_{8})P_{m_{1},m_{2}}, 
\eeq
with the factor $(-)^{m_{1}+m_{2}}$ in the  square coefficient taking into account 
  the position of the stack on the  $(\pi,\pi)$
point.

In this case,  there is a consequent change  in the reflection coefficients of $\tilde{\mathcal{M}}$
\bea 
\tilde{\mathcal{M}} &=& - 2 \frac{\a' }{v_{2}}\frac{N}{2}\left(\hat{V}_{8} - \hat{S}_{8}\right)(-)^{m_{1} + m_{2}}   \left( \frac{1 + (-)^{m_{1}} + (-)^{m_{2}} -  (-)^{m_{1} +  m_{2}}}{2}\right) P_{m_{1},m_{2}} \nn \\
 &=& - 2 \frac{\a' }{v_{2}}\frac{N}{2}\left(\hat{V}_{8} - \hat{S}_{8}\right)  \left( \frac{ - 1 + (-)^{m_{1}} + (-)^{m_{2}} +  (-)^{m_{1} +  m_{2}}}{2}\right) P_{m_{1},m_{2}}, 
\eea
since the coefficients in front of the propagators needs to be the geometrical
means of those of the other two transverse amplitudes.

The vacuum open string amplitude $\mathcal{M}$ in this last case then reads
\beq
\mathcal{M} = -  \frac{N}{2}\left(\hat{V}_{8} - \hat{S}_{8}\right)\left( - W_{2n_{1},2n_{2}}+ W_{2n_{1} + 1,2n_{2}}+ W_{2n_{1}, 2n_{2}+ 1} +  W_{2n_{1}+1, 2n_{2}+ 1 } \right),
\eeq
From this   expression and the Annuls vacuum amplitude  (\ref{A00B}) 
we can read $N^{2}/2 + N/2$ states for the gauge bosons so that  the gauge group
is $USp(16)$.
Clearly one can obtain a generic gauge group $SO(N) \times USp(2M)$ with $N + 2M = 16$
by displacing two stacks of branes, one  on a traditional O-7 plane and the other
on an inverted O-7 one.   More in general, one can obtain also 
   unitary gauge groups if some of the branes are not on top of orientifold planes, 
as discussed in the section dedicated to the Wilson Lines.

\vspace{.2 cm}

To summarise, the presence of a non-vanishing NSNS $B_{ij}$ has the effect of inverting
tension and charge of some of the orientifold planes.  On  branes that
 are on top of   reverted planes live a  $USp$ gauge group, while when  branes
 are on top of traditional O-planes, (with
negative tension and charge),  $SO$ gauge groups emerge \cite{Bianchi:1991eu}.

All these configurations are connected continuously by varying the internal component
of the vector open-string background, i.e. by moving the branes on different fixed
points through continuous Wilson lines.
 The total tension and charge of the orientifold planes has been lowered by the 
$B$ background and so the total rank of the gauge group.

For a generic $T^{d}$ compactification with a non-vanishing $B_{ij}$ of rank $r$,
 the total rank of the gauge group is lowered to $2^{4 - r/2}$.
 All these effects  were first observed in the context of rational toroidal compactification
 and named  originally  \emph{Discrete Wilson lines} \cite{Bianchi:1990tb}. We will come back
on the subject in the next chapter when we will  employ a similar mechanism
 for breaking supersymmetry without generating a vacuum energy \cite{Angelantonj:2004cm}.

\vspace{3 cm}


\newpage

\chapter{Mechanisms for Supersymmetry breaking}

\section{Brane Supersymmetry Breaking}
\everypar{\hspace{-.6cm}}
Let us start with $D =10$ uncompactified spacetime and study the supersymmetry breaking effect
induced by the presence of $\bar{D9}$ anti-branes in the background.
 Suppose we have $n$ $D9$ branes $(+,+)$ and $\bar{n}$  $\bar{D9}$ $(+,-)$,
where the first sign denotes the tension of the object while the second its charge.
 
The relative signs for tensions and charges between  D-branes and orientifold planes
  is measured by $\tilde{\mathcal{M}}$, the transverse amplitude for closed strings
 to propagate between a  D-brane and an O-plane \cite{bsbII}

\beq
\tilde{\mathcal{M}} =  \frac{2}{2}\left[ (n + \bar{n})\e_{NS}\hat{V}_{8} - (n  - \bar{n})\e_{R}\hat{S}_{8}\right]. \label{Mzoo}
\eeq
The reversed sign in front of $\bar{n}$ in the R sector  with respect to the corresponding type I amplitude,
takes into account  the opposite charge of the  $\bar{D9}$ with respect to a $D9$. 

 With the different choices for the signs $\e_{NS}$ and $\e_{R}$ one can take
into account the presence of all  possible kinds of O9-planes in the background.

Type I corresponds to $\e_{NS} = \e_{R} = -$,  with conventional $O9$ $(-,-)$ planes, while
the choice $\e_{NS} = \e_{R} = +$ corresponds  to  $O9_{-}$ planes $(+,+)$,    for $\e_{NS} = + $ and $\e_{R} = -$ 
 we have $\bar{O9}_{-}$  $(+,-)$  and finally  $\e_{NS} = - $ and $\e_{R} = +$ gives $\bar{O9}$ anti-orientifold planes  $(-,+)$.

 It is mandatory to cancel the overall RR charge in order not to violate gauge invariance,
while the total tension (NSNS tadpole) of the objects in the background can be non-vanishing
 with the effect of breaking supersymmetry on the branes, since as previously discussed
in the presence of supersymmetry all tadpoles automatically vanish \cite{Martinec:1986wa,Friedan:1985ge}.

The transverse Klein bottle  $\tilde{\mathcal{K}}$, being   proportional  to the square of the tension
and charge of the planes, has  the same expression
 for every kind
of O-planes in the background
\beq
 \tilde{\mathcal{K}} = \frac{2^{5}}{2}(V_{8} - S_{8}), \label{Kzoo}
\eeq 
and therefore despite supersymmetry is broken or not on the branes by an uncancelled
NSNS tadpole, it is exact  in the bulk at the tree level, since the closed string amplitudes
remain the same.

Due to eq. (\ref{Mzoo}) and eq.(\ref{Kzoo}), the   transverse Annulus is 
\bea 
\tilde{\mathcal{A}} &=&  \frac{2^{-5}}{2}\left[ (n + \bar{n})^{2}V_{8} - (n - \bar{n})^{2} S_{8}\right]\nn \\
 &=& \frac{2^{-5}}{2}  \left[( n^{2} +  \bar{n}^{2})(V_{8} - S_{8}) + n\bar{n}(V_{8} + S_{8}) \right],
\eea
 which displays the opposite RR charge between branes and anti-branes in the
negative sign for $\bar{n}$ in the coefficient in front of
  $S_{8}$.


A change of the choice of proper time in the cylinder diagram gives the loop Annulus diagram,
and via a  modular $S$ transformation on the characters  $(V_{8} + S_{8})(i/t) = (O_{8} - C_{8})(it)$, one recovers
\beq
\mathcal{A} = \left[\left( \frac{n^{2}}{2}  +  \frac{\bar{n}^{2}}{2}\right)(V_{8} - S_{8}) + n\bar{n}(O_{8} - C_{8}) \right], \label{Atachyon}
\eeq
which shows that the spectrum of open strings stretching between a D9 and anti-D9
 has a reversed GSO projection \cite{tachyonopen}, and therefore contains the open string tachyon.
The presence of a tachyonic excitation 
 signals the instability of the system due to the mutual attraction
  of the opposite charged objects.

The loop open string M\"obius amplitude  is as usual obtained from (\ref{Mzoo}) form a $P$ modular tranformation and reads
\beq
\mathcal{M} = \frac{\e_{NS}}{2}(n + \bar{n})\hat{V}_{8} -  \frac{\e_{R}}{2}(n - \bar{n})\hat{S}_{8}.
\eeq
 In the  absence of anti-branes, $\bar{n} = 0$, we have a tachyon free spectrum 
  as shown by eq. (\ref{Atachyon}).
  RR tadpole cancellation requires  the conventional type I choice $\e_{R} = -$, 
  therefore there are two
possibilities for the choice of the sign    $\e_{NS}$.     $\e_{NS} = -$  corresponds to conventional O9 planes and therefore
to type I superstring.  Relaxing  the NSNS tadpole with the choice  $\e_{NS} = +$ 
 corresponds to the presence of  $\bar{O9}_{-}$ $(+,-)$  planes in the background.
This configuration is not supersymmetric and the gauge boson has $N^{2}/2 + N/2$ states,
which corresponds to a $USp(32)$ gauge group, while the massless open string fermion
remains in the anti-symmetric representation \cite{Sugimoto:1999tx}.

The presence of a non-vanishing total tension in the background corresponds to
a non vanishing vacuum energy, which curves   the spacetime.
Among the solutions of the tadpole-corrected string equations of motion, there are 
 spacetimes with an $SO(1,8)$ isometry group and  a warping metric on the ninth dimension \cite{Dudas:2000ff}.

It is worth to stress again   that both in the case of the presence of $\bar{D9}$ anti-branes and
for the choice of reversed tension $\bar{O9}_{-}$ plane, supersymmetry is broken 
on the branes i.e. in the open spectrum \cite{Sugimoto:1999tx,bsb,bsbII}, while the closed spectrum in the bulk
 remains supersymmetric at the tree level,
  since $\mathcal{T}$ and $\mathcal{K}$ 
 are  identical to the type I ones.

However, this is a tree level feature, since  higher genus diagrams, starting from $g = 3/2$,
 are expected to transmit the breaking of supersymmetry also 
  to the closed string sector.

\vspace{3 cm}

\subsection{Breaking Supersymmetry with a Vanishing Vacuum Energy}
\everypar{\hspace{-.6cm}}
The breaking of supersymmetry is, in one way or in another, connected
to the issues of background redefinition and to the issues  related to the  
  effects of   quantum string fluctuations on  the  background  dynamics.
 One  might  avoid  at least  the problem of background redefinition
 if the mechanism that induces the breaking of supersymmetry
 has the virtue of not generating a vacuum energy.  At the  perturbative
  level this would imply 
that every single vacuum diagram, or their  sum, cancels order by order \cite{bg}.

Although a proper technology for computing  arbitrarily high-genus diagrams
 is presently  missing, it is at times possible to check 
whether this higher   order corrections to the vacuum energy are vanishing or not.

In the following we will discuss a novel mechanism for  supersymmetry breaking \cite{Angelantonj:2004cm},
 that does not  generate a perturbative vacuum energy.
  Differently  to  what discussed in the previous sections,
 in this non-supersymmetric class of  backgrounds there is a cancellation
 of the NSNS tadpole,  since
 an uncancelled dilaton tadpole   generates a non-vanishing
 vacuum energy  already at $g = 1/2$  disk and crosscap diagrams.
  
  The mechanism in question  originates from the interactions between D-branes and both standard O-planes    
and exotics $ O_{-}$ ones, with reversed tension and charged,
 whose presence is induced by a nonvanishing
$B_{ij}$ NSNS discrete modulus, as discussed in the previous chapter. 
 One of the main features of the mechanism
 is a  \emph{deconstructions}
of   D-branes in their NS and R elementary constituents that  gives  the possibility of building
 configurations
 respecting Fermi-Bose degeneracy. For a symmetric splitting of the branes we will
show that the $g = 1$ open string amplitudes are vanishing, and
the vacuum energy does not  receive one-loop contributions.

Moreover, at higher genus, although an explicit computation of the vacuum diagram
 is out of reach, we will show that the vacuum energy does not receive contributions
 as a consequence of higher genus generalisations of the well known
 Abstrusa Jacobi Aequatio (\ref{Abstrusa}).

Finally, we will study the stability of  these configurations of D-branes and O-planes,
 by computing the one loop potential for open string moduli (Higgs fields),  that control
D-brane positions along compact directions.
 Unfortunately,  as it is typically the case for non-supersymmetric backgrounds,
there are instability directions in the potential, and the configuration
 with vanishing perturbative vacuum energy is not a true minimum but
 actually  a saddle point in the moduli space.

\section{A Prototype Six-Dimensional Example}\label{A Prototype Six-Dimensional Example}
\everypar{\hspace{-.6cm}}

In order to illustrate the supersymmetry breaking mechanism at work
 we start by  considering a toroidal compactification in a rational CFT point.
 Let us choose a   $T^{4}$ compactification  of the type IIB superstring on a rigid torus at the SO(8) enhanced symmetry point.
The rational point is reached by choosing  for  the background metric $G_{ij}$  the $SO(8)$  
 Cartan matrix

\bea
 G_{ij}  = 
 \left( \begin{array}{cccc} 
2 & -1 & 0 & 0  \\ 
-1 & 2 & -1 & -1  \\ 
0 & -1 & 2 & 0 \\
0 & -1 & 0 & 2 \end{array} \right),    
\eea
and for the $B_{ij}$ NSNS background   the  adjacency  matrix
\bea
 B_{ij}  = 
 \left( \begin{array}{cccc} 
0 & 1 & 0 & 0  \\ 
-1 & 0 & 1 & 1  \\ 
0 & -1 & 0 & 0 \\
0 & -1 & 0 & 0 \end{array} \right).    
\eea

The four dimensional lattice sum in the torus amplitude for the above choices 
reduces to the sum of  characters relative to the $SO(8)$ conjugacy classes,
the right factor in the following torus relation

\beq
\label{torus4.2}
\mathcal{T} = |V_8 - S_8 |^2 \, \left( |O_8|^2 + |V_8 |^2 + |S_8 |^2 + |C_8|^2 \right) \,.
\eeq

Its open string descendants  were constructed long ago in  \cite{Bianchi:1990yu}   and exhibit all the main properties of orientifold constructions  \cite{Angelantonj:2002ct,Dudas:2000bn,Dabholkar:1997zd}.

 For  rational  compactifications  the non trivial effects of a discrete $B_{ij}$
 modulus in connection to  open string Wilson lines,  were first noticed in \cite{Bianchi:1990tb} and  
 named  in this context  \emph{discrete Wilson lines}.
 In this case the possibility of obtaining  either orthogonal or simplectic gauge groups
 are related to the subtleties of the $P$ modular transformation in the M\"obius amplitude,
defined by eq. (\ref{PMoebius}) and (\ref{P'Moebius}).

A world-sheet parity projection of the spectrum in (\ref{torus4.2}) 
 amounts to introducing the Klein-bottle amplitude 
\beq
\label{discreteKleind}
\mathcal{K} = {\textstyle{1\over 2}} (V_8 - S_8)\,  ( O_8 + V_8 + S_8 + C_8)\,,
\eeq
which describes  the presence of O-9 planes in the background. 
 The $\Omega$-invariant six-dimensional massless excitations comprise an $\mathcal{N} =(1,1)$ supergravity multiplet coupled to four vector multiplets. 

In the transverse channel
\beq
\label{discreteKleint}
\tilde\mathcal{K} = {2^4 \over 2} (V_8 - S_8 ) \, O_8 
\eeq
develops non-vanishing NSNS and RR tadpoles that require the introduction of D-branes to compensate the tension and charge of  the O9-planes.

In the absence of Wilson lines, {\it i.e.} for $N$ coincident D-branes, the transverse-channel annulus amplitude is given by
\beq\label{discreteannt}
\tilde\mathcal{A} = {2^{-4}\over 2}\, N^2\, (V_8-S_8)\, (O_8 + V_8 + S_8 + C_8)\,,
\eeq
since  the closed string states that flow in the tube are actually those in  (\ref{torus4.2}), 
  that satisfy  reflection conditions
at the boundaries of the cylinder,  that identify the holomorphic parts of the characters
with the anti-holomorphic ones.  

The above amplitude  together with (\ref{discreteKleint}) implies the transverse-channel M\"obius amplitude
$$
\tilde\mathcal{M} = - N \, (\hat V_8 - \hat S_8 ) \, \hat O_8 \,.
$$
However, this is not the only possible choice for $\tilde\mathcal{M}$. The standard hatted characters for the internal lattice contribution decompose with respect to $SO (4)\times SO (4)$ according to
\footnote{In the M\"obius amplitudes, both in the loop open string and
in the tree closed string channels, the coefficients in front of the power of $q$ have an
alternate sign due to the real part of the modulus of the surface.
 Since the character $\hat{V}_{4}\hat{V}_{4}$ begins with massive states at  the first closed string level,
it is natural to attribute to it a minus sign in order to pair the sign of all its
 massive modes with those contained in $\hat O_4 \hat O_4$.}

$$
\hat O_8 = \hat O_4 \hat O_4 - \hat V_4 \hat V_4 \,.
$$

But following  \cite{Bianchi:1990tb,Bianchi:1990yu,Bianchi:1990yu} one may introduce discrete Wilson lines to modify the $ SO (4)\times SO (4)$ decomposition according to
$$
\hat O _8 ' = \hat O_4 \hat O_4 + \hat V_4 \hat V_4 \,,
$$
and write the alternative M\"obius amplitude
$$
\tilde\mathcal{M} \, ' = - N\, (\hat V_8 - \hat S _8 ) \, \hat O_8 ' \,.
$$
The two different choices reflect the sign ambiguity of the reflection coefficients
in the transverse M\"obius amplitude  and, since affect only \emph{massive} modes, they lead
  to the same  tadpole cancellation condition $N=16$ in the transverse-channel.
 However, a different choice for the  above SO(4) decomposition    \emph{does affect} the open-string spectrum,
 since the corresponding $P$ transformation is also modified.

 In fact,  for the $SO(4)$ characters, $P$ interchanges $\hat O_4$ and $\hat V_4$,
\beq
\label{Ptransf}
P:\qquad \hat O _8 \to - \hat O_8 \,, \quad{\rm but}\quad \hat O_8 ' \to + \hat O _8 '
\,.
\eeq
Therefore, the loop-channel annulus amplitude
$$
\mathcal{A} = {\textstyle{1\over 2}} \, N^2 \, (V_8 - S_8) \, O_8 
$$
has two consistent (supersymmetric) projections
$$
\mathcal{M} = + {\textstyle{1\over 2}} \, N \, (\hat V_8 - \hat S_8 ) \, \hat O_8 \,,
$$
and
$$
\mathcal{M}\, ' =  -{\textstyle{1\over 2}} \, N \, (\hat V_8 - \hat S_8 ) \, \hat O_8 ' \,:
$$
the former yields a $USp(16)$ gauge group while the latter yields an $SO(16)$ gauge group,
as is clear by counting the multiplicity of states for  $V_8$ in the two case :  $(N^{2}/2 + N/2)$ for $\mathcal{M}$,  while $(N^{2}/2 - N/2)$ choosing $\mathcal{M} '$. 

Notice, however, that one could have well
 decided to modify the internal lattice only in the NS or only in R sector of the D-brane \cite{Angelantonj:2004cm}
$$
\tilde\mathcal{M} \, '' = - N \, \left(\hat V_8 \, \hat O_8 - \hat S_8 \, \hat O_8 ' \right) \,.
$$
Tadpole conditions are again preserved, but in the direct channel 
$$
\mathcal{M}\, '' = {\textstyle{1\over 2}} \, N \,\left(\hat V_8 \, \hat O_8 + \hat S_8 \, \hat O_8 ' \right)
$$
breaks supersymmetry and yields a USp(16) gauge group with fermions in the antisymmetric 120-dimensional (reducible) representation. However, a non-vanishing cosmological constant emerges at one loop, as a result of Fermi-Bose asymmetry in the open-string sector.
In order to obtain a restoration of Fermi-Bose degeneracy  we  employ two different
 stacks of branes $N$ and $M$ that support different sign choices for the reflection coefficient
 in opposite way for the \emph{massive} NS and R states

\bea
\tilde\mathcal{M} &=& - \left[ \hat V _8 \, \left( N \,\hat O_8 + M\, \hat O_8 ' \right)
-\hat S_8 \, \left( N \,\hat O_8 '+ M\, \hat O_8  \right) \right]\,.
 \nonumber \\
& =& - \left\{ (N+M) \, (\hat V_8 - \hat S_8) \, \hat O_4 \hat O_4 +
\left[ (N-M)\, \hat V_8 - (-N+M)\,  \hat S_8\right] \, \hat V_4 \hat V_4 \right\}.
\label{discreteMoebt}
\eea
As a consequence,  the transverse Annulus amplitude needs to have the following form, in order for the states
flowing in the transverse M\"obius to have as  reflection coefficients the geometrical mean
 between those of the Klein and the Annulus itself
\bea
\tilde\mathcal{A} &=& {2^{-4}\over 2}\, \left\{ (N+M)^2 \, (V_8 - S_8 ) \,  
\left( O_4 O_4 + \ldots \right) \right.
\nonumber \\
&& \left. + \left[ (N-M)^2 \, V_8 - (-N+M)^2\, S_8 \right] \left( V_4 V_4 + \ldots \right) 
\right\}\,,
\label{annulus}
\eea

where the dots correspond to a right combination of sums of the SO(4) decomposition of the $ V_8, S_8, C_8 $
 internal characters allowed to flow as well in $\tilde\mathcal{A}$. 
  In fact, the reflection  conditions at the boundary
 of the cylinder   identify the holomorphic and the anti-holomorphic
  parts of the  states  contained in $|O_8|^2, |V_8 |^2, |S_8 |^2, |C_8|^2 $ 
 in the torus amplitude (\ref{torus4.2}). 

Through a $P$  modular transformation one recovers the loop M\"obius amplitude
\beq \label{Moebius4.2}
\mathcal{M} = {\textstyle{1\over 2}}\, \left[ \hat V_8 \left( N \,\hat O_8 - M\, \hat O_8 ' \right)
-\hat S_8 \, \left( -N \,\hat O_8 '+ M\, \hat O_8  \right) \right] \,,
\eeq
 which allows to obtain the  still missing combinations of characters in $\tilde\mathcal{A}$,
via the loop channel constraint imposing for the  M\"obius amplitude
 to be the  $\Omega$ projection of the annulus.
 
 The change of proper time from the cylinder to the annulus
 implies the use of an  S modular transformation,
 on the  SO(4) characters $\{ O_4 , V_4 , S_4 , C_4 \}$ given by 
$$
S = {1\over 2} \left( \matrix{ +1 & +1 & +1 & +1 \cr
+1 & +1 & -1 & -1 \cr
+1 & -1 & -1 & +1 \cr
+1 & -1 & +1 & -1 \cr} \right).
$$







The complete form of the transverse Annulus that matches the loop channel constraint 
 is
\bea \label{annulus44}
\tilde\mathcal{A} &=& {2^{-4}\over 2}\, \{ ( N + M)^2 \, (V_8 - S_8 ) \,  
\left( O_4 O_4 + V_4 O_4 + S_4 S_4 + C_4 S_4 \right) \nn \\
&+& \left[ (N - M)^2 \, V_8 - (-N + M)^2\, S_8 \right] \left( V_4 V_4 + O_4 V_4 + C_4 C_4 + S_4 C_4 \right)
\} 
\eea

that indeed in the direct channel gives
$$
\mathcal{A} = {\textstyle{1\over 2}} \, (V_8 - S_8 ) \left[ (N^2 + M^2)\, (O_4 O_4 + V_4 V_4 )
+ 2 NM\, (O_4 C_4 + V_4 S_4) \right]\,.
$$
This amplitude satisfies the request since indeed its unoriented states flow in $\mathcal{M}$.

Once again, it is worth to stress that this deformation 
  involves only the signs for the reflection coefficients for massive states
flowing in the transverse M\"obius and therefore the total tadpoles,
whose contributions arise only from \emph{massless} closed string states
 flowing in the transverse diagrams, are identical to the supersymmetric case
and read
$$
N+M =16.
$$

\vspace{.1 cm}


From the loop open string amplitudes one can read that the
 massless excitations on the D-branes 
 comprise  six-dimensional gauge bosons and 
four scalars in the adjoint of the Chan--Paton group
$$ 
G_{CP} =  USp(N) \otimes  SO(M)
$$
and non-chiral fermions in the anti-symmetric representation for the symplectic gauge
group and in the symmetric one for the orthogonal gauge group.

For a generic choice of numbers $N$ and $M$ for the two stacks of branes,
 the configuration gives rise to a one loop non vanishing vacuum energy 
 generated by the  M\"oebius amplitude  $\mathcal{M}$, since  $\mathcal{A}$,  $\mathcal{K}$  and  $\mathcal{T}$     
  all vanish, being identical to the supersymmetric case.

But for a splitting of the 16 original branes in two identical sets

$$
N=M = 8\,,
$$

 also   $\mathcal{M}$ vanishes being in this case proportional to $V_8 - S_8$, which 
is zero as a consequence of the Jacobi Abstrusa identity. 

In this last interesting configuration, the brane excitations give rise
to the gauge group
$$ 
G_{CP} =  USp(8) \otimes SO(8)
$$
with fermions in the representations
$(27\oplus 1,1)\oplus (1,35\oplus 1)$.

The presence of the two singlet fermions is crucial for a consistent coupling of the non-supersymmetric matter sector to the supersymmetric bulk supergravity \cite{emilian} . Although the massless D-brane excitations are as in \cite{AngelantonjHR} , the deformations we have employed here are different, and have the nicer feature of yielding an exact Fermi-Bose degenerate massless and massive spectrum for both the closed-string sector (that is still supersymmetric) and for the non-supersymmetric open-string sector, thus preventing any non-vanishing contribution to the one-loop vacuum energy. As we shall see, this guarantees that the contributions from genus three-half surfaces, and reasonably from generic genus-$g$ Riemann surfaces, vanish as well.

\section{Deforming away from the rational point}\label{section Deforming away from the rational point}
\everypar{\hspace{-.6cm}}

  We  turn our attention to a generic point in the closed string moduli space,
 away from the rational point, and study the supersymmetry breaking mechanism 
  for a generic  $T^{2}$  compactification with a non vanishing  $B_{ab} ={\alpha '\over 2} \epsilon_{ab}$
  discrete background \cite{Angelantonj:2004cm}.
 We consider    the presence of D7 branes that are pointlike in the target torus
 and  O7  planes to neutralise the overall NSNS and RR charges.
   The planes, pointlike in the compact space, induce the projection  by $\Omega \, I_2 \, (-1)^{F_{\rm L}}$ on both the closed and open string spectra,  where $I_2$ denotes the inversion of the two compact coordinates and $(-1)^{F_{\rm L}}$ is  the left-handed fermion index,  needed in order that the projector squares to the identity \cite{dpSO,Sen:1996vd}.

 The torus amplitude is
\beq \label{torusirr}
 \mathcal{T} = |V_8 - S_8|^2 \, \Lambda_{(2,2)} (B)\,,
\eeq
  while  the Klein bottle amplitude
$$
 \mathcal{K} \, = {\textstyle{1\over 2}} \, (V_8 - S_8) \, W_{2n_1 , 2n_2}
$$
involves only even windings, due to the presence of a non-vanishing (quantised) $B_{ab}$ background. 
 In the transverse channel
\beq \label{kleintirr}
 \tilde{\mathcal{K}} \, = {2^{3} \over 2}\, {\alpha '  \over R_1 R_2} \, (V_8 - S_8) \, \left( {1 + (-1)^{m_1}+ (-1)^{m_2} - (-1)^{m_1 +m_2} \over 2} \right) ^2 \, P_{(m_1 , m_2)}
\eeq
neatly displays the geometry of the orientifold planes: three standard $O$ planes are sitting at the fixed points $(0,0)$, $(\pi R_1 , 0)$ and $(0,\pi R_2)$, while an $O^{-}$ plane
with reversed tension and charge is sitting at the fourth fixed point $(\pi R_1 , \pi R_2)$, 
 as already discussed in  sec \ref{sectionDiscreteModuli} of the previous chapter.

On the other hand, the transverse-channel annulus amplitude
\beq \label{anntirr}
\tilde{\mathcal{A}} \, = {2^{-5} \over 2} \,  {\alpha '  \over R_1 R_2}\, \left[ \left( N + (-1)^{m_1 + m_2} M \right)^2 V_8 - \left( (-1)^{m_1 + m_2} N + M \right)^2 S_8 \right] \, P_{(m_1 , m_2)}
 \eeq
encodes all the relevant informations on the geometry of the D-branes. Notice that here we have decomposed the supersymmetric  D-branes of the type I superstring into their elementary NS and R constituents

The corresponding transverse-channel M\"obius amplitude can be unambiguously determined from (\ref{kleintirr})\ and (\ref{anntirr}) , and reads
\bea 
\tilde{\mathcal{M}} &=& - {1\over 2}\,  {\alpha '  \over R_1 R_2} \, \left[ \left( N + (-1)^{m_1 + m_2} M \right) \hat V_8 - \left( (-1)^{m_1 + m_2} N + M \right) \hat S_8 \right] \nn \\
&& \times  \left( {1 + (-1)^{m_1}
+ (-1)^{m_2} - (-1)^{m_1 +m_2} \over 2} \right) \, P_{(m_1 , m_2)} \,.\label{moebtirr}
\eea
The tadpole conditions are as in the supersymmetric case, and require
$$ 
N+M =16\,.
$$
Finally, the direct channel annulus 
$$
\mathcal{A} = (V_8 - S_8) \, \left[  {\textstyle{1\over 2}} (N^2 + M^2) \,W_{(n_1 , n_2)} +
NM \, W_{(n_1 +{1\over 2} , n_2 + {1\over 2})} \right]
$$
and M\"obius-strip

\bea
\tilde{\mathcal{M}} &=& -{\textstyle{1\over 2}} \Big[ \left( N\, \hat V _8 - M\, \hat S_8 \right) 
\left( W_{(2n_1 , 2 n_2)} + W_{(2n_1 +1 , 2n_2)} + W_{(2n_1 , 2n_2 +1)} - W_{(2n_1 +1 , 2n_2 +1 )} \right) \nn \\
&+&  \left(M\, \hat V _8 - N\, \hat S_8 \right) \left( -W_{(2n_1 , 2 n_2)} + W_{(2n_1 +1 , 2n_2)} + W_{(2n_1 , 2n_2 +1)} + W_{(2n_1 +1 , 2n_2 +1 )}  \right) \Big] \nn \\  \label{mobiusnonsusy}  
\eea
amplitudes  yield an eight-dimensional massless spectrum comprising
gauge bosons and pairs of massless scalars in the adjoint representation of $G_{\rm CP} = {\rm SO} (N) \otimes {\rm USp} (M)$, and non-chiral fermions in the (reducible) representations
$({1\over 2} N (N+1) ,1) \oplus (1,{1\over 2} M (M-1))$. 

As to the contributions of the four one-loop amplitudes to the vacuum energy, $\mathcal{T}$, $\mathcal{K}$ and $\mathcal{A}$ vanish identically as a result of Jacobi's {\it aequatio abstrusa}.
Furthermore, if the branes are split in two identical sets, {\it i.e.} if $N=M$, also $\mathcal{M}$ does not give any contribution to the cosmological constant. As we shall see in the next section, this also provides strong clues that all higher-genus vacuum amplitudes vanish.

\vspace{3 cm}

\section{Higher-genus amplitudes}\label{Higher-genus amplitudes}
\everypar{\hspace{-.6cm}}
Until now we have shown how acting with (suitable) discrete Wilson lines in the open unoriented sector and splitting the sixteen D-branes into two identical sets
leads to a vanishing one-loop contribution to the vacuum energy. Of course, this is not enough, since higher-order corrections might well spoil this result. We shall now provide arguments that actually this is not the case: at any order in perturbation theory, no contributions to the vacuum energy are generated, if the branes are separated in equal sets, {\it i.e.} if $N=M$. 
This is obvious for closed Riemann surfaces, both oriented and unoriented, of arbitrary genus, since the closed-string sector is not affected by the deformation and therefore has the same properties as in the supersymmetric type I string case

\begin{figure}
[ptb]
\begin{center}
\includegraphics[scale=1,height=5cm]{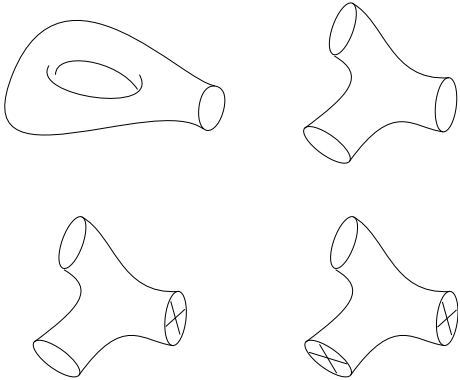}
\caption{Genus three-half Riemann surfaces with boundaries.}
\label{highgenus}
\end{center}
\end{figure}


When boundaries are present, one has to be more careful, since the non-supersymmetric deformation might well induce non-vanishing contributions to the vacuum energy. Our claim, however, is that these are always multiplied by a numerical coefficient proportional to $(M-N)$, that vanishes 
for our choice of brane displacement \cite{Angelantonj:2004cm}. Let us substantiate this statement with a closer look at a genus three-half amplitude. Among the surfaces with boundaries depicted in the figure let us concentrate on the one with two cross-caps and one hole. Similarly to the one-loop case, there is a particular choice for the period matrix $\Omega_{\alpha\beta}$ for which this surface describes a tree-level three-closed-string interaction diagram, weighted by the product of disc ($\mathcal{B}_i$) and cross-cap 
($\mathcal{C}_j$) one-point functions of closed states, that can be read from the transverse-channel Klein-bottle, annulus and M\"obius-strip amplitudes. More precisely, a formal expression for the one-disc--two-cross-caps amplitude is
\beq \label{scattering}
\mathscr{R}_{[0,1,2]} = \sum_{i,j,k} \mathcal{C}_i \, \mathcal{C}_j \, \mathcal{B}_k \, \mathscr{N}\,{}_{ij}{}^{k}\,
\mathscr{V}\,{}_{ij}{}^k \  \Omega_{\alpha\beta} \,,
 \eeq

where $[h,b,c]$ counts the number of holes $h$, boundaries $b$ and cross-caps $c$, 
$ \mathscr{N}\,{}_{ij}{}^k$ are the fusion rule coefficients and $\mathscr{V}\,{}_{ij}{}^k (\Omega_{\alpha\beta} )$ is a complicated function of the period matrix encoding the kinematics of the three-point interaction among states $i$, $j$ and $k$. This amplitude is expected to vanish in the supersymmetric (undeformed) case, and this requirement imposes  some relations among the functions $\mathscr{V}\,{}_{ij}{}^k (\Omega_{\alpha\beta} )$ that we have not defined explicitly \footnote{These are the analogues of the Jacobi identity $V_{8} = S_{8}$ for one-loop theta constants, that garanties the vanishing of the one-loop vacuum energy in the
ten-dimensional superstring.}

For instance, for the supersymmetric solution with transverse amplitudes   (\ref{discreteMoebt})
 and (\ref{annulus44})

\bea
\tilde{\mathcal{A}} &=& {2^{-4}\over 2}\, \big[ (N+M)^2 \, (V_8 - S_8 ) \,  
\left( O_4 O_4 + V_4 O_4 + S_4 S_4 + C_4 S_4 \right) \nn \\
&+& (N-M)^2 \, (V_8 -  S_8) \left( V_4 V_4 + O_4 V_4 + C_4 C_4 + S_4 C_4 \right) 
\big]
\eea

and
\beq
\tilde{\mathcal{M}} = - (N+M) \, (\hat V_8 - \hat S_8) \, \hat O_4 \hat O_4 - (N-M)\, (\hat V_8 -   
\hat S_8) \, \hat V_4 \hat V_4 \,,
\eeq
the genus three-half amplitude takes the form

\bea
 \mathscr{R}_{[0,1,2]} &=& {\textstyle{1\over 4}} \, (N+M) \, \left[ \mathscr{ V}\,{}_{111} + 
3 \mathscr{ V}\,{}_{133} + \mathscr{ V}\,{}_{122} + \mathscr{ V}\,{}_{144} +2\mathscr{ V}\,{}_{234} \right] \nn \\
&+& {\textstyle{1\over 4}} \, (N-M) \, \left[
\mathscr{ V}\,{}_{122} + \mathscr{ V}\,{}_{144} +2\mathscr{ V}\,{}_{234} \right] \,,\label{thsusy}
\eea
where the relative numerical coefficients of the $ \mathscr{V}$'s take into account the combinatorics of diagrams with given external states. The indices $1,2,3,4$ refer to the four characters $V_8\, O_4 O_4$,
$V_8 \, V_4 V_4$, $-S_8 \, O_4 O_4$ and $-S_8 \, V_4 V_4$, that identify the only states  with a non-vanishing $\mathcal{C}_i$, as can be read from eq. (\ref{discreteKleint}) , and
their non-vanishing fusion rule coefficients, all equal to one, are $\mathscr{ N}\,{}_{111}$, $\mathscr{ N}\,{}_{122}$, $\mathscr{ N}\,{}_{133}$, $\mathscr{ N}\,{}_{144}$ and $\mathscr{ N}\,{}_{234}$, and, both in $\mathscr{ V}\,$ and in $\mathscr{ N}\,$, we have lowered the indices using the diagonal metric $\delta_{kl}$, since all characters in this model are self conjugate.

For a supersymmetric theory this amplitude is expected to vanish independently of brane locations, and thus the condition $\mathscr{ R}_{[0,1,2]} = 0$ amounts to the two constraints
 \bea \label{relations}
 \mathscr{V}\,{}_{111} + 3 \mathscr{ V}\,{}_{133} &=& 0 \,, \nn \\
 \mathscr{V}\,{}_{122} + \mathscr{V}\,{}_{144} +2 \mathscr{V}\,{}_{234} &=& 0 \,.
 \eea

Turning to the non-supersymmetric open sector in eqs. (\ref{annulus44}) and (\ref{discreteMoebt}) , one finds instead
\bea
\mathscr{R}_{[0,1,2]} &=& {\textstyle{1\over 4}} \, (N+M) \, \left[ \mathscr{V}\,{}_{111} + 
3 \mathscr{V}\,{}_{133} + \mathscr{V}\,{}_{122} + \mathscr{V}\,{}_{144} +2\mathscr{V}\,{}_{234} \right]\nn \\
&-& {\textstyle{1\over 2}} \, (N-M) \, \left[
\mathscr{V}\,{}_{122} - \mathscr{V}\,{}_{144} \right] \,,\label{thnonsus}
\eea
since now $B_4 = -N +M$ has a reversed sign. Using eq. (\ref{thnonsus}) the non-vanishing contribution to the genus three-half vacuum energy would be
$$
\mathscr{R}_{[0,1,2]} = - {\textstyle{1\over 2}} \, (N-M) \, \left[
\mathscr{V}\,{}_{122} - \mathscr{V}\,{}_{144} \right]
$$
that however vanishes if $N=M$.

Similar considerations hold for the other surfaces in figure \ref{highgenus}, that are simply more involved since more $\mathscr{V}\,{}_{ij}{}^k$ terms contribute to them. In all cases, however, one can show that all the potential contributions are multiplied by the breaking coefficients $N-M$, that vanish for $N=M$.

It is not hard to extend these observations to the case of higher-genus amplitudes with arbitrary numbers of handles, holes and cross-caps, given a choice of period matrix that casts them in the form
\beq \label{generalampl}
\mathscr{R}\,_{[h,b,c]} = \sum_{\{ n_i \} , \{m_j \} } \, \prod_{i=1}^c \, \prod_{j=1}^b \, \mathcal{C}_{n_i} \, B_{m_j} \, \mathscr{N}\,\,{}^{[h]}{}_{n_1 \ldots n_c | m_1 \ldots m_b} \,
\mathscr{V}\,\,{}^{[h]}_{n_1 \ldots n_c | m_1 \ldots m_b} (\Omega_{\alpha \beta}) \,,
\eeq

where $\mathscr{V}\,\,{}^{[h]}_{n_1 \ldots n_c | m_1 \ldots m_b}$ is a complicated expression encoding the $h$-loop interaction of $b+c$ closed strings, $ \mathscr{N}\,\,{}^{[h]}{}_{n_1 \ldots n_c | m_1 \ldots m_b}$ are generalised Verlinde coefficients \cite{yassen} ,
while the sum is over the closed-string states $\Phi_\ell$ with disc and cross-cap one-point  functions $\mathcal{B}_\ell$ and $\mathcal{C}_\ell$, respectively. 
The expression (\ref{generalampl}) naturally descends from the definition of higher-genus Verlinde coefficients \cite{yassen}
$$
\mathscr{ N}\,\,{}^{[h]}{}_{n_1 \ldots n_c | m_1 \ldots m_b} = \sum_k {S_{n_1 k} \over S_{0k}}
\ldots {S_{n_c k} \over S_{0k}} {S_{m_1 k} \over S_{0k}} \ldots {S_{m_b k} \over S_{0k}}
{1\over (S_{0k})^{2(h-1)}} \,,
$$ 
and from the fusion algebra
$$
\mathscr{ N}\,\,{}_i \, \mathscr{ N}\,\,{}_j = \sum_k \mathscr{ N}\,\,{}_{ij}{}^k \, \mathscr{ N}\,\,{}_k \,.
$$
For instance, an amplitude $\mathscr{ R}\,\,{}_{[0,4,0]}$ can be represented as a combination of two three-closed-string interaction vertices
\bea
\mathscr{ R}\,\,_{[0,4,0]} &=& \sum_{i,j,k,l} \, B_i \, B_j \, B_k \, B_l \, \sum_m \mathscr{ N}\,\,{}_{ij}{}^m\, \mathscr{ N}\,\,{}_{mkl}\,\, \mathscr{V}\,\,^{[0]}{}_{ijkl} \, (\Omega_{\alpha\beta} )\nn \\
&=&  
\sum_{i,j,k,l} \, B_i \, B_j \, B_k \, B_l \, \sum_n \mathscr{ N} \,\,{}_{ik}{}^n \, \mathscr{ N}\,\,{}_{jnl}
\,\, \mathscr{V}\,\,^{[0]}{}_{ijkl} \, (\Omega_{\alpha\beta} ) \nn \\
&=& \sum_{i,j,k,l} \, B_i \, B_j \, B_k \, B_l \, \sum_n \sum_p \sum_q 
{S_{ip}S_{kp}S_{np}^\dagger \over S_{0p}} \, 
{S_{jq}S_{nq}S_{lq} \over S_{0q}}  \,\, \mathscr{V}\,\,^{[0]}{}_{ijkl} \, 
(\Omega_{\alpha\beta} ) \nn \\
&=& \sum_{i,j,k,l} \, B_i \, B_j \, B_k \, B_l \, \sum_p {S_{ip} S_{kp} S_{jp} S_{lp}
\over (S_{0p})^2}   \,\, \mathscr{V}\,\,^{[0]}{}_{ijkl} \, (\Omega_{\alpha\beta} )\nn \\
&=& \sum_{i,j,k,l}  \, B_i \, B_j \, B_k \, B_l \, \mathscr{ N}\,\,{}^{[0]}_{ijkl}
\,\, \mathscr{V}\,\,^{[0]}{}_{ijkl} \, (\Omega_{\alpha\beta} ) \,,
\eea
and similarly for other surfaces.

Again, the only differences in the amplitudes (\ref{generalampl})  with respect to the supersymmetric case, where these amplitudes are supposed to vanish, are present in terms containing at least one $B_\ell = N-M$, and thus vanish if $N=M$.

All these considerations are not special to the rational case, and can be naturally extended to irrational models.


\vspace{3 cm}

\subsection{One-loop effective potential and Higgs masses}
\everypar{\hspace{-.6cm}}

 In the previous sections  we have presented 
  a new non-supersymmetric class of open-string vacua
 where different discrete Wilson lines are introduced in
 the NS and R sectors.
 In this class of models with vanishing NS-NS and, of course R-R tadpoles,
 supersymmetry is broken on the branes while it is exact at tree level  in the bulk.
  If the branes are separated into two identical sets
 the closed and open-string spectra have an exact Fermi-Bose degeneracy,
 and thus the one-loop contribution to
 the cosmological constant vanishes identically.
  In section \ref{Higher-genus amplitudes}  we gave some qualitative arguments
 suggesting that higher-order perturbative contributions
 from surfaces with increasing numbers of holes,
 crosscaps and boundaries vanish as well. 

Of course, String Theory calculations should be at
 all compatible with Field Theory results,
 where only massless states are taken into account.
In the case at hand, it is easy to verify that
 the one-loop contribution to the vacuum energy vanishes
 identically also in Field Theory as a result of
 exact Fermi-Bose degeneracy of the massless degrees of freedom.
 However, non vanishing contributions do emerge at
 two loops, and possibly at higher loops.

This is not in principle inconsistent with our String Theory results.
In fact, while at the one-loop level the vanishing
 of the Field Theory amplitudes for the massless fields is a necessary,
 though not sufficient, condition for the vanishing
 of the corresponding String Theory amplitudes,
 this is not the case at higher loops. In fact, 
 one-loop amplitudes only involve a free propagation
 of the spectrum at each mass level and thus
 cancellations can only occur among states of equal mass,
 while at higher loops
interactions must be taken into account.
 In Field Theory these exist only among
 massless fields, while in String Theory
 non-trivial interactions also exist 
among massless and massive excitations,
 and these do contribute to the cancellation 
of higher-genus amplitudes. In the Field Theory limit 
 massless-massive interactions decouple and
 non-vanishing contributions to the vacuum energy may emerge.
 Of course, this can be checked only after complete
 \emph{quantitative} expressions of higher-genus
 vacuum amplitudes in String Theory are known.

\vspace{1 cm}

 We turn now to address the problem of the stability \cite{higgsfield} 
of the configuration of branes and orientifold planes.
 To this end we study the displacement of $N$ and $M$ branes
 in the eight-dimensional model of section \ref{section Deforming away from the rational point},
 by studying the effect of
 Wilson lines, introduced in sec. \ref{section Wilson Lines}
 at pag.  \pageref{section Wilson Lines}.

We consider a non vanishing background value for the internal components of the
 gauge vector field in the Cartan subalgebra
\beq
\langle A_{i} \rangle = diag \left(\frac{a_{i}}{R_{i}},...,\frac{a_{i}}{R_{i}},\frac{-a_{i}}{R_{i}},...\frac{-a_{i}}{R_{i}},\frac{b_{i}}{R_{i}},....,\frac{b_{i}}{R_{i}},\frac{-b_{i}}{R_{i}},...,\frac{b_{i}}{R_{i}}\right),\eeq
 with $i = 8,9$ along the $T^{2}$ directions.
 This $16 \times 16$ matrix  encodes the splitting 
   of the two stacks $N$ and $M$, of eight branes each, into
  two sets $(N, \bar{N})$ and  $(M, \bar{M})$, each containing  four branes.

 This background matrix corresponds to a pure gauge configuration for the gauge potential,
  obtained by  a U(1) gauge transformation
\beq
A_{i} = -i \Lambda^{-1}\p_{i}\Lambda,  
\eeq
with $\Lambda$ given by
\beq
 \Lambda = diag \left(e^{ ia_{i}y^{i}/R_{i}},...,e^{ia_{i}y^{i}/ R_{i}},e^{-ia_{i}y^{i}/R_{i}},...,e^{-ia_{i}y^{i}/R_{i}}          ,e^{ib_{i}y^{i}/R_{i}},....,e^{ib_{i}y^{i}/R_{i}} , e^{-ib_{i}y^{i}/R_{i}}   ,..., e^{-ib_{i}y^{i}/R_{i}}\right).
\eeq 

As shown in section  \ref{section Wilson Lines},  the presence of  non vanishing Wilson lines
 manifest itself  in the transverse Annulus amplitudes with the appearance 
 of phases in the lattice sum, that neatly encode the brane displacement along the
two circles on the internal $T^{2}$
\bea
N &\rightarrow &  N e^{2\pi i a_{1}m_{1}}e^{2\pi i a_{2}m_{2}} +  \bar{N} e^{- 2\pi i a_{1}m_{1}}e^{- 2\pi i a_{2}m_{2}},  \nn \\
M &\rightarrow &  M e^{2\pi i b_{1}m_{1}}e^{2\pi i b_{2}m_{2}} +  \bar{M} e^{- 2\pi i b_{1}m_{1}}e^{- 2\pi i b_{2}m_{2}}.  \label{replacement} 
\eea
These changes in the reflection coefficients in turn induce a modification 
in the M\"obius  amplitude, which,
 being the only amplitude responsible for the breaking of supersymmetry, 
 generates a  one-loop effective potential for the  Higgs fields $a$ and $b$
\beq
 \mathscr{V} = - \int_{0}^{\infty}\frac{dl}{\hat{\eta}^{8}(\frac{1}{2}+il)} \label{potentialM}
 \mathcal{V}(l,a,b). 
 \eeq           
This potential is obtained by 
 replacing (\ref{replacement}) in the M\"obius amplitude (\ref{mobiusnonsusy})
\beq
 \mathcal{V} = -  \frac{4\a'}{R_{1}R_{2}} \hat{V}_{8}\cdot \{ \mathscr{U}(a_{1},a_{2}) -  \mathscr{U}(b_{1},b_{2})\}. \label{potentialmm}
\eeq
  The above expression   takes into account  the condition $N = \bar{N} = M = \bar{M} = 4$
  and the identity $V_{8} = S_{8}$, 
   consequence  of the Jacobi Abstrusa Aequatio (\ref{Abstrusa})
 and the vanishing of $\theta_{1}(0|\tau) = 0$. 

\vspace{.2 cm}

The function $\mathscr{U}(a_{1},a_{2})$  is given by the following lattice sum

\beq
\mathscr{U}(a_{1},a_{2}) = \sum _{m_{1},{m_{2}} \in \mathbb{Z}}  
e^{2\pi i a_{1}m_{1}}e^{2\pi i a_{2}m_{2}}(1 - (-)^{m_{1} + m_{2}})e^{-2\pi l
\frac{\a'}{4}\frac{m_{1}^{2}}{R_{1}^{2}}}e^{-2\pi l \frac{\a'}{4}\frac{m_{2}^{2}}{R_{2}^{2}}}. \label{U}
\eeq

In the expression  (\ref{potentialM}) for the one-loop potential  we have recollected  the correct  overall
 negative sign that we always  drop  for $g = 1$ amplitudes
 in discussing supersymmetric vacua, since in this case all the amplitudes are vanishing
and thus an overall sign does not matter.
 However, we still discard  an overall normalisation coefficient in (\ref{potentialmm}),  irrelevant for the
study of the behaviour of the potential.

 The function $\mathscr{U}$   appearing in the integrand function (\ref{potentialmm}),  
can be expressed in terms  of Jacobi Theta functions
\beq
\theta\left[^{a}_{b}\right](z|\tau) = \sum_{n \in \mathbb{Z}}q^{\frac{1}{2}(n+a)^{2}}e^{2\pi i(z+b)(n+a)} \qquad q = e^{2\pi i\tau}, \label{Thetasum}
\eeq
as follows
\beq
 \mathscr{U}(a_{1},a_{2}) = \theta\left[^{0}_{0}\right](a_{1}|\tilde{\tau}) \theta\left[^{0}_{0}\right](a_{2}|\tilde{\tau})                           - \theta\left[^{0}_{\frac{1}{2}}\right](a_{1}|\tilde{\tau}) \theta\left[^{0}_{\frac{1}{2}}\right](a_{2}|\tilde{\tau}). \label{subpotential}
\eeq

Where  $ \tilde{\tau} = \frac{i l}{2} \frac{\a '}{R^{2}}$ and,
  for simplicity, we have chosen equal radii  $R_{1}=R_{2}=R$ for the squared $T^{2}$.\\

  We want to look for equilibrium configurations  of  D-branes and O-planes
 where the net classical  forces, given by exchange of closed strings
 in the transverse $g = 1$ diagrams,  vanish.
  The stationary points as a function of the Higgs fields $a$ and $b$  are
given by the solutions of

\beq
\frac{\p}{\p a_{1}}\mathcal{V}=\frac{\p}{\p a_{2}}\mathcal{V}=\frac{\p}{\p b_{1}}\mathcal{V}=\frac{\p}{\p b_{2}}\mathcal{V}=0, 
\eeq

that through eq. (\ref{potentialmm}) and (\ref{subpotential})  translates into the following condition
\beq
\frac{\p}{\p \z_{i}} \mathscr{U}(\z_{1},\z_{2})=\frac{\p}{\p \z_{i}} \theta\left[^{0}_{0}\right](\z_{1}|\tilde{\tau}) 
\theta\left[^{0}_{0}\right](\z_{2}|\tilde{\tau})- \frac{\p}{\p \z_{i} } 
\theta\left[^{0}_{\frac{1}{2}}\right](\z_{1}|\tilde{\tau}) \theta\left[^{0}_{\frac{1}{2}}\right](\z_{2}|\tilde{\tau}) = 0. \label{condizionen}
\eeq
From the definition (\ref{Thetasum}) of the Jacobi functions as a power series in $q$
\beq
\frac{\p}{\p \z} \theta\left[^{0}_{0}\right](\z|\tilde{\tau}) = \frac{\p}{\p \z}\sum_{n \in \mathbb{Z}}q^{n^{2}/2}e^{2\pi i \z}
= 2\pi i \sum_{n \in \mathbb{Z}}n \sin(2\pi \z n)q^{n^{2}/2} \label{thetader0}
\eeq
 one obtains  that  eq. (\ref{condizionen}) is satisfied  for $\z =  0,\frac{1}{2}$. 

Similarly

\beq
\frac{\p}{\p \z} \theta\left[^{0}_{\frac{1}{2}}\right](\z|\tilde{\tau}) =
\frac{\p}{\p \z} \theta\left[^{0}_{0}\right](\z + 1/2|\tilde{\tau}) =
 \frac{\p}{\p \z}\sum_{n \in \mathbb{Z}}q^{n^{2}/2}e^{2\pi i (\z + 1/2)}
= 2\pi i \sum_{n \in \mathbb{Z}}n \sin 2\pi (\z + 1/2) nq^{n^{2}/2} \label{thetader1/2}
\eeq
  vanishes for  $\z =  0,\frac{1}{2}$.

\begin{figure}  
\begin{center} 
\includegraphics[scale=1, height=8cm]{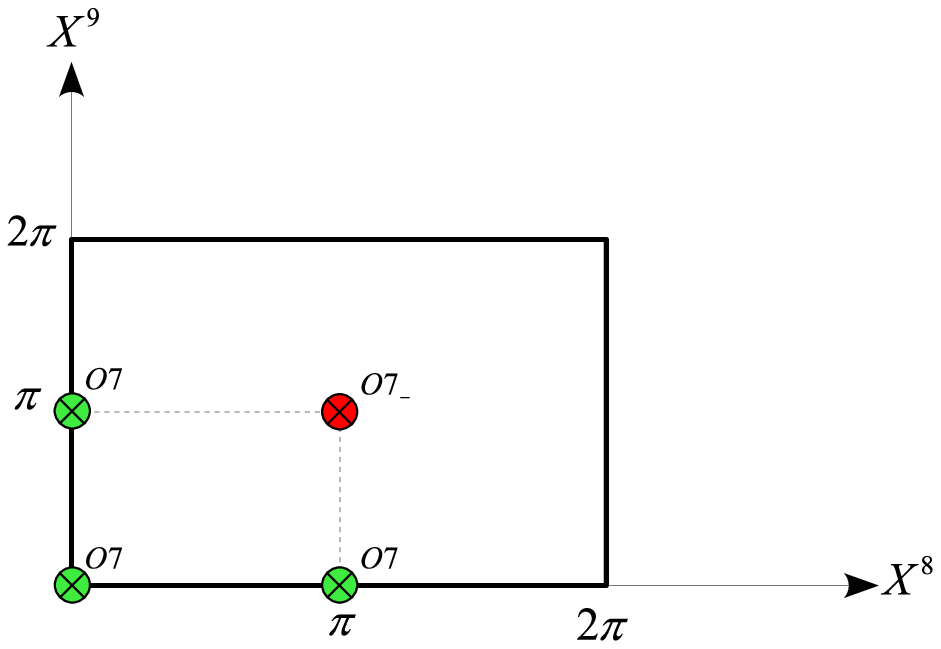}    
\caption{Three $O7_{+}$ planes (in green) with negative tension occupy three of the four  fixed point of the involution $I_{2}$,
while an exotic   $O7_{-}$ plane (in red) with \emph{positive} tension and charge is located in the 
fourth fixed point.}                       
\label{O7distributionII}
\end{center}
\end{figure}

We have  therefore  $16$  equilibrium configurations for $a_{1}$, $a_{2}$, $b_{1}$, $b_{2}$ 
either $0$ or $\frac{1}{2}$, corresponding  to all the possible displacements of the  
 four stacks of  D-branes  $N$, $\bar{N}$ ,$M$, $\bar{M}$
on the three $O$-planes located  on the two circles of $T^{2}$ at $(0,0)$, $(0,\pi R)$,  $(\pi R,0)$  
and  on the inverted plane   $\bar{O}_{-}$  in  $(\pi R, \pi R)$, (see figure \ref{O7distributionII}).

By writing  the vanishing-argument Theta constant in the usual notation
\beq
  \theta\left[^{0}_{0}\right](0|\tilde{\tau}) = 
\theta_{3}(\tau), \qquad \theta\left[^{0}_{\frac{1}{2}}\right](0|\tau) = \theta_{4}(\tau), 
\eeq
 and by using the periodicity property
\beq
\theta\left[^{0}_{\frac{1}{2}}\right]\left(\frac{1}{2}|\tau\right) = \theta\left[^{0}_{0}\right](0|\tau),
\eeq

 one can obtain the   values  of   $\mathscr{U}(\z_{1},\z_{2})$
 at the stationary points
 
\bea
\mathscr{U}(0,0) &=&  - \ \ \mathscr{U}\left(\frac{1}{2},\frac{1}{2}\right) = \theta^{2}_{3}(\tilde{\tau}) -  \theta^{2}_{4}(\tilde{\tau}) \nn \\  
\mathscr{U}\left(\frac{1}{2},0\right) &=&  \mathscr{U}\left(0,\frac{1}{2}\right) = 
\theta_{3}(\tilde{\tau})\theta_{4}(\tilde{\tau}) - \theta_{4}(\tilde{\tau})\theta_{3}(\tilde{\tau})=0.\label{Ustationary}
\eea

Recalling from eq. (\ref{potentialmm}) that the one loop potential is proportional to
\beq
\mathcal{V} \sim \mathscr{U}(a_{1},a_{2}) - \mathscr{U}(b_{1},b_{2}), 
\eeq

 it is then easy to check with the help of (\ref{Ustationary})  that,  besides the configuration of
interest (0,0,0,0), the following points have a vanishing one loop energy
\beq
(1/2,0,1/2,0),(0,1/2,0,1/2), (1/2,0,0,1/2),(0,1/2,0,1/2), (1/2,1/2,1/2,1/2).
\eeq

Whence the following configurations
\beq
(1/2,0,0,0),(0,1/2,0,0),(1/2,1/2,1/2,0),(1/2,1/2,0,1/2),
\eeq
have a non vanishing vacuum energy, proportional to 
\beq
\Lambda \sim - \frac{\a'}{R^{2}}\int_{0}^{\infty} \frac{dl \hat{V}_{8}}{\hat{\eta}_{8}}\left( \theta^{2}_{3}(\tilde{\tau}) -  \theta^{2}_{4}(\tilde{\tau})\right). \label{CC}
\eeq
  In the above expression because of tadpole cancellation,
  the divergent contribution from the constant term   in the $q = e^{-2\pi l}$
 theta expansion must be discarded.
  This term  corresponds to the tadpole contribution
 from the M\"obius strip, responsible for the  divergence in the 
 integral  in the region $l \rightarrow \infty$.

\vspace{.2 cm}

Finally, in the remaining stationary  points
\beq
(0,0,1/2,0),(0, 0,0,1/2),(1/2,0,1/2,1/2),(0,1/2,1/2,1/2),
\eeq
the vacuum energy is again non vanishing and opposite in sign
  to (\ref{CC})
\beq
\Lambda \sim + \frac{\a'}{R^{2}}\int_{0}^{\infty} \frac{dl \hat{V}_{8}}{\hat{\eta}_{8}}\left( \theta^{2}_{3}(\tilde{\tau}) -  \theta^{2}_{4}(\tilde{\tau})\right).
\eeq

\vspace{2 cm}

In order to study  the nature  of the 16 equilibrium  configurations 
 one considers  the Hessian matrix  from the
one loop  potential. 
 The off-diagonal elements of the Hessian of $\mathscr{U}(\z_{1},\z_{2})$
are given by
\beq
\frac{\p^{2}}{\p a_{1} \p a_{2}} \mathscr{U}(a_{1},a_{2})= \frac{\p}{\p a_{1}} 
\theta\left[^{0}_{0}\right](a_{1}|\tilde{\tau}) \frac{\p}{\p a_{2}}   
\theta\left[^{0}_{0}\right](a_{2}|\tilde{\tau})- \frac{\p}{\p a_{1} } 
\theta\left[^{0}_{\frac{1}{2}}\right](a_{1}|\tilde{\tau}) \frac{\p}{\p a_{2}}   
\theta\left[^{0}_{\frac{1}{2}}\right](a_{2}|\tilde{\tau})
\eeq
 which are zero at the stationary points as a consequence of eq.(\ref{thetader0}) and 
 (\ref{thetader1/2}).

Therefore  in the stationary points  the Hessian is  already in a diagonal form.

 For the  non-vanishing  diagonal elements we have for example  
\beq
\frac{\p^{2}}{\p a_{1}^{2}} \mathscr{U}(a_{1},a_{2})=\frac{\p^{2}}{\p a_{1}^{2}}      
\theta\left[^{0}_{0}\right](a_{1}|\tilde{\tau}) \theta\left[^{0}_{0}\right](a_{2}|\tilde{\tau})-
 \frac{\p^{2}}{\p a_{1}^{2}} \theta\left[^{0}_{\frac{1}{2}}\right](a_{1}|\tilde{\tau}) 
\theta\left[^{0}_{\frac{1}{2}}\right](a_{2}|\tilde{\tau}) \label{hessian}, 
\eeq
  which implies that
\beq
 \frac{\p^{2}}{\p a_{2}^{2}} \mathscr{U}(a_{1},a_{2}) = \frac{\p^{2}}{\p a_{1}^{2}} \mathscr{U}(a_{2},a_{1}).
\eeq

With the  definition $f(\z_{1},\z_{2}) =  \frac{\p^{2}}{\p \z_{1}^{2}} \mathscr{U}(\z_{1},\z_{2})$,  
one can write the  Hessian matrix $\mathbb{H}$ in the simple form
\beq
 \mathbb{H} = diag\big(f(a_{1},a_{2}), f(a_{2},a_{1}), - f(b_{1},b_{2}), - f(b_{2},b_{1})\big) \label{hessian2}.
\eeq

Moreover, the periodicity property of the Jacobi Theta functions
 implies also that
\bea
f(0,0) &=& - f\left(\frac{1}{2},\frac{1}{2}\right) =: \l   \nn \\
f\left(\frac{1}{2},0\right) &=& - f\left(0,\frac{1}{2}\right) =: \m. \label{hessianf}
\eea

Therefore  we can write the $4\times 4$  $\mathbb{H}$ matrix at the   16 stationary points
 already in a diagonal form, in terms of the  only  two independent eigenvalues $\l$ and $\m$
\bea 
\l &=& \frac{\p^{2}}{\p a_{1}^{2}} \theta \biggl[{}^{0}_{0}\biggr](0|\tilde{\tau}) 
\theta\left[{}^{0}_{0}\right](0|\tilde{\tau})- \frac{\p^{2}}{\p a_{1}^{2}} 
\theta\biggl[{}^{0}_{\frac{1}{2}}\biggr](0|\tilde{\tau}) \theta\left[^{0}_{\frac{1}{2}}\right](0|\tilde{\tau}) =
 2\pi \p_{\tilde{\tau}}  \left(\theta_{3}^{2}(\tilde{\tau}) -  \theta_{4}^{2}(\tilde{\tau})\right) \nn\\
\m & = & \frac{\p^{2}}{\p a_{1}^{2}}\theta\left[^{0}_{0}\right]\left(\frac{1}{2}|\tilde{\tau}\right) 
\theta\left[^{0}_{0}\right](0|\tilde{\tau})- \frac{\p^{2}}{\p a_{1}^{2}} 
\theta\left[^{0}_{\frac{1}{2}}\right]\left(\frac{1}{2}|\tilde{\tau}\right) 
\theta\left[^{0}_{\frac{1}{2}}\right](0|\tilde{\tau})\nn\\
 &=& 4\pi i \theta_{3}(\tilde{\tau}) \theta_{4}(\tilde{\tau})\p_{\tilde{\tau}}ln\left(\frac{\theta_{4}}{\theta_{3}} (\tilde{\tau}) \right). \label{hessian3} 
\eea

We are not particularly interested in the values of the two independent eigenvalues,
rather in checking  the sign of the eigenvalues of the Hessian matrix.

From the relations (\ref{hessian2}) and  (\ref{hessianf}) it
 is easy to check that 14 of the 16 stationary points are saddle points.
The list unfortunately includes also all the points with vanishing vacuum energy
and in particular the configuration of interest, that corresponds to the point
$(0,0,0,0)$, where $\mathbb{H} = diag\big(f(0,0), f(0,0),-f(0,0), -f(0,0)\big) =
 diag\big(\l, \l, -\l, -\l \big)$, clearly a saddle point.

  The two points where the one-loop potential is extremum are
 $(1/2,1/2,0,0)$ and  $(0,0,1/2,1/2)$ since $\mathbb{H}$ is given by  $diag\big(\l, \l, \l, \l \big)$
 for the first, while   $diag\big(-\l, -\l, -\l, -\l \big)$
 for the second.
The vacuum energy is not vanishing in the true minimum for the one-loop potential,
and is  proportional to
\beq
|\Lambda| \sim  \frac{\a'}{R^{2}} | \int_{0}^{\infty}
 \frac{dl \hat{V}_{8}}{\hat{\eta}_{8}}\left( \theta^{2}_{3}(\tilde{\tau})
 -  \theta^{2}_{4}(\tilde{\tau})\right)|.                           
\eeq


 An explicit computation of the above integral shows a runaway
behaviour in the radii $R$.

  Besides the instability in the radius, the above value
 is naturally related to the string scale $\a'$
 and therefore  by far to high to represent
the correct order of  magnitude
 of the observed Cosmological Constant. 
  It is also fear to remind that this
is a  one-loop value and, one should 
compute contributions from higher genus vacuum diagrams,
  in order to analyse 
  in  a systematic fashion the quantum effects of
supersymmetry breaking and single out the true vacuum.

\newpage

\section{Breaking Supersymmetry in the Bulk: The Scherk-Schwarz Mechanism}
\everypar{\hspace{-.6cm}}

There is a different way to break supersymmetry through  compactification,
the Scherk-Schwarz mechanism  \cite{scherk}.
 In field theory it corresponds to
 assign  periodic boundary conditions
 to the fermionic fields along a compact cycle up to an R-symmetry of the Lagrangian.
This produces a shift on the Fourier modes with respect to the bosons ones 
  thus breaking partially or totally the   supersymmetry.
 The easiest example corresponds to anti-periodic conditions,
 since the fermionic fields appear always as bilinears in every
Lagrangian, and therefore this choice is compatible with the symmetries of
the Action.
  This last case being  quite similar  to what we have discussed
 for the world-sheet spinors
in the NSR superstring but this time in the target space.
 
 A Fourier expansion for an anti-periodic field gives half integer modes, thus
  anti-periodic fermions on a circle have a tower of Kaluza-Klein
 states with masses
\beq
M^{2}_{m} = \frac{(m + 1/2)^{2}}{R^{2}}, \qquad m \in \mathbb{Z}. \label{masskkss}
\eeq
This choice gives therefore a mass to \emph{all} the  fermions, 
  controlled by the compactification radius $R$, with the 
result of breaking supersymmetry of the original uncompactified Lagrangian.

In string theory there are by far a larger spectrum of possibilities.
  Already at the closed string level \cite{kounnas} there is  the possibility of  mechanisms
that work not only on the Kaluza-Klein momenta quantum number $m$ as in  (\ref{masskkss}),
but also  on the closed string winding numbers $n$   or on 
 both $m$ and $n$ at the same time. This  latter interesting 
 option will be  described in section \ref{Asymmetric Shifts}, where a computation of the one-loop
potential induced by this mechanism shows interesting quantum effects
of the breaking of supersymmetry on the closed string
 geometrical moduli.

  In the presence of  D-branes in the background there are even further
possibilities, since depending on the geometry of the branes with respect
to the SS cycle, where the twisted periodicity conditions
 for the fermions are imposed,  it is possible either
  to connect a bulk gravitational non-supersymmetric
 sector to a gauge supersymmetric one 
 or to break supersymmetry also on the branes \cite{ADSi,adds,higgsfield}.

 A detailed study of the effects of the SS mechanism on general configurations
of intersecting branes will be the subject of the next chapter \cite{Angelantonj:2005hs}, where
  it will be stressed how this mechanism can be used to remove
undesidered non-chiral massless  states in standard model-like spectra.

Here, we shall focus on the effects of the SS mechanism on the closed string spectrum,
 and in particular, to the study of the  quantum effects induced
 by this  breaking of supersymmetry by computing one-loop effective potentials.

\subsection{Scherk-Schwarz on a Circle}
 \label{sectionSSCircle}
\everypar{\hspace{-.6cm}}
 We consider type IIB compactified on a circle $S^{1}$ of radius $R$ with 
 anti-periodic boundary conditions for the fermions. As a consequence the fermionic modes
have half-integer momentum quantum numbers, therefore naively one would expect 
the following Torus amplitude
\beq
 \mathcal{T} = \int_{\mathcal{F}}\frac{d^{2}\tau}{\tau_{2}^{6} (\eta \bar{\eta})^{8}}\left(|V_{8}|^{2} + |S_{8}|^{2} \right)\sqrt{\tau_{2}}\sum_{m,n} \Lambda_{m,2n} 
              -\int_{\mathcal{F}}\frac{d^{2}\tau}{\tau_{2}^{6} (\eta \bar{\eta})^{8}} \left( V_{8}\bar{S}_{8}  +  \bar{V}_{8}S_{8} \right)\sqrt{\tau_{2}}\sum_{m,n} \Lambda_{m + 1/2,2n},\label{SSincompletem} 
\eeq
 a modification of the circle torus amplitude   (\ref{toruscircle}), that takes into accounts
for the different momenta quantum number of the bosons and the fermions. 
 We recall the form of the lattice sum, first introduced in (\ref{Lambdalatticesum}),
  that takes into account the discreetness of the left and right momenta along the circle coordinate

\beq
\sum_{m,n}\Lambda_{m,n}(R) = \sum_{n \in \mathbb{Z}} e^{-\pi \tau_{2} \a'\left(\frac{m^{2}}{R^{2}} +  \frac{(nR)^{2}}{(\a')^{2}} \right)}e^{2\pi i\tau_{1} mn}.  
\eeq

However, (\ref{SSincompletem}) is not the complete answer for the torus amplitude,
 since  this expression is not modular invariant. To see this
 one can rewrite   (\ref{SSincompletem}) in the equivalent way

\bea
 \mathcal{T} &=& \left(|V_{8}|^{2} + |S_{8}|^{2} \right) \Lambda_{m,2n}(R) 
              - \left( V_{8}\bar{S}_{8}  +  \bar{V}_{8}S_{8} \right) \Lambda_{m + 1/2,2n}(R),  \nn \\ 
&=& \frac{1}{2} |V_{8} - S_{8}|^{2} \Lambda_{m,n}(2R) +  \frac{1}{2} |V_{8} + S_{8}|^{2} (-)^{m} \Lambda_{m,n}(2R),\label{SSincompletem2}
\eea
where to go from the first to the second line we  doubled the radius $R \rightarrow 2R$,
 (the meaning of this doubling  will be explained in the next section), and the circle compactification lattice sum
 is given by  
\beq
 \sum_{m,n}(-)^{m} \Lambda_{m,n}(R) =  \sum_{m, n \in \mathbb{Z}} (-)^{m} e^{-\pi \tau_{2} \a'\left(\frac{m^{2}}{R^{2}} +  \frac{(nR)^{2}}{(\a')^{2}} \right)}e^{2\pi i\tau_{1} mn} \label{latticephasem}.
\eeq
While the first term in  (\ref{SSincompletem}) is clearly  modular invariant, being  $1/2$ of the circle torus amplitude(\ref{toruscircle}),
the second terms is not  invariant under a generic modular transformation.
 The character $V_{8} + S_{8}$ 
  under  an $S$ and $T$ transforms as 
 \beq
|V_{8}+S_{8}|\stackrel{S}{\leftrightarrow} |O_{8}-C_{8}| \stackrel{T}{\leftrightarrow}|O_{8}+C_{8}|.\label{charinv}, 
\eeq
  thus being  only invariant under the subgroup
 of $PSL(2,\mathbb{Z})$ generated by $T$ and $ST^{2}S$.

\vspace{.2 cm}
The lattice sum (\ref{latticephasem}) is clearly invariant under a $T: \tau_{1} \rightarrow \tau_{1} +1$, and
after a Poisson resummation (\ref{Poisson}) over $m$, it can be rewritten as 

\bea
\sum_{m,n}(-)^{m} \Lambda_{m,n} &=& \frac{R}{\sqrt{\a ' \tau_{2}}} 
\sum_{\m,n}   e^{-\frac{\pi R^{2}}{\a ' \tau_{2}}|n\tau+ \m+ 1/2|^{2}} \nn \\ 
  &=& \frac{R}{\sqrt{\a ' \tau_{2}}} \sum_{\m,n}   e^{-\frac{\pi R^{2}}{\a ' 4\tau_{2}}|2n\tau+ 2\m+ 1|^{2}} \nn \\
 &=& \frac{R}{\sqrt{\a ' \tau_{2}}} \sum_{p \in  2\mathbb{Z} + 1}
 \sum_{c,d} e^{-\frac{\pi R^{2}p^{2}}{\a ' 4\tau_{2}}|2c\tau+ 2d + 1|^{2}}. \label{resm} 
\eea

In the last line $p$ is the biggest integer contained in both the numbers $2n$ and $2\m + 1$.

Recall that a modular transformations acts on  the modulus $\tau$   through  $2 \times 2$   $M_{c d}$ matrices

\bea
M_{c d}= \left(\ba{cc} a &  b \\ 2c & 2d+1  \ea \right),
\eea

  in the following way  
\beq
  \tau \rightarrow \frac{a\tau+b}{2c\tau+2d+1}.
\eeq
 As a consequence  the  immaginary part $\tau_{2}$ transforms as
\beq
 \tau_{2} \rightarrow \frac {\tau_{2}}{|2c\tau+2d+1|^{2}} = M_{cd}\tau_{2}.\label{imazione}
\eeq
Therefore in the last line of (\ref{resm}) one recognises the action of a generic
 matrix of this form with the last two entries, being two numbers
 one even and the other odd, coprime.

Clearly  this constraint on the   two entries of  $M_{cd}$ suggests
that
the lattice sum (\ref{latticephasem})
\beq
\sqrt{\tau_{2}} \sum_{m,n}(-)^{m} \Lambda_{m,n} =  \frac{R}{\sqrt{\a '}} \sum_{p \in  2\mathbb{Z} + 1}
 \sum_{c,d} e^{-\frac{\pi R^{2}p^{2}}{\a ' 4 M_{c,d}\tau_{2}}}, \label{latticeMcd}
\eeq
cannot be invariant under the full modular group.

Actually the set $\{ M_{c d} \}$ gives a  representation of a subgroup $\Gamma_{0}[2]$ of the 
 modular group, called a congruence subgroup,
 since the last two entries of these matrixes are equal mod(2).
The lattice sum written in the form  (\ref{latticeMcd})
 exhibits clearly
   an invariance under  $\Gamma_{0}[2]$, since it contains a full \emph{orbit}
 of the parameter $\tau$ under these class of matrixes.
It turns out that  $\Gamma_{0}[2]$ is generated by $T$ and  $ST^{2}S$, 
the same two generators under which the character $V_{8} + S_{8}$ is shown 
by (\ref{charinv}) to be invariant. 
  
Of course this last observation also suggests that, to obtain a complete
 modular invariant torus amplitude, one has to include the missing orbits,
 namely the matrixes with the last two entries $(2c + 1 , 2d)$  and those
 with $(2c + 1 , 2d + 1)$.

 These matrixes are contained in two different lattice sums
\beq 
\sum_{\m,n} =  e^{-\frac{\pi R^{2}}{\a ' \tau_{2}}| (n + 1/2)\tau+ \m |^{2}}, \label{latticeS}
\eeq
and
\beq 
\sum_{\m,n} =  e^{-\frac{\pi R^{2}}{\a ' \tau_{2}}|(n + 1/2)\tau+ (\m+ 1/2)|^{2}}, \label{latticeTS}
\eeq

The second of the previous lattice sums can be easily obtained from the first
after a $T: \tau \rightarrow \tau + 1$, while 
  an  $S: \tau \rightarrow - 1/ \tau$ transformation on  the first sum gives
\bea 
\sum_{\m,n}   e^{-\frac{\pi R^{2} |\tau|^{2}}{\a ' \tau_{2}}| -(n + 1/2)\tau  - \m  |^{2}}
&=& \sum_{\m,n}   e^{-\frac{\pi R^{2}}{\a ' \tau_{2}}|-(n + 1/2)  + \m \tau |^{2}}, \nn \\
&=& \sqrt{\tau_{2}}\frac{\sqrt{\a'}}{R} \sum_{m,n}(-)^{m} \Lambda_{m, n}.
\eea
    The  last line of the previous equation  can be obtained by comparing  (\ref{latticephasem}) 
  with  the second line, 
 after the change of names $n \leftrightarrow - \m$.  

Moreover, the  lattice sum in (\ref{latticeS}) corresponds to
\beq
\sum_{\m,n}   e^{-\frac{\pi R^{2}}{\a ' \tau_{2}}| (n + 1/2)\tau+ \m |^{2}} =  \sqrt{\tau_{2}}\frac{\sqrt{\a'}}{R} \sum_{m,n} \Lambda_{m, n + 1/2}
\eeq 
as it is clear again by the first line of (\ref{latticephasem}).

While for  the   (\ref{latticeTS}) we have
\beq
\sum_{\m,n}   e^{-\frac{\pi R^{2}}{\a ' \tau_{2}}| (n + 1/2)\tau+ \m + 1/2 |^{2}} =  \sqrt{\tau_{2}}\frac{\sqrt{\a'}}{R} \sum_{m,n} (-)^{m} \Lambda_{m, n + 1/2}.
\eeq 

To summarise we have shown that one needs three lattice sums in order to obtain a complete
 set of  orbits of the full modular group, which are related as follows
\beq
  \sqrt{\tau_{2}} \sum_{m,n} (-)^{m} \Lambda_{m, n} \stackrel{S}{\leftrightarrow}
\sqrt{\tau_{2}} \sum_{m,n}  \Lambda_{m, n + 1/2} \stackrel{T}{\leftrightarrow}
\sqrt{\tau_{2}} \sum_{m,n} (-)^{m}  \Lambda_{m, n + 1/2}
\eeq
where the lattice sum on the left of the previous relation  is $T$ invariant, while
the one on the right is $S$ invariant.
 
  An identical relation holds  between the  characters
\beq
|V_{8}+ S_{8}|\stackrel{S}{\leftrightarrow} |O_{8}-C_{8}| \stackrel{T}{\leftrightarrow}|O_{8}+C_{8}|.
\eeq

Altogether, the invariant modular torus amplitude is therefore

\bea
 \mathcal{T} &=& \frac{1}{2}\int_{\mathcal{F}}\frac{d^{2}\tau}{\tau_{2}^{11/2} (\eta \bar{\eta})^{8}} 
  \bigg( |V_{8} - S_{8}|^{2}  \sum_{m,n} \Lambda_{m,n} 
   + |V_{8} + S_{8}|^{2} \sum_{m,n} (-)^{m} \Lambda_{m,n} \nn \\
  &+& |O_{8} - C_{8}|^{2}\sum_{m,n} \Lambda_{m,n + 1/2} 
 + |O_{8} + C_{8}|^{2} \sum_{m,n} (-)^{m}\Lambda_{m ,n + 1/2} \bigg) \label{SScompletem} 
\eea

 After the rescaling $R \rightarrow R/2$ one can then obtain the 
 missing part of  (\ref{SSincompletem}) and write the complete modular invariant
torus amplitude for type IIB compactified on a circle with  anti-periodic
 boundary conditions for the spacetime fermions   

\bea
 \mathcal{T} &=& \left(|V_{8}|^{2} + |S_{8}|^{2} \right)\Lambda_{m,2n} 
              - \left( V_{8}\bar{S}_{8}  +  \bar{V}_{8}S_{8} \right) \Lambda_{m + 1/2,2n} \nn \\
 &+&  \left(|O_{8}|^{2} + |C_{8}|^{2} \right)\Lambda_{m,2n + 1}
   - \left( O_{8}\bar{C}_{8}  +  \bar{O}_{8}C_{8} \right) \Lambda_{m + 1/2,2n + 1},
\eea
where to lighten the notation we have omitted the integration over the torus modulus
  and the contribution from the transverse bosons as well.

We see that the request of modular invariance introduces 
 states in the spectrum with inverted GSO projection with respect to  the type II ones.
 In particular there is an infinite tower of NSNS scalars of the form 
 \beq
e^{ip_{L}X_{L} + ip_{R}X_{R}}|0 \rangle,
 \eeq
with quantum numbers $(0,2n + 1)$.
Since the conformal weights of this operator are $(\frac{\a'}{2}P^{2}_{L},\frac{\a'}{2}P^{2}_{R})$,
 for a given value of the radius $R$ this states become relevant if
\beq  
\frac{\a'}{2}(2n + 1)^{2}\frac{R^{2}}{\a'^{2}} < 1 \label{tachyonregion}
\eeq
i.e. if $R < \frac{1}{2n + 1}\sqrt{2\a'}$. Therefore there is a critical radius $R_{c} \sim \sqrt{\a'}$,
below which the lowest NSNS scalar $(m = 0, n = 1)$ becomes tachyonic and  the potential  diverges.

  



\vspace{4 cm }
\subsection{Scherk-Schwarz as a freely Acting Orbifold: Momentum Shift}
\everypar{\hspace{-.6cm}}

In the torus amplitude (\ref{SScompletem}) we had  to add
  the last two terms in order to achieve modular invariance.
  The need for  introducing new states can also be  understood in the general
framework of orbifold constructions \cite{Dixon:1985jw,Dixon:1986jc}.
 To this aim let us first rewrite eq.(\ref{SScompletem}),

\beq
 \mathcal{T} = \frac{1}{2}\int_{\mathcal{F}}\frac{d^{2}\tau}{\tau_{2}^{6} (\eta \bar{\eta})^{8}} 
  \left[ |V_{8} - S_{8}|^{2} \sqrt{\tau_{2}} \sum_{m,n} \Lambda_{m,n} 
   +\left(1 + S + TS \right)\circ \left( |V_{8} + S_{8}|^{2} \sqrt{\tau_{2}} \sum_{m,n} (-)^{m} \Lambda_{m,n}\right)\right], \label{SSm}
  \eeq
where we stressed the modular transformations that connect the various contributions.

Besides the last two terms that complete a closed modular orbit, the second term indeed
acts as a projection on the type IIB spectrum that selects even $m$ quantum numbers for the bosons
and odd ones for the fermions.

A shift operation on the circle $X \rightarrow X + \pi R$, translates on the left and right mover
circle coordinates as
\bea
 X_{R} &\rightarrow&  X_{R} + \pi R/2, \nn \\
 X_{L} &\rightarrow&  X_{L} + \pi R/2,
\eea
which is usually denoted by $A_{1}$-shift.
This  translation is realised on a closed string state $|m,n\rangle$  by the following operator
\beq
e^{ip_{R} \pi R/2 + i p_{L} \pi R/2 }|m,n\rangle = e^{i\left(\frac{m}{R} - \frac{nR}{\a'}\right) \pi R/2 
      + i \left(\frac{m}{R} + \frac{nR}{\a'}\right) \pi R/2 }|m,n\rangle = (-)^{m}|m,n\rangle.
\eeq
If we denote by $\d_{m}$ the above operation on the states, we see that the second term 
 in (\ref{SSm}) is obtained from the first by the action of the operator $\d_{m}(-)^{F}$,
and  $(-)^{F} = (-)^{F_{R} + F_{L}}$  gives a minus sign
to the fermionic NSR characters. 

\vspace{.2 cm}

Therefore by halving the circle torus amplitude and adding the second term in (\ref{SSm})
 one is actually inserting a projector into the trace that computes  this vacuum diagram \cite{bd,ADSi}
\beq
\mathcal{T} =  \int_{\mathcal{F}}\frac{d^{2}\tau}{\tau_{2}}Tr\left(\frac{1 + \d_{m}(-)^{F}}{2}\right)
q^{\frac{1}{4}L_{0}}\bar{q}^{\frac{1}{4}\bar{L}_{0}} \label{torustracem}.
\eeq
However, this is not the end of the story since the above expression encodes only the
first two terms  of the complete torus amplitude (\ref{SSm}). Actually
 by selecting invariant states under the action of  $\d_{m}(-)^{F}$ through the projector 
in (\ref{torustracem}) one is identifying points on the $S^{1}$ that have a distance of $\pi R$,
 therefore  string satisfying the following boundary conditions 
\beq
X(\s + \pi, \tau) = X(\s + , \tau) + \frac{n}{2}\cdot 2\pi R
\eeq  
are actually closed due to the  $\d$  identification. These \emph{twisted} 
closed strings represent the missing states  that complete the full torus amplitude
 that, as shown by eq.  (\ref{SScompletem}), have indeed half-integer winding numbers.

Finally, as we have seen at the beginning of the previous section, one actually 
  recovers the correct Scherk-Schwarz modes that follow by imposing anti-periodicity for
the fermions after halving the radius $R \rightarrow R/2$ of this orbifold description.
This is simply explained by the fact that after the identification by  $\d$,
the original circle is mapped in a circle with half of the original radius, that indeed corresponds
to the \emph{physical} compact part of the target space.   

\vspace{3 cm}

\subsection{One-loop effective potentials from Scherk-Schwarz Breaking}\label{potentialconventionalSS}
\everypar{\hspace{-.6cm}}
In a circle compactification with the Scherk-Schwarz mechanism  supersymmetry is broken
 by the asymmetry in the Kaluza-Klein excitations between spacetime bosons and fermions
and it is formally recovered in the decompactification limit $R \rightarrow \infty$.
However, from the torus amplitude and from  the analysis of the dependence of the mass
 on the radius $R$ for the NSNS scalar contained in $|O_{8}|^{2}$, we already know (see (\ref{tachyonregion}))
the emergence of a tachyonic excitation in the region $R < \sqrt{\a'}$.
It is thus natural to ask about the behaviour of the one-loop potential 
  amplitude as a function of $R$ \cite{Itoyama:1986ei,Ghilencea:2001bv}
\beq
  \mathscr{V} = - \frac{1}{2}\int_{\mathcal{F}}\frac{d^{2}\tau}{\tau_{2}^{6} (\eta \bar{\eta})^{8}} 
  \left(1 + S + TS \right)\circ \left( |V_{8} + S_{8}|^{2} \sqrt{\tau_{2}} \sum_{m,n} (-)^{m} \Lambda_{m,n}\right),
  \eeq
given by the last three terms of the torus amplitude (\ref{SSm}) with a crucial global
minus sign that we usually discard in writing the (vanishing) amplitudes for supersymmetric vacua.

\begin{figure}
[ptb]
\begin{center}
\includegraphics[scale=1,height=5cm]{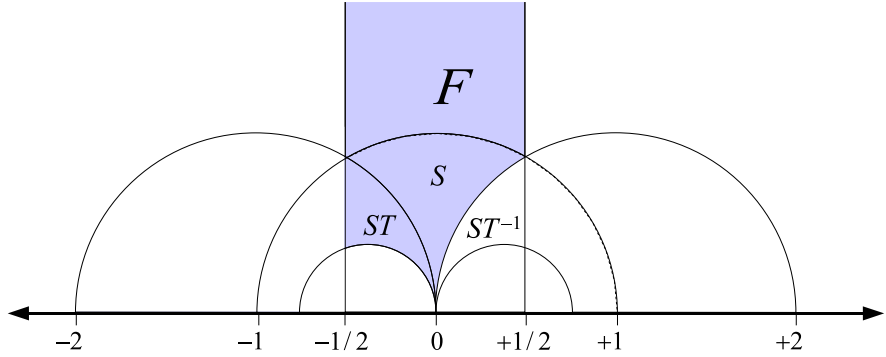}
\caption{ The region in light blue corresponds to the fundamental domain of  $\Gamma_{0}[2]/T$:  $\mathcal{F} \cup S(\mathcal{F}) \cup  ST(\mathcal{F})$.}\label{congruencefd}
\end{center}
\end{figure}

 The lattice sum after a Poisson resummation can be written as in  (\ref{latticeMcd}),
\bea
\mathscr{V} &=& - \frac{R}{2\sqrt{\a '}} \sum_{p \in  2\mathbb{Z} + 1} \sum_{c,d}
\int_{\mathcal{F}}\frac{d^{2}\tau}{\tau_{2}^{6} (\eta \bar{\eta})^{8}} 
  \left(1 + S + TS \right)\circ \left( |V_{8} + S_{8}|^{2} e^{-\frac{\pi R^{2}p^{2}}{\a ' 4 M_{c,d}\tau_{2}}}          \right) \nn \\
&=& - \frac{R}{2\sqrt{\a '}} \sum_{p \in  2\mathbb{Z} + 1} 
\int_{\bigcup_{c,d} M_{c,d}(\mathcal{F})}\frac{d^{2}\tau}{\tau_{2}^{6} (\eta \bar{\eta})^{8}} 
  \left(1 + S + TS \right)\circ \left( |V_{8} + S_{8}|^{2} e^{-\frac{\pi R^{2}p^{2}}{\a ' 4 \tau_{2}}}          \right) \nn  \\
&=& - \frac{R}{2\sqrt{\a '}} \sum_{p \in  2\mathbb{Z} + 1} 
\int_{\Gamma_{0}[2](\mathcal{F})}\frac{d^{2}\tau}{\tau_{2}^{6} (\eta \bar{\eta})^{8}} 
  \left(1 + S + TS \right)\circ \left( |V_{8} + S_{8}|^{2} e^{-\frac{\pi R^{2}p^{2}}{\a ' 4 \tau_{2}}}          \right),
\eea
and thus one can unfold the integration domain from the fundamental region $\mathcal{F}$
 into the region $\bigcup_{c,d} M_{c,d}(\mathcal{F})$ which is the image under 
 $\Gamma_{0}[2]/T$ of  $\mathcal{F}$ shown in fig.\ref{congruencefd}.

Moreover by using the decomposition 
 $\left(1 + S + TS \right)\circ \Gamma_{0}[2]$
 for the full modular group, one can reduce the integration domain 
 to the half strip $\mathcal{S} = \{ |\tau| > 1, \  -1/2 <\tau_{1} \le 1/2, \   0 \le \tau_{2} \}$. 
\beq
\mathscr{V}    - \frac{R}{2\sqrt{\a '}} \sum_{p \in  2\mathbb{Z} + 1} 
\int_{-1/2}^{1/2}d\tau_{1} \int_{0}^{\infty} \frac{d\tau_{2}}{\tau_{2}^{6} (\eta \bar{\eta})^{8}} 
  \left( |V_{8} + S_{8}|^{2} e^{-\frac{\pi R^{2}p^{2}}{\a ' 4 \tau_{2}}}\right).
\eeq
After this unfolding of the integration domain one then
 is able  to perform  the integral over  $\tau$  
\beq
\mathscr{V} =   - \frac{R}{2\sqrt{\a '}} \sum_{p \in  2\mathbb{Z} + 1} 
 \int_{0}^{\infty} \frac{d\tau_{2}}{\tau_{2}^{6}} 
   e^{-\frac{\pi R^{2}p^{2}}{\a ' 4 \tau_{2}}}
\int_{-1/2}^{1/2}d\tau_{1} \left|\frac{\theta_{2}^{2}}{\eta^{12}}\right|^{2}.
 \eeq

  By  expanding the modular functions  in a $q$-power series expansion 
\beq
 \frac{\theta_{2}^{4}}{\eta^{12}}(\tau) = \sum_{N = 0}^{\infty}d_{N}q^{N} \label{sviluppo},
\eeq
  we have
\beq
 \frac{|\theta_{2}^{4}|^{2}}{|\eta|^{24}}(\tau) = \sum_{N = 0}^{\infty}
 \sum_{N' = 0}^{\infty}d_{N}d_{N'} q^{N} \bar{q}^{N'}, 
\eeq
and therefore the $\tau_{1}$-integration gives
\bea
\int_{-\frac{1}{2}}^{\frac{1}{2}}d\tau_{1}
\frac{|\theta_{2}^{4}|^{2}}{|\eta|^{24}}(\tau)
  &=& \sum_{N = 0}^{\infty}
 \sum_{N' = 0}^{\infty}d_{N}d_{N'} e^{-2\pi \tau_{2}(N+N')}
   \int_{-\frac{1}{2}}^{\frac{1}{2}}d\tau_{1}  e^{-2\pi i\tau_{1}(N-N')} \nn \\
  &=&  \sum_{N = 0}^{\infty}\sum_{N' = 0}^{\infty}d_{N}d_{N'} e^{-2\pi
    \tau_{2}(N+N')}\delta_{N-N'}  =  \sum_{N = 0}^{\infty}d_{N}^{2}e^{-4\pi \tau_{2}N}.
\eea

The one-loop potential is therefore given by
\beq
\mathscr{V} =   - \frac{R}{2\sqrt{\a '}} \sum_{p \in  2\mathbb{Z} + 1} 
  \sum_{N = 0}^{\infty} d_{N}^{2}   \int_{0}^{\infty} \frac{d\tau_{2}}{\tau_{2}^{6}} 
   e^{-\frac{\pi R^{2}p^{2}}{\a ' 4 \tau_{2}}- 4\pi \tau_{2}N}
 \eeq

that can be divided into two contributions, one from the massless closed string states $N = 0$ 
 and the other from the massive states $N > 0$
\beq
\mathscr{V} =   - \frac{64R}{2\sqrt{\a '}} \sum_{p \in  2\mathbb{Z} + 1} 
      \int_{0}^{\infty} \frac{d\tau_{2}}{\tau_{2}^{6}} 
   e^{-\frac{\pi R^{2}p^{2}}{\a ' 4 \tau_{2}}} 
   - \frac{R}{2\sqrt{\a '}} \sum_{p \in  2\mathbb{Z} + 1} 
  \sum_{N = 1}^{\infty} d_{N}^{2}   \int_{0}^{\infty} \frac{d\tau_{2}}{\tau_{2}^{6}} 
   e^{-\frac{\pi R^{2}p^{2}}{\a ' 4 \tau_{2}}- 4\pi \tau_{2}N}. \label{potentialm3}
 \eeq

The first integral can be calculated with the help of the formula
\beq
 \int_{0}^{\infty} dx x^{n} e^{- \mu x} = n!/\mu^{n + 1}
\eeq 
so that the contribution $\mathscr{V}_{0}$ from the massless states is given by  \cite{Itoyama:1986ei}
\beq
 \mathscr{V}_{0} =  - \left(\frac{\sqrt{\a'}}{R} \right)^{9} \frac{2^{15} \cdot 4!}{ \pi^{5}} \sum_{p \in  2\mathbb{Z} + 1}\frac{1}{p^{10}} =  - \left(\frac{\sqrt{\a'}}{R} \right)^{9}
 \frac{3 \cdot 256 \cdot 1023}{ \pi^{5}} \z(10),  
\eeq
where we have computed the series in terms of  the zeta-Riemann function 
\beq
 \sum_{p \in  2\mathbb{Z} + 1}\frac{1}{p^{10}} = \frac{2^{10} - 1}{2^{10}} \z(10) = \frac{1023}{1024} \z(10).
\eeq

The contribution from the massive closed string excitations, the second integral in (\ref{potentialm3}),
is computed instead by the use of the following formula
\beq
\int_{0}^{\infty} dx x^{n -1}e^{-Ax -B/x} = 2 \left( \frac{B}{A} \right)^{n/2} K_{n}( 2 \sqrt{AB}), 
\eeq

so that 
\beq
 \mathscr{V}_{M} = -  32 \left(\frac{\sqrt{\a'}}{R} \right)^{4} \sum_{p \in  2\mathbb{Z} + 1} 
  \sum_{N = 1}^{\infty} d_{N}^{2} \frac{N^{5/2}}{p^{5}}K_{5}\left( \frac{R}{\sqrt{\a'}} 2\pi p \sqrt{N}\right). \label{massivem}\eeq

\begin{figure}
[ptb]
\begin{center}
\includegraphics[
height=2.4751in,
width=4.6613in
]%
{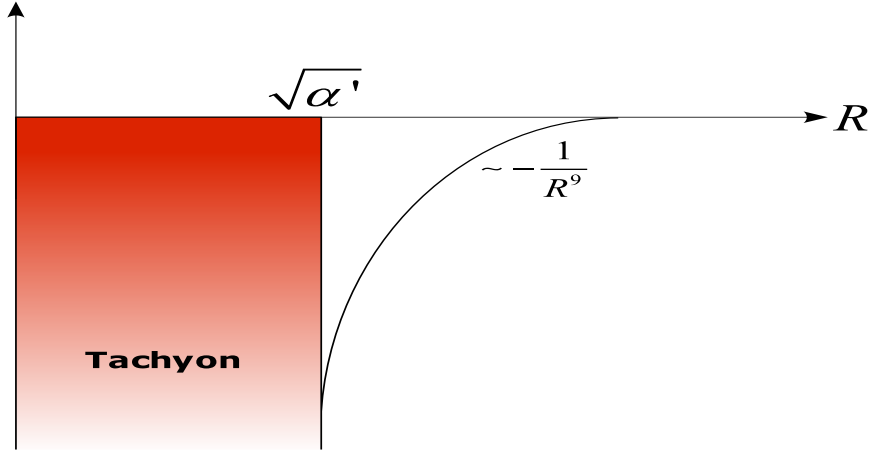}%
\label{MShift}
 \caption{The one-loop potential in the compactification radius $R$ resulting from a momentum
 Scherk-Schwarz mechanism. In red the tachyonic region $R < \sqrt{\a'}$ where the potential is divergent.}
\end{center}
\end{figure}

For large argument the Bessel function has the following behaviour
\beq
K_{n}(x) \sim \frac{e^{- x}}{\sqrt{x}},
\eeq
therefore for large  radius  $R$, the massive modes give  exponentially suppressed contributions
 to the potential, and the leading order is given by the massless contributions, so that the potential
  goes to zero with a $-1/R^{9}$ behaviour, as shown in fig. \ref{MShift}

From the analysis of the spectrum, obtained by reading the modular invariant torus amplitude,
we  have seen the appearance of closed string tachyon in the region $R < \sqrt{\a'}$, eq.(\ref{tachyonregion}).
For these values of the radius the torus integral has an IR divergence $\tau_{2} \rightarrow \infty$,
as follows from the general structure of this amplitude
\beq    
 \int_{0}^{\infty}\frac{d\tau_{2}}{\tau_{2}^{11/2}}\sum_{n} c_{N} e^{- \a' M_{n}^{2}\tau_{2}},
\eeq
that diverges in the presence of a tachyonic excitation.

Here we recover the same region of divergence for the potential, by considering that the 
number of closed string states for large $N$  grows as $ d_{N}^{2} \sim  \frac{e^{2\pi \sqrt{N}}}{N^{11/4}}$,
 while the behaviour of the Bessel functions  for large $N$ is
 $K_{5} \sim  \frac{1}{N^{1/4}}e^{-\frac{2\pi Rp}{\sqrt{\a'}}\sqrt{N}}$.
Therefore the sum over $N$ in the contribution (\ref{massivem}) from the massive states to the potential converges only
if $R/\sqrt{\a'} \ge 1$\footnote{For the critical value of the radius $R = \sqrt{\a'}$, where the
 lowest twisted $NS \otimes NS$ scalar
becomes massless, the potential is finite since,  for  large $N$, the generic term in (\ref{massivem}) 
 goes as $\sim 1/N^{13/4}$ and therefore the series in (\ref{massivem}) converges.}.

\vspace{3 cm}

\subsection{Asymmetric Shifts and One-Loop potentials in various dimensions}\label{Asymmetric Shifts}

\everypar{\hspace{-.6cm}}
A second possible Scherk-Schwarz mechanism consists in a breaking of the Fermi-Bose degeneracy
 by performing a projection on the closed string  winding numbers $n$.
 
This second option if  restricted to a pure closed string theory
 is  connected to the  mechanism  on the momenta  $m$, 
 by a T-duality. 
 Although in the  presence of D-branes in the background the two options
are rather distinct, giving rise to quite different and interesting effects, 
a subject discussed in the last chapter,
  here we want to focus only on the quantum effects induced by
  these mechanisms  emerging from
 the computation of  one-loop closed string potentials.

\vspace{.2 cm}

 The winding shift mechanism can be obtained by the 
following asymmetric shift

\bea
 X_{R} &\rightarrow&  X_{R} - \pi /2R, \nn \\
 X_{L} &\rightarrow&  X_{L} + \pi /2R,
\eea
usually called a $A_{3}$ shift
which  translates on the states 
\beq
e^{- ip_{R} \pi R/2 + i p_{L} \pi/2R }|m,n\rangle = e^{ - i\left(\frac{m}{R} - \frac{nR}{\a'}\right) \pi/2R 
      + i \left(\frac{m}{R} + \frac{nR}{\a'}\right) \pi/2R }|m,n\rangle = (-)^{n}|m,n\rangle.
\eeq
Notice, as already  anticipated, that
  after a T-duality along the circle coordinates
\bea 
 X_{R} &\rightarrow& -  X_{R}, \nn \\
 X_{L} &\rightarrow&  X_{L},
\eea
we recover the  $A_{1}$  momentum  shift discussed in the previous section. 

\begin{figure}
[ptb]
\begin{center}
\includegraphics[
height=2.4751in,
width=4.6613in
]%
{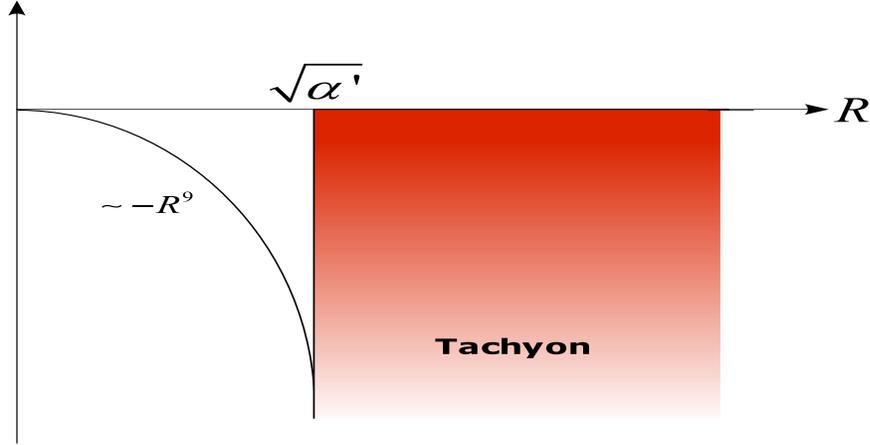}%
\label{A3Shift}
 \caption{The one-loop potential in the compactification radius $R$ resulting from a winding
 Scherk-Schwarz mechanism. In red the tachyonic region $R > \sqrt{\a'}$ where the potential is divergent.}
\end{center}
\end{figure}

 The one-loop potential for winding shift can thus  be obtained from the momentum shift potential
by the
 replacement $R \rightarrow \sqrt{\a'}/R$, and its behaviour is displayed in fig. \ref{A3Shift} 

\vspace{1 cm}

A third interesting option is a mechanism  involving  momenta $m$ and winding $n$ quantum numbers
 at the same time, as the following    selfdual asymmetric shift 
\bea
 X_{R} &\rightarrow&  X_{R} +  \pi R/2 - \pi \a'/2R , \nn \\
 X_{L} &\rightarrow&  X_{L} + \pi R/2  + \pi \a'/2R. \label{asymmetricdualshift} 
\eea
which corresponds to $A_{1}\circ A_{3}$ and is called $A_{2}$ shift.

 The level matching condition requires that one has 
to act simultaneously on  pairs of 
coordinates\footnote{
 In fact for \ref{asymmetricdualshift} acting on  a \emph{single} coordinate,
   the level matching condition in the twisted sector would yield
 $p^{2}_{L} - p^{2}_{L} + N_{1/2} - \bar{N}_{1/2} =
  2(m + 1/2)(n + 1/2) + N_{1/2} - \bar{N}_{1/2} = 0$, which has no solutions.}.

\vspace{1 cm}

\subsection{Tachyonic regions in the moduli space for  a $T^{2d}$ compactification}
\everypar{\hspace{-.6cm}}
\label{tachyonicregions}



\vspace{.4 cm}

We want to study   a general $A_{2}(-)^{F}$ asymmetric 
 $T^{2d}$ compactification, and  how  the appearance  of a  tachyonic excitation in the  spectrum  
 depends  on  $d$ and  the compactification moduli \cite{inpreparation}.

 The modular invariant  torus amplitude reads

\bea
\mathcal{T} &=& (|V_{8}|^{2} + |S_{8}|^{2})\frac{1 + (-)^{(\vec{m} + \vec{n})\vec{\e}}}{2} 
\Lambda_{\vec{m}, \vec{n}}^{(2d)} +
(V_{8}\bar{S}_{8} + \bar{V}_{8}S_{8})\frac{1 - (-)^{(\vec{m} + \vec{n})\vec{\e}}}{2} 
\Lambda_{\vec{m}, \vec{n}}^{(2d)} \nn \\
  &+& (|O_{8}|^{2} + |C_{8}|^{2})\frac{1 + (-)^{d + (\vec{m} + \vec{n})\vec{\e}}}{2}\Lambda_{\vec{m} + \vec{1}/2, \vec{n}  + \vec{1}/2}^{(2d)}
- ( O_{8}\bar{C}_{8} + \bar{O}_{8} C_{8})\frac{1 - (-)^{d + (\vec{m} + \vec{n})\vec{\e}}}{2}\Lambda_{\vec{m} + \vec{1}/2, \vec{n} + \vec{1}/2}^{(2d)}.\nn \\ \label{torus2d}
 \eea

 The zero-mode lattice sum $\Lambda_{\vec{m}, \vec{n}}^{(2d)}$, defined in (\ref{dlattice}),  encodes the contribution to the vacuum energy from   left and right moving   momenta (\ref{PLPR}), which depend on the background moduli $G_{ij}$ and $B_{ij}$.   $\vec{\e}$ in eq. (\ref{torus2d})  is the $2d$ unitary vector,
 $\e_{i} = 1, \ i = 1,\ldots, 2d$.  

Let us consider first the case of a diagonal  $G_{ij}$ and vanishing  $B_{ij}$,  describing 
a torus which is the product of $2d$ circles with radii $R_{i}$,   $i = 1, \ldots 2d$. 
For this special case the generalised Scherk-Schwarz projection in (\ref{torus2d})
 corresponds to the action of the asymmetric shift defined in  (\ref{asymmetricdualshift})
 along the compact coordinates, since
\beq
  e^{ip_{L, i}\left(\pi R^{i}/2 + \pi\a'/2R^{i} \right) + ip_{R, i}\left(\pi R^{i}/2 - \pi\a'/2R^{i}\right)}|\vec{m},\vec{n}\rangle = (-)^{(\vec{m} + \vec{n})\cdot \vec{\e}}|\vec{m},\vec{n}\rangle.
\eeq
 One thus  expects
 in  this region of the  moduli space  the one-loop potential to  enjoy the symmetry $R^{i} \rightarrow \a'/ R^{i}$.

Let us check  whether there are  relevant (tachyhonic) states in the 
twisted spectrum, described in the two last terms of (\ref{torus2d}).  The NSNS scalars
\beq
e^{ip_{L, i}X^{i}_{L} +  ip_{R, i}X^{i}_{R}}|0 \rangle \label{NSNSscalar}
\eeq
 have conformal weights $ (h, h) = (\frac{\a'}{2}p^{2}_{L}, \frac{\a'}{2}p^{2}_{R})$, with
\beq
p_{L, i} = \frac{m_{i} + 1/2}{R^{i}} + (n_{i} + 1/2 )\frac{R^{i}}{\a'}, \qquad p_{R, i} = \frac{m_{i} + 1/2 }{R^{i}} -  (n_{i} + 1/2) \frac{R^{i}}{\a'},
\eeq
 and $\vec{m}$ and $\vec{n}$  such that  the level matching condition
\beq
p^{2}_{L} - p^{2}_{R} = 4\left(\vec{m} + \frac{\vec{1}}{2} \right)\left(\vec{n} + \frac{\vec{1}}{2} \right) = 0,
\eeq
is satisfied.

The conformal weights thus satisfy the inequalty  
\beq
h  =  \frac{\a'}{2}p^{2}_{L} = \frac{\a'}{2}p^{2}_{R}    \ge d/2,
\eeq
with the minimum  reached for $R^{i} = \sqrt{\a'}$.

It follows that  the non-supersymmetric closed string spectrum obtained by the $A_{2}(-)^{F}$
mechanism for a factorizable $T^{2d}$ compactification is free of relevant modes 
if $2d \ge 4$, which means that the one-loop potential in this case is  finite
for all the values of the radii $R^{i}$.
 In   section \ref{computationfourtorus} we will consider the lowest dimensionality for the 
 target torus where this property occurs by  explicitly  compute the
 one-loop  potential 
and  study  its behaviour.

However, this appealing property of a tachyon-free spectrum  for $2d \ge 4$
 is restricted only for a factorisable $T^{2d}$ and 
 disappears whenever one switches on  off-diagonal metric
 componets  and/or a  $B_{ij}$ background.

In fact, to observe the presence of a  tachyon region for the $T^{4}$ compactification  it is enough
to switch on a $B_{ij} = b\e_{ij}$ along   the first two  directions of the torus.
The zero modes in the presence of the NSNS modulus  are then given by ( see eq. \ref{PLPR}) 
\beq
p_{L, i} = \frac{1}{R^{i}}(m_{i} +  b\e_{ij} n^{j}) + n_{i}\frac{R^{i}}{\a'} \qquad p_{R, i} = \frac{1}{R^{i}}(m_{i} +  b\e_{ij} n^{j}) - n_{i}\frac{R^{i}}{\a'},
\eeq
for the two coordinates where the  $b$ field  is not zero, while 
\beq
p_{L, i} = \frac{m_{i}}{R^{i}} + n_{i}\frac{R^{i}}{\a'} \qquad p_{R, i} = \frac{m_{i}}{R^{i}}  - n_{i}\frac{R^{i}}{\a'},
\eeq 
 for the other two coordinates.

 The NSNS scalars (\ref{NSNSscalar}) have therefore conformal weights given by
\bea
h &=& \frac{\a'}{2}\bigg[ \frac{1}{R^{2}_{1}}\left(m_{1} + \frac{1}{2} - b\left(n_{2} + \frac{1}{2} \right)\right)^{2}+ \left(n_{1} + \frac{1}{2}\right)^{2}R^{2}_{1} \nn \\
&+& \frac{1}{R^{2}_{2}}\left(m_{2} + \frac{1}{2} + b\left(n_{1} + \frac{1}{2} \right)\right)^{2}+ \left(n_{2} + \frac{1}{2}\right)^{2}R^{2}_{2} \nn \\
&+& \frac{1}{R^{2}_{3}}\left(m_{3} + \frac{1}{2} \right)^{2}+ \left(n_{3} + \frac{1}{2}\right)^{2}R^{2}_{3}
+ \frac{1}{R^{2}_{4}}\left(m_{4} + \frac{1}{2} \right)^{2}+ \left(n_{4} + \frac{1}{2}\right)^{2}R^{2}_{4}
\bigg].
\eea

 For $b = 1$ and $ \vec{m} = \vec{n} = \vec{0}$ we have

\beq
h = \frac{\a'}{8}\left(\frac{R^{2}_{1} + R^{2}_{2} + R^{2}_{3} + R^{2}_{4}}{(\a')^{2}} + \frac{1}{R_{3}^{2}}
+ \frac{1}{R_{4}^{2}} \right) \ge \frac{1}{2},\label{beffect} 
\eeq

thus showing the existence of  the tachyonic  region $ 1/2 \le h < 1$ in  the moduli space.

The reason behind  the apparence of a tachyonic region for a nonvanishing  $b$ 
 is a consequence
of the effect   of a  \emph{simultaneous} shift
 along two or more directions on  the target torus. 

Recall that in the specific case of a $T^{2}$ compactification, the background data $G_{ij}$
and  $B_{ij}$ can be organised to form two complex numbers
 encoding  the shape and the volume of the $T^{2}$,  the complex structure $U$ and the K\"ahler class $T$.

\beq
   b = T_{1}, \qquad  \sqrt{detG} = T_{2}, 
\eeq  
and
\bea
 G_{ij}  = 
\frac{T_{2}}{U_{2}} \left( \begin{array}{cc} 
1 & U_{1}   \\ 
U_{1}  & |U|^{2} 
 \end{array} \right). \label{metrictwotorus}    
\eea

\begin{figure}
[ptb]
\begin{center}
\includegraphics[scale=1,height=5cm]{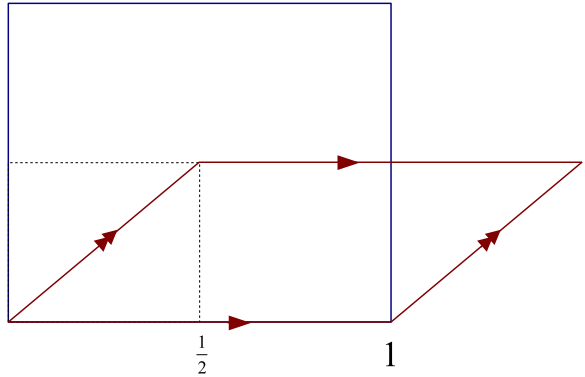}
\caption{After the identification by an $A_{1}$-shift a squared  $T^{2}$ with modulus $U = iU_{2} = i R_{2}/R_{1}$ becomes a tilted 
two-torus with modulus $U' = 1/2 + iU/2$.}
\label{A1shiftonT2}
\end{center}
\end{figure}

 The action of a \emph{simultaneous} shift on  two 
internal coordinates   changes the shape of the target $T^{2}$.
In fact, it is not too difficult to show that after the identification induced by a $A_{1}$-shift   
 a squared $T^{2}$ with modulus $U = iU_{2} = i R_{2}/R_{1}$ becomes a tilted 
two-torus with modulus $U' = 1/2 + i R_{2}/2R_{1}$, as shown in figure \ref{A1shiftonT2}.

An $A_{3}$-shift produces an identical effect on the $T^{2}$ K\"ahler class, since a T-duality along
one of the two coordinates interchanges $U$ with $T$, and at the same time
 turns an $A_{1}$-shift into an $A_{3}$-shift and vice-versa.

To summarise, if one takes a squared torus with a vanishing $b$,
 with K\"ahler class   $T = iT_{2} = iR_{1}R_{2}$,
 and modes out by an $A_{3}$ shift, the quotient space will have a K\"ahler class   
 given by  $T' = 1/2 + iT_{2}/2 = 1/2 +  iR_{1}R_{2}/2$.
Therefore it becomes clear that by turning on a $b = 1/2$ on the quotient space
  one is  able to cancel
the effect of the $A_{3}$-shift and vice-versa\footnote{The  string spectrum for a toroidal compactification
 is invariant under
 the transformation $b \rightarrow b + 1$ (Peccei-Queen symmetry), and thus a background with $b = 1$
  is equivalent to  one with  a vanishing $b$.}

  Actually, in (\ref{beffect})  we have observed that a $b = 1$ background cancels the effects
 of the $A_{2}$-shift along two coordinates, thus producing a tachyonic region
for small values of the corresponding radii.  
  The fact that it occours for $b = 1$ \emph{and not} for $b = 1/2$  is  the consequence of our 
 description of  states for  a tilted torus compactification
 in terms of (a projection of)  states of a squared torus compactification.
 In fact,  the quotient by  $A_{2} = A_{1} \cdot A_{3}$  halves  $T_{2} \rightarrow T_{2}/2$, the volume of the $T^{2}$.
 By  working  with the  double  of the   $T$ modulus
  one expects to see the cancelation  of the $A_{2}$ shift for
 $b = 1$ and not $b = 1/2$.

In order to see the same effects from a different point of view, let us  recall the form of the potential
For a $T^{2}$ case 
 
\beq
 \mathscr{V} \sim  -{1\over 2} \left( 1 + S + TS   \right)\circ \left(   |V_8 + S_8|^2 \,\sum_{\vec m , \vec n} (-)^{(\vec m + \vec n ) \cdot \vec \epsilon} \Lambda_{\vec m , \vec n}(b, R_{1}, R_{2})\right),\label{potentialb}
\eeq
 where in the above expression we omit the $\tau$ integration and the contributions to
the potential from the transverse coordinates.
  
In  the mass formula

\beq
 M^{2} = \frac{(m_{1} - bn_{2})^{2}}{R_{1}^{2}} + \frac{(m_{2} + bn_{1})^{2}}{R_{2}^{2}} + n_{1}^{2}\frac{R_{1}^{2}}{(\a')^{2}} + n_{1}^{2}\frac{R_{2}^{2}}{(\a')^{2}},
\eeq

we choose $b = 1$, the potential can be rewritten as
\beq
 \mathscr{V} =  -{1\over 2} \left( 1 + S + TS   \right)\circ \left(   |V_8 + S_8|^2 \,\sum_{\vec m'\vec n } (-)^{ \vec m' \cdot \vec \epsilon} \Lambda_{\vec m' , \vec n}(0, R_{1}, R_{2})\right),\label{potentialbb}
\eeq
 after  the substitution in (\ref{potentialb})  
\bea
m_{1} - n_{2} &=& m_{1}' \nn \\
m_{1} + n_{2} &=& m_{2}'.
\eea

 Eq. (\ref{potentialbb})   corresponds to the potential  from a  $A_{1}(-)^{F}$ breking mechanism, i. e. a
simultaneous Scherk-Schwarz involving only Kaluza-Klein momenta.  The effects of the $A_{3}$ winding 
shift have been cancelled by  the background value $b = 1$. 

\vspace{2 cm}

\subsection{Computation of the one-loop potential from asymmetric Scherk-Schwarz $T^{2}$ compactification}

\everypar{\hspace{-.6cm}}

\label{computationduetoro}




 We now turn  to study the simplest case of a compactification on a $T^{2}$
 and compute the one-loop potential generated by a  $A_{2}(-)^{F}$ 
 supersymmetry breaking mechanism \cite{inpreparation}.
 In the next section we will compute the one-loop potential for a $T^{4}$ compactification,
 a quite interesting case, since as discussed in the previous section,
 for vanishing values of the compact NSNS moduli, the spectrum
 turns out to be tachyon-free, and the potential has a stable AdS minimum 
 when the four $T^{4}$ radii are stabilised  at  the string length scale $\sqrt{\a'}$.  

 The  present  $T^{2}$ case, although plagued by a tachyonic region in the compactification
 radii, is interesting because one is able to compute the
 potential  in the presence of    \emph{all the  NSNS geometrical  moduli} \cite{inpreparation}.

The left and right moving  momenta for a generic toroidal compactification are given by eq. (\ref{PLPR}), and
  in the $T^{2}$ case it is convenient
to encode the background data $G_{ij}$ and $B_{ij}$  into two complex numbers
$U$ and $T$, called the complex and Kh\"aler structures for the two-torus, in terms
of which   $b = T_{1}$, $\sqrt{detG} = T_{2}$ and 
  
\bea
 G_{ij}  = 
\frac{T_{2}}{U_{2}} \left( \begin{array}{cc} 
1 & U_{1}   \\ 
U_{1}  & |U|^{2} 
 \end{array} \right). \label{metrictwotorus2}    
\eea

One can then check that the squares of the left and right momenta that appear
in the torus amplitude can be written  in terms of $U$ and $T$ as follows

\bea
p^{2}_{L} = p^{T}_{L}G^{-1}p_{L} &=& \frac{1}{T_{2}U_{2}}|Um_{1} - m_{2} - T(n_{1} + Un_{2})|^{2}, \nn \\  
p^{2}_{R} = p^{T}_{R}G^{-1}p_{R} &=& \frac{1}{T_{2}U_{2}}|Um_{1} - m_{2} - \bar{T}(n_{1} + Un_{2})|^{2},
\eea
with the  lattice sum  given as usual by
\beq
\sum_{\vec{m},\vec{n}}\Lambda_{\vec{m},\vec{n}} = \sum_{\vec{m},\vec{n}}q^{\frac{\a'}{4}p^{2}_{R}} 
\bar{q}^{\frac{\a'}{4}p^{2}_{L}}.
\eeq
  We consider  the action of  the operator  $V_{\vec{m},\vec{n}}$ 
 on the two internal momenta $\vec{m}$ and winding numbers $\vec{n}$
 
 
\beq
V_{\vec{m},\vec{n}}|\vec{m},\vec{n}\rangle  = (-)^{\vec{m} + \vec{n}}|\vec{m},\vec{n}\rangle.\label{Vmn}
\eeq 

The modular invariant torus  amplitude therefore reads
\footnote{Recall that the last  term  in (\ref{torusA2}) is obtained as a $T$ transformation from the previous one,
  the  minus sign  in front of this last term comes from a $T$ on  the $\tau_{1}$-dependent term in the lattice sum. 
 For  a single circle one would have $e^{2\pi i (m + 1/2)( n + 1/2)\tau_{1}}$
 that gives a factor $i$ after a $T$ transformation.}

\bea
\mathcal{T} &=& {\textstyle{1\over 2}} \sum_{\vec m , \vec n} \, \Bigl[ |V_8 - S_8 |^2 \Lambda_{\vec m , \vec n} + |V_8 + S_8|^2 \, (-)^{(\vec m + \vec n ) \cdot \vec \epsilon} \Lambda_{\vec m , \vec n} \nn \\&+&  |O_8 - C_8|^2 \Lambda_{\vec m +{1\over 2}, \vec n+{1\over 2}} -
|O_8 + C_8|^2 \, (-)^{(\vec m + \vec n ) \cdot \vec \epsilon}
\Lambda_{\vec m +{1\over 2}, \vec n+{1\over 2}} \Bigr] \label{torusA2} \\
&=&  {1\over 2} \sum_{\vec m , \vec n} \,  |V_8 - S_8 |^2 \Lambda_{\vec m , \vec n}
 + {1\over 2} \left( 1 + S + TS   \right)\circ \left(   |V_8 + S_8|^2 \,\sum_{\vec m , \vec n} (-)^{(\vec m + \vec n ) \cdot \vec \epsilon} \Lambda_{\vec m , \vec n}\right).\nn 
\eea

 and  generates the one-loop potential

\beq 
\mathscr{V} = -  \int_{\mathscr{F}} {d^2 \tau \over \tau_2^5} {{\mathcal{T}}\over |\eta|^{16}} 
= -  \int_{\tilde{\mathscr{F}}} {d^2 \tau \over \tau_2^5} \left| 
{\theta_2^4 \over \eta^{12}}\right|^2  \sum_{\vec m , \vec n}   (-)^{(\vec{m} + \vec{n} ) \cdot \vec{\epsilon}} \Lambda_{\vec m , \vec n}, \label{partialunfolding} 
\eeq

where $\tilde{\mathscr{F}} = (1 + S + ST ) \mathscr{F}$, 
\footnote{Recall the decomposition
of the full modular as $\Gamma_{0}[2]\times (S + ST)$, discussed in section \ref{sectionSSCircle}}
the region shown in figure \ref{congruencefd},
 corresponding to a   fundamental domain of $\Gamma_{0}[2]/T$.
 The integrand function    in  eq. (\ref{partialunfolding})  is invariant only
 under the congruence subgroup $\Gamma_0 [2] \subset {\rm PSL}(2;\mathbb{Z})$,
 whose generic element  has  the form

\bea 
 \left( \begin{array}{cc} 
a & b   \\ 
 2c  & 2d + 1 
 \end{array} \right),  \qquad    a (2d+1) - 2cb = 1.  
\eea

 After a  Poisson resummation over the pair of Kaluza-Klein momenta $m_1$ and $m_2$
     in   the lattice sum,  the potential can be rewritten in the following form \cite{Dixon:1990pc} 
\beq
\mathscr{V} = -  T_2  \int_{\tilde{\mathscr{F}}} {d^2 \tau \over \tau_2^6} \left| {\theta_2^4 \over \eta^{12}}\right|^2 \, \sum_{\{M\}} e^{{i\pi\over 2} (1- T) \det (M) } \, \exp \left\{ - {\pi T_2 \over 4 \tau_2 U_2} \left| \left( \matrix{ 1 & U \cr} \right) \, M \, \left( \matrix{\tau \cr 1 \cr} \right) \right|^2 \right\}
\eeq
with $M$ an integral matrix of the form
$$
M= \left( \matrix{ 2n_1 & 2\ell_1 + 1 \cr 2 n_2 & 2 \ell_2 + 1 \cr} \right) \,.
$$
 In order to unfold the integration domain from the region $\tilde{\mathscr{F}}$ into the
half strip $\mathcal{S}$, where  the computation of the $\tau$
integral becomes possible, 
one needs to decompose the space of integral matrices  into orbits of the congruence subgroup
 $\Gamma_{0}[2]$.
 In this way, the sum over the set of generic matrices can be restricted to contributions from a single representative of each independent orbit and correspondingly the integration domain is  enlarged
from  $\tilde{\mathscr{F}}$ to the image of this region under the
 action of $\Gamma_{0}[2]$.

 In the case at hand one has to distinguish between two classes of orbits. Those generated by
$$
M_0 = \left( \matrix{ 0 & 2p + 1 \cr 0 & 2 q + 1 \cr} \right) \,, \qquad p,q\in\mathbb{Z}
$$
with vanishing determinant, and the non-degenerate orbits whose representatives can be cast into the form
$$
M_1 = \left( \matrix{ 2 k & 2p + 1 \cr 0 & 2 q + 1 \cr} \right) \,, \qquad 2k > 2p+1 >0\,, \quad q\in\mathbb{Z} \,.
$$
Actually, the matrices $M_0$ are invariant under the $T$-modular transformation, and thus generate orbits of $\Gamma_0 [2]/T$. As a result, the integration domain unfolds in this case
 into the half-strip $\mathcal{S} = \Gamma_0 [2] (\tilde{\mathscr{F}}) /T$,
 while the contribution of the non-degenerate matrices
 has to be integrated over two copies of the
 upper-half plane $\mathbb{C}_+= \Gamma_0 [2] (\tilde{\mathscr{F}})$.

After several pages of  rather tricky computations 
 we were able to obtain
the expressions for the one-loop effective potential \cite{inpreparation}
$$
\mathscr{V} = \mathscr{V}_0 +  \mathscr{V}_1 
$$

The degenerate orbits  yield
\bea
\mathscr{V}_0 &=& - {1\over 2 (4 \pi^2 \alpha')^4  T_2^4} \, \left[ (2^{10} + 1) E_5 (U) - 2^5 E_5 (U/2) - 2^5 E_5 (2U) \right] \nn \\
 &-&  {2^8\,T_2\over (2 \pi)^3  {\alpha'}^4} 
\sum_{N=1}^{\infty} \sum_{p,q\in \mathbb{Z}} \, N^5\, d_N^2 \, { K_5 (y) \over y^5 }
\eea

where the argument of the special Bessel function $K_5$ is $y= 2\pi \sqrt{{NT_2 \over U_2} |2p+1 + U(2q+1)|^2} $, $d_{N}$ counts, as usual, the total degeneracy of string oscillators of mass $N$
$$
{V_8 + S_8 \over \eta^8} = {\theta_2^4 \over \eta^{12}} = \sum_{N=1}^{\infty} d_N \, q^N \,,
$$
and $E_5 (U)$ is the Eisenstein series
$$
E_{2k} (U) = {\Gamma (k) \over 2 \pi^k}\,  {\sum_{m,n}}^{\, \prime} {U_2^k \over |m + Un|^{2k}} \,,
$$
where the primed sum does not include the point $(m,n)=(0,0)$.

Similarly the non-degenerate orbits contribute to the potential with 
\beq
\mathscr{V}_1 = - {(T_2 U_2)^{-4}\over  (2\pi )^{25/2} \,{\alpha'}^4}
 \sum_{\ell , q\in\mathbb{Z}} \sum_{N =0}^\infty \sum_{k=1}^\infty 
 {d_N\, d_{|N+\ell k|} \over (2q+1)^9} \, e^{-i\pi (2q+1) \left[ (T_1 -1) k - (U_1 -1) \ell\right]}
\,x^{9/2} \, K_{9/2} (x) 
\eeq
where now the argument of the special Bessel function $K_{9/2}$ is 
\beq
x = 2\pi \sqrt{ T_2 U_2 (2q+1)^2 \left[ N + {\textstyle{1\over  4}} \left( {U_2 \over T_2} \ell + {T_2 \over U_2} k\right)^2 \right]} \,. 
\eeq

\vspace{3 cm}

\subsection{Computation of the one-loop potential for asymmetric shift  in a
$T^{4}$ compactification}

\everypar{\hspace{-.6cm}}
\label{computationfourtorus}

In this section we consider the computation for the one-loop potential generated by the
asymmetric Scherk-Schwarz supersymmetry breaking mechanism $A_{2}(-)^{F}$ acting on
 a factorisable $T^{4}$. As shown in section \ref{tachyonicregions},  in this region of the moduli space, which corresponds to a diagonal constant metric background $G_{ij}$ and a vanishing $B_{ij}$, the closed string spectrum is
free of any tachyonic excitation and, as we will show in the following, the potential
 has a unique negative minimum  for $R^{i} = \sqrt{\a'}$.
 This is a quite interesting result since it shows the existence of a stable  AdS minimum,
 as the effect of moduli stabilisation induced by one-loop quantum effects.
 However, as 
 already discussed  in section  \ref{tachyonicregions}, by turning on off-diagonal components
      of the background metric or the NSNS $B$ field one recovers tachyonic regions in the
moduli space that spoil this result. A mechanism that may project out such moduli
 and the study of the tachyon condensation\footnote{ for 
  various approaches to the  study of  closed string tachyon
 condensation see for example  \cite{Headrick:2004hz,Fradkin:1985fq,Fradkin:1985ys,Tseytlin:2000mt,Harvey:2000na,Vafa:2001ra,Dine:2003ca}} are presently under investigation. 

The starting point is the expression for the quantum potential 
 
\beq \label{formpot}
 \mathscr{V} = \int_{\mathscr{F}} {d^2 \tau \over \tau_2^2}\, {1\over \tau_2^4 \, |\eta |^{16}} \, \left( 1 + S + TS \right) \circ \left( \left|{\theta_2^4 \over \eta^4}\right|^2 \, \prod_{i=1}^4 \, \tau_2^2 \, \sum_{m_i,n_i} \, (-)^{m_i} (-)^{n_i} \, \Lambda_{m_i,n_i} 
\right).
\eeq
 
In order to  compute the $\tau$-integral we will follow a different path from the $T^{2}$
case, and use  an ad hoc trick   \cite{trickpotential}. 
The reason is that in the $T^{4}$ case it  is rather difficult, if not impossible,
 to find a decomposition of the set of matrices that one recovers
 after Poisson resummation of the four-dimensional lattice sum, in terms of orbits of the $\Gamma_{0}[2]$ congruence
group, (see section \ref{computationduetoro}   for a successful use of this technique in the computation
of the potential in the $T^{2}$ case).

Therefore in the following we will describe a  way-out for an apparent no-go that appears at one stage of
the computation.
  
As usual, to evaluate $\mathscr{V}$ one needs to  unfold the fundamental region 
\begin{plain}
$$
 \mathscr{F} = \{ z\in \mathbb{C} \, : \quad -{\textstyle{1\over 2}} < {\rm Re} (z) < {\textstyle{1\over 2}} \,, \quad |z| >1 \},
$$
\end{plain}
into the half-strip
\beq
\label{strip}
 \mathscr{S} = \{ z\in \mathbb{C} \, : \quad -{\textstyle{1\over 2}} < {\rm Re} (z) < {\textstyle{1\over 2}} \,, \quad 0 < {\rm Im}(z) < \infty \} \,.
 \eeq

Although the integral on $ \mathscr{S}$ extends till the dangerous region $\tau_2 \to 0$, its convergence is ensured after a Poisson re-summation of the lattice contribution over the momenta
\beq \label{momenta}
\sum_{m, n} (-)^{m} (-)^{n} \,  \Lambda_{m, n} = {R \over \sqrt{\tau_2\, \alpha '}} \sum_{\mu ,n} (-)^n \, e^{-{\pi R^2 \over \alpha ' \tau_2} |n \tau +\mu+{1\over 2}|^2} \,,
\eeq
or over the windings
\beq \label{windings}
\sum_{m, n} (-)^{m} (-)^{n} \,  \Lambda_{m , n} = \sqrt{\alpha ' \over R^2 \tau_2} \sum_{m,\nu} (-)^m \, e^{-{\pi \alpha ' \over R^2 \tau_2} |m \tau +\nu+{1\over 2}|^2} \,.
 \eeq
These two expressions are suited to study the $R\to \infty$ and $R\to 0$ regions, respectively. Since the asymmetric shift we are using  is invariant under T duality we expect identical behaviours in the two opposite regions.

  Let us resum over the momenta $m$, and notice that eq. (\ref{momenta})  can be conveniently rewritten (by pulling out a factor ${1\over 2}$ from the absolute value) so that only even $n$ and odd $\mu$ contribute to it. Thus, denoting by  $p \in\mathbb{Z}$ the maximum integer contained in both $\mu$ and $n$, so that
$$
\mu = (2d+1) \, p \,, \qquad n = 2c\, p\,, \qquad {\rm with}\quad d\in\mathbb{Z},\ c\in\mathbb{N} \,,
$$
one arrives at the expression
\bea
\label{congr}
\sum_{m,n} (-)^{m+n} \, \Lambda_{m,n} &=& {R\over \sqrt{\alpha ' \tau_2}} 
\sum_{p \in 2\mathbb{Z} +1} \sum_{c \in \mathbb{N} ,d \in \mathbb{Z}}
 \, (-)^c \, e^{-{\pi R^2 p^2 \over 4\alpha ' \tau_2} |2c\tau + 2d +1|^2} \nn \\
&=& {R\over \sqrt{\alpha ' \tau_2}} \sum_{p \in 2\mathbb{Z} +1} \sum_{c \in \mathbb{N} ,d \in \mathbb{Z}}\, (-)^c \, e^{-{\pi R^2 p^2 \over 4\alpha '} (M_{cd} \tau_2 )^{-1} }\,,
\eea
where we have introduce the matrix
\begin{plain}
$$
M_{cd} = \left( \matrix{ a & b \cr 2c & 2d +1\cr} \right) \in {\rm PSL} (2,\mathbb{Z}) \,,
$$
whose projective action on the Teichm\"uller parameter is as usually given by
$$
\tau \to M_{cd}\, \tau = {a \tau + b \over 2c \tau + 2d+1} \,.
$$
\end{plain}
Actually, the set of matrices $M_{cd}$ is a representation of the congruence subgroup $\Gamma_0 [2]$  of the modular group. Recall that for a generic $n\in \mathbb{N}$, the congruence subgroup $\Gamma_0 [n]$ is represented by matrices whose third entrance is a multiple of $n$ and the fourth one is co-prime to it. 

After (\ref{congr}) is plugged into eq. (\ref{formpot}), the resulting expression differs from the more conventional Scherk-Schwarz one computed in section \ref{potentialconventionalSS}  by an alternating sign $(-)^c$, whose origin can be traced to the stringy nature of the deformation we are now employing that simultaneously affects momenta and windings. Notice that this sign depends explicitly on $M_{cd}$, and thus its presence in the integrand could obstruct the unfolding of the fundamental domain. A careful analysis of the entire expression shows, however, that this sign disappears after the change of variable
\beq \label{changevar}
\tau \to M_{cd}\, \tau
 \eeq
is performed. To prove this important result, we decompose $M_{cd}$ in terms of the generators $ST^2S$ and $T$ of $\Gamma_0 [2]$:

\beq \label{decomp}
M_{cd} = (ST^2S)^{k_1} \, T^{\ell_1} \, \ldots \, (ST^2S)^{k_N} \, T^{\ell_N} \,,
 \eeq
for some choice of positive or null integers $k_i$ and $\ell_i$, and with the standard representation
\begin{plain}
$$
S= \left( \matrix{0 & 1 \cr -1 & 0\cr} \right)\,, \qquad T=\left( \matrix{1 & 1 \cr 0 & 1 \cr} \right) \,,
$$
\end{plain}
for the ${\rm PSL} (2,\mathbb{Z})$ generators. A direct calculation shows that one-half of the entry 
$(M_{cd})_{21}$ equals the number of times the $ST^2S$ generator appears in the decomposition (\ref{decomp}) plus an \emph{even integer number} $2\mathcal{N}$ 
\begin{plain}
$$
M_{cd} = \left( \matrix{a & b \cr 2c & 2d +1\cr} \right) =
(ST^2S)^{k_1} \, T^{\ell_1} \, \ldots \, (ST^2S)^{k_N} \, T^{\ell_N}
= \left( \matrix{ a & b \cr 2(k_1 + \ldots + k_N + 2\mathcal{N} ) & 2d+1 \cr} \right) \,.
$$
\end{plain}
The change of variable (\ref{changevar}) then results into a series of $ST^2S$ and $T$ transformations on the integrand. While the latter leaves both $V_8 + S_8$ and the lattice contributions invariant, the former fully compensates the alternating sign $(-)^c$. In fact, although under a single $ST^2S$ transformation
$V_8 + S_8$ is invariant, the three remaining lattice contributions
$$
\left[ \sqrt{\tau_2} \sum_{m,n} (-)^{m+n} \Lambda_{m,n} (\tau ) \right]^3 \stackrel{ST^{2}S}{\rightarrow}
- \left[ \sqrt{\tau_2} \sum_{m,n} (-)^{m+n} \Lambda_{m,n} (\tau ) \right]^3
$$
get a minus sign, that for a generic $M_{cd}$ yields the required factor $(-)^{k_1 + \ldots + k_N} = (-)^c$.  Therefore, (\ref{changevar})  maps (\ref{formpot}) into
\bea \label{residual}
 \mathscr{V} &=& {R\over \sqrt{\alpha '}} \int_{\cup_{c,d} M_{c,d} ( \mathscr{F})} \, {d^2\tau \over \tau_2^{9/2} \, |\eta|^{16}}  \nn \\
&\times&  \left( 1 + S + TS \right) \circ \left[ \left| {\theta_2^4 \over \eta^4} \right|^2 \prod_{j=2}^4 \, \sum_{m_j, n_j} (-)^{m_j + n_j} \Lambda_{m_j,n_j}  \sum_{p \in 2 \mathbb{Z} +1} \, e^{- {\pi R^2 p^2 \over 2\alpha' \tau_2}} \right] \,.
\eea

Since the integrand is invariant under $T$ modular transformations, $\tau_2 ' = M_{cd}\, \tau_2$ does not depend on $a$ and $b$, and since

\begin{plain}
$$
T^m M_{cd} = \left( \matrix{ a+mc & b + md \cr 2c & 2d+1 \cr}\right) \,,
$$
\end{plain}
we can always shift the images $M_{cd} (\mathscr{F})$ inside the half-strip $\mathscr{S}$, so that
the integration domain is now
$$
\bigcup_{c,d} M_{cd} (\mathscr{F}) = \Gamma_0 [2] (\mathscr{F}) /T \subset \mathscr{S} \,.
$$
At this point we use the residual modular transformations in (\ref{residual}) to unfold the fundamental domain into $\mathscr{S}$
$$
(1 + S + ST ) \circ \Gamma_0 [2] (\mathscr{F}) /T = \Gamma ( \mathscr{F}) /T = \mathscr{S} \,,
$$
where we have made use of the important decomposition 
$$
(1 + S + TS+ \ldots + T^{N-1} S) \circ \Gamma_0 [N] = \Gamma \,.
$$
As a result, the potential reads
$$
 \mathscr{V} = {2 R\over \sqrt{\alpha '}} \int_0^\infty {d\tau_2 \over \tau_2^{9/2}} \int_{-1/2}^{1/2} d\tau_1 \, \left|{\theta_2^4 \over \eta^{12}}\right|^2 \, \prod_{j=2}^4 \, \sum_{m_j, n_j} (-)^{m_j + n_j} \Lambda_{m_j,n_j}  \sum_{p \in 2 \mathbb{N} +1} \, e^{- {\pi R^2 p^2 \over 2\alpha' \tau_2}} \,.
$$
Using the explicit expression for the lattice contributions
$$
\prod_{j=2}^4 \, \sum_{m_j , n_j} (-)^{m_j + n_j} \Lambda_{m_j ,n_j} =
\sum_{\vec m , \vec n} (-)^{(\vec m + \vec n) \cdot \vec \epsilon} \, e^{2\pi i \tau_1 \vec m \cdot \vec n} \, e^{-\pi \alpha' \tau_2 \left( {|\vec m|^2 \over R^2} + |\vec n|^2 {R^2 \over \alpha '{}^2} \right)} \,,
$$
where now $\vec m$ and $\vec n$ are three-dimensional vectors, and the expansion 
$$
{\vartheta_2^4 \over \eta^{12}} = \sum_{N=0}^\infty \, d_N \, q^N
\,,
$$
the $\tau_1$ integration yields the familiar level-matching condition, so that we are left with
$$
 \mathscr{V} = {2 R\over \sqrt{\alpha '}} \sum_{p \in 2\mathbb{N}+1}\, \sum_{\vec m , \vec n} 
 (-)^{(\vec m + \vec n) \cdot \vec \epsilon} \, \sum_{N=0}^\infty \, d_N \, d_{N+\vec m \cdot \vec n} \, \, \Theta (N + \vec m \cdot \vec n )
\int_0^\infty {d\tau_2 \over \tau_2^{9/2}}  \, e^{-{\pi R^2 p^2 \over 4 \alpha ' \tau_2} - \tau_2 h} ,
$$
where $\Theta (x)$ is the step function
$$
\Theta (x) = \cases{1 & for $x\ge 0$ \cr 0 & for $x<0$ \cr}\,,
$$
and we have defined
$$
h = h(N,R,\vec m , \vec n ) =  \pi\left( 4 N + \alpha '\left| {\vec m \over R} + {\vec n \, R \over \alpha '} \right|^2  \right)
\,.
$$

After the last change of variable $\tau_2 =1/x$, the integral becomes of the form
$$
\int_0^\infty dx\, x^{n-{1\over 2}} \, e^{-r x - {s\over x}} = (-)^n \sqrt{\pi}\, {\partial^n \over \partial r^n} \left( {e^{-2\sqrt{rs}}\over \sqrt{r}}\right)
\,, \qquad {\rm with}\ {\rm Re} (r),\ {\rm Re} (s) >0 \,,
$$
with $n=3$, $s= h$ and $r= {1\over 4} \pi R^2 p^2/\alpha '$, and thus we arrive at the final expression
\begin{plain}
$$
\eqalign{
 \mathscr{V} =& {2^5 \, 5!! \over \pi^3} \left({\sqrt{\alpha '} \over R} \right)^6 \, \sum_{\vec m , \vec n} 
 (-)^{(\vec m + \vec n) \cdot \vec \epsilon} \, \sum_{N=0}^\infty \, d_N \, d_{N+\vec m \cdot \vec n} \, \, \Theta (N + \vec m \cdot \vec n ) 
\cr
& \times \sum_{p \in 2\mathbb{N}+1} \left[ {1\over p^7} + 
{f\over p^6} + {\textstyle{2 \over 5}}  {f^2 \over p^5} +
{\textstyle{1\over 15}} {f^3 \over p^4} \right] \, e^{- pf}
\cr}
$$
with $f=R \sqrt{ \pi h / \alpha'}$. The contribution from the massless states, with $N=0$, $\vec m = \vec n =0$, can be exactly evaluated 
$$
\mathscr{V}_0 = {5 !! \, 2^5 \, c_0^2 \over \pi^3} \, \left( {\sqrt{\alpha '}\over R} \right)^6 \sum_{p \in 2  \mathbb{N}+1} {1\over p^7} = 
{5 !! \, 2^6 \, 127 \over \pi^3} \, \left( {\sqrt{\alpha '}\over R} \right)^6 \, \zeta (7) \,,
$$
and determines the large-radius behaviour of $\mathscr{V}$. For generic $R$, the remaining terms can be expressed in terms of poly-logarithms 
$$
{\rm Li}_n (x) = \sum_{p=1}^\infty {x^p \over p^n} \,,
$$
\end{plain}
so that the full potential can be re-expressed as 
\bea
\label{finalpot}
\mathscr{V} &=& {5 !! \, 2^6 \, 127 \over \pi^3} \, \left( {\sqrt{\alpha '}\over R} \right)^6 \, \zeta (7) \nn \\
&+&
{2^4 \, 5!! \over \pi^3} \, \left( {\sqrt{\alpha '} \over R} \right)^6 \sum_{\{ \vec m , \vec n \} \not= \{0,0\}} (-)^{(\vec m +\vec n ) \cdot \vec \epsilon} \, \sum_N \, d_N \, d_{N+\vec m \cdot \vec n} \, \Theta (N+\vec m \cdot \vec n ) \nn \\
&\times& \Biggl[  {\rm Li}_7 (e^{- f})  +  f \, 
{\rm Li}_6 (e^{- f}) + {\textstyle{2\over 5}}  \, f^2 \,
{\rm Li}_5 (e^{-  f}) \nn \\
&+& {\textstyle{1\over 15}} \, f^3 \,
{\rm Li}_4 (e^{- f}) - {\rm Li}_7 (-e^{-\pi f})  - f \, 
{\rm Li}_6 (-e^{-  f}) \nn \\
& -& {\textstyle{2\over 5}}  \, f^2 \,
{\rm Li}_5 (-e^{- f}) 
- {\textstyle{1\over 15}} \, f^3 \,
{\rm Li}_4 (-e^{- f}) \Biggr]\,.
\eea

To determine the $R\to 0$ behaviour, one can notice that our starting expression is invariant under the T-duality $R \to \alpha ' /R$, and thus one would expect
$$
 \mathscr{V} \sim {5 !! \, 2^6 \, 127 \over \pi^3} \, \left( {R\over \sqrt{\alpha '}} \right)^6 \, \zeta (7) \,.
$$
Although this result can be obtained from the final expression for the potential (\ref{finalpot})
  after a careful summation of an infinite number of terms, it can be more easily derived if we started with a winding re-summation, as in (\ref{windings}).

\vspace{.3 cm}

 Due to the absence of a tachyonic excitation in the moduli space of a  factorisable $T^{4}$,
 the potential is a finite continuous negative function of the four radii $R^{i} / i = 1,\ldots, 4$.
 
Let us study its  behaviour as a function of a single radius $R$.
  The constraints from  self-duality
 on the behaviour of the potential $\mathscr{V}(\a'/ R) = \mathscr{V}(R)$
  implies   that if  there was a maximum(minimum) $\tilde{R} \ne \sqrt{\a'}$,
then 
\beq
 \mathscr{V}'\left(\frac{\a'}{\tilde{R}}\right) =
 - \frac{\a'}{\tilde{R}^{2}}\mathscr{V}'( \tilde{R}),
\eeq
 the dual point $\a'/\tilde{R}$ would be a maximum(minimum),
 a possibility incompatible with the self-duality of $\mathscr{V}$.

Therefore the only extremum can exist for the self dual radius $R = \sqrt{\a'}$,
which is indeed a  minimum of the one variable function $\mathscr{V}(\cdot \ \ , R)$,
  since the potential is a negative function with no divergences that vanishes 
 for $R \rightarrow 0$ and $R \rightarrow \infty$ \cite{trickpotential}.

\begin{figure}
[ptb]
\begin{center}
\includegraphics[scale=1,height=5cm]{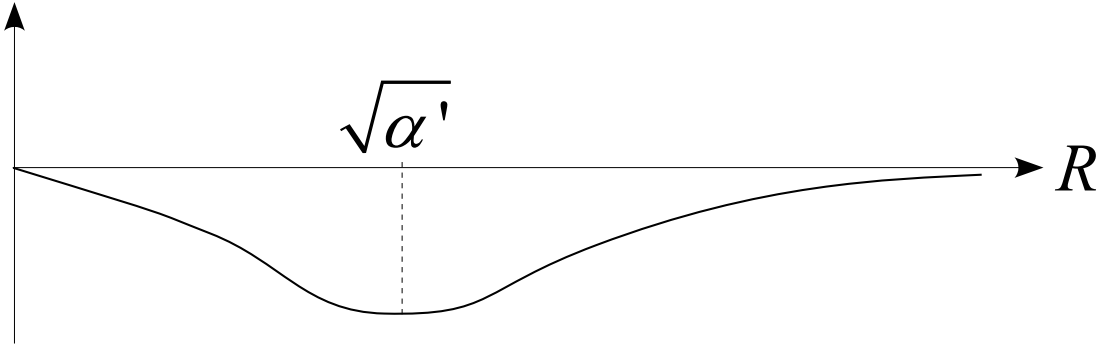}
\caption{The one-loop potential  
 is a continuous  negative function with a global minimum for  $R_{i}  = \sqrt{\a'}$.}
\label{MNshift}
\end{center}
\end{figure}

By following the same line of thought one therefore recognises that 
 $\mathscr{V}$ is a negative function with a global minimum
in $R_{i} = \sqrt{\a'}$, as shown in figure \ref{MNshift}.

\newpage
--------------------------------------------------------------------------------------------------------------
\newpage


\chapter{Genus-One  Amplitudes for intersecting Branes}
\everypar{\hspace{-.6cm}}
In this chapter we will discuss some of the main features of toroidal compactifications
with intersecting branes \cite{douglas}.
This quite interesting class of string backgrounds lead to four-dimensional
chiral spectra, and eventually offer diverse mechanisms for breaking supersymmetry \cite{wittenso,bachas,penta}.
They have been object of intense investigation  in  orientifold constructions
\cite{intersectingi,intersectingii,intersectingiii,intersectingiv,intersectingv}
 and in its T-dual version as magnetic backgrounds \cite{bachas,us,ralphi}.
 The  aim here is to introduce the main aspects that are relevant
for the discussion  of the effects  of  the Scherk-Schwarz mechanism 
 on D-branes at angles \cite{Angelantonj:2005hs}, presented in the next chapter.
 We will focus  on the construction of one-loop amplitudes, 
  on computations of  tadpole contributions from the transverse diagrams
  and on the analysis of the spectra of
 closed and open string excitations.

Since, as we shall explain, configurations of intersecting branes on one side and magnetised D-branes
 on the other are connected via T-duality,
in the following we will switch from one picture to the other
 following our convenience.

\vspace{1 cm}

\section{World-Sheet  Action and Boundary Conditions for Open Strings on Magnetic Backgrounds}
 \everypar{\hspace{-.6cm}}
  
 Open strings couple in a  natural way 
  through their charged endpoints to a magnetic background $F_{ij}$ \cite{tseytlin,aboob}.
   The $U(1)$-invariant
 coupling for the bosonic part of the open string action 
 in  conformal gauge is
\beq
S = \frac{1}{4\pi \a'} \int d\s d\tau \p^{\a}X^{\mu} \p_{\a}X_{\mu}  
 - q_L \int d\tau  A_{i}(X)\p_{\tau}X^{i}(0,\tau) -   q_R \int d\tau A_{i}(X)\p_{\tau} X^{i} (\pi,\tau), 
\eeq
where $A_{i}$ is the vector potential $F_{ij} = \p_{i}A_{j}  - \p_{j}A_{i}$.


 In the simplest case  of  constant
 magnetic field  along two directions  $F_{i j} =  F\e_{i j}$
  $i,j = 1,2$, 
 the open string boundary conditions for the  $q_{L}$ charged  left end-point are modified as follows
\bea
\p_{\s}X^{1}(0,\tau) - 2\pi \a'  q_{L} F \p_{\tau}X^{2}(0,\tau) &=& 0, \nn \\
\p_{\tau}X^{1}(0,\tau) + 2\pi \a' q_{L} F \p_{\s}X^{2}(0,\tau) &=& 0.    \label{mixed} 
\eea

 Similar conditions hold for the right  endpoint,
  after the replacement $q_L \rightarrow -q_R$.

The  constant $F_{ij}$ along two
 directions on a D-p brane thus 
 mixes   the boundary conditions for the two  open-string 
coordinates that couple to the background.
 
 By performing a T-duality, say along the 2-direction 
$X^{2} \rightarrow \tilde{X^2}$,
 one can achieve a more geometrical picture 
for the meaning of these 
 boundaries conditions.
  T: $X^{2} \rightarrow \tilde{X^2}$  
interchanges the two world-sheet coordinates
 for this direction  ($\tilde{X}^{2}(\s,\tau) = X^{2}(\tau,\s))$
  and thus $\p_{\s}X^{2}\stackrel{T}{\leftrightarrow} \p_{\tau}\tilde{X}^{2}$ and
 $\p_{\tau}X^{2}\stackrel{T}{\leftrightarrow} \p_{\s}\tilde{X}^{2}$,
   so that the  boundary conditions in the T-dual picture read
\bea
\p_{\s}\left(X^{1}(0,\tau) -    2\pi \a' q_L F \tilde{X}^{2}(0,\tau)\right) &=& 0 \nn \\
\p_{\tau} \left(X^{1}(0,\tau) +    2\pi \a'  q_L F \tilde{X}^{2}(0,\tau)\right) &=& 0. \nn \label{Tbc} \\
\eea

\begin{figure}  
\begin{center} 
\includegraphics[scale=1, height=9cm]{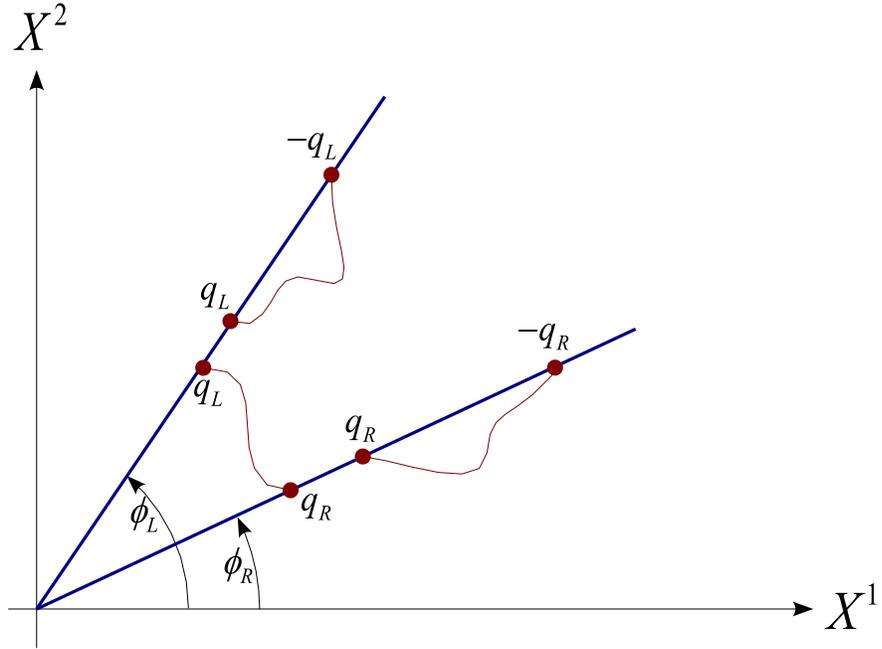}
\caption{After a $T$-duality along one of the two coordinates of the two torus,
  magnetised D-branes become intersecting (D-1)-branes. 
 The left endpoint of the open string lives on one of such  branes,
 rotated by an angle $tg(\phi_L )= -  2\pi \a'  q_L F$,
 while   the right endpoint lives on a brane rotated by  $tg(\phi_R )= 2\pi \a' q_R F$.
  Thus open strings with a net total charge stretch between two distinct
  branes, while dipole open string with  $ q_L = -  q_R $ live on a single
  brane, rotated by  $tg(\phi)= 2\pi \a' q F$.}
\label{intersectingTdual}
\end{center}
\end{figure}

 From  the conditions in (\ref{Tbc}) it follows that the
left endpoint of the open string lives on a D-(p-1) brane,
 rotated by an angle $tg(\phi_L )= -  2\pi \a'  q_L F$.
Similar conditions apply to the right 
open string endpoint which is constrained to
 live on a \emph{distinct}
 D-(p-1) brane rotated by  $tg(\phi_R )= 2\pi \a' q_R F$, as far as $ q_L \ne -  q_R $.
  Therefore a neatly charged open
 string is effectively stretched
 between two D-(p-1) branes intersecting under
 the non trivial angle $\phi = |\phi_L|+ |\phi_R| = 
  tg^{-1}( 2\pi \a'q_L F) +  tg^{-1}( 2\pi \a'q_R F)$.
 While if   $q_L = -q_R$ the two angles coincide
 and thus  both
 the endpoints live on a single rotated
  D-(p-1) brane, as shown in fig. \ref{intersectingTdual} .

\vspace{1 cm}

 In order to compute the spectrum
  of open strings on magnetised branes one
  needs to  find the eighenfunctions of the
 wave operator satisfying the mixed conditions (\ref{mixed}).
 
   By using  complex coordinates
 $X^{\pm} = \frac{1}{\sqrt{2}} \left( X^{1}\pm iX^{2} \right)$
  one is able to disentangle the boundary conditions

\bea
\p_{\s}X^{+}(0,\tau) - 2i \pi \a'  q_{R} F \  \p_{\tau}X^{+}(0,\tau) &=& 0, \nn \\
\p_{\tau}X^{+}(\pi,\tau) + 2\pi \a' q_{L} F \  \p_{\s}X^{+}(\pi,\tau) &=& 0, \label{magnetisedbc}
\eea
where  $X^{-} = (X^{+})^*$.

The eighenfunctions of the world-sheet wave
 operator $\p^{\a}\p_{\a}$  satisfying (\ref{magnetisedbc})
for a \emph{charged} open string   $(q_{L}+ q_{R} \ne 0)$  are

  \beq
 \phi_{n}(\s,\tau) = \frac{1}{\sqrt{|n-\d|}}e^{-i\tau(n-\d)}\cos [ (n-\d)\s + \Phi_L  ],
\eeq
 $\d  = (\phi_L + \phi_R)/ \pi$ being the shift
 on the vibrational frequency  depending on the angle
  $(\phi_L + \phi_R)$
 (in $\pi$ unites) between the two branes.

The solution for a charged classical vibrating 
 string $ q_{L}+ q_{R} \ne 0$ of $(\p^{2}_{\s} - \p^{2}_{\tau})X^{+} = 0$ 
is then a general combination of fundamental harmonics $\phi_{n}$
\beq 
X^{+}= x^{+} + i\left(\sum_{n \ge 1} a_n \phi_{n} - \sum_{n \ge 0} b^{*}_n \phi_{-n}\right).\label{modeexpansionB}
\eeq

After canonical quantisation one finds  the
 usual harmonic oscillator commutator relations between 
the operators  $a_n$ and $b_n$, 
so that  $a_n$ creates states whose masses
 are  lowered by $-\d$, while  the states created by $b_n$ 
 have  masses shifted by 
 $+ \d$.

  Also the centre of mass operators $x^{+}$
  and $x^{-}$ do not commute anymore
\beq
[ x^{+},  x^{-}] = \frac{\pi \a '}{\phi_L + \phi_R}. \label{noncom}  
\eeq

Since the Hamiltonian does not depend on these operators
 one can diagonalise at the same time the energy
 with say $x^{+}$, so that the spectrum is $x^{-}$
  degenerate.

  This is the stringy analogous  of the 
the Landau degeneracy for a point charge
 in a uniform magnetic field.

Moreover, the  absence of a momentum in the normal mode expansion (\ref{modeexpansionB})
can be explained by the fact that in the $T$-dual
 picture the centre of mass of the open string
 is stuck at the brane intersection,
 with a quantum mechanical uncertainty (\ref{noncom}), proportional
 to a string effective length. 

\vspace{1 cm}  

For open dipole strings  the frequencies are not shifted $\d = (\phi_{L} + \phi_{R})/\pi = 0$,
  but   the  zero modes in the coordinate expansion are modified as follows \cite{aboob}
\beq
X^{+} = \frac{ x^{+} + p_{-}(\tau - i \ tg(\Phi))(\s - \frac{\pi}{2}) }{\sqrt{1 + tg^{2}(\Phi)}} + \ldots , \label{dipoleexpansion}
\eeq

 the ellipsis referring  to the contributions
 from the  unshifted  oscillators.

In the T-dual picture with intersecting branes
on a compact space 
  the scaling  factor that appears
  in the zero mode in (\ref{dipoleexpansion})  
 takes simply into account the length of the rotated brane
  on the compact space, as we will discuss in the following.

\vspace{.5 cm}

Moving to the coupling of the  world-sheet fermions  to the magnetic background
\beq
S_F  = \frac{i}{4\pi \a '}\int d\s d\tau \bar{\psi}^{\mu}\gamma^{\a}\p_{\a}\psi_{\mu} -\frac{iq_L}{4}
\int d\tau \bar{\psi}^{\nu}(0)\gamma^{0} \psi_{\mu}(0) -\frac{iq_R}{4}
\int d\tau \bar{\psi}^{\nu}(\pi)\gamma^{0} \psi_{\mu}(\pi),
\eeq

 the
boundary conditions read

\bea
(\psi^{1}_{L} - Fq_L \psi^{2}_{R}) \d \psi^{1}_{L} - (\psi^{1}_{R} + Fq_L \psi^{2}_{L})  \d \psi^{1}_{R}  \nn \\ +
(\psi^{2}_{L} + Fq_L \psi^{1}_{R}) \d \psi^{2}_{L} - (\psi^{2}_{R} - Fq_L \psi^{1}_{L}) \d \psi^{2}_{R} = 0,
\eea
for  the left charged endpoint and similarly for the right endpoint
 after the  replacement $q_L \rightarrow - q_R$. 

Notice how the magnetic background induces a mixing  between $\psi^1$  and  $\psi^2$
  with respect to  the usual boundary conditions on the
type I world-sheet fermions

\beq
\psi^{\mu}_{L} \d \psi^{\mu}_{L} - \psi^{\mu}_{R} \d \psi^{\mu}_{R} = 0.  
\eeq

In  a complex basis $\psi^{\pm} = \frac{1}{\sqrt{2}}(\psi^{1} \pm i \psi^{2})$
one is able to resolve the mixing between the two coordinates
\bea
(1 \mp iF q_L )\psi^{\pm}_{L}(0) &=& (1 \pm iF q_L )\psi^{\pm}_{R}(0) \nn \\ 
(1 \pm iF q_R )\psi^{\pm}_{L}(\pi) &=& (-)^{a}(1 \mp iF q_R )\psi^{+}_{R}(\pi),
\eea
where in the above equations $a = 0$ for the  R sector while $a = 1$ for the NS one.

The eighenfunction of the two dimensional Dirac operator which solves 
 the above boundary conditions for the R sector are
\begin{displaymath}
 \psi^{R \pm}_{n} = \left( \begin{array}{c} 
 \psi^{\pm}_{n, L} \\ 
 \psi^{\pm}_{n, R}    \end{array} \right) = \frac{1}{\sqrt{2}} \left( \begin{array}{c} 
 e^{-[i(n \pm \d)(\tau + \s) \pm \Phi_L ]} \\ 
 e^{-[i(n \pm \d)(\tau - \s) \mp  \Phi_L ]}\end{array} \right),   
\end{displaymath}
while in the NS sector are
\begin{displaymath}
 \psi^{NS \pm}_{n} = \left( \begin{array}{c} 
 \psi^{\pm}_{n, L} \\ 
 \psi^{\pm}_{n, R}    \end{array} \right) = \frac{1}{\sqrt{2}} \left( \begin{array}{c} 
 e^{-[i(n - \frac{1}{2} \pm \d)(\tau + \s) \pm \Phi_L ]} \\ 
 e^{-[i(n -  \frac{1}{2}  \pm  \d)(\tau - \s)  \mp  \Phi_L ]}\end{array} \right).   
\end{displaymath}

The generic solution of the equations of motion is thus
 given by an expansion
 in the normal modes  eighenspinors
 listed above. 
  After the canonical anticommutators
 are imposed on the oscillation modes,
 one can construct the zero mode Virasoro operator

\beq
  L_0 = - \sum_{\mathbb{Z}+ a/2}(n - \frac{a}{2}   + \d ): d_{-n}^{-} d_{n}^{+}: + \Delta (a),
\eeq
 with $a = 0$ for the R sector while  $a = \frac{1}{2}$ for the NS one.

An important quantity in order to recover    the open string spectrum
  is the $\Delta (a)$ normal order constant that we  will  compute  in the next section.

\begin{figure}  
\begin{center} 
\includegraphics[scale=1, height=5cm]{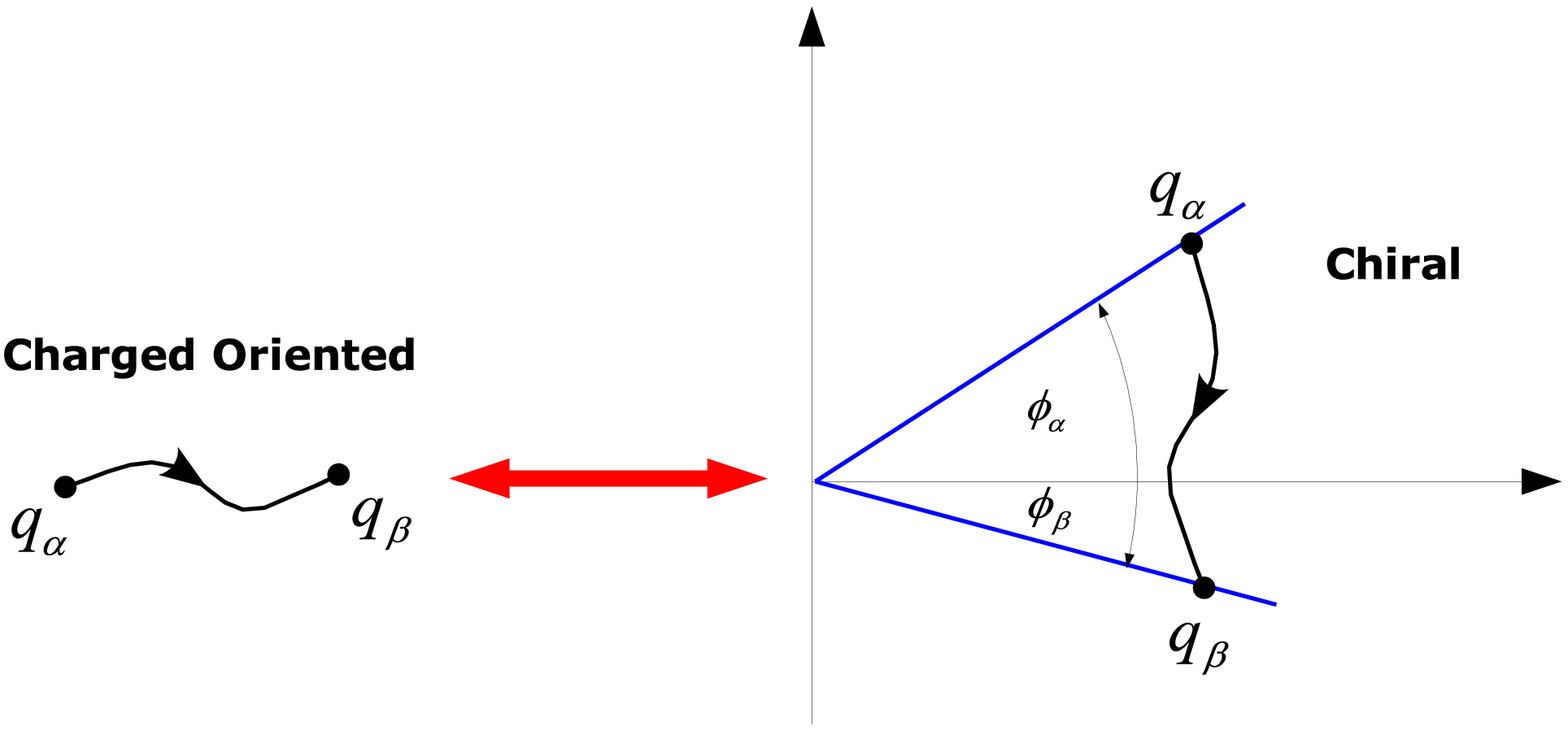}  
\caption{}
\label{}
\end{center}
\end{figure}

\begin{figure}  
\begin{center} 
\includegraphics[scale=1, height=5cm]{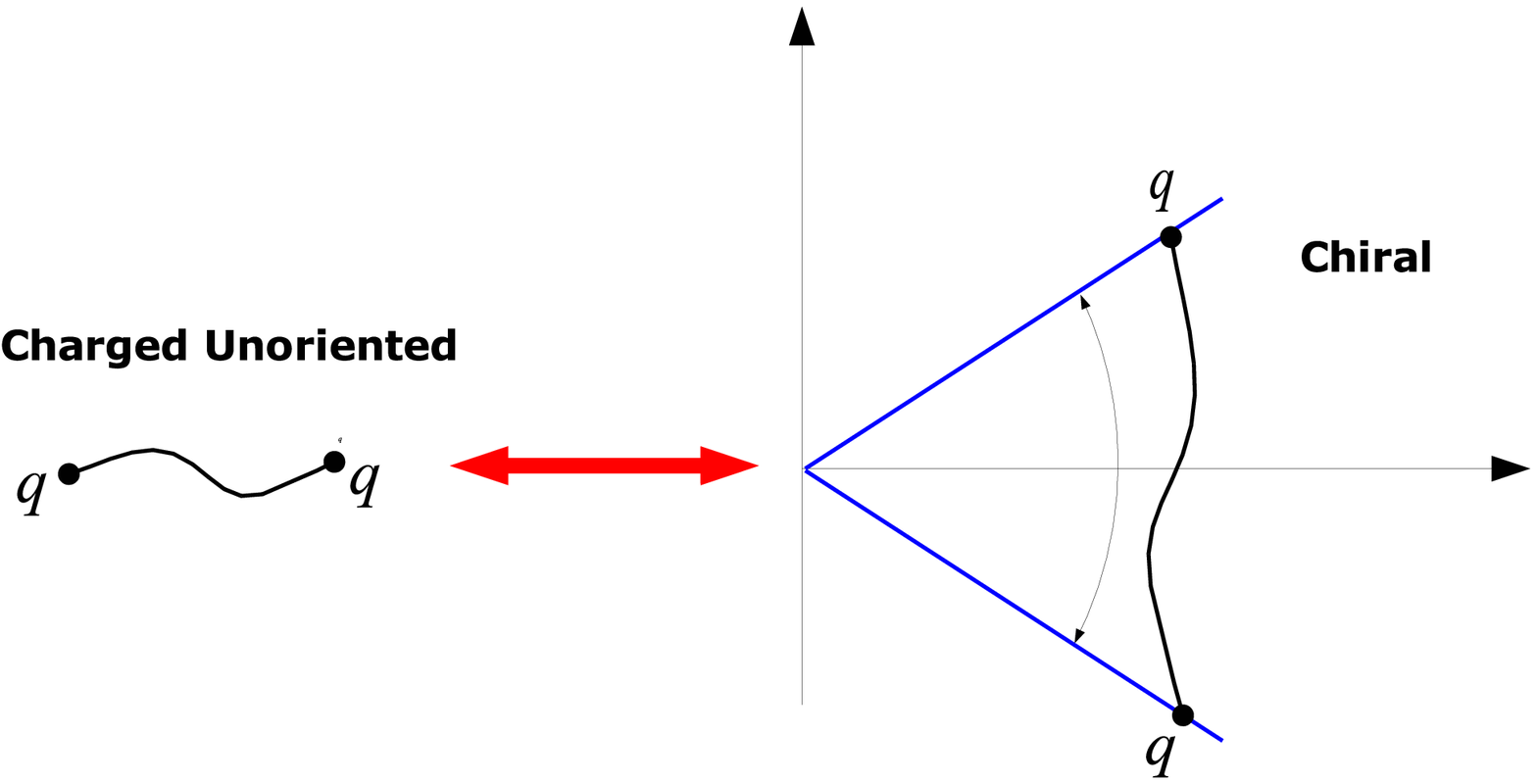} 
\caption{}
\label{}
\end{center}
\end{figure}

\begin{figure}  
\begin{center} 
\includegraphics[scale=1, height=5cm]{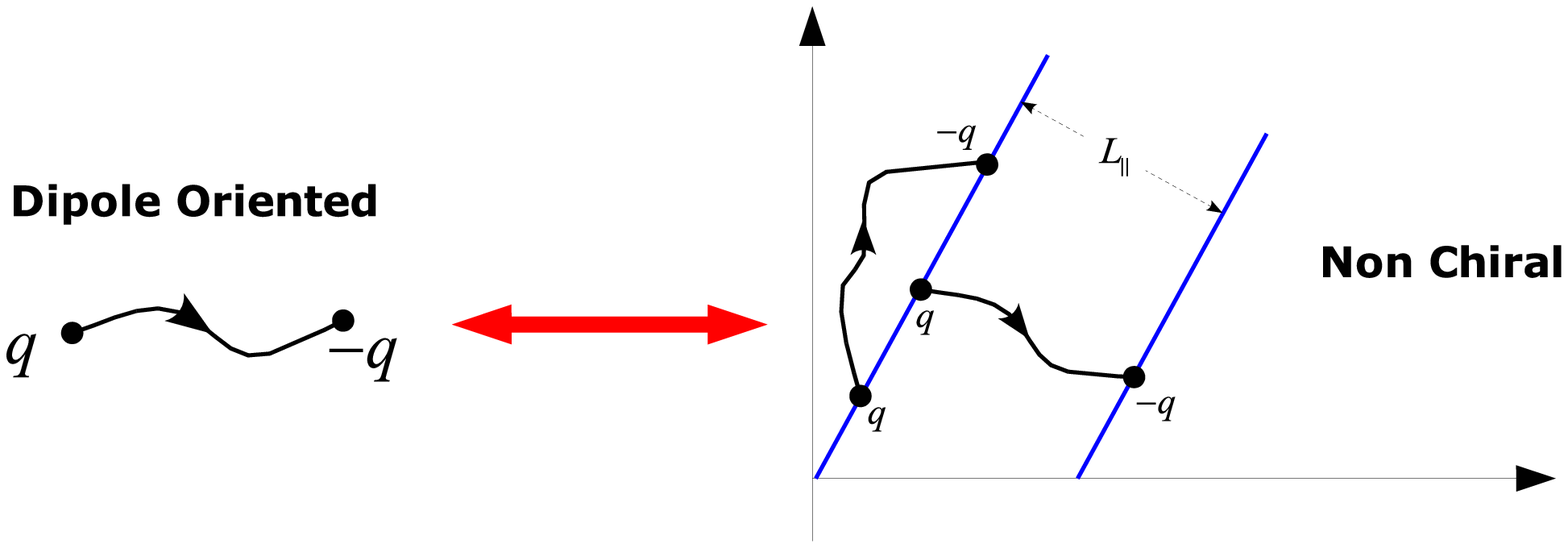} 
\caption{}
\label{}
\end{center}
\end{figure}

\vspace{3 cm}

\subsection{Shifts on The Zero Point Energies}

 \everypar{\hspace{-.6cm}}
 As shown in the previous section, both in the R and NS sectors the frequencies 
of the open string oscillators are shifted by
 the amount $\d = \phi_{ab}/ \pi$ due to the 
interaction with the magnetic background.

 The  zero point energies for the   $SO(8)$  characters
   can be  computed by regularising  
 the  divergent sums  from normal ordering
   world-sheet modes.

For the two coordinates $X_{\pm}$
 one can make the  sum 
  $\sum_{n=1}^{\infty}(n+ \d) $
 to converge, by inserting a  dumping  factor $e^{-\e(n+ \d)}$ 
 and look at the finite
$\e$-independent part of the  expansion
 
\beq
 \sum_{n=1}^{\infty}(n+ \d)e^{-\e(n+ \d)} = \frac{1}{\e^{2}} -\frac{1}{12} - \frac{\d}{2} - \frac{\d^{2}}{2} +O(\e).
\eeq 

   Thus the vacuum shift due to $X_{\pm}$ is given by
 \beq  
\Delta_{X_{\pm}} = \frac{1}{2}\left(   -\frac{2}{12} - \d - \d^{2} \right) =  -\frac{2}{24} - \frac{\d}{2} - \frac{\d^{2}}{2}.
\eeq 

For  the two world-sheet fermions  
 in the  Ramond sector,  the contribution is  opposite to that computed for the bosonic coordinates 
\beq  
\Delta_{\psi _{R \pm} } =  +\frac{2}{24} + \frac{\d}{2} +\frac{\d^{2}}{2},
\eeq 
 as one might  expect  by the integer-mode expansion  and by world-sheet supersymmetry.

  The NS contribution instead
gives 
\beq  
\Delta_{\psi _{NS \pm} } =  -\frac{2}{48}  +\frac{\d^{2}}{2}.
\eeq

\vspace{1 cm}
Taking into account both the shifts in 
 the masses and in the zero point energies we can compute 
the characters for open strings living on the constant magnetic background

\bea 
 tr_{X_{\pm}}q^{L_{0}} &=& q^{- \left(\frac{2}{24} - \frac{\d}{2} - \frac{\d^{2}}{2}\right)}\cdot \frac{1}{\prod^{\infty}_{n=1}(1- q^{n -\d})(1- q^{n +\d -1})}, \nn  \\
  tr_{ \psi_{R} \pm}\left(\mathcal{P}^{\pm}_{GSO}  q^{ L_{0}} \right) &=& 2q^{+ \left(\frac{2}{24} + \frac{\d}{2} + \frac{\d^{2}}{2}\right)   } \cdot \frac{ \prod^{\infty}_{n=1}\left( (1+ q^{n -\d})(1+ q^{n +\d -1})\pm  (1- q^{n -\d})(1- q^{n +\d -1})\right)}{2}, \nn  \\
tr_{\psi_{NS}\pm} \left(\mathcal{P}^{\pm}_{GSO}  q^{ L_{0}} \right) &=& q^{- \left(\frac{2}{48}  - \frac{\d^{2}}{2}\right)}\cdot  \frac{\prod^{\infty}_{n=1}\left(  (1+ q^{n - \frac{1}{2} -\d})(1+ q^{n -\frac{1}{2}+\d })\pm   (1+ q^{n - \frac{1}{2} -\d})(1+ q^{n -\frac{1}{2}+\d })\right)}{2}. \nn \\     \label{magchar}
\eea 

The different choices of 
GSO projections in eq. (\ref{magchar}) give
 the $SO(2)$  characters that encode the effects 
 induced by the constant magnetic background on the spectrum
\bea
O_2 (\d) &=&  tr_{\psi_{NS}\pm } \left(\mathcal{P}^{+}_{GSO}  q^{ L_{0}} \right), 
\qquad V_2 (\d) =  tr_{\psi_{NS}\pm } \left(\mathcal{P}^{-}_{GSO}  q^{ L_{0}} \right), \nn  \\  
C_2 (\d) &=&  tr_{\psi_{R}\pm } \left(\mathcal{P}^{+}_{GSO}  q^{ L_{0}} \right),
\qquad S_2 (\d) =  tr_{\psi_{R}\pm } \left(\mathcal{P}^{-}_{GSO}  q^{ L_{0}} \right). \nn
\eea

These characters can also  be written in terms of Jacobi theta functions 

\bea
O_2 (\d) &=& q^{\frac{\d^{2}}{2}} \frac{\theta_{3}(\d \tau|\tau) + \theta_{4}(\d \tau|\tau)}{2 \eta},
\qquad V_2 (\d) = q^{\frac{\d^{2}}{2}} \frac{\theta_{3}(\d \tau|\tau) - \theta_{4}(\d \tau|\tau)}{2 \eta},     \nn  \\  
C_2 (\d) &=&  q^{\frac{\d^{2}}{2}} \frac{\theta_{2}(\d \tau|\tau) - i\theta_{1}(\d \tau|\tau)}{2 \eta}, 
\qquad S_2 (\d) =  q^{\frac{\d^{2}}{2}} \frac{\theta_{2}(\d \tau|\tau) + i\theta_{1}(\d \tau|\tau)}{2 \eta},                                                                          \label{thetamagn}
\eea 

while  the  
 contributions from  the two bosonic
  coordinates coupled to  the magnetic background is given by

\beq
 \chi(\d) =   \frac{i \eta}{ q^{\frac{\d^{2}}{2}}\theta_{1}(\d \tau|\tau)}.\label{bosmagn}  
\eeq

\vspace{3 cm}

\section{A first Look at the Spectrum at Branes Intersections}
 \everypar{\hspace{-.6cm}}

\begin{figure}  
\begin{center} 
{\includegraphics[scale=1, height=8cm]{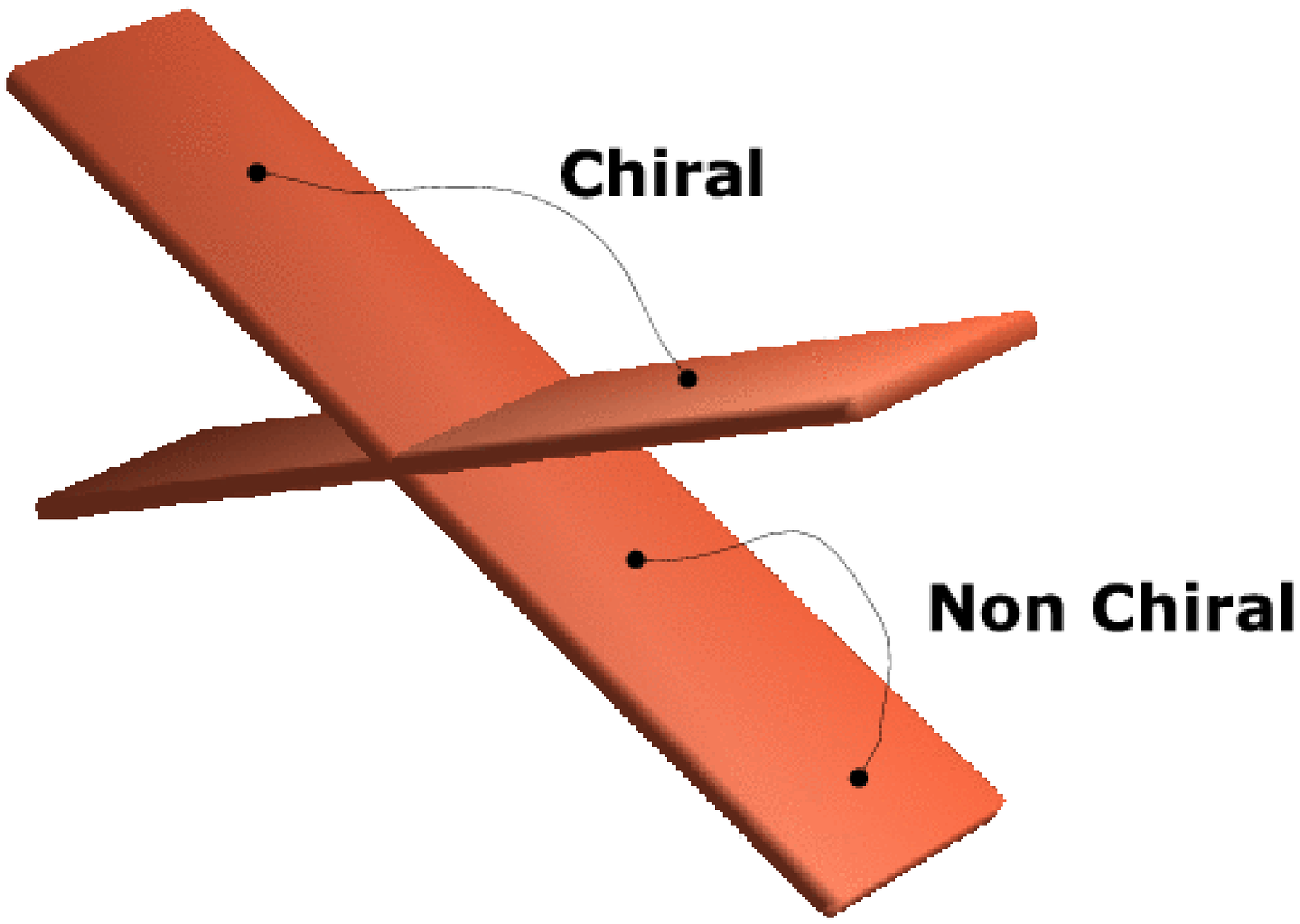}}
\caption{  .}
\label{intersecting2D}
\end{center}
\end{figure}

In this section we want to give a first look at
 the spectrum of open strings   living at the 
intersection of two D1-branes forming an angle $\phi_{ab} = \pi\d$ 
 ($ |\d|<1$), as in fig. (\ref{intersecting2D}).

 To this end,  we decompose the ten-dimensional characters
according to the breaking of the ten dimensional 
Lorentz group $SO(1,9) \rightarrow SO(1,7) \times SO(2)$

\beq   
(V_8 (\d) - S_8 (\d))\chi(\d) = \left( V_6 (0) O_2(\d) +  O_6 (0) V_2(\d) - S_6 (0) S_2(\d) - C_6 (0) C_2(\d) \right)
\chi(\d), \label{decomposition}
\eeq 

where the internal characters are 
 given in (\ref{thetamagn}) and (\ref{bosmagn}).

\vspace{1 cm}

The essential features of the open-string spectra
  are that states with 
opposite internal helicity couple  differently
 to the magnetic background. 
Their modes 
are shifted in opposite ways and,  thus at the intersecting points
  only one chirality  
remains  massless,
while  states of opposite chirality 
acquire a mass.

 In particular, this is the case
 for  compactifications  to four dimensions, being
 one of the interesting feature of the 
open string spectra from intersecting branes
configurations.  

The spectrum at the brane intersection
is encoded in the decompositions of the characters 
given in eq. (\ref{decomposition}).  

The eight dimensional vector becomes massive, since
 $V_6(0)O_2(\d)\chi(\d) \sim 6 q^{ \d/ 2}$, 
while   $ O_6(0) V_2(\d) \chi(\d)  \sim q^{- \mid \d \mid  +\frac {\mid \d \mid}{2}}$,
  the lowest scalar state is thus  tachyonic.

Its presence signals the instability
of the configuration of the two 
intersecting branes that  
 break all the supersymmetries.
 In fact the two branes are each separately 1/2 BPS
 but for a generic intersection angle they preserve 
 two  different combinations of  supercharges, thus  breaking all the original
 type II  supersymmetries.

In this situation there is not a cancellation
between gravitational attraction and RR repulsion
between the two branes and the system is not a
true vacuum, where quantum fluctuations might be
considered   under control. Indeed tree-level exchanges
of closed states will certainly drive the
system to a  new configuration, maybe through  brane   
recombination  where, after  a  tachyon condensation process \cite{tachyonopen},
the system can reach  a nearby \emph{true vacuum}.

\begin{figure}  
\begin{center} 
{\includegraphics[scale=1, height=5cm]{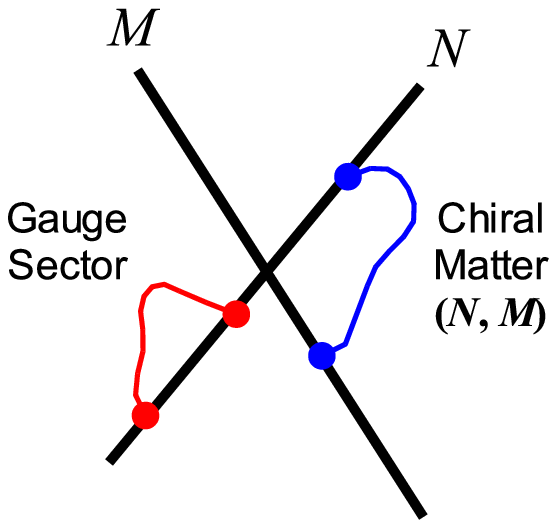}}
\caption{  .}
\label{intersecting2DD}
\end{center}
\end{figure}

However, when the two branes intersects on
a higher dimensional torus $T^d$, for 
$d \ge 4$,   particular configurations of 
intersecting branes exist  where some amount of the original
  supersymmetry is preserved
and  the   scalar excitations  return to be
massless, being part of a matter supermultiplet.

\vspace{.3 cm}

The ten-dimensional  character $S_{8}$
   decomposes into the sum of two 
eight dimensional characters  with opposite chirality  ($S_6$), ($C_6$).

Due to the magnetic background, the
 states in one of the two eight-dimensional characters
become all massive,  while those in  the other 
one remain massless.
Which of the two internal helicity characters acquires a
mass 
depends on the sign of the relative angle 
between the two D-branes. 
This can be clearly  seen by the
 leading term  expansions for the internal characters
\beq  
 C_{6}(0) S_2(\d) \chi(\d)  \sim  q^{-\frac{\d }{2} +
 \frac {\mid \d \mid}{2}}  \qquad   
  S_{6}(0) C_2(\d) \chi(\d)  \sim  q^{\frac { \d }{2} +\frac {\mid \d \mid}{2}}.
\eeq

 \vspace{1 cm}
Therefore the character that contains massless states supplies half of the degree of freedom
 of an eight dimensional Weyl spinor. 
The remaining half is given by the mirror sector, open strings with left end right
endpoints interchanged that experience an \emph{opposite shift} $ - \d$ in their mode
expansion.
 The presence of a mirror sector is due to the introduction of orientifold
planes,  as usual  necessary  in order to
cancel the total RR charge and get
  an anomaly-free massless spectrum.

Not to break translational  invariance 
in the extended dimensions, the  D-branes  transverse 
directions need to be compact.
Since the  D-branes are sources of RR fluxes on
their transverse directions and the flux lines
cannot escape at infinity on the compact space,
the presence of an opposite charged object that can  sink 
 the  Faraday  lines of the D-branes is necessary to avoid a violation
of the equations of motion, that are a version of Gauss
low for  charged extended objects.

Orientifold planes in turns require the presence of
mirror D-branes, whose open-string
excitations have left and right endpoints
interchanged  with respect to the original ones.
Their normal modes are therefore subjected to an opposite shift.

 It turns out that in  the mirror sector there are the missing degrees of freedom
 to form a one chirality   
   eight-dimenisonal Weyl  spinor.

Therefore at the brane 
intersection  we have the number of degrees of freedom
of an eight-dimensional   Weyl spinor, accompanied by (scalar) tachyonic excitations.  

\vspace{3 cm}

\section{Magnetised Branes Topological Invariants  and Intersecting Branes Wrapping Numbers}

 \everypar{\hspace{-.6cm}}

In order to take into account  all the open string sectors
   excitations of the magnetised branes wrapping a torus,
 one still need a generalisation of the Dirac quantisation condition
 between electric and magnetic charges. In the T-dual picture this condition
 translates into a statement on the geometry of
 the rotated-branes configurations.
 
 \vspace{.5 cm} 
 
 Let us consider a constant magnetic field $F$
 on a two torus $T^2$, whose coordinates on the
 two canonical cycles are  $x \sim x + L_1$ and $y \sim y + L_2$.
One can write the $U(1)$ potential locally as $A = (0, \frac{1}{2}Fx)$,
 but globally, in order
 to respect the periodicity along the two cycles, two different coordinate  patches are needed. 
   
\begin{figure}  
\begin{center} 
\includegraphics[scale=1, height=5cm]{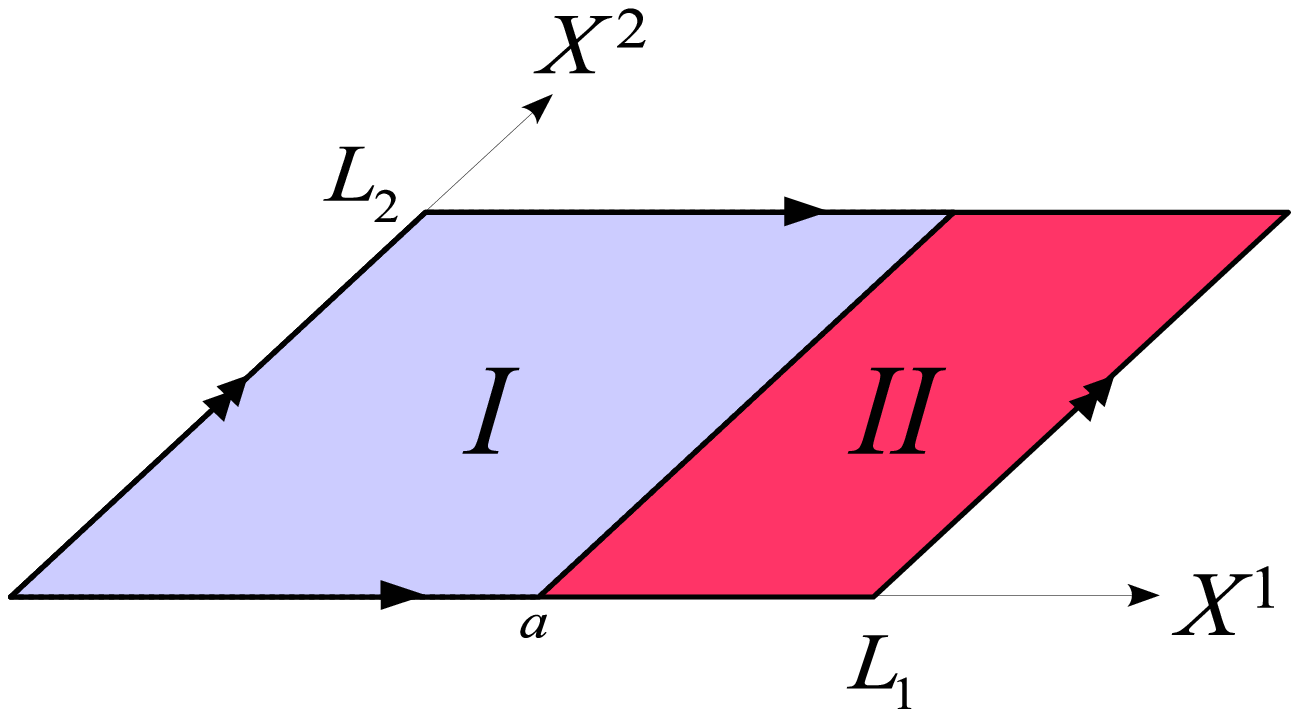}  
\caption{}
\label{regions}
\end{center}
\end{figure}

  As an example, one can take the following two regions as in fig. \ref{regions},
    $0 \le x < a$  with $A_I = (0, \frac{1}{2}Fx)$, and 
   $a \le x < L_1$ with  $A_{II} = (0, \frac{1}{2}F(x - L_1) )$.

 The  magnetic field on the torus is therefore a non trivial $U(1)$ gauge bundle
 ($F \ne 0$ in cohomology), and
 as a result   a non vanishing  magnetic field
 corresponds to the presence of magnetic charges, (see  (\ref{Diracquantpoint}) 
and the related discussion).  

  The first Chern number associated to the non trivial bundle is  
 a non vanishing integer
\beq
\frac{1}{4\pi^{2}}\int_{T^{2}} F = m.  
\eeq

 For a constant magnetic field on a square $T^2$ the above condition gives
\beq
\frac{1}{4\pi^{2}}  F L_1 L_2  = m.
\eeq

 If one considers a magnetised  brane wrapping $n$ times the $T^2$
 the firs Chern number needs to be modified to 
\beq
\frac{n}{4\pi^{2}}\int_{T^{2}} F = m,\label{Chernclass}  
\eeq
since due to the wrapping number $n$,
 the brane volume  is $n$ times the $T^2$ volume.

 A T-duality along the $y$ direction $L_2 \rightarrow  \a'/L_2$ 
 transforms the  magnetised  D-p brane
into a D-(p-1)brane  rotated by an angle $tg\phi = 2\pi \a' F$,
 and  the relation (\ref{Chernclass})
 into 

\beq
 tg\phi = \frac{m}{n}\frac{L_2}{L_1},
\eeq

which shows that the integer $m$ corresponds
 to the number of times the brane wraps around the
 vertical cycle, while $n$ corresponds the the horizontal wrapping number \cite{ralphi}.





\vspace{3 cm}

\section{One-Loop Amplitudes for Intersecting Branes Wrapping a Two Torus}

\begin{figure}  
\begin{center} 
\includegraphics[scale=1, height=5cm]{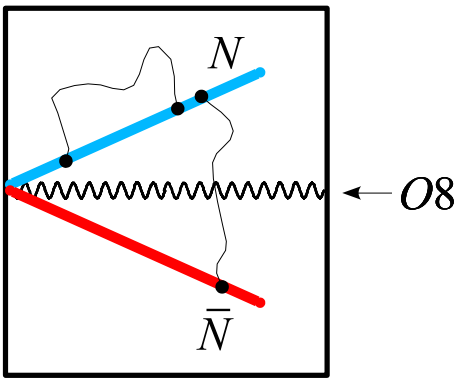}  
\caption{A portion of a stack of $N$ branes is represented in blue, while in red is represented its
image under the mirror orientifold plane (horizontal wavy line).}
\label{T2intersecting}
\end{center}
\end{figure}

 \everypar{\hspace{-.6cm}}
 We consider   the configuration of  intersecting branes
 wrapping   a two-torus $T^{2}$ shown in fig. \ref{T2intersecting}.

  An O8 plane wraps one of the
two homology cycles, while
 a stack 
$N_{\a}$ and its orientifold image  $\bar{N}_{\a}$   form 
 an angle  $\phi_{\a}$   (respectively $-\phi_{\a}$) with the $O8$-plane.

The presence of the O8 plane, necessary to cancel the total RR charge,
 induces a     $\Omega I$ 
  symmetry on both  the closed
 and open string spectrum,  $I$ being a reflection along $X^{2}$, the orthogonal direction
 to the O-plane and $\Omega$ the world-sheet parity.

 
  The $I$ reflection
  needs to act in a  crystallographic way
 on the target two torus,
 it is not then  difficult to see that
 this is the case only  when
the real part $U_{1}$ of the $T^{2}$ complex structure
  is vanishing or equals to  $1/2$.

Therefore the  off-diagonals entries of the
 metric in the canonical  homology basis,
  given by $U_{1}$  in  eq.(\ref{metrictwotorus}),
 in the presence of the O8-plane
 become  discrete moduli with the possible values $U_{1} = 0,1/2$.
This is the T-dual manifestation of what we have already
  encountered for
the $B_{ij}$ field in the presence of O7 planes as discussed
in sec. \ref{sectionDiscreteModuli}. In fact,
after a T-duality along one of the two internal coordinates, 
the dimensions of the D-branes
and O-planes are increased or lowered, depending
 whether the duality is  along directions  orthogonal or parallel  to these objects.
 The $T^{2}$ complex structure $U$ and  the Kh\"aler class $T$ 
 get  also interchanged and, in particular, $b$ plays the role 
 here played by $U_{1}$. 
At the moment we consider the case $U_{1} = 0$,
 corresponding to  a square $T^{2}$, we will consider
 in the next chapter  also   the  case of   a $U_{1} = 1/2$
 \emph{tilted} target torus, 
  which however  presents only few slightly differences.

\vspace{1 cm}

The closed string  spectrum for the background we are
considering  is encoded in the one-loop amplitudes  
$\frac{1}{2}\mathcal{T} + \mathcal{K}$

\bea
\mathcal{T} &=& (V_8 - S_8)(\bar{V_8  } - \bar{C_8})\Lambda_{ \vec{m}  \vec{n}}, \nn \\
\mathcal{K} &=& \frac{1}{2}(V_8 - S_8)P_{\vec{m}}W_{\vec{n}}.  \label{TKintersecting}
\eea

 A change of proper time in the second of eqs. (\ref{TKintersecting})
 gives the   transverse Klein bottle amplitude 
$\tilde{\mathcal{K}}$, proportional to the 
square of the  tension and the charge of the $O8$ plane 

\beq
 \tilde{\mathcal{K}} =  \left(2^{2}\  \sqrt{\frac{L_1}{L_2}}\right)^{2}  (V_8 - S_8) W_{\vec{2n}}P_{\vec{2m}}. \label{kbintersecitng}
\eeq


Turning to the open  string amplitudes, the brane excitations
are encoded in
  $\mathcal{A}$ and   $\mathcal{M}$.
 
\begin{figure}  
\begin{center} 
\includegraphics[scale=1, height=5cm]{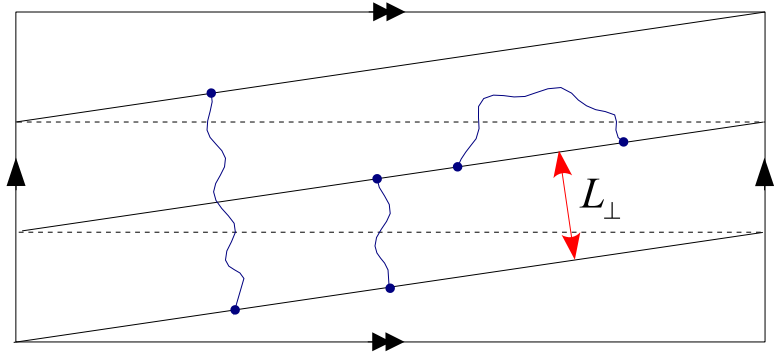}  
\caption{Open strings living on a rotated brane wrapping the $T^{2}$.
 The Kaluza-Klein momenta are proportional to $1/L_{||}$, where $L_{||}$ is the length
of the brane, while the windings are multiple of $L_{\perp}$, the minimal distance
between two consecutive wrappings of the brane on the $T^{2}$.}
\label{Lorthogonal}
\end{center}
\end{figure}

The Annulus is given by
\bea
&\mathcal{A}& = N_{\a}\bar{N}_{\a}( V_8 - S_8 )(0)P_{m}(L_{||})  W_{n}(L_{\perp}) \nn  \\
 &+& \left( \frac{N_{\a}^{2}}{2}( V_8 - S_8 )\left(2 \Phi_{\a}\tau / \pi  \right) +  
     \frac{\bar{N}_{\a}^{2}}{2}( V_8 - S_8 ) \left(2 \Phi_{\a}\tau / \pi  \right)  \right)
      \frac{ 2k_{\a}\omega_{\a} \cdot i\eta}{\theta_{1}\left( 2 \Phi_{\a}\tau / \pi \mid \tau  \right)},
\label{anellomagn} \eea

where   the first line gives the contribution 
from dipole string that live on  rotated branes,
whose Kaluza-Klein momenta are scaled by the
effective length of the brane $L_{||}$, and  winding
are possible between  parallel wrappings of 
the branes, $L_{\perp}$ being
the minimal distance   between two consecutive
 D-brane wrapping
\footnote{For a D-brane that forms an angle $\phi$ with one of the horizontal  cycles of  $T^{2}$
\beq
tg \phi = \frac{\omega}{\kappa}\frac{L_{2}}{L_{2}},\label{quantcond}
\eeq
 with the integers  $\omega$ and $\kappa$ being the horizontal and vertical wrapping numbers.

The length of the brane is given by Pitagora's  theorem
\beq
L_{||} = \sqrt{ (\omega L_{1})^{2} + (\kappa L_{2})^{2}},
\eeq

while the distance $L_{\perp}$  between two consecutive brane wrappings is given by
\beq
L_{\perp} = \frac{L_{1}L_{2}}{L_{||}}.
\eeq
Using the quantisation condition (\ref{quantcond}) one can also obtain the following useful
relations
\beq
 L_{||} = \frac{\omega L_{1}}{\cos \phi} =  \frac{\kappa L_{2}}{\sin \phi}, 
\qquad L_{\perp} = \frac{L_{2}}{\omega}\cos \phi = \frac{L_{1}}{\kappa} \sin \phi.
\eeq

 The zero modes contribution to the mass spectrum for dipole strings 
 is written in terms of the two geometrical data $L_{||}$ and $L_{\perp}$
\beq
M^{2}_{Z.M.} = \left(\frac{m}{L_{||}}\right)^{2} + \frac{1}{\a'^{2}}(nL_{\perp})^{2}.
\eeq},  (see fig. \ref{Lorthogonal}).

In the second line  of (\ref{anellomagn})  is displayed the contribution 
from the unoriented charged open strings and 
their image under the mirror O8 plane.
 In this sector open strings stretch
between  branes and their orientifold images.

The multiplicity factor $2 \k_{\a}\omega_{\a}$
is the intersection number $I_{\a \bar{\a}}$ \footnote{The intersection number between to 
 stacks $N_{\a}$ and $N_{\b}$ of  D-branes on a square $T^{2}$
 is a topological invariant $I_{\a \b} = k_{\a}\omega_{\b} - k_{\b}\omega_{\a}$,
 that depends on the two pairs of wrapping numbers $(\omega_{\a}, \k_{\a})$
 and  $(\omega_{\b}, \k_{\b})$ of the two stacks.}, that 
corresponds to the number of intersections  between
 the brane and its image.
 Due to the possibility of multiple intersections of two lines
 on a compact space, 
  replicas of  chiral matter sectors  emerge naturally from 
 intersecting branes configurations.

\vspace{1cm}

 In the  M\"obius only  the unoriented states in the last two terms of (\ref{anellomagn}) can flow

\beq
\mathcal{M} =    -\left( \frac{N_{\a}}{2}(\hat{V}_8 -  \hat{S}_8 )\left(2 \Phi_{\a}\tau / \pi  \right) +  
     \frac{\bar{N}_{\a}   }{2}( \hat{V}_8 -  \hat{S}_8 ) \left(2 \Phi_{\a}\tau / \pi  \right)  \right)
          \frac{ 2k_{\a}\cdot i\hat{\eta}}{\hat{\theta} _{1}\left( 2 \Phi_{\a}\tau / \pi \mid \tau  \right)}.
 \label{moebius5}
\eeq

\begin{figure}  
\begin{center} 
\includegraphics[scale=1, height=8cm]{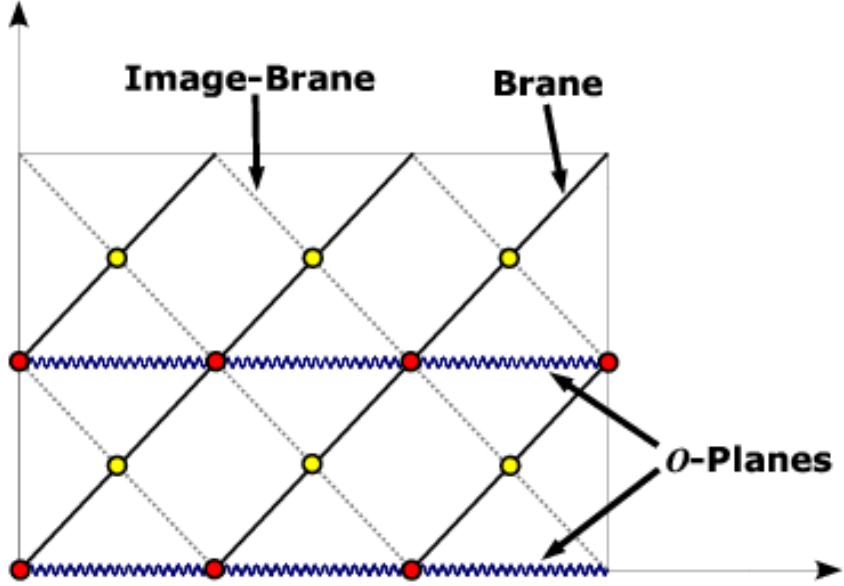}  
\caption{ Multiple intersections between a brane and its image with respect an orientifold plane.
 If the brane has integer wrapping numbers $(\omega, \kappa)$ and its image $(\omega, -\kappa)$,
the total number of intersections is given by $2\omega \kappa$.
 $2 \kappa$ of such points (in red) belong also to the pair of O-planes,
 while the remaining $2(\omega -1)\kappa$ intersection points (in yellow) fall outside of the O-planes.}
\label{intersectionsOD}
\end{center}
\end{figure}

Notice the  multiplicity factor $2 k_{\a}$, corresponding  to 
the number of intersection points between    
the brane and its image that belong  also to 
the pair of O-8 planes,
  the two fixed circles
  $y=0$ and $y= L_2 /2$ under the involution $I$,
  (see fig. \ref{intersectionsOD}).

\vspace{1 cm}
 The gauge sector is given by the massless excitations
of the  oriented dipole strings in the annulus amplitude, the first term in (\ref{anellomagn}).
After a dimensional reduction of the original SO(8) characters we have
\beq
\mathcal{A}^{Or}_{0} = N_{\a}\bar{N}_{\a}( V_6 O_2(0) + O_6 V_2(0) - S_6 C_2(0) - C_6 S_2(0)), \label{orientedA}
\eeq

where the multiplicity $N_{\a}\bar{N}_{\a}$ in front of $V_{8}$
 corresponds to the dimension of the adjoint representation of 
the unitary group $U(N_{\a})$.

The reason why the gauge group is unitary
 follows from the fact that   the vectors and the gauginos
  correspond to oriented  dipole string excitations that
cannot flow in the M\"obius.
Only in the limit of uncharged open strings, corresponding to a vanishing
angle between the stacks of branes and the O-plane, these states
become unoriented and one recovers the previously discussed
 configurations of
  D-branes on top of  O-planes, with  $SO(N)$ or $USp(2M)$ gauge groups.

\vspace{.5 cm}

From eq. (\ref{orientedA}) one can read the physical degrees of freedom of
 an   eight-dimensional gauge boson vector,  contained
 in   $V_6 O_2(0)$,
    two massless scalars  from $O_6 V_2(0)$ and 
two opposite chirality gauginos from   $S_6 C_2(0)$ and   $C_6 S_2(0)$.
Altogether, these massless excitations
  correspond to a  $\mathcal{N} = 1$ $d=8$ super Yang Mills  $U(N_{\a})$    vector multiplet.

\vspace{1 cm}

The non-supersymmetric matter lives at the intersection points between
  branes  and   image branes, and they come
 in a number of
  families which equals the intersection number.
 The matter states are given by   fermionic 
massless states  
and  tachyonic excitations.
  In the present case of a single stack $N_{\a}$ and its
 orientifold image $\bar{N}_{\a}$, the open string stretching between those objects
  have \emph{equal charge}
at their endpoints and thus are
  \emph{unoriented}.  These states  are given by  the last two terms in the Annulus 
 (\ref{anellomagn}) and are the only contributions
  to the  M\"obius amplitude  (\ref{moebius5})


\bea
&&\mathcal{A}^{Un}_{0} + \mathcal{M}_{0} = \nn \\ && \left[ \left(\frac{N_{\a}^{2}}{2} - \frac{N_{\a}}{2}\right) \frac{2\omega_{\a}k_{\a} + 2k_{\a}}{2} + 
\left(\frac{N_{\a}^{2}}{2} + \frac{N_{\a}}{2}\right) \frac{2\omega_{\a}k_{\a} - 2k_{\a}}{2} \right] 
\left(O_6  V_2 (2\phi_{\a}) - C_6 S_2 (2\phi_{\a})\right) \nn \\
 &+& \left[ \left(\frac{\bar{N}_{\a}^{2}}{2} - \frac{ \bar{N}_{\a}}{2}\right) \frac{2\omega_{\a}k_{\a} + 2k_{\a}}{2} + 
\left(\frac{\bar{N}_{\a}^{2}}{2} + \frac{\bar{N}_{\a}}{2}\right) \frac{2\omega_{\a}k_{\a} - 2k_{\a}}{2}\right] \left(O_6 V_2 (-2\phi_{\a}) - S_6 C_2 (-2\phi_{\a})\right). \nn
\eea

\vspace{.5 cm}

 From the above amplitudes one can read 
 $2\omega_{\a}k_{\a}$ 
families of chiral spinors and 
(tachyonic) scalars,
 each for every  intersection between the brane and  its orientifold image.

$A_{\a} = (2\omega_{\a}k_{\a} + 2k_{\a})/2$ of these
families carry an antisymmetric representation of
the gauge group $U(N_{\a})$,  (see fig. \ref{intersectionsOD} ).

Whenever a stack $N_{\a}$   wraps more than one time 
the horizontal one-cycle  ($\omega_{\a} > 1$),
a number given by
$S_{\a}=  (2\omega_{\a}k_{\a} - 2k_{\a})/2$ of intersection points of the total   
 $2\omega_{\a}k_{\a}$  fall outside the orientifold planes and
 $S_{\a}$ families  carry a \emph{symmetric} 
representation of the $U(N_{\a})$ gauge group.

\vspace{1 cm}
  In a more general situation,  where more than one
 rotated stack are present,  the open strings stretching
between two different stacks $N_{\a}$ and $N_{\b}$
  have different charge at their endpoints and
therefore are oriented. 
These states cannot flow
  in the M\"obius,  
 thus  come in  \emph{bifundamental} representations
 of the  gauge group $U(N_{\a}) \times U(N_{\b})$.

The presence of tachyonic excitations
at the intersection points  
signals the instability of the configuration
that breaks all the supersymmetries.

To compute the  fate of the final configuration is in
general quite problematic 
in string theory due to the dificulties 
connected to the issues of background 
redefinition in a first quantised  formulation.
Nonetheless, various old and new suggestions have
been given to these problems
\cite{Fischler:1986ci,Fischler:1986tb,Dudas:2004nd}.
 At the end of the day
tachyonic excitations might be very welcome,
since they could trigger gauge symmetry 
breaking through  a process of tachyon condensation  
which also induces 
brane recombination processes. 

\vspace{3 cm}

\section{Tadpole cancellation conditions}

 \everypar{\hspace{-.6cm}}

In order to compute  the various tadpole contributions 
one  needs to extract the coefficients in front of the
 closed string massless states   flowing 
in the  cylinder, Klein bottle and M\"obius strip 
tree-level diagrams.

The characters  
in the presence of the magnetic background
 have been written in terms of Jacobi theta function  in (\ref{thetamagn}),
therefore to go to the transverse channel one 
 can use the well known behaviour of the theta functions
 under an S modular transformation

\beq
\theta\left[^{\a}_{\b}\right]\left(\frac{\z}{\tau}|-\frac{1}{\tau}\right) = (-i \tau)^{\frac{1}{2}}e^{2\pi \a\b} e^{i\pi \frac{\z^{2}}{\tau}} \theta\left[^{-\b}_{\a}\right](\z|\tau). 
\eeq

From  the one loop Annulus amplitude (\ref{anellomagn}), and  with the help of the previous relation,
one can compute the transverse Annulus amplitude

\bea
\tilde{\mathcal{A}} &=& 2^{-5}\left(\frac{L_{||}}{L_{\perp}}N_{\a}\bar{N}_{\a}(V_8 - S_8)W_{n}(L_{\perp})P_{m}(L_{||})\right) \nn \\ &+&  2^{-4}\left(\frac{N_{\a}^{2}}{2}(V_8 - S_8)(2\phi_{\a}/\pi |l) + \frac{\bar{N}_{\a}^{2}}{2}(V_8 - S_8)(-2\phi_{\a}/\pi |l) \right) \frac{2k_{\a}\omega_{\a} \eta}{\theta_{1}(2\phi_{\a}/\pi |l)} \label{cylind}.
\eea

 From    the  expressions for the theta
functions one can obtain the leading terms in the expansions
 of  the  closed string   characters 
 
\bea
O_2 (\phi/\pi |l) &\sim& 1
\qquad V_2(\phi/\pi |l)\sim \cos(\phi)\nn  \\  
C_2 (\phi/\pi |l)     &\sim& e^{-i \frac{\phi}{2}} 
\qquad S_2 (\phi/\pi |l) \sim e^{i \frac{\phi}{2}},                                                                          \label{expansions}  
\eea 
 and
\beq 
\theta_{1}(\phi/\pi |l) \sim 2sin(\phi). \label{expansion}
\eeq

These leading terms  in (\ref{expansions}) and (\ref{expansion})
are the ingredients that one needs  
in order to  extract
the tadpole  contributions from the various diagrams.

 Since closed-string modes are not shifted by 
the magnetic field, the effects 
of the $U(1)$ background   enter only in modifying the  reflection coefficients
in the transverse amplitudes.
In particular for this reason one does not need to care
about   zero point energy shifts in the closed
string  characters, since they cancel each-other as in the standard (unmagnetised) case.

\vspace{1 cm}

The Annulus amplitude  contribution to the NSNS tadpole
is given by the  coefficient
in front of the NSNS massless states, contained in the character $V_6$ in
(\ref{cylind})
\beq
 V_8(2\phi_{\a}/\pi |l) = V_6(0 |l) O_2(2\phi_{\a}/\pi |l) +  O_6(0 |l) V_2(2\phi_{\a}/\pi |l). 
\eeq

It corresponds to  $\mathcal{B}^{2}$,  the
 square of the one point function for the emission of the dilaton $\phi$ or $g_{\m}^{\m}$
  from a disk.
  \
  By using the (\ref{expansions}) and 
(\ref{expansion}) one then obtains

\beq
 \mathcal{B}^{2}
 =  2^{-5}\left[\frac{L_{||}}{L_{\perp}}N_{\a}\bar{N}_{\a}
+ \left(\frac{N_{\a}^{2}}{2} + \frac{\bar{N}_{\a}^{2}}{2}\right)\frac{k_{\a} \omega_{\a}}{sin\phi_{\a}\cos\phi_{\phi_{\a}}} \right] \label{boundarycoef5}.
\eeq

  As expected, the above expression can be written as e perfect square 
by using the relation between
the lengths $L_{||}$ and $L_{\perp}$ and 
the angle $\phi_{\a}$.

\beq
 \mathcal{B}^{2} = \left(2^{-3}\sqrt{ \frac{L_{||}}{L_{\perp}}}(N_{\a}+ \bar{N}_{\a})\right)^{2}\label{boundarycoeff}.
\eeq

\vspace{.5 cm}

The transverse M\"obius amplitude is given by
\beq
\tilde{\mathcal{M}} =  - \left(\frac{N_{\a}}{2}(\hat{V}_8 - \hat{S}_8)(\phi_{\a}/\pi |l) + \frac{\bar{N}_{\a}}{2}(\hat{V}_8 - \hat{S}_8)(- \phi_{\a}/\pi |l) \right) \frac{2k_{\a} \hat{\eta}}{\hat{\theta}_{1}(2\phi_{\a}/\pi |l)},
\eeq
where  to lighten the notation in the above expression, we did not explicitly  write the real part  of the
 M\"obius proper time .

The contribution from this diagram to the 
NSNS tadpole is therefore
\beq
\tilde{\mathcal{M}} = -2 ( N_{\a}+ \bar{N}_{\a}) \frac{L_{||}}{L_{2}}
\eeq

that is consistently equal to  
 $- 2\sqrt{   \mathcal{B}^{2}\cdot \mathcal{C}^{2}}$,
 twice  the square-root of the product
  between the amplitude $\mathcal{B}$ for the emission of massless states 
 from an  O-plane $\mathcal{C}$ and  those from a D-brane (\ref{boundarycoef5}).

 The total NSNS tadpole contributions from the transverse diagrams  reads
\bea
\tilde{\mathcal{K}}_{0}+ \tilde{\mathcal{A}}_{0}+\tilde{\mathcal{M}}_{0} &=&  \mathcal{C}^{2}
 + \mathcal{B}^{2}  - 2 \mathcal{C}\mathcal{B} = \left(  \mathcal{B} - \mathcal{C} \right)^2 \nn \\
&=& \left( 2^{-3}\sqrt{\frac{L_{||}}{L_{\perp}}}(N_{\a} + \bar{N}_{\a}) - 2^{2}\sqrt{\frac{L_{1}}{L_{2}}}\right)^2. \nn
\eea
NSNS tadpole cancellation condition  therefore asks for
\beq
\sqrt{\frac{L_{||}}{L_{\perp}}}(N_{\a} + \bar{N}_{\a}) =  2^{5}\sqrt{\frac{L_{1}}{L_{2}}}, \label{tadpoleNS}
\eeq

 that, with the help of the relations between the
 effective length of the brane and the angle eq. (\ref{tadpoleNS}),   can be rewritten as 
\beq
\omega_{\a}(N_{\a} + \bar{N}_{\a}) \sqrt{1 + tg(\phi_{a})^2 } =  2^{5}.\label{NSNStadpolemagn}
\eeq

In the present case of a $T^{2}$ compactification there is not a  way to cancel the NSNS tadpole
  (\ref{NSNStadpolemagn}) for a non vanishing angle $\phi_{\a} \ne 0$,
 and thus the configuration  breaks all the supersymmetries\footnote{Recall that  supersymmetry implies
the vanishing of all the tadpoles \cite{Martinec:1986wa} \cite{Friedan:1985ge}}.
 However, this is not the case starting for an higher dimensional $T^{d}$ compactifications 
  for $d \ge 4$, 
 where both  BPS and not BPS configurations of
 intersecting branes do exist that cancel the NSNS tadpole.

After a T-duality along the
vertical coordinate  of the two torus, the l.h.s. of eq.  (\ref{NSNStadpolemagn}), corresponds 
to the Dirac Born Infeld action \cite{bidirac,tseytlin} for a stack $N_{\a}$
of  magnetised D9-branes wrapping $\omega_{\a}$ times the 
two torus
\beq
 \omega_{\a}( N_{\a} + \bar{N}_{\a}) \sqrt{det( \1 +  \mathcal{F}) } = 2\times 2^{4}.
\eeq

This relation of course  
 corresponds again  to a condition for the cancellation
between the tension of the stack of  magnetised branes
and the tension of the O9 planes,  equal to $16$ for each of the two planes, \footnote{
O-p planes correspond to the loci of fixed points under the orientifold involution
 $I_{\perp}\Omega$, a mix between a world-sheet involution $\Omega$ and a target-space one 
 $I_{\perp}$ along orthogonal directions respect to the O-p planes. Therefore 
 for $p < 9$ the planes always come in pairs, each carrying a tension
 that equals $16$ times those of D-p branes.  $16$ is the number of \emph{physical}
branes  corresponding  to the rank of the gauge group $SO(32)$ in type I vacua.} 
  $tg(\phi_{a}) =  2\pi \alpha'\mathcal{F}$,   $\mathcal{F}$ being
 the value of the constant magnetic field on the two-cycle

\begin{displaymath}
 \mathcal{F} = 
2\pi \alpha' \left( \begin{array}{cc} 
0 & F \\ 
-F & 0 \end{array} \right).    
\end{displaymath}

\vspace{1 cm}

Turning  to the RR tadpole, 
one needs  to take into account the contributions 
 from the decomposition of the RR ten dimensional characters
  in $\tilde{\mathcal{A}}$ and $\tilde{\mathcal{M}}$
\beq
S_8(2\phi_a /\pi |l)   =   S_6(0|l) C_2 (2\phi_a /\pi |l) + S_6(0|l) C_2 (2\phi_a / \pi |l).
\eeq

 By using the $q$-expansions in (\ref{expansions}) and (\ref{expansion}),  
 it is clear that different internal helicities \footnote{
Here helicity is referred to the massless states  contained in the same characters
when they appear in the loop open string amplitudes. In the present case the characters
actually describe closed string RR states propagating in the transverse diagrams.}
  
 will introduce   different phases in their couplings to the magnetic background.
\vspace{.2 cm}
 
The massless state from  $S_6$ in $\tilde{\mathcal{A}}$
  gives the following contribution
\beq
\mathcal{B}^{2}_{S_6} = 2^{-6}\frac{L_{||}}{L_{\perp}} \left( N_{\a}e^{-i\phi_{\a}} + \bar{N}_{\a}e^{ i\phi_{\a}}\right)^2,
\eeq
while the  contribution from  $C_6$  is
\beq
\mathcal{B}^{2}_{C_6} = 2^{-6}\frac{L_{||}}{L_{\perp}} \left( N_{\a}e^{i\phi_{\a}} + \bar{N}_{\a}e^{-i\phi_{\a}}\right)^2
\eeq
Both the contributions can be expressed by  a single formula
\beq
\mathcal{B}^{2}_{RR} = 2^{-6}\frac{L_{||}}{L_{\perp}} \left( N_{\a}e^{2i\phi_{\a}\l} + \bar{N}_{\a}e^{-2i\phi_{\a}\l}\right)^2
\eeq
where $\l = \pm \frac{1}{2}$ is the  helicity of the internal 
character  (in the open string sense), that couples in different ways
to the magnetic background. 
 This formula generalises nicely  to higher-dimensional (factorised)
 torus compactifications as we will see in the next chapter. 

Taking also into account the contributions from
 the Klein bottle and the M\"obius amplitudes,
 the RR tadpole cancellation  condition reads
\beq
\sqrt{\frac{L_{||}}{L_{\perp}}} \left( N_{\a}e^{2i\phi_{\a}\l} + \bar{N}_{\a}e^{-2i\phi_{\a}\l}\right) =
 2^{5} \sqrt{\frac{L_{1}}{L_{2}}} \label{RRtad}.
\eeq

 The real part of the above equation gives
\beq
\omega_{\a}\left( N_{\a} + \bar{N}_{\a} \right) = 2^5, \label{RRcharg}
\eeq
 while the imaginary part just  asks for
 an equal number of branes and image branes
\beq
 N_{\a} - \bar{N}_{\a} = 0.
\eeq
 After having imposed this last condition 
eq. (\ref{RRcharg}) asks for a cancellation
  between the RR charge of the stack of branes
  and the RR  of orientifold plane

\beq
\omega_{\a} N_{\a} = 16.
\eeq

 It is worth to notice that an horizontal wrapping
 $\omega_{\a}> 1$
 for the D8 stack  increases its effective charge
  by $\omega_{\a}$ times its original value.
 This has a simple explanation  after a T-duality
 along the vertical coordinate of the two torus, where
 the D8 branes gets mapped  into  $\omega_{\a}$ copies of D9 branes or,
 equivalently, to a D9  wrapping $\omega_{\a}$-times  the two torus.

\newpage
--------------------------------------------------------------------------------------------------------------
\newpage

\chapter{Scherk-Schwarz Breaking and Intersecting Branes}
\everypar{\hspace{-.6cm}}

In this chapter we analyse the effects of the Scherk-Schwarz mechanism
on configurations of intersecting branes \cite{Angelantonj:2005hs}.
After a discussion on the conditions under which  a D-brane wrapping a torus  
feels the breaking or  remains supersymmetric, we will show how the  Scherk-Schwarz  mechanism
can be successfully employed to give a tree-level  mass to non-chiral fermionic
excitations.
 
This offers a solution for a longstanding problem of intersecting branes vacua
since non-chiral massless fermions, ubiquitous in open string spectra 
of intersecting branes, are certainly not present in the standard model.

Finally, we will describe the effects of the Scherk-Schwarz mechanism on $D-7$  intersecting branes
 wrapping  a $T^{4}$.











\section{Scherk-Schwarz and M-theory breaking}
\everypar{\hspace{-.6cm}}
 In the following we shall review the main features of type IIB orientifolds with supersymmetry broken via momentum or winding deformations. Actually, to emphasise the geometry of these orientifolds, we focus on the more conventional momentum-deformed closed-string spectrum and study the $\Omega$ and $\Omega I$ orientifolds, where $\Omega$ is the world-sheet parity and $ I$ is an inversion of the compact coordinate. The latter orientifold better described within type IIA is the T-dual description of type IIB with winding shifts.  

The deformed nine-dimensional spectrum of the closed IIA and IIB oriented strings is encoded in the one-loop partition function
\bea\label{niness}
\mathcal{T} &=& {\textstyle{1\over 2}} \biggl[ |V_8 - S_8 |^2 \, \Lambda_{m,n} + |V_8 + S_8 |^2\, (-1)^m \, \Lambda_{m,n} \nn \\
&+ & |O_8 - C_8 |^2 \, \Lambda_{m,n+{1\over 2}} + |O_8 + C_8 |^2 \, (-1)^m \,\Lambda_{m,n+{1\over 2}}\biggr] 
\eea

where $\Lambda_{m,n}$ is a $(1,1)$-dimensional Narain lattice. Indeed, all the fermions have acquired a mass proportional to the inverse radius, and a twisted tachyon is present if $R<2 \sqrt{\alpha '}$. 

The Klein bottle amplitudes associated to the two orientifolds $\Omega$ and $\Omega ' = \Omega I$ can be straightforwardly determined from \ref{niness}\ and read
\footnote{As recently shown in \cite{emilian2}\ one has the additional option of symmetrising the R-R sector while acting simultaneously on the tower of Kaluza-Klein states with order-two shifts.}

$$
\mathcal{K} = {\textstyle{1\over 2}} \, (V_8 - S_8 ) \, P_{2m} \,,
$$
and
$$
\mathcal{K} \,\,' = {\textstyle{1\over 2}} \, (V_8 - S_8) \, W_n + {\textstyle{1\over 2}} \, (O_8 - C_8 ) \, W_{n+{1\over 2}} \,.
$$
The former amplitude clearly spells out the presence of O9 planes while the latter involves conjugate pairs of O8 and $\overline{{\rm O}8}$ planes sitting at the two edges of the segment $S^1 / I$. 

More interesting is the open-string spectrum associated to these orientifolds. In the first case, it consists of open strings stretched between D9 branes. In particular, these branes wrap the compact direction and thus their excitations are affected by the Scherk-Schwarz deformation. Indeed, the corresponding annulus and M\"obius-strip amplitudes
$$
\mathcal{A} = {\textstyle{1\over 2}} \, N^2\, \left( V_8 \, P_{2m} - S_8 \, P_{2m+1} \right) \,,
$$
and 
$$
\mathcal{M} = - {\textstyle{1\over 2}} \, N \, \left( \hat V_8 \, P_{2m} - \hat S_8 \, P_{2m+1} \right) \,,
$$
clearly reveal that the fermions are now massive and supersymmetry is spontaneously broken. In the second case, the D8 branes are transverse to the compact direction along which the Scherk-Schwarz deformation takes place, and thus the open-string spectrum is unaffected. However, the brane configuration has to respect the symmetries we are gauging, and therefore the pairs of image (anti-)branes have to be displaced at points diametrically opposite. In equations
$$
\mathcal{A}\,\, ' = \left[ {\textstyle{1\over 2}} \left( N^2 + M^2 \right) \, (V_8 - S_8)  + M\, N \, (O_8 - C_8) \right] \left( W_n
+ W_{n+{1\over 2}} \right) \,,
$$
and
$$
\mathcal{M} \,\, '= -{\textstyle{1\over 2}} \, (N+M) \, \hat V _8 \, \left( W_n + W_{n+{1\over 2}} \right) 
+ {\textstyle{1\over 2}}\, (N-M) \, \hat S_8 \, \left( W_n - W_{n+{1\over 2}} \right) \,.
$$
As expected the fermions are still massless (at tree level), while supersymmetry is broken only by the simultaneous presence of branes and anti-branes. 

In conclusion, we can summarise our results as follows \cite{ADSi}: if the Scherk-Schwarz deformation connects points on the brane then supersymmetry is broken in the open-string sector, otherwise image branes have to be properly added in order to respect the gauged symmetry and the fermions stay massless. 
 
\section{Wilson lines on magnetised branes}
\everypar{\hspace{-.6cm}}
In this section we review some known facts about magnetised (or intersecting) branes 
\cite{intersectingi,intersectingii,intersectingiii,intersectingiv,intersectingv} and comment on the role of Wilson lines and/or brane displacements\footnote{See \cite{standardxiv}\ for a detailed analysis of Higgsing in intersecting brane vacua, and 
\cite{marianna}\ for related issues.}. This will give us the opportunity to present our notation and introduce some simple deformations that however will play an important role in the forthcoming sections. 

\subsection{Preliminaries on intersecting branes: notation and conventions}
\label{preliminaries}
\everypar{\hspace{-.6cm}}
Let us focus our attention on the simple case of eight-dimensional reductions with D8 branes extending along one generic direction of the $T^2$. Lower-dimensional reductions are quite simple to study, and will be discussed in the following sections. For simplicity, we take the torus to be rectangular with horizontal and vertical sides of length $R_1$ and $R_2$, respectively. Each set of parallel branes is then identified by the pair of integers $(q, k)$ that correspond to the number of times the branes wrap the horizontal and vertical sides of the $T^2$, the two canonical one-cycles. In turn, these numbers determine the oriented angle $\phi$ between the horizontal axis of the torus and the brane itself via the relation
\beq \label{dirac}
\tan \phi = {k R_2 \over q R_1} \,,
\eeq
and it is positive (negative) if starting from the horizontal axis we move counter-clock-wise
(clock-wise) towards the brane. By simple trigonometry arguments, we can then determine the effective length of the brane
$$
L_\| = \sqrt{(q R_1)^2 + (k R_2)^2}\,,
$$
and the distance between consecutive wrappings on the $T^2$
$$
L_\perp = {R_1 R_2 \over L_\|} \,.
$$
Using the quantisation condition (\ref{dirac}), $L_\|$ and $L_\perp$ can be conveniently written as
$$
L_\| = {q R_1 \over \cos\phi} = {kR_2 \over \sin\phi}\,, \qquad \quad 
L_\perp = {R_2 \over q} \cos\phi = {R_1 \over k} \sin\phi \,.
$$

\begin{figure}  
\begin{center} 
\includegraphics[scale=1, height=8cm]{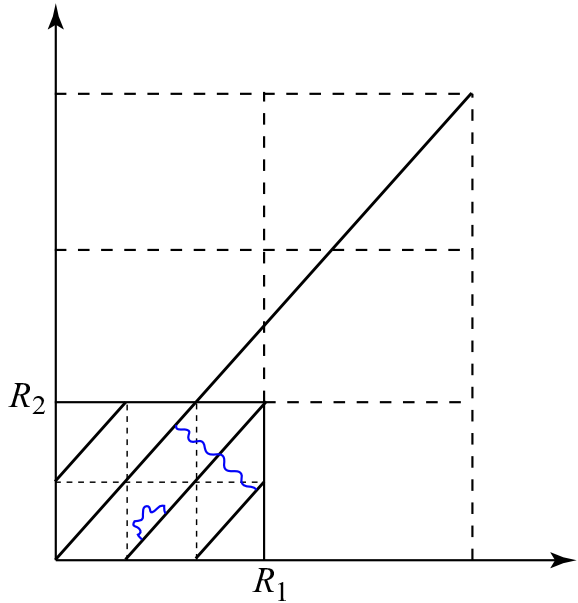}  
\caption{ Zero-mode spectrum on a rotated brane with wrapping numbers $(2,3)$.}
\label{zeromode}
\end{center}
\end{figure}

Given these two quantities it is then easy to determine the zero-mode spectrum of open strings stretched on this rotated brane \cite{bachas,intersectingi,intersectingii,intersectingiii,intersectingiv,intersectingv}
$$
M^2_{\rm z.m.} = \left( {m \over L_\|}\right)^2 + {1\over \alpha'{}^2} \left( n L_\perp\right)^2 \,.
$$
It simply consists of Kaluza-Klein states along the compact direction of the brane and of winding states between two portions of the brane in the elementary cell of the $T^2$, as indicated in figure \ref{zeromode}.

If orientifold planes and/or other branes are present, one has to consider new sectors corresponding to open-strings stretched between two different branes. These new sectors are particularly appealing since they typically support chiral matter at the intersection loci, while the masses of the string excitations now depend on the relative angle $\phi_\alpha - \phi_\beta$, and the precise dependence changes in the NS and R sectors. 

As usual, it is convenient to summarise the complete spectrum of string excitations in the annulus, and eventually M\"obius-strip, partition function. To this end let us consider the simple case of an $\Omega \mathscr{R}$ orientifold, with  $\Omega$ the standard world-sheet parity and $\mathscr{R}:\ z\to \bar z$ an anti-conformal involution acting on the complex coordinate $z\equiv y_1 + i y_2$ of the $T^2$. This operation, a symmetry of the IIA string, introduces two pairs of horizontal O8 planes, passing through the points $y_2 =0$ and $y_2 = {1\over 2} R_2$. Introducing a stuck of $N_\alpha$ coincident D8 branes with wrapping numbers $(q_\alpha,k_\alpha)$ breaks in general the orientifold symmetry, unless a suitable stack of $N_\alpha$ image branes with wrapping numbers $(q_\alpha,-k_\alpha)$ is also added \cite{intersectingi,intersectingii,intersectingiii,intersectingiv,intersectingv}. The annulus amplitude then consists of different sectors corresponding to strings stretched between a pair of (image-)branes and to strings stretched between a brane and its image:

\bea
\mathcal{A} &=& N_\alpha \bar N_\alpha \, (V_8 - S_8) [{\textstyle{0\atop 0}}]\, P_m (L_{\|\alpha}) W_n (L_{\perp\alpha}) \nn \\
&+& {\textstyle{1\over 2}} \left( N^2_\alpha\, (V_8 - S_8 ) [{\textstyle{\alpha\bar\alpha\atop 0}}] +  \bar N ^2_\alpha \, (V_8 - S_8) [{\textstyle{\bar\alpha \alpha\atop 0}}]
\right) \, {2 q_\alpha k_\alpha \over \Upsilon_1 [{\textstyle{\alpha\bar\alpha\atop 0}}]}
\eea

\begin{figure}  
\begin{center} 
\includegraphics[scale=1, height=5cm]{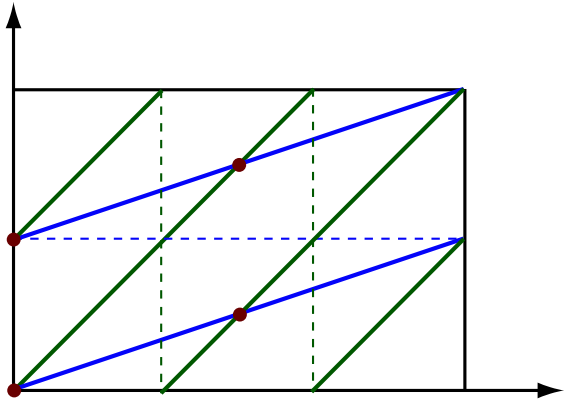}  
\caption{Two branes with wrapping numbers (2,1) and (3,2) intersecting four times on a $T^2$.}
\label{  }
\end{center}
\end{figure}

Here 
$$
P_m (L_{\|\alpha}) = \eta^{-1}\, \sum_m q^{{\alpha ' \over 2} (m/L_{\|\alpha} )^2} \quad {\rm and}\qquad
W_n (L_{\perp\alpha} ) = \eta^{-1} \sum_n q^{{1\over 2\alpha'} (nL_{\perp\alpha} )^2 } 
$$
denote the momentum and winding lattice sums, while we have endowed the $SO(8)$ characters with additional labels, reflecting the mass-shift of the string oscillators induced by the non-trivial intersection angle. In particular, for this $T^2$ reduction
$$
(V_8 -S_8) [{\textstyle{\alpha\beta \atop \gamma\delta}}] = V_6 \, O_2 (\zeta ) + O_6 \, V_2 (\zeta ) -  S_6 \, S_2 (\zeta) - C_6\, C_2 (\zeta)\,,
$$
with 
$$
\zeta = {1\over \pi} \left[ (\phi_\alpha - \phi_\beta ) {i\tau_2 \over 2} + \phi_\gamma - \phi_\delta \right]\,,
$$
$\tau_2$ being the proper time of the annulus and of the M\"obius-strip, and we use a barred index to define the angle $\phi_{\bar\alpha} = - \phi_\alpha$ of the image brane, whose wrapping numbers are $(q_{\bar\alpha}, k_{\bar\alpha})=(q_\alpha , - k_\alpha )$. The $SO(2)$ characters are defined by

\bea
O_2 (\zeta ) &=& {e^{2i\pi \zeta} \over 2\eta} \, \left[ \theta_3 (\zeta |\tau) + \theta_4 (\zeta |\tau) \right], \qquad
V_2 (\zeta ) = {e^{2i\pi \zeta} \over 2\eta} \, \left[ \theta_3 (\zeta |\tau) - \theta_4 (\zeta |\tau) \right]\,, \nn \\
 \qquad
S_2 (\zeta ) &=& {e^{2i\pi \zeta} \over 2\eta} \, \left[ \theta_2 (\zeta |\tau) +i \theta_1 (\zeta |\tau) \right]\,, \qquad
C_2 (\zeta ) = {e^{2i\pi \zeta} \over 2\eta} \, \left[ \theta_2 (\zeta |\tau) -i \theta_1 (\zeta |\tau) \right]\,, \nn
\eea

with the argument of the theta functions now depending on the relative angle, and $\tau$ the Teichm\"uller parameter of the double covering torus, {\it i.e.} $\tau = i\tau_2 /2$ for the annulus amplitude and $\tau = (1+i\tau_2)/2$ for the M\"obius-strip amplitude. Finally,
$$
\Upsilon_1 [{\textstyle{\alpha\beta \atop \gamma\delta}}] = {e^{2i\pi \zeta}\, \theta_1 (\zeta |\tau) \over i \eta}\,,
$$
encodes the contribution of the rotated world-sheet bosonic coordinates. Notice the multiplicative factor $2 q_\alpha k_\alpha$ in the unoriented sector. It has a simple geometrical interpretation in terms of the number of times a brane and its image intersect on the $T^2$. More generally, two branes with wrapping numbers $(q_\alpha , k_\alpha)$ and $(q_\beta , k_\beta)$ intersect a number of times given by (see fig. 2) 
$$
I_{\alpha\beta} = q_\beta k_\alpha - q_\alpha k_\beta  \,,
$$
its sign determining the chirality of massless fermions living at the intersection points
\cite{intersectingi,intersectingii,intersectingiii,intersectingiv,intersectingv}.

\begin{figure}  
\begin{center} 
\includegraphics[scale=1, height=5cm]{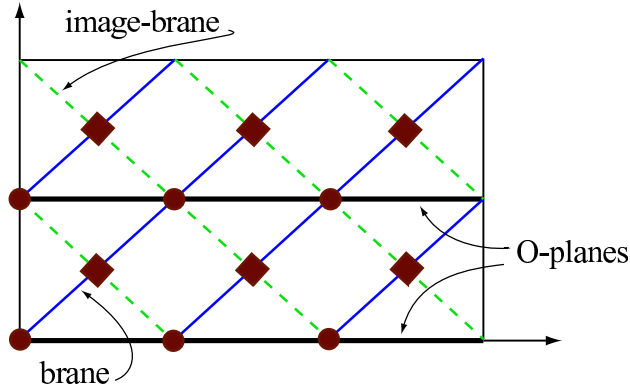}  
\caption{A brane (solid line) and its image (dashed line). The intersections denoted with a dot lie
 on the orientifold planes while those denoted with a diamond do not, and thus do not contribute to the M\"obius amplitude.}
\label{branes-oplanes}
\end{center}
\end{figure}

To conclude this brief review on intersecting branes, the M\"obius-strip amplitude
$$
\mathcal{M} = - {\textstyle{1\over 2}} \left( N_\alpha \, (\hat V_8 - \hat S_8) [{\textstyle{\alpha\bar\alpha\atop 0}}] + \bar N_\alpha \, (\hat V_8 - \hat S_8) [{\textstyle{\bar\alpha \alpha\atop 0}}] \right) \, {K_\alpha \over \hat\Upsilon_1 
[{\textstyle{\alpha\bar\alpha\atop 0}}] }
$$
receives contributions only from those $K_\alpha = 2 k_\alpha$ intersections that live on the orientifold planes (see fig. \ref{branes-oplanes}). As a result, the massless spectrum will in general contain both symmetric and antisymmetric representations of the unitary $ U(N_\alpha)$ gauge group


\vspace{1 cm}

\subsection{Wilson lines and brane displacements}
\label{Wilson}
\everypar{\hspace{-.6cm}}

As in the more conventional case of space-filling or point-like branes, one has the option of deforming the previous spectrum by introducing suitable Wilson lines \cite{Bianchi:1991eu}\ or by displacing the branes in the space transverse to their world volume \cite{pcj}. This simply amounts to the deformed zero-mode mass spectrum 
$$
M^2_{\rm z.m.} (a,c) = \left( {m\over L_\|} + a \right)^2 + {1\over \alpha'{}^2} (w L_\perp + c)^2 \,.
$$
It is then clear that for arbitrary values of $a$ and $c$ the open-string excitations are massive, while massless states emerge if
$$
a = {\mu \over L_\|} \,, \qquad c = \nu L_\perp\,,
$$ 
for $\mu$ and $\nu$ integers.
The latter condition simply reflects a symmetry under a rigid translation of the brane by an integer multiple of the distance between two consecutive wrappings. 

The corresponding modifications of the annulus partition function are then quite natural. Considering for simplicity the case of an orthogonal displacement $c=\delta L_\perp$ of $M_\alpha$ branes, that still have the same wrapping numbers, one has
\bea
\mathcal{A} &=& (V_8 - S_8 ) [{\textstyle{0\atop 0}}] \,P_{m} (L_\|) [ (N_\alpha \bar N_\alpha + M_\alpha \bar M_\alpha ) W_n (L_\perp) \nn \\
&+&  N_\alpha \bar M_\alpha W_{n-\delta} (L_\perp ) + \bar N_\alpha M_\alpha W_{n+\delta} (L_\perp)] \nn \\
&+& {\textstyle{1\over 2}} \bigg[ \left( N_\alpha^2 + M_\alpha^2 + 2 N_\alpha M_\alpha \right) (V_8 - S_8 ) [{\textstyle{\alpha\bar\alpha\atop 0}}] \nn \\
&+& \left( \bar N_\alpha^2 + \bar M_\alpha^2 + 2 \bar N_\alpha \bar M_\alpha \right) (V_8 - S_8 ) [{\textstyle{\bar\alpha \alpha \atop 0}}] \bigg] \,
 {2 q_\alpha k_\alpha  \over \Upsilon_1 [{\textstyle{\alpha\bar\alpha\atop 0}}]} 
\eea

with an obvious deformation of the M\"obius-strip amplitude.
It is then clear that for arbitrary $\delta$ the original gauge group $U(N_\alpha + M_\alpha)$ is broken to $U(N_\alpha ) \times  U(M_\alpha)$ while the (anti-)symmetric representations decompose into the sum of (anti-)symmetric representations of each group factor plus additional bi-fundamentals. As expected, for $\delta$ integer new massless vectors emerge from strings stretched between overlapping wrappings of the $N_\alpha$ and $M_\alpha$ branes and the gauge symmetmery is consequently enhanced to the original $ U(N_\alpha + M_\alpha)$.

\vspace{1 cm}
\subsection{Branes vs antibranes: an intriguing $\pi\!$uzzle}
\begin{figure}  
\begin{center} 
\includegraphics[scale=1, height=5cm]{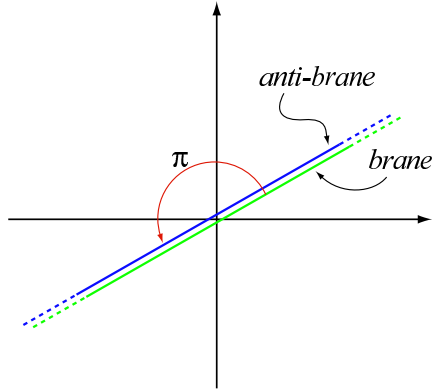}  
\caption{A brane and its anti-brane differing by a $\pi$ rotation.}
\label{antibrane}
\end{center}
\end{figure}

\everypar{\hspace{-.6cm}}
The formalism of intersecting branes here reviewed may hide some ambiguities. It is well known in fact that anti-branes (with positive tension and negative R-R charge) are nothing but regular branes (with positive tension and positive R-R charge) that have undergone a $\pi$ rotation, as shown in fig. \ref{antibrane}. On a compact space, if a brane has wrapping numbers $(q,k)$ and angle $\phi$ its conjugate partner has opposite wrapping numbers $(-q , -k)$ and an angle $\phi +\pi$. Despite branes and anti-branes satisfy the same quantisation condition 
\ref{dirac} and induce similar mass shifts in the spectrum of string excitations, their relative $\pi$ angle plays a crucial role in selecting the correct GSO projection.

Let us consider in fact the spectrum of open strings stretched between two branes with angles $\phi_\alpha$ and $\phi_\beta$. As already observed, it is encoded in the partition function
$$
{(V_8 - S_8) [{\textstyle{\alpha\beta\atop 0}}] \over \Upsilon_1 [{\textstyle{\alpha\beta \atop 0}}]} \sim {V_6 O_2 (\zeta ) + O_6 V_2  (\zeta ) -S_6 S_2 (\zeta ) - C_6 C_2  (\zeta)
\over \theta_1  (\zeta |\tau) \, q^{{1\over \pi}(\phi_\alpha - \phi_\beta)}} \,,
$$
with $\zeta = (\phi_\alpha - \phi_\beta)\tau/\pi$, where we have omitted irrelevant numerical factors.  After rotating the $\alpha$ brane, say, by an angle $\pi$ and converting it to an anti-brane, the partition function obviously reads
\beq \label{annbab}
{(V_8 - S_8) [{\textstyle{\alpha'\beta\atop 0}}] \over \Upsilon_1 [{\textstyle{\alpha'\beta \atop 0}}]} \sim {V_6 O_2 (\zeta_\pi ) + O_6 V_2  (\zeta_\pi ) -S_6 S_2 (\zeta_\pi ) - C_6 C_2  (\zeta_\pi) \over \theta_1  (\zeta_\pi  |\tau) \, q^{{1\over \pi} (\phi_\alpha - \phi_\beta) +1}} \,,
\eeq
with now $\zeta_\pi = (\phi_\alpha - \phi_\beta+\pi)\tau/\pi$. From the very definition of theta functions
$$
\theta [{\textstyle{a\atop b}}] (z|\tau) = \sum_n q^{{1\over 2} (n + a)^2} \, e^{2i \pi (n+a)(z+b)}
\,,
$$
one can then deduce the periodicity properties

\bea
\theta_3 (\zeta_\pi |\tau ) &=& + q^{- {\phi\over \pi}-{1\over 2}} \, \theta_3 (\zeta |\tau), 
\qquad
\theta_2 (\zeta_\pi |\tau) = + q^{- {\phi\over \pi}-{1\over 2}} \, \theta_2 (\zeta |\tau )\nn \\
\theta_4 (\zeta_\pi |\tau) &=& - q^{- {\phi\over \pi}-{1\over 2}} \, \theta_4 (\zeta |\tau)\,,
\qquad
\theta_1 (\zeta_\pi |\tau) = - q^{- {\phi\over \pi}-{1\over 2}} \, \theta_1 (\zeta |\tau)\,,
\eea

\begin{plain}
that, in turn,  induce a non-trivial reshuffling of the SO(2) characters
$$
\eqalign{
O_2 (\zeta_\pi ) =& V_2 (\zeta) \, q^{-{\phi\over \pi}+{1\over 2}}\,,
\cr 
V_2 (\zeta_\pi ) =& O_2 (\zeta) \, q^{-{\phi\over \pi}+{1\over 2}}\,,
\cr}
\qquad
\eqalign{
S_2 (\zeta_\pi ) =& C_2 (\zeta) \, q^{-{\phi\over \pi}+{1\over 2}}\,,
\cr 
C_2 (\zeta_\pi ) =& S_2 (\zeta) \, q^{-{\phi\over \pi}+{1\over 2}}\,.
\cr}
$$
As a result, the GSO projection in (\ref{annbab}) changes to
$$
{(V_8 - S_8) [{\textstyle{\alpha'\beta\atop 0}}] \over \Upsilon_1 [{\textstyle{\alpha'\beta \atop 0}}]} \sim {V_6 V_2 (\zeta ) + O_6 O_2  (\zeta ) -S_6 C_2 (\zeta ) - C_6 S_2  (\zeta) \over \theta_1  (\zeta |\tau) \, q^{{1\over \pi} (\phi_\alpha - \phi_\beta) }} \,,
$$
\end{plain}
that indeed pertains to open strings stretched between pairs of branes and anti-branes \cite{tachyonopen}.

It is evident, that the formalism introduced in sub-sections \ref{preliminaries}   and
 \ref{Wilson}   is well tailored to study any kind of brane if care is used in dealing with apparently innocuous $\pi$ angles. To avoid ambiguities,  in this paper we adopt the convention to use the term brane when $q$ is positive, and anti-brane when $q$ is negative. This is clearly suggested by their contribution to the R-R tadpoles. Then according to the sign of their vertical wrapping number $k$, they can induce positive or negative charges for lower-degree gauge potentials, and consequently transmute into lower-dimensional branes (for $k>0$) or anti-branes (for $k<0$).

\section{Freely acting orbifolds, Scherk-Schwarz deformations and intersecting branes}
\label{freely}
\everypar{\hspace{-.6cm}}
We can now turn to the study of the effects freely acting shifts have on rotated and intersecting branes. For simplicity we shall focus to the case of a $\mathbb{Z}_2$ shift, although similar considerations can be straightforwardly extended to the more general case. 
\begin{plain}
On a given $T^2$ one has in principle to distinguish among the three cases of shifts acting along the horizontal axis only, along the vertical axis only or along both the horizontal and vertical axis. It turns out that it is enough to consider one case, since the others can be unambiguously determined. To this end, let us consider the shift

$$
\delta :\quad  \left\{
\eqalign{
y_1 &\to y_1 + {\textstyle{1\over 2}} R_1 \,,
\cr
y_2 & \to y_2\,,
\cr}\right.
$$
where $(y_1,y_2)$ are the natural coordinates on a rectangular torus with sides of length $R_1$ and $R_2$. With respect to the natural reference frame adapted to the brane this $\mathbb{Z}_2$ shift decomposes as
$$
\delta :\quad \left\{
\eqalign{
\bar y_1 &\to \bar y_1 + {\textstyle{1\over 2}} R_1 \cos\phi \,,
\cr
\bar y _2 &\to \bar y _2 - {\textstyle{1\over 2}} R_1 \sin\phi \,,
\cr} \right.
$$
with $\bar y_1$ and $\bar y _2$ labelling the directions longitudinal and transverse to the brane, respectively. This shift maps in general points on the branes to points in the bulk unless ($\mu\in\mathbb{Z}$)
$$
y_2 + \mu R_2 = \tan\phi \left( y_1 + {\textstyle{1\over 2}} R_1 \right) \quad \Rightarrow
\quad
\mu R_2 = {k R_2 \over 2 q} \quad \Rightarrow \quad k\in 2 \mathbb{Z}\,,
$$
\end{plain}
that is to say that the shifted point still belongs to the line identified by the brane, modulo
 ${\rm SL} (2,\mathbb{Z})$ identifications of the lattice.

\begin{figure}  
\begin{center} 
\includegraphics[scale=1, height=5cm]{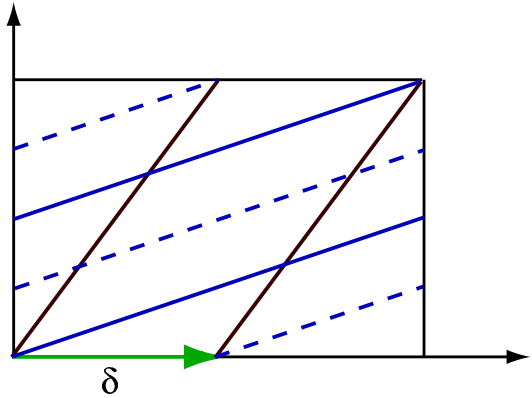}  
\caption {The two classes of branes in the presence of a freely acting horizontal shift $\delta$. The dashed brane is the image under $\delta$.}
\label{ssbreak}
\end{center}
\end{figure}


It is then evident that we can identify two equivalence classes of branes, as indicated in figure 
\ref{ssbreak}: those with odd $k$ and those with even $k$. In the former case the shift is not a symmetry unless we introduce image branes with the same angle but separated by a distance $- {\textstyle{1\over 2}} R_1 \sin\phi$. In the latter case the shifted points still belong to the brane and thus the given configuration is already symmetric under the action of $\delta$. Similarly, for shifts along the vertical axis (or along the diagonal of the torus) only for branes with $q$ even (or with $k$ and $q$ both odd) the image points still belong to the brane world volume.
This is somewhat reminiscent of what happens when we act with freely acting shifts along the world-volume of the brane and or along directions transverse to it.

Given these observations, it is straightforward to write down the associated vacuum amplitudes. Since in this toy model we are modding out by $\Omega \mathscr{R}$, our parent theory is the type IIA superstring compactified on a $T^2$  
$$
\mathcal{T} = (V_8 - S_8)(\bar V _8 - \bar C_8) \, \left[ \Lambda_{2m_1,n_1} (R_1) \Lambda_{m_2,n_2} (R_2) +
\Lambda_{2m_1,n_1+{1\over 2}} (R_1) \Lambda_{m_2,n_2} (R_2)\right] \,,
$$
where each $\Lambda (R)$ denotes the Narain lattice for a circle of radius $R$ and momenta and windings specified. The associated direct-channel Klein-bottle amplitude reads
$$
\mathcal{K} = {\textstyle{1\over 2}} (V_8 - S_8 ) \, P_{2m_1} (R_1) W_{n_2} (R_2)  \,.
$$

\begin{figure}  
\begin{center} 
\includegraphics[scale=1, height=5cm]{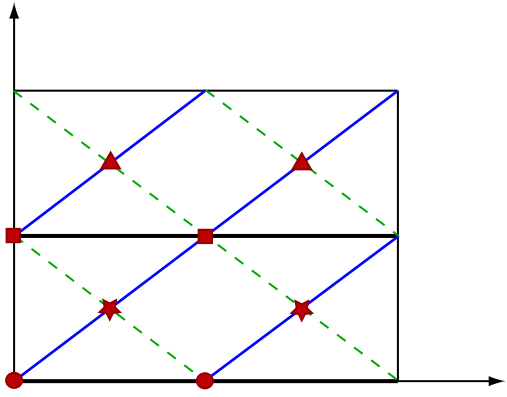}  
\caption {A (1,1) brane and its images under $\Omega \mathscr{R}$ and $\delta$. The intersections marked with alike symbols are identified under $\delta$.}
\label{counting}
\end{center}
\end{figure}

\begin{plain}
Moving to the open-string sector, for $k_\alpha$ odd, the direct-channel annulus and M\"obius-strip amplitudes now read
$$
\eqalign{
\mathcal{A} =& N_\alpha\bar N_\alpha \, (V_8 - S_8) [{\textstyle{0\atop 0}}] P_m \left[ W_{n} + W_{n+{1\over 2}} \right]
\cr
&+ {\textstyle{1\over 2}} \,
\left[ N^2_\alpha \, (V_8 - S_8) [{\textstyle{\alpha\bar\alpha \atop 0}}]
+ \bar N^2_\alpha \, (V_8 - S_8) [{\textstyle{\bar\alpha \alpha \atop 0}}]  \right] \, {2 I_{\alpha \bar\alpha} \over \Upsilon_1 [{\textstyle{\alpha\bar\alpha \atop 0}}]} \,,
\cr}
$$ 
and
$$
\mathcal{M} = - {\textstyle{1\over 2}} \left[N_\alpha \, (\hat V_8 - \hat S_8) 
[{\textstyle{\alpha\bar\alpha \atop 0}}] + \bar N_\alpha \,
(\hat V_8 - \hat S_8 ) [{\textstyle{\bar \alpha \alpha \atop 0}}]
 \right] {K_\alpha \over \hat\Upsilon_1 [{\textstyle{\alpha\bar\alpha \atop 0}}]
}
\,.
$$

Notice the important difference with the standard case: the annulus amplitude unambiguously reflects the presence of image branes under the action of the horizontal shift both by the presence of dipole strings with shifted windings, and by the doubling of the number of families for unoriented strings due to the doubling of local intersections.
As for the M\"obius-strip amplitude, instead, it is not modified since the effective number of intersections sitting on the O-planes is not affected. This counting of effective intersections is clearly depicted in figure \ref{counting} in the case of (1,1) branes, where points marked with the same symbol are identified under $\delta$. 

Moving to the case of branes with even $k_\alpha$, one is not required any longer to introduce brane images, these branes being invariant under $\delta$, and the annulus and M\"obius-strip amplitudes read
$$
\eqalign{
\mathcal{A} =& N_\alpha\bar N_\alpha \, (V_8 - S_8) [{\textstyle{0\atop 0}}] P_{2m} W_n
\cr
+& {\textstyle{1\over 2}} \left[ N^2_\alpha \, (V_8 -S_8) [{\textstyle{\alpha\bar\alpha \atop 0}}]
 + \bar N ^2_\alpha \, (V_8 - S_8)  [{\textstyle{\bar\alpha \alpha \atop 0}}]
\right] \, {I_{\alpha\bar\alpha} \over 2 \Upsilon_1  [{\textstyle{\alpha\bar\alpha \atop 0}}]
 } \,,
\cr}
$$
and 
$$
\mathcal{M} = - {\textstyle{1\over 2}} \left[ N_\alpha \, (\hat V_8 - \hat S_8 ) 
[{\textstyle{\alpha\bar\alpha \atop 0}}]  + \bar N_\alpha  \, (\hat V _8 - \hat S_8) 
[{\textstyle{\bar \alpha \alpha \atop 0}}] \right] {K_\alpha \over 2 \hat \Upsilon_1 
 [{\textstyle{\alpha\bar\alpha \atop 0}}] }\,.
$$
\end{plain}

\begin{figure}  
\begin{center} 
\includegraphics[scale=1, height=5cm]{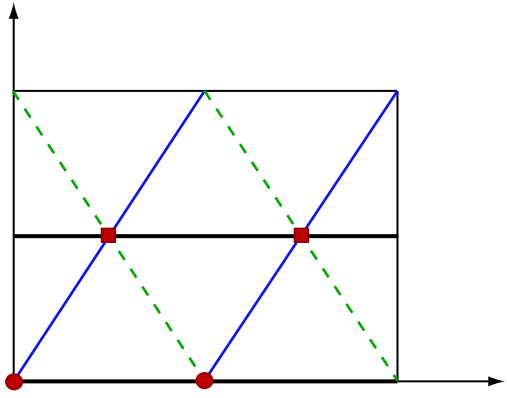} 
\caption {A (1,1) brane and its images under $\Omega \mathscr{R}$ and $\delta$. The intersections marked with alike symbols are identified under $\delta$.}
\label{countingeven}
\end{center}
\end{figure}

In contrast to the previous case, the multiplicities in the annulus and M\"obius-strip amplitudes have now been halved, since $\delta$ identifies pairs of intersection points, as depicted in figure \ref{countingeven}.

We can now straightforwardly generalise the open-string amplitudes to the case of a more interesting $T^6= T^2 \times T^2\times T^2$ compactification with the shift acting along  the horizontal axis $y_1$ of the first $T^2$. For simplicity we shall focus our attention to the case of factorisable D6 branes or, in the T-dual language, to commuting magnetic backgrounds, although more general choices turned out to be very promising in stabilising closed-string moduli  \cite{tristan},\cite{trevigne}. In this simple case, D6 branes on a $T^2 \times T^2\times T^2$ are then identified by their angles $\phi^\Lambda_\alpha$ ($\Lambda = 1,2,3$ labelling the three two-tori) and wrapping numbers $(q_\alpha^\Lambda \,,\, k^\Lambda_\alpha)$ related, as usual, by
$$
\tan \phi^\Lambda_\alpha = {k^\Lambda_\alpha \, R^\Lambda_2 \over q^\Lambda_\alpha \, R^\Lambda_1} \,,
$$
where $R^\Lambda_1$ and $R^\Lambda_2$ denote the sizes of the horizontal and vertical sides of the $\Lambda$-th $T^2$, respectively. 
We then split the generic index $\alpha$ into the pair $a$ and $i$ labelling respectively $n_{\rm o}$ different stacks of $N_a$ branes ($a=1,\ldots,n_{\rm o}$) with $k^1_a$ odd and $n_{\rm e}$ different stacks of $N_i$ branes ($i=1,\ldots,n_{\rm e}$) with $k^1_i$ even. The annulus amplitude is then

\bea
\label{torann}
\mathcal{A} &=& \sum_{a=1}^{n_{\rm o}} N_a \bar N_a \, (V_8 - S_8) [{\textstyle{0\atop 0}}] \, P_m^1 \left[ W_n^1 + W_{n+{1\over 2}}^1\right] P_m^2 W_n^2  P_m^3 W_n^3 \nn \\
& +& \sum_{i=1}^{n_{\rm e}} N_i \bar N_i \, (V_8 - S_8) [{\textstyle{0\atop 0}}] \, P_{2m}^1 W_n^1 P_m^2 W_n^2 P_m^3 W_n^3 \nn \\
&+& {\textstyle{1\over 2}} \sum_{a=1}^{n_{\rm o}} \left( N_a^2 (V_8 - S_8) [{\textstyle{a\bar a\atop 0}}]
+ \bar N_a^2 (V_8 - S_8)
[{\textstyle{\bar a  a \atop 0}}] \right) \, 
{2 I_{a\bar a}  \over \Upsilon_1 [{\textstyle{a\bar a\atop 0}}]} \nn \\
&+& {\textstyle{1\over 2}} \sum_{i=1}^{n_{\rm e}} \left( N_i^2 (V_8 - S_8)
[{\textstyle{i\bar\imath\atop 0}}] + \bar N_i^2 (V_8 - S_8) 
[{\textstyle{\bar\imath i\atop 0}}] \right) \, 
{I_{i\bar\imath} \over 2 \,\Upsilon_1 [{\textstyle{i\bar\imath\atop 0}}]} \nn \\
&+& \sum_{{a,b=1 \atop b<a}}^{n_{\rm o}} \left( N_a \bar N_b (V_8 - S_8 ) 
[{\textstyle{a b\atop 0}}] + \bar N_a N_b (V_8 - S_8 ) [{\textstyle{\bar a \bar b\atop 0}}]\right)\,
{ 2\, I_{a b} \over \Upsilon_1 [{\textstyle{a b\atop 0}}]} \nn \\
&+& \sum_{{a,b=1 \atop b<a}}^{n_{\rm o}} \left( N_a  N_b (V_8 - S_8 ) 
[{\textstyle{a\bar b\atop 0}}] + \bar N_a \bar N_b (V_8 - S_8 ) 
[{\textstyle{\bar a b\atop 0}}] \right) \,
{ 2\, I_{a\bar b} \over \Upsilon_1 [{\textstyle{a\bar b\atop 0}}]} \nn \\
&+& \sum_{{i,j=1 \atop j<i}}^{n_{\rm e}} \left( N_i \bar N_j (V_8 - S_8 ) 
[{\textstyle{i j\atop 0}}]
+ \bar N_i N_j (V_8 - S_8 ) [{\textstyle{\bar\imath \bar\jmath\atop 0}}]\right) \,
{ I_{ij} \over 2\, \Upsilon_1 [{\textstyle{ij\atop 0}}]} \nn \\
&+& \sum_{{i,j=1 \atop j<i}}^{n_{\rm e}} \left( N_i  N_j (V_8 - S_8 ) 
[{\textstyle{i\bar\jmath\atop 0}}] + \bar N_i \bar N_j (V_8 - S_8 ) 
[{\textstyle{\bar\imath j\atop 0}}] \right) \,
{ I_{i\bar\jmath} \over 2\, \Upsilon_1 [{\textstyle{i\bar\jmath\atop 0}}]} \nn \\
&+& \sum_{a=1}^{n_{\rm o}} \sum_{i=1}^{n_{\rm e}} \left( N_a \bar N_i \, (V_8 -S_8) 
[{\textstyle{a i\atop 0}}]+ \bar N_a N_i \, (V_8 - S_8) 
[{\textstyle{\bar a \bar\imath\atop 0}}] \right) \, 
{I_{a i} \over \Upsilon_1 [{\textstyle{a i \atop 0}}]} \nn \\
&+& \sum_{a=1}^{n_{\rm o}} \sum_{i=1}^{n_{\rm e}} \left( N_a N_i \, (V_8 -S_8) 
[{\textstyle{a\bar\imath \atop 0}}]+ \bar N_a \bar N_i \, (V_8 - S_8) 
[{\textstyle{\bar a i\atop 0}}] \right) \, {I_{a\bar\imath} \over \Upsilon_1 [{\textstyle{a\bar\imath \atop 0}}]}
\eea

To lighten the notation, we omit here and in following similar expressions the ``effective radius'' dependence of the various momentum and winding lattice contributions. In particular, $P_m^\Lambda$ ($W_n^\Lambda$) is a short-hand notation for $P_m (L_{\alpha\|}^\Lambda)$ ($W_n (L_{\alpha\perp}^\Lambda)$), where $L_{\alpha\|}^\Lambda$ ($L_{\alpha\perp}^\Lambda$) is the total length (the transverse distance among the consecutive wrappings) of the $\alpha$-th stack of branes in the $\Lambda$-th torus.
The M\"obius-strip amplitude instead is a simple combination of the previous ones and reads
\begin{plain}
$$
\eqalign{
\mathcal{M} =& - {\textstyle{1\over 2}} \sum_{a=1}^{n_{\rm o}}
\left(N_a \, (\hat V_8 - \hat S_8) [{\textstyle{a\bar a\atop 0}}] + \bar N_a \,
(\hat V_8 - \hat S_8 ) [{\textstyle{\bar a a\atop 0}}] \right) {K_a \over \hat\Upsilon_1 
[{\textstyle{a\bar a\atop 0}}]}
\cr
&- {\textstyle{1\over 2}} \sum_{i=1}^{n_{\rm e}}
\left( N_i\, (\hat V_8 - \hat S_8 ) [{\textstyle{i\bar\imath\atop 0}}]
+ \bar N_i \, (\hat V _8 - \hat S_8) [{\textstyle{\bar \imath i\atop 0}}] 
\right) {K_i \over 2\, \hat \Upsilon_1 [{\textstyle{i \bar\imath\atop 0}}]}
\,.
\cr}
$$
\end{plain}
Here we have adapted our notation to the case of multiple $T^2$'s. As already stated we append an index $\Lambda$ to angles and wrapping numbers relative to the $\Lambda$-th torus, while now
$$
I_{\alpha\beta} = \prod_{\Lambda =1}^3 \, \left( q_\beta^\Lambda \, k_\alpha^\Lambda - q_\alpha^\Lambda \, k_\beta^\Lambda \right)
$$
and
$$
K_\alpha = \prod_{\Lambda = 1,2,3} \, 2 \, k^\Lambda_\alpha
$$
count the total number of intersections in the $T^6$, and those sitting on the O6 planes. Moreover, the contribution of the world-sheet bosons is  
$$
\Upsilon_1 \, [{\textstyle{\alpha\beta \atop \gamma\delta}}]
= \prod_{\Lambda=1,2,3} \, {\theta_1 (\zeta^\Lambda |\tau) \over i\eta (\tau)} \,
e^{2i \pi \zeta^\Lambda} \,,
$$ 
with
$$
\zeta^\Lambda = {1\over \pi}\left[
\left( \phi^\Lambda_\alpha - \phi^\Lambda_\beta \right) \tau + \left( \phi_\gamma^\Lambda - \phi_\delta^\Lambda \right) \right] \,.
$$
The contribution of the world-sheet fermions is more involved and requires an $SO (8)\to SO (2) \times SO (2) \times SO (2) \times SO (2)$ breaking of the original characters. This is a consequence of the fact that, in the T-dual language of magnetised backgrounds, fields with different helicities couple differently to the magnetic fields. More explicitly
\begin{plain}
$$
\eqalign{
V_8 \, [{\textstyle{\alpha\beta \atop \gamma\delta}}] = V_2 &\, \left[
O_2 (\zeta^1 ) \, O_2 (\zeta^2) \, O_2 (\zeta^3) + 
O_2 (\zeta^1 ) \, V_2 (\zeta^2) \, V_2 (\zeta^3) \right.
\cr
&\left.+ 
V_2 (\zeta^1 ) \, V_2 (\zeta^2) \, O_2 (\zeta^3) +
V_2 (\zeta^1 ) \, O_2 (\zeta^2) \, V_2 (\zeta^3) \right]
\cr
+ O_2 &\, \left[
V_2 (\zeta^1 ) \, O_2 (\zeta^2) \, O_2 (\zeta^3) + 
V_2 (\zeta^1 ) \, V_2 (\zeta^2) \, V_2 (\zeta^3) \right.
\cr
&\left. +
O_2 (\zeta^1 ) \, V_2 (\zeta^2) \, O_2 (\zeta^3) +
O_2 (\zeta^1 ) \, O_2 (\zeta^2) \, V_2 (\zeta^3) \right] \,,
\cr
S_8 \, [{\textstyle{\alpha\beta \atop \gamma\delta}}] = S_2 &\, \left[
S_2 (\zeta^1 ) \, S_2 (\zeta^2) \, S_2 (\zeta^3) +  
S_2 (\zeta^1 ) \, C_2 (\zeta^2) \, C_2 (\zeta^3) \right.
\cr
&\left. + 
C_2 (\zeta^1 ) \, S_2 (\zeta^2) \, C_2 (\zeta^3) + 
C_2 (\zeta^1 ) \, C_2 (\zeta^2) \,S_2 (\zeta^3) \right]
\cr
+ C_2 &\, \left[
C_2 (\zeta^1 ) \, S_2 (\zeta^2) \, S_2 (\zeta^3) + 
C_2 (\zeta^1 ) \, C_2 (\zeta^2) \, C_2 (\zeta^3) \right.
\cr
&\left. +
S_2 (\zeta^1 ) \, S_2 (\zeta^2) \, C_2 (\zeta^3) +
S_2 (\zeta^1 ) \, C_2 (\zeta^2) S_2 (\zeta^3) \right] \,.
\cr}
$$
\end{plain}
As usual, the massless spectrum can be extracted expanding $\mathcal{A}$ and $\mathcal{M}$. Aside from the full $\mathscr{N} = 4$ super-Yang-Mills multiplet in the adjoint representation of the Chan-Paton gauge group, one typically gets tachyonic excitations and (non-)chiral fermions in various representations. The spectrum of chiral fermions can be easily computed and is encoded in table 1, where, as usual, $A_\alpha$ and $S_\alpha$ denote respectively the anti-symmetric and symmetric representations of the $\alpha$-th factor in the gauge group. As for non-chiral fermions, these emerge whenever branes are parallel in a given $T^2$. Their multiplicity is then given by  the number of times the corresponding branes intersect in the remaining tori. For instance, if the branes of type $\alpha$ and $\beta$ are parallel in the first $T^2$, then one finds 
$$
I^{\rm non\ chiral}_{\alpha\beta} = \prod_{\Sigma=2,3} \left( q_\beta^\Sigma \, k^\Sigma_\alpha - q_\alpha^\Sigma \, k^\Sigma_\beta \right)
$$
non-chiral fermions in the representation $(N_\alpha , \bar N_\beta )$.

\begin{center}
\begin{table}[h]
\hskip5.5cm\begin{tabular}{|c c|}
\hline
 Rep. &   multiplicity \\ 
\hline
 $A_a$ & $\quad\quad {1\over 2} (2\, I_{a\bar a} \pm K_a)$ \\
$S_a$ & $\quad\quad {1\over 2} (2\, I_{a\bar a} \mp K_a)$ \\
$A_i$ & $\quad\quad {1\over 4} (I_{i\bar\imath} \pm K_i)$ \\
$S_i$ & $\quad\quad {1\over 4} (I_{i\bar\imath} \mp K_i)$\\
$(N_a , N_b)$ & $\quad\quad 2\, I_{a\bar b}$\\
$(N_a , \bar N_b)$ & $\quad\quad 2\, I_{a b}$ \\
$(N_i , N_j)$ & $\quad\quad {1\over 2}\, I_{i\bar\jmath}$\\
$(N_i , \bar N_j)$ & $\quad\quad {1\over 2}\, I_{ij}$\\
$(N_a , N_i)$ & $\quad\quad I_{a\bar\imath}$\\
$(N_a , \bar N_i)$ & $\quad\quad I_{ai}$\\
\hline
\end{tabular}\\
\caption{ Spectrum of four-dimensional chiral fermions. The sign of $I_{\alpha\bar\beta}$ determines the chirality of the four-dimensional spinors, while the two signs in the multiplicity of (anti-)symmetric representations refer to branes ($q>0$) or to anti-branes ($q<0$), respectively.}
\end{table}
\end{center}

Turning to the transverse channel, the massless tadpoles can be extracted as usual from the leading terms in $\tilde{\mathcal{K}}$, $\tilde{\mathcal{A}}$ and $\tilde{\mathcal{M}}$. Introducing the combinations

\beq \label{radii}
{\bf R} = \prod_{\Lambda =1}^3 \, \sqrt{ R_1^\Lambda \over R_2^\Lambda}
\,,
\qquad {\bf L}_\alpha = \prod_{\Lambda = 1}^3 \, \sqrt{ L^\Lambda_{\|\, \alpha}
\over L^\Lambda_{\perp\,\alpha}}\,,
\eeq
the NS-NS dilaton tadpole reads
\beq \label{nstadpole}
2\, \sum_{a=1}^{n_{\rm o}} \, {\bf L}_a \, \left( N_a + \bar N_a \right) + \sum_{i=1}^{n_{\rm e}} \, {\bf L}_i \, \left( N_i + \bar N_i \right) = 2^5 \, {\bf R}\,.
\eeq
Notice that, after a T-duality along the vertical axis, the left-hand-side is nothing but the Dirac-Born-Infeld Action
$$
{\bf L_\alpha} = \prod_{\Lambda=1,2,3} \, \sqrt{ R_1 R_2 \, {\rm det}\,\left[ q^\Lambda_\alpha \, \left( {\bf 1} + F^\Lambda_\alpha \right) \right]} \,,
$$
with $F^\Lambda$ a two-by-two antisymmetric matrix, whose only independent entry is the magnetic field background $H^\Lambda$. 
The eight different R-R tadpoles can be straightforwardly obtained from (\ref{nstadpole})\  
after a multiplication of the Chan-Paton multiplicities by a phase depending on the scalar product of the rotation angles $\phi^\Lambda_\alpha$ and the helicities $\eta^\Lambda = \pm {1\over 2}$ of the internal spinors. With our convention such that $S_2$ has helicity $+{1\over 2}$ and $C_2$ has helicity $-{1\over 2}$, they read 
\beq\label{rrtadpole}
2\, \sum_{a=1}^{n_{\rm o}} \, {\bf L}_a \, \left( N_a \, e^{2i \phi_a \cdot \eta}
+ \bar N_a \, e^{-2i \phi_a \cdot \eta} \right) + \sum_{i=1}^{n_{\rm e}} \, {\bf L}_i \, \left( N_i \, e^{2i\phi_i \cdot \eta}  + \bar N_i \, e^{-2i\phi_i \cdot \eta}\right) = 2^5 \, {\bf R}\,,
\eeq
with the internal product $\phi_\alpha \cdot \eta = \sum_{\Lambda =1}^3 \, \phi_\alpha^\Lambda \, \eta^\Lambda$. This single R-R tadpole actually encodes couplings to different forms as dictated by the (T-dualised) Action
$$
S \sim \int \, \sum_p \, C_{p+1} \, e^{i F} \sim \int \, C_{10} + i \, C_8 \wedge F - {\textstyle{1\over 2}}\, C_6 \wedge F \wedge F - {\textstyle{1\over 6}} \,i \,C_4 \wedge F \wedge F \wedge F \,. 
$$
Each factor, in turn, receives contributions from several terms according to which internal $T^2$ the magnetic field points to.
For instance, the real part of the tadpole (\ref{rrtadpole})\ yields the conditions 
\begin{plain}
$$
\eqalign{
\left[
\sum_{a=1}^{n_o} \, 2 \,  q^1_a \, q^2_a \, q^3_a \, N_a +
\sum_{i=1}^{n_e} \, q^1_i \, q^2_i \, q^3_i \, N_i\right] 
\sqrt{R^1_1 \, R^2_1 \, R^3_1 \over R_2^1 \, R_2^2 \, R_2^3}=& 16 \,
\sqrt{R^1_1 \, R^2_1 \, R^3_1 \over R_2^1 \, R_2^2 \, R_2^3} \,,
\cr
\left[
\sum_{a=1}^{n_o} \, 2 \,  q^1_a \, k^2_a \, k^3_a \, N_a +
\sum_{i=1}^{n_e} \, q^1_i \, k^2_i \, k^3_i \, N_i \right] 
\sqrt{R^1_1 \, R^2_2 \, R^3_2 \over R_2^1 \, R_1^2 \, R_1^3}=& 0\,,
\cr
\left[ \sum_{a=1}^{n_o} \, 2 \,  k^1_a \, q^2_a \, k^3_a \, N_a +
\sum_{i=1}^{n_e} \, k^1_i \, q^2_i \, k^3_i \, N_i \right]
\sqrt{R^1_2 \, R^2_1 \, R^3_2 \over R_1^1 \, R_2^2 \, R_1^3}=& 0\,,
\cr
\left[
\sum_{a=1}^{n_o} \, 2 \,  k^1_a \, k^2_a \, q^3_a \, N_a +
\sum_{i=1}^{n_e} \, k^1_i \, k^2_i \, q^3_i \, N_i \right]
\sqrt{R^1_2 \, R^2_2 \, R^3_1 \over R_1^1 \, R_1^2 \, R_2^3}=& 0\,.
\cr}
$$
\end{plain}
After the vertical axis of the three $T^2$'s are properly T dualised, one can then read the couplings with the ten-form potential $C_{10}$ and with the three six-form potentials $C^\Lambda_6$, whose six indices point to the four non-compact space-time directions and to the two compact coordinates of the $\Lambda$-th torus. The imaginary part of the tadpole (\ref{rrtadpole}) is proportional to $N_\alpha - \bar N_\alpha$ and thus vanishes identically. This is fully consistent with the fact that the $C_8$ and $C_4$ forms are projected out by the orientifold involution and thus do not belong to the physical spectrum.

\subsection{Breaking supersymmetry in the dipole-string sector}
\everypar{\hspace{-.6cm}}
After we have learned how shifts act on rotated branes, we can now proceed to the study of the combined effect of $\delta$ and $(-1)^F$. As is well known the (freely-acting) orbifold generated by $\delta \, (-1)^F$ is a convenient string description of Sherck-Schwarz deformations  that in general are responsible for giving masses to fermions, and thus for breaking supersymmetry. Following the lines of section \ref{freely} we expect that the space-time fermion index has a non trivial effect only when the shift $\delta$ moves points along the brane, that is only when $k_i$ is an even number. In this case the annulus amplitude reads
\bea
\mathcal{A} &=& N_i \bar N_i  \, \big( V_8 [{\textstyle{0\atop 0}}]
\, P_{2m} - S_8 [{\textstyle{0\atop 0}}]
\, P_{2m+1} \big) \, W_n \nn \\
&+& {\textstyle{1\over 2}} \left( N^2_i \, (V_8 - S_8) [{\textstyle{i\bar\imath \atop 0}}]
+ \bar N ^2_i (V_8 - S_8 ) [{\textstyle{\bar\imath i\atop 0}}]
\right) \, {I_{i\bar\imath} \over 2\, \Upsilon_1 [{\textstyle{i\bar\imath \atop 0}}]}
\label{openss}
\eea

As is spelled out by this direct-channel amplitude, the adjoint fermions have now got a mass 
$$
m_{1/2} \sim L_\|^{-1} = {\cos \phi_i \over q_i R_1} = {\sqrt{1 + \alpha ' {}^2 H_i^2} \over q_i R_1}
= {1 \over q_i^2 \, R_1^2} \sqrt{ q_i^2 \, R_1^2 + \alpha ' {}^2 {k_i^2 \over R_2^2}}\,,
$$ 
of the order of the TeV,
and, consequently, supersymmetry is broken in the neutral dipole-string sector. In the transverse channel
\begin{plain}
$$
\eqalign{
\tilde{\mathcal{A}} =& \, 2^{-6} \, {L_{\| i} \over L_{\perp i}} \, N_i \bar N_i \, \left[ ( V_8 - S_8) 
[{\textstyle{0 \atop 0}}] \,\tilde W_{n}
+ (O_8 - C_8) [{\textstyle{0\atop 0}}]
 \,\tilde W_{n+{1\over 2}} \right] \, \tilde P_m
\cr
&+ 2^{-5} \, \left[ N^2_i \, (V_8 - S_8) [{\textstyle{0 \atop i\bar\imath}}]
 + \bar N ^2_i \, (V_8 - S_8) [{\textstyle{0 \atop \bar\imath i}}]
  \right]\,
{I_{i\bar\imath} \over2\, \Upsilon_1 [{\textstyle{0 \atop i\bar\imath}}]} \,,
\cr}
$$
\end{plain}
the twisted closed-string R-R states are massive, and the only massless tadpoles one is to worry about are those for the untwisted R-R fields in $S_8$.

The generalisation to higher-dimensional tori is also straightforward. The only modification is in the dipole sector of the $N_i$ branes that now would read
\beq
\label{nonsusyann}
\sum_{i=1}^{n_{\rm e}} M_i \bar M_i \, \left( V_8 \, [{\textstyle{0\atop 0}}] \, P_{2m}^1
\, - S_8 \, [{\textstyle{0\atop 0}}] \, P_{2m+1}^1 \right)
 W_n^1 P_m^2 W_n^2 P_m^3 W_n^3 \,.
\eeq
It reflects itself in the transverse-channel contribution
$$
2^{-6} \, 
\sum_{i=1}^{n_{\rm e}} \, ({\bf L}^\Lambda _i)^2\,  
N_i \bar N_i \, \Bigl[ ( V_8 - S_8) [{\textstyle{0\atop 0}}] 
\,\tilde W_{n}^1 
 + (O_8 - C_8) [{\textstyle{0\atop 0}}] \,\tilde W_{n+{1\over 2}}^1 \Bigr] \, \tilde P_m^1 
\tilde W_n^2 \tilde P_m^2 \tilde W_n^3 \tilde P_m^3 \,,
$$
and thus, also in this case, the massless NS-NS and R-R tadpoles (\ref{nstadpole})\ and (\ref{rrtadpole})\ are not affected.

\vspace{1 cm}
\subsection{Lifting non-adjoint non-chiral fermions}
\everypar{\hspace{-.6cm}}
\begin{plain}
As we have remarked in previous subsections, non-chiral fermions arise not only in the dipole-string sector, where open strings in the adjoint representation end on the same stuck of branes. In fact, it may well happen that although branes intersect at non-trivial angles 
in some tori, they are actually parallel in one $T^2$. This is typically reflected by a vanishing intersection number $I_{\alpha \beta} =0$ for the given pair of $\alpha$-type and $\beta$-type branes, and in turn by an undetermined ${0\over 0}$ expression in the annulus amplitude. For concreteness, let us suppose that the $\alpha$ and $\beta$ branes are parallel in the $\Sigma$-th  torus. By definition, this implies that their angles are the same
$\phi^\Sigma_\alpha = \phi^\Sigma_\beta$ and thus also the corresponding wrapping numbers coincide
$$
q^\Sigma_\alpha \equiv q^\Sigma_\beta \,, \qquad 
k^\Sigma_\alpha \equiv k^\Sigma_\beta \,.
$$
The internal $SO(2)$ characters corresponding to the first $T^2$ have then vanishing argument, and can be combined with the space-time $SO(2)$ little group to reconstruct a full $SO(4)$ symmetry, whose spinor representation is vector-like from the four-dimensional viewpoint. 
As a result, in the $N_\alpha \bar N_\beta$ sector of the annulus amplitude
$$
\eqalign{
\mathcal{A} =& \left( N_\alpha \bar N_\beta \, (V_8 - S_8) [{\textstyle{\alpha \beta \atop 0}}]
+ \bar N_\alpha N_\beta \, (V_8 - S_8) [{\textstyle{\bar\alpha\bar\beta \atop 0}}] \right) 
{I_{\alpha \beta} \over \Upsilon_1  [{\textstyle{\alpha \beta \atop 0}}]}
\cr
=& \left( N_\alpha \bar N_\beta \, (V_8 - S_8) [{\textstyle{\alpha \beta \atop 0}}]
+ \bar N_\alpha N_\beta \, (V_8 - S_8) [{\textstyle{\bar\alpha\bar\beta \atop 0}}] \right) 
\prod_{\Lambda =1}^3 \, {(k^\Lambda_\alpha q^\Lambda_\beta - 
k^\Lambda_\beta q^\Lambda_\alpha )\, i\eta \, e^{2 i (\phi^\Lambda_\alpha - \phi^\Lambda_\beta)\tau} \over \theta_1 ({1\over \pi} (\phi^\Lambda_\alpha - \phi^\Lambda_\beta)\tau |\tau )}
\cr}
$$

the contribution from the $\Sigma$-th torus is clearly ill defined since both $k^\Sigma_\alpha q^\Sigma_\beta - k^\Sigma_\beta q^\Sigma_\alpha$ and $\theta_1 (0|\tau)$ vanish. Actually, one has to be very careful in cases like this, since if on the one hand it is true that parallel branes do not have any longer the tower of Landau levels, it is evident on the other hand that open strings stretched between such branes have now non-trivial zero modes, and thus their contribution has to be taken into account. In practice, this amounts to the substitution
$$
{(k^\Sigma_\alpha q^\Sigma_\beta - k^\Sigma_\beta q^\Sigma_\alpha )\, i\eta  \over \theta_1 (0 |\tau )} \to P_m (L^\Sigma_{\| \, \alpha} ) \, W_n (L^\Sigma_{\perp\, \alpha} ) \,.
$$
It is then clear that, acting with the $\delta (-1)^F$ deformation on the $\Sigma$-th torus,  non-chiral fermions can acquire a tree-level mass $m_{1/2} \sim 1/L_{\|\,\alpha}^\Sigma$
if their vertical wrapping number $k_\alpha^\Sigma$ is an even integer. In this case, the corresponding sector in the annulus amplitude would read
$$
N_\alpha \bar N_\beta \, \left( V_8 [{\textstyle{\alpha \beta \atop 0}}] P_{2m} (
L^\Sigma_{\| \, \alpha} ) - S_8 [{\textstyle{\alpha \beta \atop 0}}] P_{2m+1} (
L^\Sigma_{\| \, \alpha} ) \right) \, W_n (L^\Sigma_{\perp\, \alpha} ) 
\, {I^{\rm non\ chiral}_{\alpha\beta}\over \tilde\Upsilon_1 [{\textstyle{\alpha \beta \atop 0}}]} \,,
$$
where clearly the $\Sigma$-th torus does not contribute to $I^{\rm non\ chiral}_{\alpha\beta}$ and $\tilde\Upsilon_1$.
\end{plain}

\subsection{Comments on Scherk-Schwarz and orbifold basis}
\everypar{\hspace{-.6cm}}
In the previous sub-section we have studied the effect of the combined action of the space-time fermion index and momentum shift along a compact coordinates and we have identified it with the Scherk-Schwarz mechanism. While correct in spirit, this definition does not correspond, however, to the common use of the term in Field Theory, since the canonical Scherk-Schwarz deformation for a circle would lead to periodic bosons and anti-periodic fermions, a choice manifestly compatible with any low-energy effective field theory, where fermions only enter via their bi-linears. On the other hand, from eq. (\ref{niness}), rewritten more explicitly as
\begin{plain}
$$
\eqalign{
\mathcal{T} =& \left( |V_8 |^2 + |S_8 |^2 \right) \, \Lambda_{2m ,n} (R)- \left(S_8 \bar V_8 + V_8 \bar S_8 \right) \, \Lambda_{2m+1,n} (R)
\cr
&+ \left( |O_8|^2 + |C_8|^2 \right) \, \Lambda_{2m,n+{1\over 2}} (R)- \left( O_8 \bar C_8 - C_8 \bar O_8 \right) \, \Lambda_{2m+1,n+{1\over 2}} (R)\,,
\cr}
$$
\end{plain}
it is clear that bosons and fermions have {\it even} and {\it odd} momenta in the orbifold. 
It is however simple to relate the two settings: the conventional Scherk-Schwarz basis of Field Theory can be recovered letting $R^{\rm SS} \equiv \rho = {1\over 2} R$, so that\begin{plain}
$$
\eqalign{
\mathcal{T} =& \left( |V_8 |^2 + |S_8 |^2 \right) \, \Lambda_{m ,2n} (\rho) - \left(S_8 \bar V_8 + V_8 \bar S_8 \right) \, \Lambda_{m+{1\over 2},2n} (\rho) 
\cr
&+ \left( |O_8|^2 + |C_8|^2 \right) \, \Lambda_{m,2n+1} (\rho)- \left( O_8 \bar C_8 - C_8 \bar O_8 \right) \, \Lambda_{m+{1\over 2},2n+1} (\rho)\,,
\cr}
$$
\end{plain}
where bosons and fermions have indeed the correct quantum numbers.

Similar considerations apply also to the orientifolds, and in particular to the D-brane sector.
From eq. (\ref{openss}) one deduces that bosons and fermions have {\it even} and {\it odd} momenta along the brane world-volume. Therefore, to recover the standard Field Theory Kaluza-Klein spectrum, the effective length of the brane has to be halved $L_\|^{\rm SS} \equiv \lambda_\| = {1\over 2} L_\|$, similarly to the closed-string case. This has important consequences on the massless spectrum of open strings, induced by the unchanged quantisation condition (\ref{dirac}). As a result of the halving of the fundamental cell, the wrapping numbers of the various stacks of branes change accordingly. In particular
\bea\label{wrappings}
(q_a \,,\, k_a ) &\to& (\omega_a \,,\, \kappa_a ) =  (2 q_a \,,\, k_a )  \,, \nn \\
(q_i \,,\, k_i ) &\to& (\omega_i \,,\, \kappa_i ) = (q_i \,,\, {\textstyle{1\over 2}} k_i ) \,,
\eea
and this identification makes it clear that the model in (\ref{torann})\ and in (\ref{nonsusyann})\ corresponds to a conventional Scherk-Schwarz (and M-theory) deformation of a configuration of branes with wrapping numbers $(\omega_\alpha \,,\, \kappa_\alpha)$ as in (\ref{wrappings}).
Therefore, {\it non-chiral fermions from branes with even horizontal wrapping number $\omega_a$ stay massless, while those from branes with odd horizontal wrapping number $\omega_i$ get a tree-level mass proportional to $1/2\lambda_\|$}. Similar considerations can be straightforwardly generalised to the case of vertical or oblique shifts. This redefinition of the wrapping numbers also takes care of the additional factors of two and one-half in the multiplicities of the chiral sectors, so that the annulus 
\begin{plain}
$$
\eqalign{
\mathcal{A} =& \sum_{a=1}^{m_{\rm e}} \, N_a \bar N_a \, (V_8 - S_8) [{\textstyle{0\atop 0}}] \, 
\prod_{\Lambda =1}^3 \, P^\Lambda_m\, W^\Lambda_n
\cr
&+ \sum_{i=1}^{m_{\rm o}} \, N_i \bar N_i \, \left( V_8 [{\textstyle{0\atop 0}}]\, P^1_m -
S_8 [{\textstyle{0\atop 0}}] \, P^1_{m+{1\over 2}} \right) \, W^1_n \, 
\prod_{\Sigma =2,3} \, P^\Sigma_m \, W^\Sigma_n
\cr
&+ {\textstyle{1\over 2}} \, \sum_{\alpha =1}^{m_{\rm e} + m_{\rm o}} \left(
N_\alpha^2 \, (V_8 - S_8) [{\textstyle{\alpha\bar\alpha \atop 0}}] + \bar N_\alpha^2 \, (V_8 - S_8) [{\textstyle{\bar\alpha \alpha \atop 0}}] \right) \, {I_{\alpha\bar\alpha} \over \Upsilon_1 [{\alpha\bar\alpha \atop 0}]}
\cr
&+ \sum_{{\alpha,\beta=1 \atop \beta<\alpha}}^{m_{\rm e} +m_{\rm o}} \left( N_\alpha \bar N_\beta (V_8 - S_8 ) 
[{\textstyle{\alpha \beta \atop 0}}] + \bar N_\alpha N_\beta (V_8 - S_8 ) [{\textstyle{\bar \alpha \bar\beta \atop 0}}]\right)\,
{I_{\alpha \beta} \over \Upsilon_1 [{\textstyle{\alpha \beta\atop 0}}]}
\cr
&+ \sum_{{\alpha,\beta=1 \atop \beta<\alpha}}^{m_{\rm e}+m_{\rm o}} \left( N_\alpha  N_\beta (V_8 - S_8 ) 
[{\textstyle{\alpha\bar\beta\atop 0}}] + \bar N_\alpha \bar N_\beta (V_8 - S_8 ) 
[{\textstyle{\bar \alpha \beta\atop 0}}] \right) \,
{I_{\alpha\bar\beta} \over \Upsilon_1 [{\textstyle{\alpha\bar\beta\atop 0}}]}
\cr}
$$
and M\"obius-strip
$$
\mathcal{M} = - {\textstyle{1\over 2}} \sum_{\alpha =1}^{m_{\rm e} + m_{\rm o}} \left(
N_\alpha \, (\hat V_8 - \hat S_8) [{\textstyle{\alpha\bar\alpha \atop 0}}] + \bar N_\alpha \, (\hat V_8 - \hat S_8) [{\textstyle{\bar\alpha \alpha \atop 0}}] \right) \, {K_\alpha \over \hat \Upsilon_1 [{\alpha\bar\alpha \atop 0}]}
$$
\end{plain}
amplitudes clearly define a freely acting deformation of the spectrum of conventional branes with wrapping numbers $(\omega_\alpha , \kappa_\alpha)$. In these equations for $\mathcal{A}$ and $\mathcal{M}$ the intersection numbers $I_{\alpha\beta}$ and $K_\alpha$ are clearly expressed in terms of $\omega_\alpha$ and $\kappa_\alpha$, and $m_{\rm e} \equiv n_{\rm o}$ ($m_{\rm o} \equiv n_{\rm e}$) counts the number of different stacks with $\omega_a$ even ($\omega_i$ odd). Moreover, additional non-chiral fermions arising from parallel branes in a particular $T^2$, can also be given a mass through a similar deformation. For instance, in the case of a pair of $\check \imath$ and $\check \jmath$ branes parallel in the first $T^2$ and with $\omega^1_{\check \imath} \equiv \omega^1_{\check \jmath}$ odd one finds the contribution
\beq\label{nonchiral}
\left[ \left( N_{\check \imath} \bar N_{\check \jmath}  V_8 [{\textstyle{\check \imath \check\jmath \atop 0}}] + \bar N_{\check \imath}  N_{\check \jmath}  V_8 [{\textstyle{
\bar{\check \imath} \bar{\check\jmath} \atop 0}}]  \right) P^1_m - 
\left( N_{\check \imath} \bar N_{\check \jmath}  S_8 [{\textstyle{\check \imath \check\jmath \atop 0}}] + \bar N_{\check \imath}  N_{\check \jmath}  S_8 [{\textstyle{
\bar{\check \imath} \bar{\check\jmath} \atop 0}}]  \right) P^1_{m+{1\over 2}} \right] W^1_n \, {I^{\rm non\ chiral}_{\check\imath \check\jmath} \over \tilde \Upsilon_1
[{\textstyle{\check \imath \check\jmath \atop 0}}]}
\eeq
where as in the previous sub-section both $I^{\rm non\ chiral}_{\check\imath \check\jmath}$ and $\tilde\Upsilon_1 [{\textstyle{\check \imath \check\jmath \atop 0}}]$ do not include terms from the first torus, and the argument of the internal $SO(2)$ characters associated to it is automatically vanishing since $\phi^1_{\check\imath} - \phi^1_{\check\jmath} =0$. 

What happens if two pairs of branes are now parallel but in different $T^2$'s? One clearly has to deform the theory along both tori simultaneously. For definiteness, let us consider the case where  $\check \imath$ and $\check \jmath$ branes are parallel in the first $T^2$ and $\hat \imath$ and $\hat \jmath$ branes are parallel in the second $T^2$. Therefore, a deformation along the first torus of the type (\ref{nonchiral})\ will give mass to the non-chiral fermions in the representation $(N_{\check \imath}\,,\, \bar N_{\check\jmath})$
proportional to $1/\lambda^1_{\|\,\check\imath}$, while a similar deformation along the second torus will give a tree-level mass to the non-chiral fermions in the representation
$(N_{\hat \imath}\,,\, \bar N_{\hat\jmath})$ proportional to $1/\lambda^2_{\|\,\hat\imath}$.
What about the gauginos in the adjoint representation? Here the situation is slightly more involved since we have now to split the set of branes in four categories depending on their horizontal wrapping numbers along the first and second $T^2$. If we label with $\alpha_1$ the branes with both $\omega^1$ and $\omega^2$ even, with $\alpha_2$ the branes with $\omega^1$ even and $\omega^2$ odd, with $\alpha_3$ the branes with $\omega_1$ odd and $\omega_2$ even, and finally with $\alpha_4$ those with both 
$\omega_1$ and $\omega_2$ odd the neutral dipole sector in the annulus amplitude reads
\begin{plain}
$$
\eqalign{
\mathcal{A}_{\rm dipole} =& \sum_{\alpha_1} N_{\alpha_1} \bar N_{\alpha_1} \, 
(V_8 - S_8) [{\textstyle{0\atop 0}}] \, \prod_{\Lambda=1}^3 P^\Lambda_m W^\Lambda_n 
\cr
&+ \sum_{\alpha_2} N_{\alpha_2} \bar N_{\alpha_2} \, 
\left( V_8 [{\textstyle{0\atop 0}}] \, P^2_m - S_8 [{\textstyle{0\atop 0}}] \, P^2_{m+{1\over 2}} \right)\, W^2_n \, \prod_{\Sigma=1,3} P^\Sigma_m W^\Sigma_n 
\cr
&+ \sum_{\alpha_3} N_{\alpha_3} \bar N_{\alpha_3} \, 
\left( V_8 [{\textstyle{0\atop 0}}] \, P^1_m - S_8 [{\textstyle{0\atop 0}}] \, P^1_{m+{1\over 2}} \right)\, W^1_n \, \prod_{\Sigma=1,2} P^\Sigma_m W^\Sigma_n 
\cr
&+\sum_{\alpha_3} N_{\alpha_3} \bar N_{\alpha_3} \, 
\left[ V_8 [{\textstyle{0\atop 0}}] \, \left( P^1_m \, P^2_m + P^1_{m+{1\over 2}} \, P^2_{m +{1\over 2}} \right) \right.
\cr
&\left. -
S_8 [{\textstyle{0\atop 0}}] \, \left( P^1_m \, P^2_{m+{1\over 2}} + P^1_{m+{1\over 2}} \, P^2_m \right) \right] W^1_n W^2_n \, P^3_m W^3_m
\cr}
$$
\end{plain}

and clearly the $\alpha_1$ gauginos stay massless, the $\alpha_2$ gauginos get a mass proportional to $1/\lambda^2_{\|\, \alpha_2}$, the $\alpha_3$ gauginos get a mass proportional to $1/\lambda^1_{\|\, \alpha_3}$, while the $\alpha_4$ gauginos get a mass proportional to $\sqrt{1/(\lambda^1_{\|\, \alpha_4})^2+1/(\lambda^2_{\|\, \alpha_4})^2}$.
The generalisation to the case of deformations acting along the three $T^2$'s is then straightforward.

\vspace{1 cm}
\subsection{An alternative Field Theory description}
\everypar{\hspace{-.6cm}}
To support and clarify of our results we can study a simple Field Theory problem where the Scherk-Schwarz deformation acts as usual along $y_1$ but now the various fields depend on the coordinate $y=\sqrt{(\omega \, y_1)^2 + (\kappa\, y_2)^2}$. 
To this end, we analyse the Kaluza-Klein spectrum of such fields subject to boundary conditions twisted by the operator $(-1)^F$. They correspond to excitations of neutral (parallel) strings, the only sector that admits zero modes and thus the only sector that can be affected by the deformation.  As in the previous sub-section, the pair of integers $(\omega, \kappa )$ on the $T^2$ define the line on which the fields live. 

On a rectangular $T^2$, the eigenfunctions of the two-dimensional Laplace and Dirac operators are simple plane waves
\beq \label{eigenfunctions}
\Phi _{p_1 , p_2 } (y_1 , y_2) \sim e^{2i\pi (p_1 y_1 + p_2 y_2 )}\,,
\eeq
with the momenta determined by the periodicity conditions. As a result, a twist by $(-1)^F$ along the horizontal axis yields 
$$
p_1 = \left( m_1 + {\textstyle{1\over 2}} \, \Delta_{\rm F} \right) {1\over R_1}\quad {\rm and} \qquad p_2 = {m_2 \over R_2} \,,
$$
with $\Delta_{\rm F} = 0$ for space-time bosons, $\Delta_{\rm F} = 1$ for space-time fermions, and $m_1$ and $m_2$ both integers. We can now determine the eigenfunctions of fields living on the straight line with 
\beq \label{straightline}
\tan \phi = {\kappa R_2 \over \omega R_1}\,.
\eeq
These can be obtained by (\ref{eigenfunctions})\ by restricting the generic points $(y_1 , y_2)$ 
to lie on the straight line (\ref{straightline}). The corresponding Kaluza-Klein spectrum 
$$
M^2 = \left( {\cos \phi \over \omega \, R_1}\right)^2 \left[ \left( m_1 + {\textstyle{1\over 2}} \, \Delta_{\rm F} \right) \, \omega + m_2 \kappa \right]^2
$$
clearly shows that the twist $(-1)^F$ has a non-trivial effect on space-time fermions only if $\omega$ is an odd integer. 

Similar analysis can be performed in the case of twists along the vertical and diagonal directions, the only other choices compatible with the orientifold projection $\Omega \mathscr{R}$. In these cases, the would-be gauginos in the dipole-string sector are lifted in mass only if $\kappa$ is odd for vertical twist, or both $\omega$ and $\kappa$ are odd for a diagonal twist.
\vspace{1 cm}
\subsection{A deformed Standard-Model-like configuration}
\everypar{\hspace{-.6cm}}
As a simple application of Scherk-Schwarz deformations to models of some phenomenological relevance we can consider the intersecting-brane configuration of \cite{ibanez}. The Standard Model spectrum is there reproduced by four stacks of branes with gauge group
$$
G_{\rm CP} = U (3)_a \times U (2)_b \times U (1)_c \times U (1)_d \,.
$$
The additional Abelian factor is required in order to accommodate the right-handed leptons, that indeed emerge from open strings stretched between the $U(1)_c$ and the $U (1)_d$ branes. The four $U(1)$'s  are related to four global symmetries of the Standard Model: $Q_a$ is three times the barion number, $Q_d$ is minus the lepton number, $Q_c$ is twice the third component of the right-handed weak isospin, and finally $Q_b$ is a Peccei-Quinn symmetry. This lead the authors of \cite{ibanez}\ to identify the anomaly-free combination
$$
Q_{Y} = {\textstyle{1\over 6}} \, Q_a - {\textstyle{1\over 2}} \, Q_c + {\textstyle{1\over 2}} \, Q_d
$$
with the familiar hyper-charge. In order to reproduce the desired chiral spectrum the intersection numbers of the four stacks should be
\begin{plain}
$$
\eqalign{
I_{a b} &= 1 \,,
\cr 
I_{a c} &= -3 \,,
\cr
I_{b d} &= 0\,,
\cr
I_{c  d} &= -3 \,,
\cr}
\qquad \quad
\eqalign{
I_{a\bar b} &= 2 \,,
\cr 
I_{a \bar c} &= -3 \,,
\cr
I_{b \bar d} &= -3\,,
\cr
I_{c  \bar d} &= 3 \,.
\cr}
$$
\end{plain}
It is not difficult to show that such intersection numbers can be obtained from branes with wrappings
 in the following table.

\begin{center}
\begin{table}[h]
\begin{center}
\begin{tabular}{c c c c}
 \hline
 $N_\alpha$ & $(\omega^1_\alpha \,,\, \kappa^1_\alpha)$ &
$(\omega^2_\alpha \,,\, \kappa^2_\alpha)$ &
$(\omega^3_\alpha \,,\, \kappa^3_\alpha)$ \\
\hline
 3 & $({1\over \beta^1} , 0)$ & $(\omega^2_a , \epsilon \beta^2)$ & $({1\over \rho} , {1\over 2})$\\
 2 & $(\omega^1_b , -\epsilon \beta^1 )$ & $({1\over\beta^2},0)$ & $(1, {3\over 2}\rho)$\\
 1 & $(\omega^1_c , 3 \rho \epsilon \beta^1 )$ & $({1\over \beta^2} , 0)$ & $(0,1)$\\
 1 & $({1\over \beta^1}, 0)$ & $(\omega_d^2 , - \beta^2 \epsilon \rho)$ & $(1, {3\over 2}\rho )$\\
\hline
\end{tabular}
\end{center}
\caption{D6-brane wrapping numbers giving rise to a SM-like spectrum, from \cite{ibanez}.}
\label{tabella2}
\end{table}
\end{center}

With respect to ref. \cite{ibanez}, here we have followed our convention to use $\omega_\alpha$ and $\kappa_\alpha$ to denote the horizontal and vertical wrapping numbers, thus replacing the pairs $(n_\alpha \,,\, m_\alpha)$. Otherwise, $\epsilon =\pm 1$,  $\beta_i= 1 , {1\over 2}$ denotes (one minus) the real component of the complex structure that for $\Omega \mathscr{R} (-1)^{F_{\rm L}}$ orientifolds is discrete \cite{discrete}\ and takes the values zero or ${1\over 2}$, while $\rho=1,{1\over 3}$ is a parameter. The remaining $\omega$'s are not fully independent since their values enter in the assignment of the hyper-charge quantum numbers, and thus have to satisfy the constraint
$$
\omega^1_c = {\beta^2 \over 2\beta^1} \left( \omega_a^2 + 3\,\rho \, \omega_d^2 \right) \,,
$$
with $\omega^1_b$ undetermined. 

The choice 
$$
\rho = \beta^1 = \beta^2 = -\epsilon = 1\,, \quad \omega^2_a = 2\,, \quad \omega_b^1 = \omega_c^1 = 1 \,, \quad \omega_d^2 = 0\,,
$$
amounts to taking all the $\omega^1_\alpha$ to be odd (actually all equal to one), and thus
a Scherk-Schwarz deformation along the horizontal side of the first $T^2$ is enough to make all the would-be gaugino massive. Similarly, the non-chiral fermions in the $b\, d$ sector can be given a mass by deforming the third torus, where indeed the $b$ and $d$ branes are parallel. Moreover, non-chiral fermions in the $a\,\bar a$ and $d\,\bar d$ sectors are also massive since the branes are parallel in the first torus, while those in the $b\,\bar b$ and $c\,\bar c$ sectors, originating from branes parallel in the second $T^2$, can get a mass if the Scherk-Schwarz deformation acts also along the second torus. As a result, all massless non-chiral fermions can be properly lifted.

We should stress here that the wrapping numbers we have chosen, alike those suggested by the authors in \cite{ibanez}, do not satisfy the tadpole condition
$$
{3 \omega^2_a \over \rho\beta^1} + {2 \omega^1_b \over \beta^2} + {\omega^2_d \over \beta^1} = 16 \,.
$$
In principle, this failure can be overcome introducing extra hidden D6 branes with no intersection with the Standard-Model ones \cite{ibanez}. In this case, the above equation would be replaced by the more general one
$$
{3 \omega^2_a \over \rho\beta^1} + {2 \omega^1_b \over \beta^2} + {\omega^2_d \over \beta^1} + N_h \omega^1_h \omega^2_h \omega^3_h = 16 \,.
$$

\vspace{1 cm}
\subsection{Deforming a three generations Pati-Salam model}

As a second example, let us examine the model presented in \cite{fbralph}. It is a four-dimensional, three generation, left-right symmetric standard model with gauge group
$$
G_{\rm CP} = U (3)_a \times U (2)_b \times U(2)_c \times U (1)_d
\,.
$$
It can be obtained engineering four stacks of D6 branes with intersection numbers as in table \ref{tablex}. To get the correct spectrum reported for completeness in table \ref{tabley}. 


In this example where all $\omega$'s are odd, regardless of the choice of (horizontal) Scherk-Schwarz coordinate, all adjoint fermions are lifted in mass. As for the remaining non-chiral fermions, they can be made massive deforming also the second torus.

\vskip 10pt

\begin{table}[h]
\begin{center}
\begin{tabular}{c c c c}
\hline
 $N_\alpha$ & $(\omega^1_\alpha \,,\, \kappa^1_\alpha)$ &
$(\omega^2_\alpha \,,\, \kappa^2_\alpha)$ &
$(\omega^3_\alpha \,,\, \kappa^3_\alpha)$\\ 
\hline
 3 & $(1,0)$ & $(1,0)$ & $(3,1)$ \\
 2 & $(1,1)$ & $(1,1)$ & $(1,0)$ \\
 2 & $(1,1)$ & $(1,-2)$ & $(1,0)$\\
 1 & $(1,0)$ & $(1,-2)$ & $(3,1)$\\
\hline
\end{tabular}
\end{center}
\caption{ D6-brane wrapping numbers for a left-right symmetric model, from \cite{fbralph}.}
\label{tablex}
\end{table}
\vspace{2 cm}

\begin{table}[h]
\begin{center}
\begin{tabular}{c c}
\hline
 ${\rm SU} (3) \times {\rm SU} (2)_{\rm L} \times {\rm SU} (2)_{\rm R} \times U (1)^4 $ &  generations\\
\hline
$\qquad\quad(3,2,1)_{(1,1,0,0)}$ &  $\qquad$ 2 \\
$\qquad\quad(3,2,1)_{(1,-1,0,0)}$ &  $\qquad$ 1 \\
$\qquad\quad(3^*,1,2)_{(-1,0,1,0)}$ &  $\qquad$ 2 \\
$\qquad\quad(3^*,1,2)_{(-1,0,-1,0)}$ &  $\qquad$ 1 \\
$\qquad\quad(1,2,1)_{(0,-1,0,1)}$ &  $\qquad$ 3 \\
$\qquad\quad(1,1,2)_{(0,-1,0,-1)}$ &  $\qquad$ 3 \\
\hline
\end{tabular}
\end{center}
\caption{Left-right symmetric chiral massless spectrum, from \cite{fbralph}.}
\label{tabley}
\end{table}
\vspace{2 cm}

\subsection{Scherk-Schwarz deformations on a tilted torus}
\begin{figure}  
\begin{center} 
\includegraphics[scale=1, height=5cm]{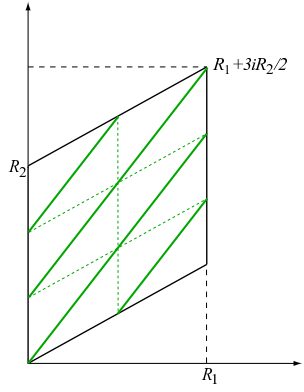} 
\caption{A (3,2) brane wrapping a tilted torus.}
\label{tilted}
\end{center}
\end{figure}

\everypar{\hspace{-.6cm}}
It is not difficult to generalise our previous results to the case of a tilted torus. 
As shown in \cite{discrete}, compatibility with the $\Omega \mathscr{R}$ projection imposes constraints on the real part of the complex structure, whose quantum deformation does not belong any more to the  physical spectrum, but nevertheless can be given a non-trivial constant background value.
 Describing magnetised or rotated branes in this tilted torus (see fig. \ref{tilted}) is quite straightforward, though some modifications are needed \cite{augusto},\cite{fbralph}. 
Firstly, if we define $(\omega , \kappa)$ as the number of times a brane wraps the two canonical cycles of the $T^2$, the quantisation condition on the angle reads
\beq\label{newangle}
\tan\phi = {(\kappa + {1\over 2} \omega ) R_2 \over \omega R_1} = {\kappa R_2 \over \omega R_1} + {R_2\over 2\, R_1} \,,
\eeq
with the additional contribution deriving from the fixed real part of the complex structure.
Moreover, the shear parameter enters both in the zero-mode sector, through a proper redefinition of the effective length of the brane
$$
L_\| = \sqrt{\left(\omega R_1\right)^2 + \left[\left(\kappa + {\textstyle{1\over 2}}\omega \right) R_2
\right]^2} \,,
$$
and in the frequencies of the string excitations that are shifted by the angle $\phi$ defined in (\ref{newangle}). As usual, for any brane with angle $\phi$ and wrapping numbers $(\omega_\alpha , \kappa _\alpha)$ one has also to introduce image branes under the action of the orientifold operator $\Omega \mathscr{R}$. They still have opposite angle $-\phi$, but now the wrapping numbers are $(\omega_\alpha , -\kappa_\alpha - \omega_\alpha )$, as a result of the tilting of the torus. Actually, all these changes are more easily taken into account after we label branes with wrapping numbers in a Cartesian basis.
In this respect, all the modifications can be summarised by noting that the wrapping numbers $(\omega_\alpha , \kappa_\alpha )$ for the case of a rectangular torus have to be replaced by  $(\omega_\alpha ,\kappa_\alpha + {1\over 2} \omega_\alpha)$ for the case of a tilted torus \cite{fbralph}.

With these modifications in mind, we can then straightforwardly generalise our previous results for a Scherk-Schwarz deformation acting along the horizontal axis to the case of a tilted torus: once more, it is the parity of $\omega_\alpha$ to discriminate between a trivial or non-trivial action on the neutral non-chiral fermions. 

\vspace{1 cm}
\section{Six-dimensional orbifold models}
\everypar{\hspace{-.6cm}}
As a second application, we consider the non-supersymmetric type IIB string compactified on the $(T^2 \times T^2)/\mathbb{Z}_2$ with the Scherk-Schwarz deformation acting along the horizontal axis of the first $T^2$. 

\vspace{1 cm}
\subsection{Preludio: the closed-string sector}
\label{preludio}
In the Scherk-Schwarz basis, the torus partition function reads \cite{ADSi}
\bea \label{SStorus}
\mathcal{T} &=& {\textstyle{1\over 2}} \Biggl[ \left( |V_8|^2 + |S_8|^2 \right) \, \Lambda_{2n_1}^{(4,4)} 
- \left( S_8 \bar V_8 + V_8 \bar S_8 \right) \, \Lambda^{(4,4)}_{m_1 + {1\over 2},2 n_1}
+ \left( |O_8|^2 + |C_8|^2 \right) \, \Lambda^{(4,4)}_{2n_1 +1}\nn \\
& -& 
\left( C_8 \bar O_8 + O_8 \bar C_8 \right) \, \Lambda^{(4,4)}_{m_1 +{1\over 2}, 2 n_1 + 1} 
+ \left( |V_4 O_4 - O_4 V_4 |^2 + |C_4 C_4 - S_4 S_4|^2 \right)\, \left| {2\eta\over\theta_2}\right|^4 \Biggr]\nn \\
&+& {\textstyle{16 \over 4}} \, \left[ \left( |Q_s + Q_c|^2 + |Q_s ' + Q_c '|^2 \right) \, \left| {\eta\over\theta_4}\right|^4
+ \left( |Q_s - Q_c|^2 + |Q_s ' - Q_c '|^2 \right) \, \left| {\eta\over\theta_3}\right|^4
\right]\,,
\eea
where $\Lambda^{(4,4)}$ is the four-dimensional Narain lattice with momenta $m_1 \,, \ldots \,, m_4$ and windings $n_1 \,, \ldots \,, n_4$. Moreover, $\Lambda_{2 n_1}^{(4,4)}$ indicates that the winding number $n_1$ is now an even integer, and similarly for the other terms. The $Q$'s are defined as usual by 
\begin{plain}
$$
\eqalign{
Q_o =& \, V_4 O_4 - C_4 C_4 \,,
\cr
Q_v =& \, O_4 V_4 - S_4 S_4 \,,
\cr
Q_s =& \, O_4 C_4 - S_4 O_4 \,,
\cr
Q_c =& \, V_4 S_4 - C_4 V_4 \,,
\cr}
\qquad
\eqalign{
Q_o ' =& \, V_4 O_4 - S_4 S_4 \,,
\cr
Q_v ' =& \, O_4 V_4 - C_4 C_4 \,,
\cr
Q_s '=& \, O_4 S_4 - C_4 O_4 \,,
\cr
Q_c '=& \, V_4 C_4 - S_4 V_4 \,,
\cr}
$$
where we have introduced additional $Q$'s that will play a role in the following. 
The Klein bottle amplitude obtained by the $\Omega \mathscr{R} (-1)^{F_{\rm L}}$ projection reads \cite{ADSi}
$$
\eqalign{
\mathcal{K} =& {\textstyle{1\over 4}} (V_8 - S_8 ) \left( P_{m}^1 W_{n}^2 P_{m}^3 W_{n}^4 + W_{2 n}^1 P_{m}^2 W_{n}^3 P_{m}^4 \right)
+ {\textstyle{1\over 4}} (O_8 - C_8 ) \, W_{2n+1}^1 P_{m}^2 W_{n}^3 P_{m}^4
\cr
& + 4 \left( Q_s + Q_c + Q_s ' + Q_c '\right) \, \left( {\eta \over \theta_4 }\right)^2 \,.
\cr}
$$
\end{plain}
After an $S$-modular transformation to the transverse channel, and adapting to our six-dimensional example the definition in (\ref{radii}) , the massless contributions to the NS-NS and R-R tadpoles
$$
\tilde{\mathcal{K}}_0 \sim  8 \, V_4 O_4 \, \left( {\bf R} + {1\over {\bf R}}\right)^2
+ 8\, O_4 V_4 \, \left( {\bf R} - {1\over {\bf R}}\right)^2
- 8 \, (C_4 C_4 + S_4 S_4) \, {\bf R}^2
$$
clearly suggest that the O7-planes stretched along the vertical axis of the two $T^2$'s come in conjugate pairs and thus do not induce any R-R charge for the corresponding eight-form potential. The configuration of O7-planes stretched along the horizontal axis has instead a non-trivial charge that has to be compensated by suitable brane configurations \cite{ADSi}.
Since we are altering the nature of some orientifold planes, we cannot expect that the 
new open-string spectrum might be obtained from the undeformed one by giving masses to the fermions in the adjoint of the $U (N_i)$ gauge group factors. In fact, this configuration would not solve any longer the untwisted R-R tadpole conditions, and thus new vacuum configurations have to be found.

\begin{figure}  
\begin{center} 
\includegraphics[scale=1, height=5cm]{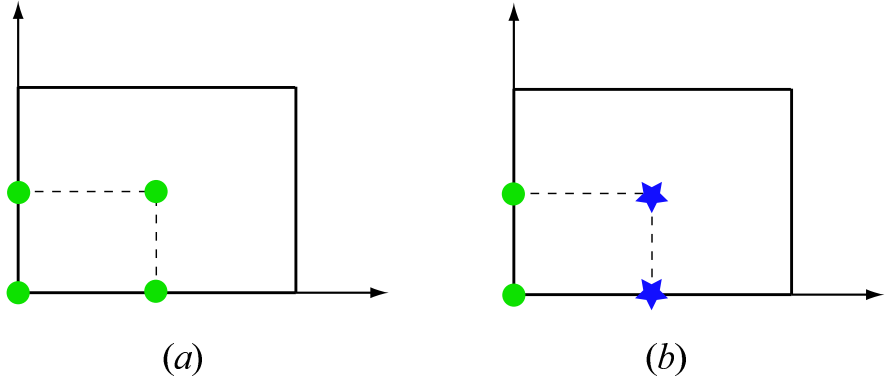} 
\caption{Fixed points for a $\mathbb{Z}_2$ orbifold without $(a)$ and with $(b)$ Scherk-Schwarz deformation.}
\label{fixed}
\end{center}
\end{figure}


Before we proceed with the construction of the open-string sector, let us pause for a moment and try to extract some useful information about the geometry of this Scherk-Schwarz-deformed $T^4/\mathbb{Z}_2$. To this end it is convenient to stare at the twisted sector in (\ref{SStorus}). The appearance of new characters $Q_{s,c}'$ and the modified multiplicity clearly spells out that the original sixteen fixed points of the geometrical $\mathbb{Z}_2$ orbifold model are now split into two sets of eight, with different GSO projections in the corresponding twisted sector. Moreover, one can also read from the terms $Q_s - Q_c$ and $Q_s' - Q_c'$ that the fixed points act differently on the internal quantum numbers. Indeed, if we call $\xi_\mu$ and $x_\mu$ ($\mu =1,\ldots , 8$) the fixed points in the two sets, the eight $\xi_\mu$ project the spectrum with respect the geometrical $\mathbb{Z}_2$ while the eight $x_\mu$ involve also the action of the Scherk-Schwarz deformation $(-1)^F$, for an overall $g\, (-1)^F$ projection, $g$ being the $\mathbb{Z}_2$ generator. This is clear both from the kind of boundary conditions we are imposing and from the structure of the modular-invariant partition function, where the projection in $Q_s' - Q_c'$ involves and additional $(-1)^F$ with respect to that in $Q_s - Q_c$. We can actually say more about the geometrical distribution of these fixed points. In our factorised $T^4 = T^2 \times T^2$ example, in fact, we have chosen to impose Scherk-Schwarz boundary conditions along the horizontal axis of the first torus. As a result, the structure of the fixed points in the second torus cannot be affected. In the first torus, instead, we have the configuration depicted in figure 9, since, as previously reminded, we are deforming the theory by changing the boundary conditions along the horizontal axis. Therefore, we can conclude that our $T^4$ is of the form $T^2_{(b)} \times T^2_{(a)}$, where $T^2_{(a)}$ and $T^2_{(b)}$ denote the undeformed and the deformed torus as in figure \ref{fixed}. The two sets of fixed points are then

\bea\label{fixedpoints}
\vec\xi &=& \bigg\{ (0,0,0,0) , (0,0,{\textstyle{1\over 2}} ,0) , (0,0,0, {\textstyle{1\over 2}}) , 
(0,0,{\textstyle{1\over 2}},{\textstyle{1\over 2}}) , \nn \\
&& (0,{\textstyle{1\over 2}},0,0), (0,{\textstyle{1\over 2}},{\textstyle{1\over 2}} ,0) , (0,{\textstyle{1\over 2}},0, {\textstyle{1\over 2}}) , 
(0,{\textstyle{1\over 2}},{\textstyle{1\over 2}},{\textstyle{1\over 2}})\bigg\} \,, \nn \\
\vec x &=&\bigg\{ ({\textstyle{1\over 2}},0,0,0), ({\textstyle{1\over 2}},0,{\textstyle{1\over 2}} ,0) , ({\textstyle{1\over 2}},0,0, {\textstyle{1\over 2}}) , 
({\textstyle{1\over 2}},0,{\textstyle{1\over 2}},{\textstyle{1\over 2}}), \nn \\
&& ({\textstyle{1\over 2}},{\textstyle{1\over 2}},0,0), ({\textstyle{1\over 2}},{\textstyle{1\over 2}},{\textstyle{1\over 2}} ,0) , ({\textstyle{1\over 2}},{\textstyle{1\over 2}},0, {\textstyle{1\over 2}}) , 
({\textstyle{1\over 2}},{\textstyle{1\over 2}},{\textstyle{1\over 2}},{\textstyle{1\over 2}})\bigg\} \,.
\eea

\vspace{1 cm}
\subsection{Intermezzo: the geometry of orthogonal branes}
\begin{figure}  
\begin{center} 
\includegraphics[scale=1, height=5cm]{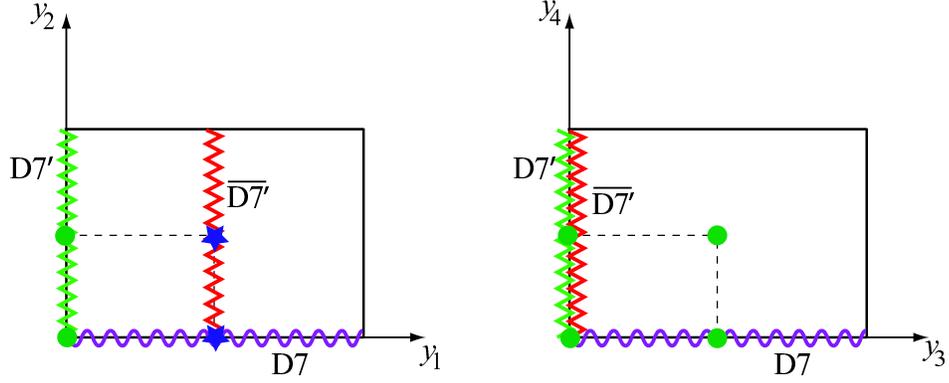} 
\caption{ The geometry of D-branes. A wavy line represents D7 branes while zig-zag lines represents ${\rm D}7'$ and $\overline{{\rm D}7'}$.}
\label{fpbranei}
\end{center}
\end{figure}

\everypar{\hspace{-.6cm}}
We can now proceed to include the D-branes, and to better appreciate the construction of the open-string sector we first review the model presented in \cite{ADSi}, with the obvious replacement of D9 and D5 with two sets of orthogonal D7 branes. To reproduce their geometry we distribute the branes as in figure \ref{fpbranei}, that, after two T-dualities along the vertical axis of the two $T^2$'s correspond indeed to space-filling D9 branes together with a set of D5 branes sitting at the fixed point $\xi_1 = (0,0,0,0)$ and a set of D5 anti-branes sitting at the fixed point $x_1 = ({1\over 2},0,0,0)$.

\begin{plain}
The direct-channel annulus in \cite{ADSi}\ is then
$$
\eqalign{
\mathcal{A} =& {\textstyle{1\over 4}} \Biggl\{ N_7^2 \left( V_8 \, P^1_m - S_8 \, P^1_{m+{1\over 2}} \right) W^2_n P^3_m W^4_n + \left( N_{7'}^2 + N_{\overline{7'}}^2 \right) (Q_o + Q_v )\, W_n^1 P_m^2 W_n^3 P_m^4
\cr
&+ 2 N_{7'} N_{\overline{7'}}\, (O_8 - C_8) W^1_{n+{1\over 2}} P^2_m W^3_n P^4_m +
R^2_7\, (V_4 O_4 - O_4 V_4 ) \left( {2\eta\over\theta_2}\right)^2
\cr
&+ \left[ R_{7'}^2 (Q_o - Q_v) + R_{\overline{7'}}^2 (Q_o ' - Q_v ') \right] \left( {2\eta \over \theta_2}\right)^2
\cr
&+ 2 \left[ N_7 N_{7'}\, (Q_s + Q_c) + N_7 N_{\overline{7'}} (Q_s ' + Q_c ' ) \right] \left( {\eta\over \theta_4} \right)^2
\cr
&+ 2 \left[ R_7 R_{7'} (Q_s - Q_c ) + R_7 R_{\overline{7'}} (Q_s ' - Q_c ' ) \right] \left( {\eta \over \theta_3}\right)^2 \Biggr\}\,.
\cr}
$$
Any single term in this expression has a clear rational. The terms in the first line simply correspond to open strings living on a given D-brane, with the important difference that D7 branes stretching along the horizontal $y_1$ axis now have a deformed non-supersymmetric spectrum. The second line is also quite standard: the first term encodes the spectrum of a brane-anti-brane system with the two ${\rm D}7'$ and $\overline{D 7'}$ branes separated along $y_1$, while the second contribution pertains to the orbifold projection on the horizontal D7 branes. Also the fourth line is quite standard: it corresponds to open strings with mixed Neumann-Dirichlet boundary conditions along the four-compact directions. The second term has different GSO projections since the strings stretch between a brane and an anti-brane. More subtle are the third and fifth lines. They both enforce the orbifold projection on open-strings, but in a different way. In fact, while the first term in the third line clearly corresponds to the conventional geometrical $\mathbb{Z}_2$ projection, the second term involves and additional minus sign in the Ramond sector, and indeed corresponds to the action of the element $g (-1)^F$. This is consistent with, and in fact dictated by, the fact that ${\rm D}7'$ branes sit on top of conventional $g$-fixed points, while the $\overline{{\rm D}7'}$ branes pass through points fixed under the action of $g (-1)^F$. Clearly, similar considerations also apply to the terms in the fifth line that describe the symmetrisation of $7\, 7'$ and $7\,\overline{7'}$ states that indeed live at the fixed points $\xi_1$ and $x_1$.
As usual, the different orbifold projections in the direct channel translate into exchanges of closed-string twisted states in the transverse channel, whose massless tadpoles clearly spell out the distribution of branes among the fixed points
$$
\tilde{\mathcal{A}}_{\rm tw} \sim 2^{-5}\, \left\{ \left[ \left(R_7 - R_{7'} \right)^2 + R^2_7 + 3\, R^2_{7'} \right]\, Q_s
+  \left[ \left(R_7 - R_{\overline{7'}} \right)^2 + R^2_7 + 3\, R^2_{\overline{7'}} \right]\, Q_s '
\right\} \,,
$$
\end{plain}
compatible with the geometry in figure \ref{fpbranei}.

\begin{figure}  
\begin{center} 
\includegraphics[scale=1, height=5cm]{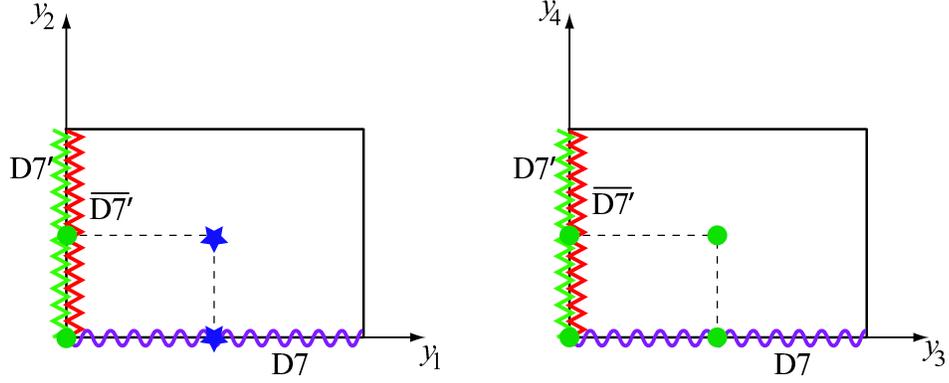} 
\caption{The geometry of D-branes. A wavy line represents D7 branes while zig-zag lines represents ${\rm D}7'$ and $\overline{{\rm D}7'}$.}
\label{fpbraneii}
\end{center}
\end{figure}


What about if we distribute branes differently? Let us consider for instance the configuration in figure \ref{fpbraneii}. Now all the branes pass through the conventional $g$-fixed points, and thus we do not expect in the amplitude terms like $Q_o' - Q_v '$. Indeed, the new annulus amplitude
\begin{plain}
$$
\eqalign{
\mathcal{A} =& {\textstyle{1\over 4}} \Biggl\{ N_7^2 \, \left( V_8 \, P^1_m - S_8 \, P^1_{m+{1\over 2}} \right) W^2_n P^3_m W^4_n + \left( N_{7'}^2 + N_{\overline{7'}}^2 \right) (Q_o + Q_v) \, W^1_n P^2_m W^3_n P^4_m 
\cr
&+ 2 N_{7'} N_{\overline{7'}} \, (O_8 - C_8) \, W^1_n P^2_m W^3_n P^4_m + R_7^2\, (V_4 O_4 - O_4 V_4 ) \left( {2\eta\over \theta_2}\right)^2
\cr
&+ \left[ \left( R_{7'}^2 + R_{\overline{7'}}^2 \right) (Q_o - Q_v)  + 2 R_{7'} R_{\overline{7'}} \,
(O_4 O_4 - V_4 V_4 - S_4 C_4 + C_4 S_4)\right] \left( {2\eta\over \theta_2}\right)^2
\cr
&+ 2 \left[ N_7 N_{7'}\, (Q_s + Q_c ) + N_7 N_{\overline{7'}}\, (Q_s '+ Q_c ')  \right] \left( {\eta\over\theta_4} \right)^2
\cr
&+ 2 \left[ R_7 R_{7'}\, (Q_s - Q_c ) + R_7 R_{\overline{7'}}\, (- O_4 S_4 + V_4 C_4 - C_4 O_4 + S_4 V_4 ) \right]  \left( {\eta\over\theta_3} \right)^2
\cr}
$$
only involves the geometrical $\mathbb{Z}_2$ projector, and yields twisted tadpoles in the transverse channel
$$
\eqalign{
\tilde{\mathcal{A}} _{\rm tw} \sim& 2^{-5} \left\{ 2 R^2_7\, Q_s ' + \left[ \left( R_7 - R_{7'} + R_{\overline{7'}} \right)^2 + R^2_7 + 3 (R_{7'} - R_{\overline{7'}} )^2 \right] \, O_4 C_4 \right.
\cr
&- \left. \left[ \left( R_7 - R_{7'} - R_{\overline{7'}} \right)^2 + R^2_7 + 3 (R_{7'} + R_{\overline{7'}} )^2 \right] \, S_4 O_4 \right\}
\cr}
$$
\end{plain}
that clearly display the distribution in figure \ref{fpbraneii}. 

Similarly, one could opt for a different distribution of branes along the compact $T^4$, that would consequently imply different projections according to the nature of the fixed points crossed by the branes. This would then reflect in twisted tadpoles compatible with the chosen geometry.

\subsection{Crescendo: rotating the branes}
\begin{figure}  
\begin{center} 
\includegraphics[scale=1, height=18cm]{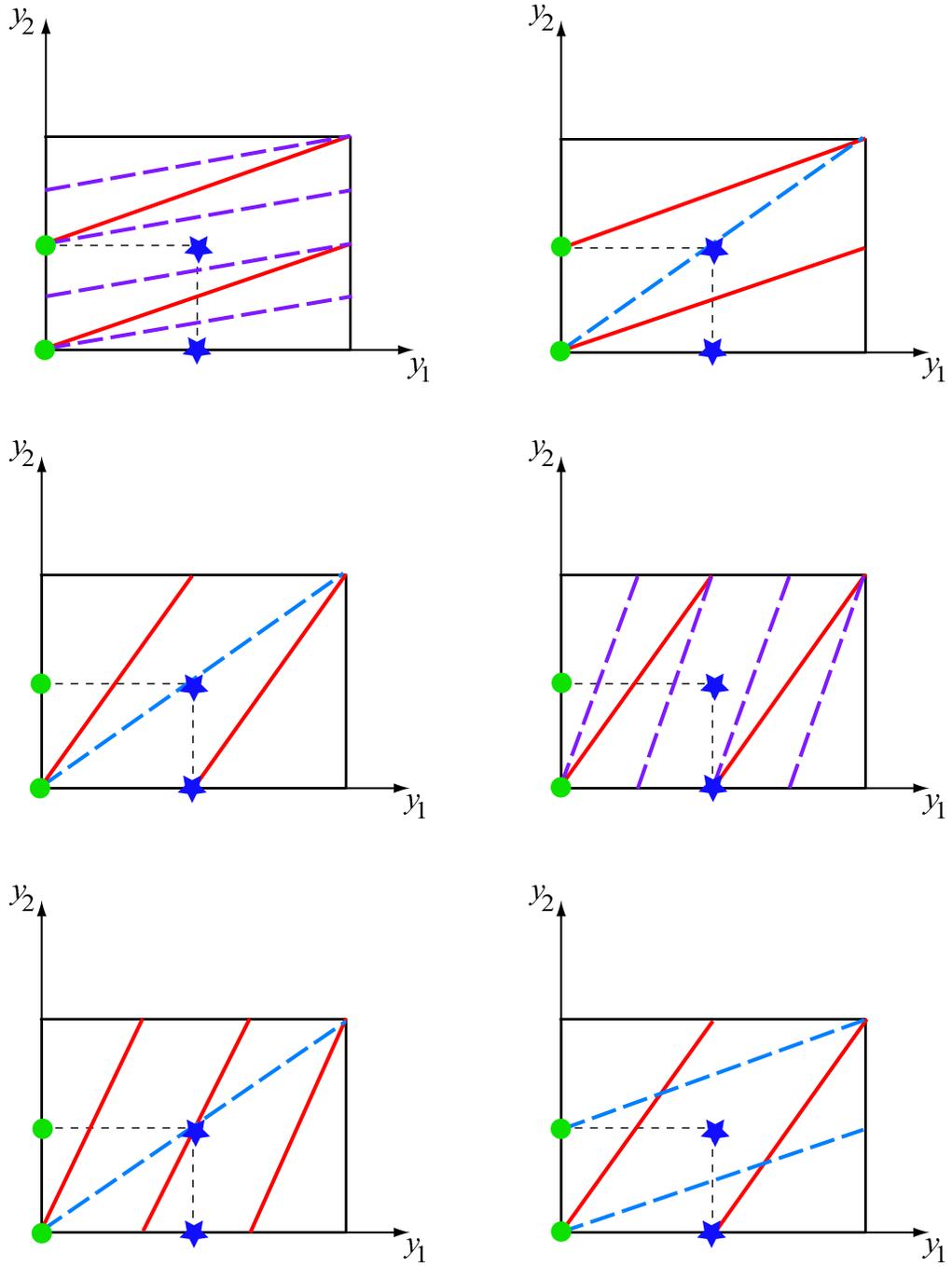}  
\caption{Branes with various wrapping numbers on a $(b)$ torus.}
\label{fprotated}
\end{center}
\end{figure}

\everypar{\hspace{-.6cm}}
We can now generalise the previous construction to the case of rotated branes, where some care is needed to identify the exact location of brane intersections. In the simple case where all branes cross the origin of the two tori, it was shown in \cite{fbralph}\ that any brane always passes through four fixed points of the $T^4/\mathbb{Z}_2$. Moreover, we can easily identify 
the four points from the wrapping numbers $(\omega_\alpha^\Lambda , \kappa_\alpha^\Lambda)$, since for a single $T^2$ branes fall into three different equivalence classes: $\omega_\alpha$ even and $\kappa_\alpha$ odd, $\omega_\alpha$ odd and $\kappa_\alpha$ even, $\omega_\alpha$ and $\kappa_\alpha$ both odd and co-prime. Focussing our attention to the first $T^2$, the only one that actually matters to identify the correct projection, one has then the cases reported in figure \ref{fprotated}.

The annulus amplitude associated to the closed-string sector in sub-section  \ref{preludio} is then
\begin{plain}
$$
\eqalign{
\mathcal{A}_{\alpha \alpha} =&  {\textstyle{1\over 2}}
\sum_{a=1}^{m_{\rm e}} N_a \bar N_a \, (Q_o + Q_v) [{\textstyle{0\atop 0}}] \, P_m^1 W_n^1 P_m^2 W_n^2 + {\textstyle{1\over 2}} \sum_{a=1}^{m_{\rm e}} N_a \bar N_a \, (Q_o - Q_v) [{\textstyle{0\atop 0}}] \, {4\over \Upsilon_2 [{\textstyle{0\atop 0}}]} 
\cr
&+ {\textstyle{1\over 2}} \sum_{i=1}^{m_{\rm o}} N_i \bar N_i \, \left( V_8  [{\textstyle{0\atop 0}}] \, P_{m}^1 - S_8 [{\textstyle{0\atop 0}}]  \, P_{m+{1\over 2}}^1 \right)  W_n^1 P_m^2 W_n^2
\cr
&+ {\textstyle{1\over 2}} \sum_{i=1}^{m_{\rm o}} N_i \bar N_i \, ( V_4 O_4 - O_4 V_4 )
[{\textstyle{0\atop 0}}] \, {4\over \Upsilon_2 [{\textstyle{0\atop 0}}]} \,,
\cr}
$$
$$
\eqalign{
\mathcal{A}_{\alpha \bar\alpha} +\mathcal{A}_{\bar \alpha \alpha} =& {\textstyle{1\over 4}} \sum_{\alpha =1}^{m_{\rm o}+m_{\rm e}} \left( N_\alpha^2 (Q_o + Q_v) [{\textstyle{\alpha\bar\alpha\atop 0}}] + \bar N_\alpha^2 (Q_o + Q_v) [{\textstyle{\bar \alpha \alpha \atop 0}}] \right) \, 
{I_{\alpha\bar\alpha}  \over \Upsilon_1 [{\textstyle{\alpha\bar\alpha\atop 0}}]}
\cr
&- {\textstyle{1\over 4}} \sum_{a=1}^{m_{\rm e}} \left( N_a^2 (Q_o - Q_v) [{\textstyle{a\bar a\atop 0}}] + \bar N_a^2 (Q_o - Q_v) [{\textstyle{\bar a a  \atop 0}}] \right) \, 
{4 \over \Upsilon_2 [{\textstyle{a\bar a\atop 0}}]} 
\cr
&- {\textstyle{1\over 4}} \sum_{i=1}^{m_{\rm o}} \left( N_i^2 (Q_o - Q_v) [{\textstyle{i\bar \imath\atop 0}}] + \bar N_i^2 (Q_o - Q_v) [{\textstyle{\bar \imath i  \atop 0}}] \right) \, 
{2 \over \Upsilon_2 [{\textstyle{i\bar \imath\atop 0}}]} 
\cr
&- {\textstyle{1\over 4}} \sum_{i=1}^{m_{\rm o}} \left( N_i^2 (Q_o '- Q_v ') [{\textstyle{i\bar \imath\atop 0}}] + \bar N_i^2 (Q_o '- Q_v ') [{\textstyle{\bar \imath i  \atop 0}}] \right) \, 
{2 \over \Upsilon_2 [{\textstyle{i\bar \imath\atop 0}}]} \,,
\cr}
$$
$$
\eqalign{
\mathcal{A}_{\alpha \beta } + \mathcal{A}_{\bar\alpha \bar \beta} =& {\textstyle{1\over 2}}
\sum_{{\alpha,\beta=1 \atop \beta<\alpha}}^{m_{\rm o}+m_{\rm e}} \left( N_\alpha \bar N_\beta (Q_o + Q_v ) 
[{\textstyle{\alpha \beta\atop 0}}] + \bar N_\alpha N_\beta (Q_o + Q_v ) [{\textstyle{\bar \alpha \bar\beta\atop 0}}]\right)\,
{I_{\alpha \beta} \over \Upsilon_1 [{\textstyle{\alpha \beta\atop 0}}]}
\cr
&+ {\textstyle{1\over 2}}
\sum_{{a , b=1 \atop b<a}}^{m_{\rm e}} \left( N_a \bar N_b (Q_o - Q_v ) 
[{\textstyle{ab\atop 0}}] + \bar N_a N_b (Q_o - Q_v ) [{\textstyle{\bar a \bar b\atop 0}}]\right)\,
{2 I_{ab} ' \over \Upsilon_2 [{\textstyle{ab\atop 0}}]} 
\cr
&+ {\textstyle{1\over 2}}
\sum_{a =1}^{m_{\rm e}} \sum_{i =1}^{m_{\rm o}}  \left( N_a \bar N_i (Q_o - Q_v ) 
[{\textstyle{ai\atop 0}}] + \bar N_a N_i (Q_o - Q_v ) [{\textstyle{\bar a \bar \imath \atop 0}}]\right)\,
{I_{ai} ' \over \Upsilon_2 [{\textstyle{ai\atop 0}}]} 
\cr
&+ {\textstyle{1\over 2}}
\sum_{{i , j=1 \atop j<i}}^{m_{\rm o}} \left( N_i \bar N_j (Q_o - Q_v ) 
[{\textstyle{ij\atop 0}}] + \bar N_i N_j (Q_o - Q_v ) [{\textstyle{\bar \imath \bar \jmath\atop 0}}]\right)\,
{I_{ij} ' \over \Upsilon_2 [{\textstyle{ij\atop 0}}]} 
\cr
&+ {\textstyle{1\over 2}}
\sum_{{i , j=1 \atop j<i}}^{m_{\rm o}} \left( N_i \bar N_j (Q_o ' - Q_v ') 
[{\textstyle{ij\atop 0}}] + \bar N_i N_j (Q_o '- Q_v ') [{\textstyle{\bar \imath \bar \jmath\atop 0}}]\right)\,
{I_{ij} '' \over \Upsilon_2 [{\textstyle{ij\atop 0}}]} 
\,,
\cr}
$$
and, finally, 
$$
\eqalign{
\mathcal{A}_{\alpha\bar\beta} + \mathcal{A}_{\bar \alpha \beta} =& {\textstyle{1\over 2}} 
\sum_{{\alpha,\beta=1 \atop \beta<\alpha}}^{m_{\rm o}+m_{\rm e}} \left( N_\alpha  N_\beta (Q_o + Q_v ) [{\textstyle{\alpha\bar\beta\atop 0}}] + \bar N_\alpha \bar N_\beta (Q_o + Q_v ) 
[{\textstyle{\bar \alpha \beta\atop 0}}] \right) \,
{I_{\alpha\bar\beta} \over \Upsilon_1 [{\textstyle{\alpha\bar\beta\atop 0}}]}
\cr
&- {\textstyle{1\over 2}}
\sum_{{a , b=1 \atop b<a}}^{m_{\rm e}} \left( N_a N_b (Q_o - Q_v ) 
[{\textstyle{a\bar b\atop 0}}] + \bar N_a \bar N_b (Q_o - Q_v ) [{\textstyle{\bar a b\atop 0}}]\right)\,
{2 I_{a\bar b} ' \over \Upsilon_2 [{\textstyle{a\bar b\atop 0}}]} 
\cr
&- {\textstyle{1\over 2}}
\sum_{a =1}^{m_{\rm e}} \sum_{i =1}^{m_{\rm o}}  \left( N_a  N_i (Q_o - Q_v ) 
[{\textstyle{a\bar \imath\atop 0}}] + \bar N_a \bar N_i (Q_o - Q_v ) [{\textstyle{\bar a i \atop 0}}]\right)\,
{I_{a\bar \imath} ' \over \Upsilon_2 [{\textstyle{a\bar \imath\atop 0}}]} 
\cr
&- {\textstyle{1\over 2}}
\sum_{{i , j=1 \atop j<i}}^{m_{\rm o}} \left( N_i N_j (Q_o - Q_v ) 
[{\textstyle{i\bar \jmath\atop 0}}] + \bar N_i \bar N_j (Q_o - Q_v ) [{\textstyle{\bar \imath j \atop 0}}]\right)\,
{I_{i\bar \jmath} ' \over \Upsilon_2 [{\textstyle{i\bar \jmath\atop 0}}]} 
\cr
&- {\textstyle{1\over 2}}
\sum_{{i , j=1 \atop j<i}}^{m_{\rm o}} \left( N_i N_j (Q_o '- Q_v ') 
[{\textstyle{i\bar \jmath\atop 0}}] + \bar N_i \bar N_j (Q_o '- Q_v ') [{\textstyle{\bar \imath j \atop 0}}]\right)\,
{I_{i\bar \jmath} '' \over \Upsilon_2 [{\textstyle{i\bar \jmath\atop 0}}]} 
\,.
\cr}
$$
Here we have used the obvious notation for the $Q [{\alpha\beta\atop\gamma\delta}]$ where the internal $SO(4)$ characters are decomposed in terms of $SO(2) \times SO (2)$ ones, and
$$
\Upsilon_2 [{\textstyle{\alpha\beta \atop \gamma\delta}}] = \prod_{\Lambda =1,2} \, {\theta_2 (\zeta^\Lambda |\tau ) \over \eta (\tau) } \, e^{2i \pi \zeta^\Lambda} 
\,,
$$
accounts for the $\mathbb{Z}_2$ orbifold generator acting on the world-sheet bosons.
Finally, $I_{\alpha\beta}$ denotes as usual the number of intersections of branes with wrapping numbers $(\omega_\alpha^\Lambda \,,\, \kappa_\alpha^\Lambda)$ and $(\omega_\beta^\Lambda \,,\, \kappa_\beta^\Lambda)$, while
$$
I'_{\alpha\beta} =  1 + \Pi (\omega_\alpha^2 - \omega_\beta^2 ) \,\Pi 
(\kappa_\alpha^2 - \kappa_\beta^2 ) 
$$
and 
$$
I''_{\alpha\beta} = \Pi (\omega_\alpha^1 +1 ) \Pi ( \omega_\beta^1 +1) \,\Pi 
(\kappa_\alpha^1 - \kappa_\beta^1 ) \, I'_{\alpha\beta} 
$$
count the number of intersections that actually coincide with the $x_p$ and $\xi_p$ fixed points in (\ref{fixedpoints}), with
$$
\Pi ( \mu ) = {\textstyle{1\over 2}} \sum_{\epsilon=0,1} e^{i\pi\epsilon \mu}\,, \qquad \mu\in\mathbb{Z}\,,
$$
a $\mathbb{Z}_2$ projector. 

\end{plain}
 
Finally, the M\"obius-strip amplitude
\bea 
\mathcal{M} &=& - {\textstyle{1\over 4}} \sum_{\alpha =1}^{m_{\rm o} + m_{\rm e}} \, 
\big[ \left(
N_\alpha \, (\hat Q _o + \hat Q_v ) [{\textstyle{\alpha \bar\alpha \atop 0}}] + \bar N_\alpha
\, (\hat Q _o + \hat Q_v ) [{\textstyle{\bar \alpha \alpha \atop 0}}] \right) \,
{K_\alpha \over \hat \Upsilon_1  [{\textstyle{\alpha \bar\alpha \atop 0}}]} \nn \\ 
&-&  \left( N_\alpha \, (\hat V_4 \hat O_4 - \hat O_4 \hat V_4 ) [{\textstyle{\alpha \bar\alpha \atop 0}}] + \bar N_\alpha \, (\hat V_4 \hat O_4 - \hat O_4 \hat V_4) [{\textstyle{\bar \alpha\alpha \atop 0}}] \right) \,{ J_\alpha \over \hat \Upsilon_2  [{\textstyle{\alpha \bar\alpha \atop 0}}]} \big]
\label{moeborb}
\eea

takes now into account also the (anti-)symmetrisation of states that live at intersections sitting on the vertical O7 planes, whose number is given by
$$
J_\alpha = \prod_{\Lambda=1,2} \, 2\, \omega_\alpha^\Lambda \,.
$$
Notice that the second line in $\mathcal{M}$ does not involve fermions. This has a simple explanation in terms of the relative R-R charges of vertical O-planes and branes whose intersections lie on them. We already commented that the absence in $\tilde{\mathcal{K}}_0$ of R-R tadpoles proportional to ${\bf R}^{-1}$ clearly spells out that the O7 planes passing through the points $(0,0)$ and $({1\over 2} R_1 , 0)$ and extending along the vertical direction are conjugates pairs. This implies that their R-R charges are equal and opposite. On the other hand the branes have positive R-R charge, and thus fermions localised at the intersections sitting on the O7 plane are (say) symmetrised while those sitting on the O7 anti-planes (equal in number) are anti-symmetrised. As a result, their net contributions to $\mathcal{M}$ is zero, consistently with the expression
 (\ref{moeborb}).

After $S$ and $P$ modular transformations to the tree-level channel the low-lying untwisted states in $\tilde{\mathcal{K}}$, $\tilde{\mathcal{A}}$ and $\tilde{\mathcal{M}}$ yield the familiar conditions
$$
\sum_{\alpha =1}^{m_{\rm e} + m_{\rm o}} \, {\bf L}_\alpha \, \left(
N_\alpha + \bar N _\alpha \right) = 32 \left( {\bf R} + {1\over {\bf R}} \right) 
$$
for the NS-NS dilaton tadpole, and 
$$
\sum_{\alpha =1}^{m_{\rm e} + m_{\rm o}} \, {\bf L}_\alpha \, 
\left( N_\alpha \, e^{2i\phi_\alpha \cdot \eta} + \bar N_\alpha \,
e^{-2i\phi_\alpha \cdot \eta} \right) = 32 \, {\bf R}
$$
for the R-R tadpole with, as usual, $\eta^\Lambda$ the chirality of the internal spinors.
In both expressions ${\bf L}_\alpha$ and ${\bf R}$ are defined as in (\ref{radii}) , with $\Lambda =1,2$ running now over the two $T^2$'s.

Much more interesting are the twisted tadpoles since, as repeatedly stated in the previous sub-sections, they clearly display the geometry of the brane configuration. After an $S$ modular transformation one then gets the massless tadpoles
\beq \label{twrri}
C_4 O_4\, : \quad \sum_{i=1}^{m_{\rm o}} \, \left( N_i - \bar N_i \right) \, P(z_p \in D_i)  =0 \,, \qquad \forall\, z_p \in \{ x\} \,,
\eeq
and
\beq\label{twrrii}
S_4 O_4 \, : \quad \sum_{a=1}^{m_{\rm e}} \, (N_a -\bar N_a ) \, P ( z_q \in D_a) +
\sum_{i=1}^{m_{\rm o}} \, (N_i - \bar N_i ) \, P (z_q \in D_i ) =0 \,, \qquad \forall \, z_q \in \{\xi\}
\eeq
where
$D_\alpha$ denotes the straight line drawn by the $\alpha$-th brane with wrapping numbers
$(\omega^\Lambda_\alpha , \kappa^\Lambda_\alpha)$, and
\begin{plain}
$$
P (w_\ell \in D_\alpha ) = 
\cases{
1 & if $w_\ell \in D_\alpha$
\cr
0 & if $w_\ell \not\in D_\alpha$
\cr}
$$
\end{plain}
equals one or vanishes depending on whether the point $w_\ell$ belongs to the line $D_\alpha$ or does not. In the simple case where all branes pass through the origin, the 
$P (w_\ell \in D_\alpha ) $ are collected in tables (\ref{table5}) and (\ref{table6}). 

\vskip 10pt

\begin{table}[ht]
\begin{center}
\begin{tabular}{c c}
\hline
 $z_p$ & $P (z_p \in D_i) $ \\
\hline
 $x_1$ & $\Pi(\kappa^1_i)\, \Pi(\omega^1_i +1)$ \\
 $x_2$ & $\Pi(\kappa^1_i)\, \Pi(\omega^1_i +1)\, \Pi(\kappa^2_i) \, \Pi(\omega^2_i +1) $ \\
 $x_3$ & $\Pi(\kappa^1_i)\, \Pi(\omega^1_i +1) \, \Pi(\kappa^2_i+1) \, \Pi(\omega^2_i) $\\
 $x_4$ & $\Pi(\kappa^1_i)\, \Pi(\omega^1_i +1) \, \Pi(\kappa^2_i+1) \, \Pi(\omega^2_i +1)$ \\
 $x_5$ & $\Pi(\kappa^1_i+1)\, \Pi(\omega^1_i +1)$\\
 $x_6$ & $\Pi(\kappa^1_i+1)\, \Pi(\omega^1_i +1)\, \Pi(\kappa^2_i) \, \Pi(\omega^2_i +1) $ \\
 $x_7$ & $\Pi(\kappa^1_i+1)\, \Pi(\omega^1_i +1)\, \Pi(\kappa^2_i+1) \, \Pi(\omega^2_i)$ \\
 $x_8$ & $\Pi(\kappa^1_i+1)\, \Pi(\omega^1_i +1)\, \Pi(\kappa^2_i+1) \, \Pi(\omega^2_i +1)$ \\\hline
\end{tabular}
\end{center}
\caption{Condition for the $\alpha$-th brane to pass through the fixed point $x_p$.}
\label{table5}
\end{table}

Turning to the twisted NS-NS tadpoles, they can be easily deduced from (\ref{twrri})  and (\ref{twrrii})  and from their analogy with the untwisted R-R tadpoles. Denoting as usual by $\eta^\Lambda = \pm {1\over 2}$ the chirality of the internal $SO(2)$ spinors, one finds
\begin{plain}
$$
\eqalign{
O_4 S_4  \,: \quad & \sum_{i=1}^{m_{\rm o}} \, \left( N_i \, e^{2 i\phi_i \cdot \eta}- \bar N_i
\, e^{2 i\phi_i \cdot \eta} \right) \, P(z_p \in D_i)  
\cr
\sim & \sum_{i=1}^{m_{\rm o}} \, N_i \, \sin ( 2 \phi_i \cdot \eta ) \, P(z_p \in D_i) 
\,, \qquad \forall\, z_p \in \{ x\}
\cr}
$$
and, similarly 
$$
O_4 C_4 \,:\quad \sum_{a=1}^{m_{\rm e}} \, N_a \, \sin ( 2\phi_a \cdot \eta)  \, P ( z_q \in D_a) +
\sum_{i=1}^{m_{\rm o}} \, N_i \, \sin ( 2 \phi_i \cdot \eta) \, P (z_q \in D_i )
\,, \quad \forall \, z_q \in \{\xi\} \,.
$$
\end{plain}
As expected, in non-supersymmetric models these tadpoles cannot be imposed to vanish, and yield additional contributions to the low-energy effective action \cite{us}.

Also in this case, additional non-chiral fermions originating from branes that are parallel in the second $T^2$ can be made massive if a Scherk-Schwarz deformation is acting also
in this torus. In this case, however, one is consequently reshuffling the fixed points in the
two sets $\xi$ and $x$ and thus some care is needed in determining the correct spectrum of the brane intersections.

\begin{table}[h]
\begin{center}
\begin{tabular}{c c}
\hline
 $z_q$ & $P (z_q \in D_i) $ \\
\hline
 $\xi_1$ & 1\\
 $\xi_2$ & $\Pi(\kappa^2_\alpha) \, \Pi(\omega^2_\alpha +1) $ \\
 $\xi_3$ & $\Pi(\kappa^2_\alpha+1) \, \Pi(\omega^2_\alpha) $ \\
 $\xi_4$ & $\Pi(\kappa^2_\alpha + 1) \, \Pi(\omega^2_\alpha +1) $ \\
 $\xi_5$ & $\Pi(\kappa^1_a + 1)\, \Pi(\omega^1_a )$\\
 $\xi_6$ & $\Pi(\kappa^1_a + 1)\, \Pi(\omega^1_a)\, \Pi(\kappa^2_a) \, \Pi(\omega^2_a +1) $\\
 $\xi_7$ & $\Pi(\kappa^1_a + 1)\, \Pi(\omega^1_a) \, \Pi(\kappa^2_a + 1) \, \Pi(\omega^2_a) $\\
 $\xi_8$ & $\Pi(\kappa^1_a + 1)\, \Pi(\omega^1_a) \, \Pi(\kappa^2_a + 1) \, \Pi(\omega^2_a +1)$\\
\hline
\end{tabular}
\end{center}
\caption{ Condition for the $\alpha$-th brane to pass through the fixed point $\xi_q$.}
\label{table6}
\end{table}

\vspace{3 cm}
\subsection{Finale: an interesting example}
\everypar{\hspace{-.6cm}}
To conclude with this deformed $T^4/\mathbb{Z}_2$ compactification let us present a simple  solution to tadpole conditions. We consider three sets of $N_\alpha$ coincident branes with wrapping numbers
\beq\label{wrapnum}
(\omega^\Lambda_1 , \kappa^\Lambda_1) = \left(
\matrix{ (1,0) \cr (1,1)\cr}\right) \,, \quad
(\omega^\Lambda_2 , \kappa^\Lambda_2) = \left(
\matrix{ (2,1) \cr (0,1)\cr}\right) \,, \quad
(\omega^\Lambda_3 , \kappa^\Lambda_3) = \left(
\matrix{ (1,2) \cr (1,-1)\cr}\right) \,, 
\eeq
as depicted in figure 13. These numbers, together with the choice $N_1 = 10$, $N_2 =12$ and $N_3 =6$ clearly satisfy the R-R tadpole conditions.

\begin{figure}  
\begin{center} 
\includegraphics[scale=1, height=5cm]{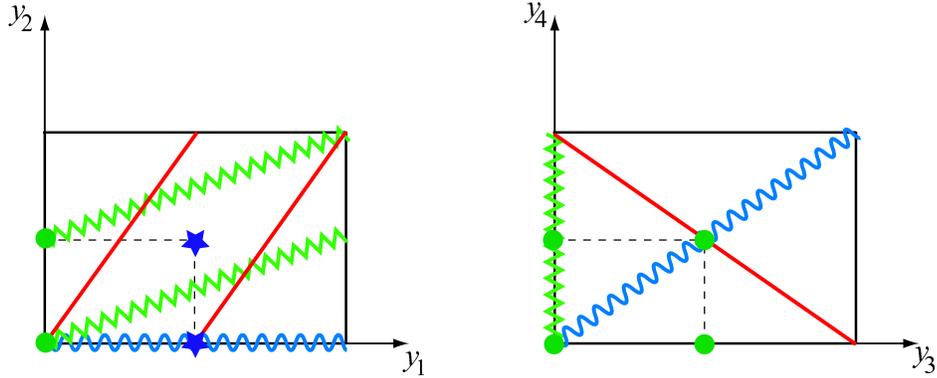}  
\caption{A simple example with three sets of branes.}
\label{fprotated2}
\end{center}
\end{figure}


At the level of toroidal compactification (i.e. before we mod out by the orbifold action and before we deform the spectrum {\it \`a la} Scherk-Schwarz) the chiral spectrum comprises left-handed fermions in the representation $2\, (10,12,1) + 5\, (1,12,6) + 4 \, (1,66,1)$ together with right-handed fermions in the representation $4\, (10,1,6 ) + 3\, (1,12,6) + 8\, (1,1,15) + 2\, (1,66,1) + 2 \, (1,78,1)$, and is clearly free of any irreducible gravitational and gauge anomaly. Before we project and deform the spectrum, let us comment about one subtlety that one meets in deriving this chiral spectrum. In particular, from table \ref{table7} one would naively deduce that the $2\, \bar 2$ sector is non-chiral since the corresponding intersection number is zero. However, since $K_2=4\not=0$,  there is a non-vanishing $N_2$-contribution in the M\"obius strip amplitude that apparently is not matched by the annulus. This is evidently not the case and the correct interpretation is as follows: despite the $N_2$ branes and their images are parallel in the second $T^2$ they intersect non-trivially in the first torus. Moreover, the non-chirality of the annulus suggests that left-handed fermions live at four intersections, and similarly do the right-handed ones, while the ``chirality'' of the M\"obius clearly states that only the left-handed fermions lie on top of the orientifold planes. As a result, one gets four copies of left-handed spinors in the antisymmetric representation of $U(12)$ together with two copies of right-handed spinors in the symmetric plus antisymmetric representations. 

\vskip 10pt

\begin{table}[h]
\begin{center}
\begin{tabular}{c  c  c  c  c  c  c}
\hline
 $\alpha\beta$ &\hspace{.4cm}& $I_{\alpha\beta}$ &\hspace{.4cm}& $I'_{\alpha\beta}$ &\hspace{.4cm}& $I'' _{\alpha\beta}$ \\
\hline
 $12$ &\hspace{.4cm}& $1$ &\hspace{.4cm}& $1$ &\hspace{.4cm}& $0$ \\
 $13$ &\hspace{.4cm}& $-4$ &\hspace{.4cm}& $2$ &\hspace{.4cm}& $2$\\
 $23$ &\hspace{.4cm}& $-3$ &\hspace{.4cm}& $1$ &\hspace{.4cm}& $0$ \\
 $1\bar 2$ &\hspace{.4cm}& $1$ &\hspace{.4cm}& $1$ &\hspace{.4cm}& $0$ \\
 $1\bar 3$ &\hspace{.4cm}& $0$ &\hspace{.4cm}& $2$ &\hspace{.4cm}& $2$\\
 $2\bar 3$ &\hspace{.4cm}& $5$ &\hspace{.4cm}& $1$ &\hspace{.4cm}& $0$\\
 $1\bar 1$ &\hspace{.4cm}& $0$ &\hspace{.4cm}& $2$ &\hspace{.4cm}& $0$\\
 $2\bar 2$ &\hspace{.4cm}& $0$ &\hspace{.4cm}& $2$ &\hspace{.4cm}& $0$\\
 $3\bar 3$ &\hspace{.4cm}& $-8$ &\hspace{.4cm}& $2$ &\hspace{.4cm}& $2$\\
\hline
\end{tabular}
\end{center}
\caption{ Intersection numbers for the three sets of branes.}
\label{table7}
\end{table}

We can now turn on the Scherk-Schwarz deformation and simultaneously project the spectrum by the geometrical $\mathbb{Z}_2$. From (\ref{wrapnum})\ one can deduce that fermions in the dipole sector of the $N_1$ and $N_3$ branes will get a mass proportional to the compactification radius, while the neutral sector of the $N_2$ branes stays supersymmetric. Moreover, the branes intersect at different fixed points, and in particular some intersections of the $N_1$ and $N_3$ branes coincide with fixed points in both sets $\{\xi_q\}$ and $\{x_p\}$. Using the explicit value of the intersection numbers in table \ref{table7}, together with the general expressions in the previous sub-section one gets the following chiral massless spectrum: right-handed fermions in the adjoint representation of the $U(12)$ gauge group and in the representations $2\, (10,1,6) + 2 \, (1,12,6) +4\, (1,1,15) + (1,66,1) + (1,78,1)$, together with left-handed fermions in the representations $(10,12,1) + 3\, (1,12,6) + 3\, (1,66,1) + (1,78,1)$. As usual, the cancellation of R-R tadpoles guarantees that the spectrum is free of irreducible gravitational and gauge anomalies, while the Green-Schwarz mechanism takes care of the reducible anomaly 
\cite{greeni},\cite{greenii}
\begin{plain}
$$
{\mathscr I}_{8} = {\textstyle{1\over 4}} \left( {\rm tr}\, F_2^2 - 2 {\rm tr}\, F_3^2 \right) \left( 
{\rm tr}\, F_1^2 + {\rm tr}\, F_3^2 - {\textstyle{1\over 2}} {\rm tr}\, R^2 \right) \,.
$$
\end{plain}

\newpage


\centerline{\bf\Huge Acknowledgments}
\everypar{\hspace{-.6cm}}

\vskip 1cm

I want to thank Carlo Angelantonj for accepting to  be the  external  adviser 
 of my PhD thesis and for the  big effort of time and energy that he has
 dedicated to me in  the last two years.
  It has been  a quite instructive and nice experience
 both from  the scientific and personal point  of view. 
 I like to thank him also for careful reading, corrections,
suggestions and comments on this manuscript.

\vspace{.2 cm}

I would like to thank my internal adviser Daniela Zanon
 for an en-joyful collaboration and for supporting
 my initiative of spending  more  than two-years of my
 PhD abroad.

 \vspace{.2 cm}

 I have been a guest at the Humboldt University of Berlin
 and at the LMU of Munich and I had a great time  
 there, thanks to the interaction and shearing  
 with the  people.  So my thanks goes to Dieter L\"ust
 for the very kind hospitality for both the two periods.
 
\vspace{.2 cm}

There are several persons that I'd like to mention
 and I will do it in a random order.

  I sheared my office with Fernando Izaurieta and Eduardo Rodrigues, 
   I thank both of them  for  being always in a such a  nice mood,
 for interesting exchanges of ideas and hours of critical and
 constructive  discussions.  
      Especially Fernando has shown  his  talent in providing
   the wonderful images  that enrich this Thesis.

\vspace{.2 cm}

  Thanks to Murad Alim for shearing the office in the latest
 times and for nice discussions at the blackboard.

\vspace{.2 cm}

 I want to thank Dan Oprisa for several lunches 
 we had   together, where we had quite often interesting
 discussions  and open-mind exchange of opinions.

\vspace{.2 cm}

Thanks to Fernando Marchesano for shearing with
me some of his time, for
  sincere exchanges of ideas and useful advises. 

 \vspace{.2 cm}

Thanks to Giuseppe Policastro for listening
 to my observations about vacuum quantum  (in)stability 
and for his enlighten comments and discussions
and for shearing some free-time.

\vspace{.2 cm}

Thanks to Prasanta Tripathy for 
 nice discussions  and sharing the department
 also  in  holidays time. 

\vspace{.2 cm}

Thanks also to Branislav Jurco
 for the  quite nice exchange of music CD 
 and for being a quite nice office-neighbour.

\vspace{.2 cm}

Thanks also to Gabriel Cardoso for his 
kindness and some useful discussion. 

\vspace{.2 cm}

 Thanks to Daniel Krefl  and Enrico Pajer
 for the nice time that we had during  Christmas
holidays in Munich  where I had a very good time
also due  to their company.

\vspace{.2 cm}

 I cannot forget Paolo Pierobon for 
 interesting discussions on  the women psychology
  and for his unforgettable  \emph{bufala}
  that I had the unexpected   privilege to eat around midnight
 in his office.

\vspace{.2 cm}

Thanks to Florian Koch  for
 interesting critical discussions about
  possible formulations of  quantum gravity.

\vspace{.2 cm}

Now I turn to my group in Milan,
 as before without following a personal criterion  of importance.

 \vspace{.2 cm}

 Thanks to Matteo Boni for all his computer help
 of these years and for having shared a 
 lot of common  experiences,  since the beginning 
of our tesi di laurea.

 \vspace{.2 cm}

I like to mention also my colleague Stefano Morisi
 for his  always nice   attitude and
   shearing  of ideas.

 \vspace{.2 cm}

 Thanks to Sergio Cacciatori, Marco Caldarelli, Alessio Celi, Dietmar Klemm,
 Luca Martucci and    Pedro Silva for various discussions, precious 
  advises  and to be always pleased
 to share their knowledge.

\vspace{.2 cm}

From conferences and  visiting periods
 I like to thank Augusto Sagnotti for
 nice discussions and advises,   Liguori Jego and Cristina Timirgaziu  for great time
  during the HEP in Crete. 
 Thanks to Cristina also    for very nice discussions during
 the String Phenomenology  conference  in Munich.

 \vspace{.2 cm} 

  Thanks to Elisa Trevigne, Gianfranco Pradisi and Pascal Anastasopoulos
    for interactions during  String Phenomenology in Munich.

\vspace{.2 cm}

Thanks  to Tristan Maillard for careful explanations at the
 blackboard about the contents of one of his papers.

\vspace{.2 cm}

 Thanks to Nikos Irges for nice collaboration and wonderful
time at Capitan Nikolaos at Falasarna.

\vspace{.2 cm}

 Since there is a good chance that in the hurry  I forgot somebody,
 I will reserve myself the opportunity 
 to improve these acknowledgements and, of course,  the rest
of this thesis with a future version. 
 As usual, I will be   very  happy to receive
 e-mail with comments and observations and to have new opportunities  
  of exchanging ideas.

\end{document}